%% file: arxiv.tex
\documentclass{siamonline1116}

\newcommand{\ja}[1]{}
\newcommand{\as}[1]{}

\usepackage{amsfonts,amsmath}
\usepackage{graphicx}
\usepackage{epstopdf}
\ifpdf
  \DeclareGraphicsExtensions{.pdf,.png,.jpg,.eps}
\else
  \DeclareGraphicsExtensions{.eps}
\fi

\usepackage{bm}

\usepackage{bbm}

\usepackage{mathtools}

\usepackage{tikz}

\usepackage{algorithmicx}
\usepackage{algpseudocode}

\usepackage{multirow}

\def\Leb{{L}}
\def\expect{{\mathbb E}}
\def\var{{\mathrm{Cov}}}
\def\argmin[#1]{{\operatorname*{arg\min}_{#1}}}
\def\eye{\mathrm{I}}

\def\frob{\mathrm{F}}

\def\euler{\mathrm{e}}
\def\ramuno{\mathrm{i}}

\def\conv{\ast}

\def\Real{{\mathbb R}}

\def\SO{{\mathrm{SO}}}

\def\mean{{\mu}}
\def\Mean{{\mean}}
\def\covar{{\Sigma}}
\def\Covar{{\covar}}

\def\projfun{{\mathcal{P}}}

\def\imag{{P}}
\def\Imag{{\imag}}
\def\imagfun{{\mathcal{P}}}

\def\summing{{I}}

\def\x{{x}}
\def\y{{y}}
\def\z{{z}}
\def\e{{e}}

\def\Z{{\z}}

\def\vol{{\x}}
\def\Vol{{\vol}}
\def\volest{{\widehat{\vol}}}
\def\volfun{{\mathcal{X}}}
\def\Volfun{{\volfun}}
\def\im{{\y}}
\def\Im{{\im}}
\def\imest{{\widehat{\im}}}
\def\imfun{{\mathcal{Y}}}
\def\Imfun{{\imfun}}
\def\noise{{\e}}
\def\Noise{{\noise}}
\def\noisefun{{\mathcal{E}}}
\def\Noisefun{{\noisefun}}
\def\coord{{\alpha}}
\def\coordest{{\widehat{\coord}}}
\def\dmap{{\beta}}
\def\dmapest{{\widehat{\dmap}}}

\def\volh{{\fourier{\vol}}}
\def\imh{{\fourier{\im}}}
\def\hh{{\fourier{h}}}

\def\basis{{Q}}
\def\volvox{{v}}

\def\vu{{\boldsymbol{u}}}
\def\vv{{\boldsymbol{v}}}
\def\vk{{\boldsymbol{k}}}
\def\vl{{\boldsymbol{l}}}
\def\vm{{\boldsymbol{m}}}
\def\vomega{{\boldsymbol{\omega}}}

\def\vi{{\boldsymbol{i}}}
\def\vj{{\boldsymbol{j}}}
\def\vc{{\boldsymbol{c}}}

\def\rot{{R}}

\def\SNR{{\mathrm{SNR}}}
\def\SNRh{{\mathrm{SNR_h}}}

\def\max{{\operatorname{max}}}
\def\min{{\operatorname{min}}}

\def\sinc{{\operatorname{sinc}}}

\def\Nsq{{N^2}}
\def\Ncu{{N^3}}

\def\per{{\mathrm{per}}}

\def\shrink{{\mathrm{(s)}}}
\def\clean{{\mathrm{(c)}}}

\newcommand{\fourier}[1]{\mathcal{F}#1}
\newcommand{\transp}{\mathrm{T}}

\newcommand{\bigO}{\mathcal{O}}

\newcommand{\TheTitle}{Structural Variability from Noisy Tomographic Projections}
\newcommand{\TheAuthors}{J. And\'{e}n and A. Singer}

\headers{Structural Variability from Projections}{\TheAuthors}

\title{{\TheTitle}\thanks{Submitted to the editors February 6th, 2018.
\funding{The authors were partially supported by Award Number R01GM090200 from the NIGMS, Simons Investigator Award, Simons Collaboration on Algorithms and Geometry from Simons Foundation, and the Moore Foundation Data-Driven Discovery Investigator Award.}}}

\author{
  Joakim And\'{e}n\thanks{Center for Computational Biology, Flatiron Institute, New York, NY %
    (\email{janden@flatironinstitute.org}).}%
  \and%
  Amit Singer\thanks{Department of Mathematics and Program in Applied and Computational Mathematics, Princeton University, NJ %
    (\email{amits@math.princeton.edu}).}
}

\usepackage{amsopn}

\ifpdf
\hypersetup{
  pdftitle={\TheTitle},
  pdfauthor={\TheAuthors}
}
\fi

\begin{document}

\maketitle

\begin{abstract}
In cryo-electron microscopy, the 3D electric potentials of an ensemble of molecules are projected along arbitrary viewing directions to yield noisy 2D images. The volume maps representing these potentials typically exhibit a great deal of structural variability, which is described by their 3D covariance matrix. Typically, this covariance matrix is approximately low-rank and can be used to cluster the volumes or estimate the intrinsic geometry of the conformation space. We formulate the estimation of this covariance matrix as a linear inverse problem, yielding a consistent least-squares estimator. For $n$ images of size $N$-by-$N$ pixels, we propose an algorithm for calculating this covariance estimator with computational complexity $\bigO(nN^4+\sqrt{\kappa}N^6 \log N)$, where the condition number $\kappa$ is empirically in the range $10$--$200$. Its efficiency relies on the observation that the normal equations are equivalent to a deconvolution problem in 6D. This is then solved by the conjugate gradient method with an appropriate circulant preconditioner. The result is the first computationally efficient algorithm for consistent estimation of the 3D covariance from noisy projections. It also compares favorably in runtime with respect to previously proposed non-consistent estimators. Motivated by the recent success of eigenvalue shrinkage procedures for high-dimensional covariance matrix estimation, we incorporate a shrinkage procedure that improves accuracy at lower signal-to-noise ratios. We evaluate our methods on simulated datasets and achieve classification results comparable to state-of-the-art methods in shorter running time. We also present results on clustering volumes in an experimental dataset, illustrating the power of the proposed algorithm for practical determination of structural variability.
\end{abstract}



\section{Introduction}
\label{sec:intro}

A single biological macromolecule often exists in a variety of three-dimensional configurations. These can be due to deformations of the molecular structure, known as conformational variability, or smaller molecules being added or removed, known as compositional variability. Since molecular structure dictates biological function, properly resolving these different configurations is of great importance in structural biology. In some cases, it is possible to isolate the different structures experimentally and subsequently image each separately in order to reconstruct its three-dimensional structure. However, this is often not possible due to the similarity in shape and size of the various configurations. In this case, the particles are imaged in a single heterogeneous sample, and their structural variability must be taken into account at the reconstruction stage.

Traditional methods such as X-ray crystallography and nuclear magnetic resonance (NMR) imaging are not well suited for this task, since both rely on aggregate measurements from the whole sample. In contrast, single-particle cryo-electron microscopy (cryo-EM) image each particle separately, and can thus potentially recover the structural variability in the sample. Unlike X-ray crystallography, cryo-EM does not require the crystallization of the sample and can handle larger molecules compared to NMR (as small as $64~\mathrm{kDa}$ \cite{khoshouei2017cryo}), making it a more flexible method. New sample preparation techniques and better detectors have recently yielded reconstructions at near-atomic resolution and the method's popularity has been steadily on the rise \cite{kuhlbrandt,amunts2014structure,liao2013structure}. The 2017 Nobel Prize in Chemistry was awarded to three pioneers of cryo-EM and the technique was named Method of the Year in 2015 by Nature Methods.

\begin{figure}
\begin{center}
\includegraphics[width=.25\textwidth]{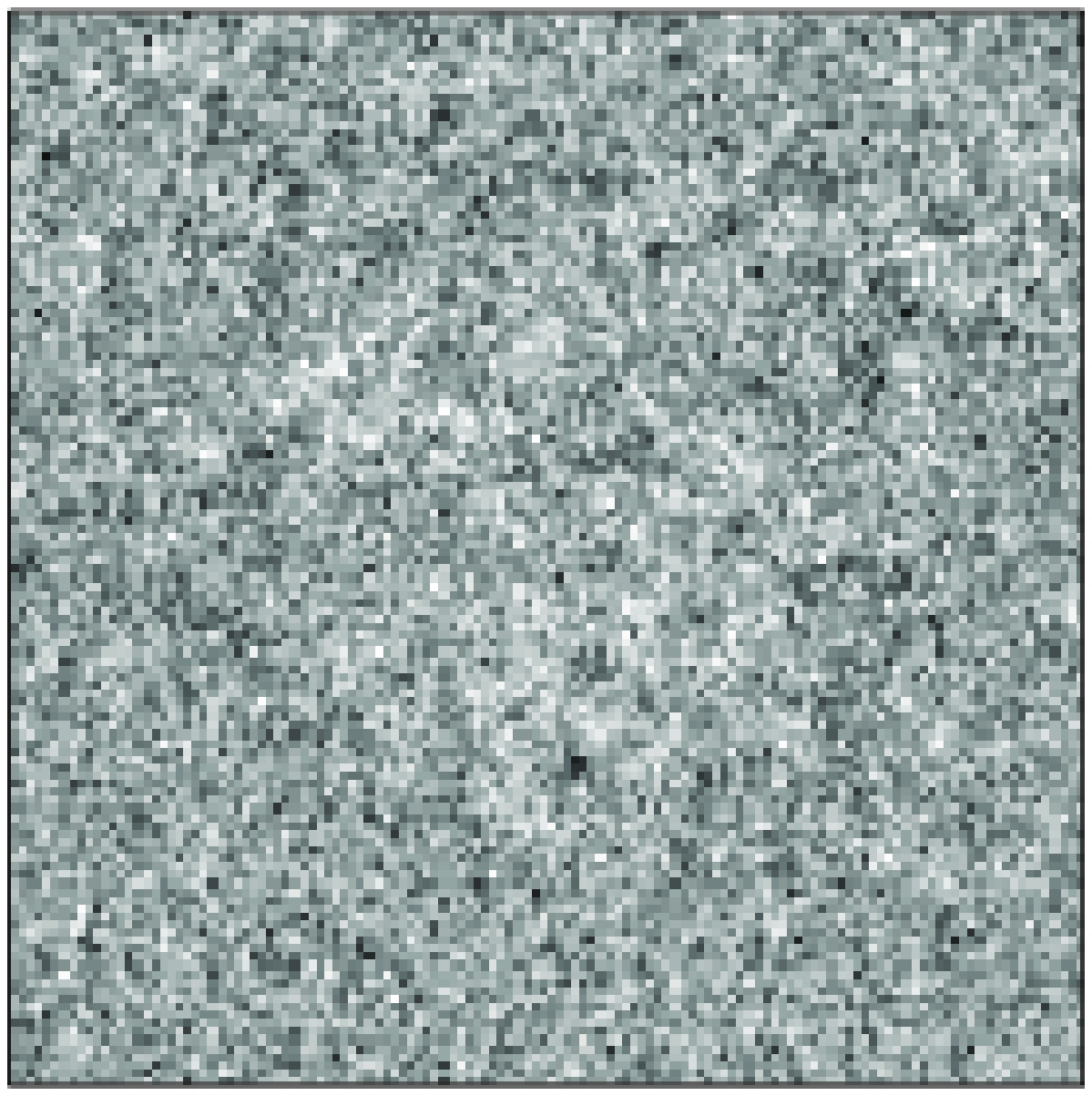}
\includegraphics[width=.25\textwidth]{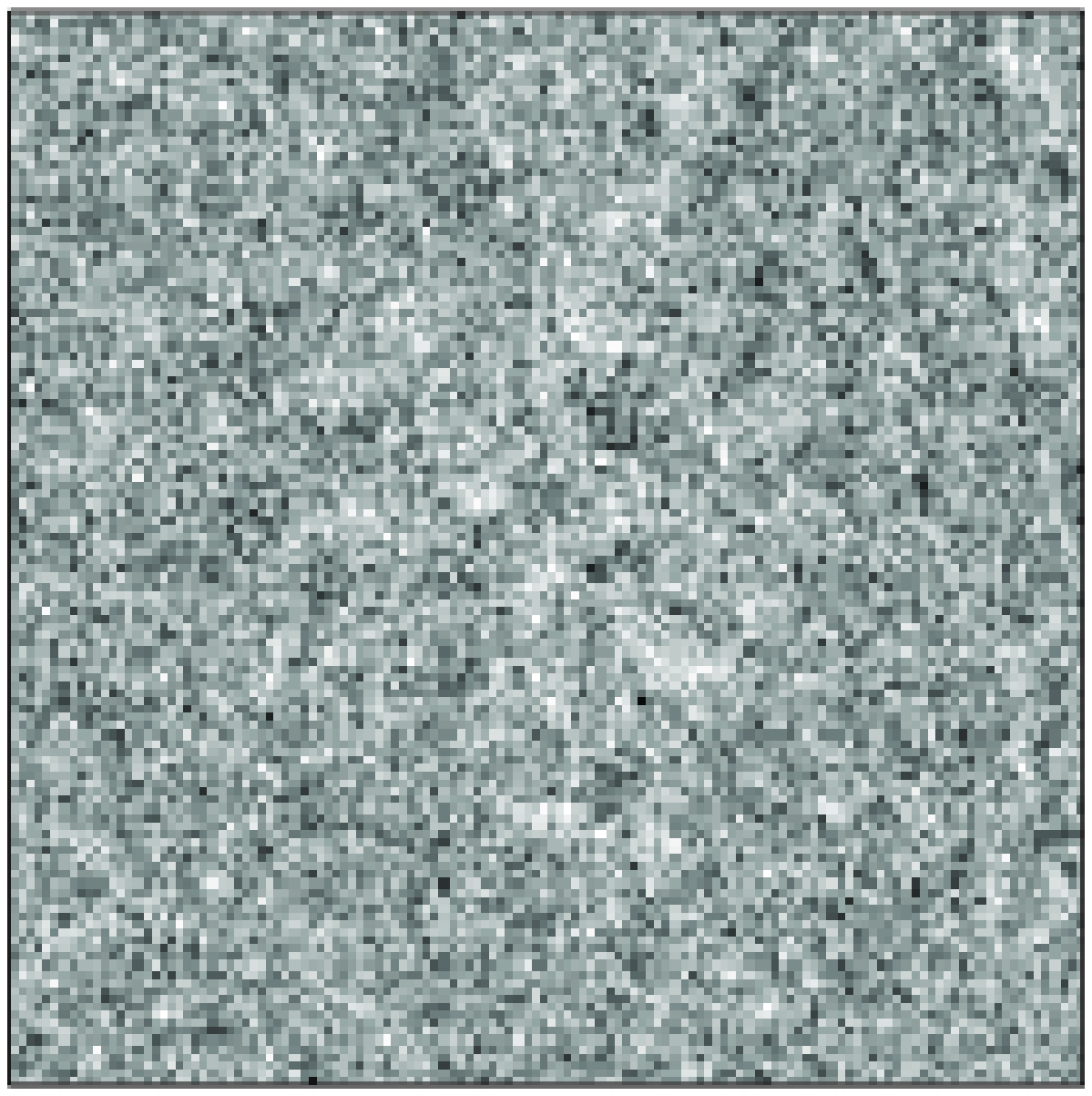}
\end{center}
\caption{\label{fig:frank70s-sample} Two sample cryo-EM images from a $10000$-image dataset depicting the 70S ribosome complex in E. Coli \cite{frank70s_10k}. Each image measures $130$-by-$130$ with a pixel size of $2.82~\mbox{\AA}$. The images depict two similar molecular structures projected in approximately the same viewing direction, but the high noise level makes it difficult to distinguish the difference in structure.}
\end{figure}

To image a set of particles using single-particle cryo-EM, the sample is frozen in a thin layer of ice and exposed to an electron beam. The transmitted electrons are then recorded, forming a set of noisy projection images, one for each particle. Images are modeled as the integral of the particle's electric potential along a particular viewing direction, followed by convolution with a point spread function and addition of noise \cite{frank}. Damaging ionization effects limit the allowable electron dose, so images are dominated by noise (see Figure \ref{fig:frank70s-sample}).

The 3D reconstruction task in single-particle cryo-EM assumes that all particles have identical structure and attempts to reconstruct that structure. As mentioned above, however, this is not always the case. The task of reconstructing the structural variability of a heterogeneous population is known as the heterogeneity problem. This variability is typically assumed to be discrete or continuous. In the case of discrete variability, each particle takes on a finite number of possible molecular configurations. This is often referred to as the 3D classification problem in single-particle cryo-EM. For continuous variability, molecular structures vary continuously, forming a smooth manifold on which each point corresponds to a distinct configuration.

The single-particle reconstruction problem in cryo-EM has been approached from many directions, bringing together ideas from statistics and tomography \cite{frank,cheng2015primer,milne2013cryo,vinothkumar2016single,marina}. The popular RELION software implements a regularized maximum-likelihood estimator using expectation-maximization \cite{scheres}. For the 3D classification problem, it fits a parametric model of discrete variability, so the number of molecular structures needs to be specified in advance. Another problem is that a high-quality initialization is often required for successful reconstruction and there is no global convergence guarantee. In particular, the expectation-maximization algorithm suffers when the number of conformations is large since more populated components take over smaller ones. Lastly, such algorithms require a significant amount of computation time, although this has recently been reduced using graphics processing units (GPUs) \cite{kimanius,brubaker}.

Another approach has focused on the covariance of the 3D volumes as represented on an $N$-by-$N$-by-$N$ voxel grid. Initial work by Liu \& Frank \cite{liu1995estimation} introduced the idea of estimating the variance of these vectors for the purpose of validating the accuracy of reconstructions. In addition, the authors discuss the possibility of quantifying conformational variability through variance estimates. Building on this, a bootstrap method for estimating the 3D variance was introduced by Penczek \cite{penczek-variance} which was later refined with applications to experimental data \cite{penczek-bootstrap,penczek-classification}. In these methods, a single dataset is resampled multiple times, each yielding a reconstruction. The variance is then calculated from the set of reconstructions.

These works touch on estimation of the entire covariance matrix, but this approach was not fully explored until later by Penczek et al. \cite{penczek-pca}. Here, the bootstrap method is used to estimate the whole covariance of the volume vectors, known as the 3D covariance matrix. Typically, the covariance is approximately low-rank, since addition or removal of a substructure is captured by a single volume vector, while deformations are often limited in spatial extent and therefore well approximated by a small number of vectors. The top eigenvectors, or ``eigenvolumes,'' of the 3D covariance thus describe the dominant modes of variability in the volumes. Projecting these eigenvolumes in the viewing direction of each image and calculating the least-squares fit yields a set of coordinates for that image. Fitting a small number of coordinates significantly reduces the noise compared to the original images. Using the coordinates, the images are then clustered and each cluster is used to reconstruct a volume using standard tomographic inversion techniques. Another advantage is that the number of clusters need not be known in advance. For $C$ volume states, the 3D covariance has rank at most $C-1$, so one plus the number of dominant eigenvalues bounds the number of clusters.

Unfortunately, the heuristic bootstrap estimator used by Penczek et al. does not come with consistency guarantees. To remedy this, an alternate approach was proposed by Katsevich et al. \cite{gene}, where the 3D covariance estimation problem is formulated as a linear inverse problem and a least-squares estimator is derived. While this estimator is consistent, its calculation involves solving a large-scale linear system, which is prohibitively expensive to invert directly for typical problem sizes. The authors therefore propose a block-diagonal approximation to the linear system in the large-sample limit which can be solved efficiently, but this is only valid for a uniform distribution of viewing angles and a fixed microscope point spread function. It is therefore of limited applicability in experimental datasets. A new approach was proposed by And\'en et al. \cite{isbi15}, where the exact linear system is solved using the conjugate gradient (CG) method \cite{hestenes}. As a result, the method is valid for non-uniform distributions of viewing directions and multiple point spread functions. However, it only converges after many iterations due to ill-conditioning and each iteration requires a separate pass through the entire dataset, resulting in long running times.

In this paper, we propose an improved version of the method of And\'en et al. \cite{isbi15} for efficient and accurate estimation of the 3D covariance matrix. Our method exploits the fact that projection followed by its dual (backprojection) is a convolution operator \cite{nufft}, also known as a Toeplitz operator. This has already resulted in efficient reconstruction techniques in MRI \cite{wajer-pruessmann,fessler-toeplitz,guerquin-kern} and cryo-EM \cite{vonesch,lanhui-firm}. The 3D covariance least-squares estimator has a similar structure, letting us pose it as a deconvolution problem in six dimensions. As a result, only one pass through the dataset is required to calculate the convolution kernel, allowing each CG iteration to be computed quickly. To reduce the number of iterations required for convergence, we employ a circulant preconditioner \cite{circulant} to improve the conditioning of the system. Our method makes mild assumptions on the distribution of viewing angles and handles image-dependent point spread functions, providing a flexible method for covariance estimation. It is a consistent estimator of the 3D covariance, but unlike the methods of Katsevich et al. \cite{gene} and And\'en et al. \cite{isbi15}, it can be applied efficiently to a wide range of data.

The proposed algorithm has computational complexity $\bigO(nN^4+\sqrt{\kappa} N^6 \log N)$ for $n$ images, where $\kappa$ is the condition number of the preconditioned convolution operator and is typically in the range $1$--$200$. This outperforms the algorithm of Katsevich et al. \cite{gene}, which has computational complexity of $\bigO(nN^6+N^{9.5})$ \cite{gene}. It similarly outperforms the method of And\'en et al. \cite{isbi15}, which has a complexity of $\bigO(\sqrt{\kappa^\prime} n N^7)$, whose condition number $\kappa^\prime$ is of the order of $5000$. The computational complexity is also lower compared to the covariance matrix estimation method introduced by Liao et al. \cite{liao-kaczmarz}, which uses a block Kaczmarz method. Although the paper does not provide an explicit computational complexity of the algorithm, its complexity is at least $\bigO(TN^{10})$, where $T$ is the number of iterations (typically around $20$).

We also introduce a modified covariance estimator based on eigenvalue shrinkage, which lowers the estimation error in the high-dimensional regime. This technique is based on prior work for high-dimensional covariance estimation, where eigenvalue shrinkage methods have been shown to consistently outperform other approaches \cite{donoho-gavish,donoho-gavish-svd,dobriban}. These ideas are most relevant when dealing with vectors of dimensionality comparable to the number of samples, which is often the case for cryo-EM. As a result, we can accurately estimate the 3D covariance at lower signal-to-noise ratios than is possible for the conventional least-squares estimator.

We evaluate the proposed algorithms on simulated datasets, showing their ability to deal with high noise levels, non-uniform distribution of viewing angles, and optical aberrations. In particular, we find that the number of images $n$ necessary to obtain a given covariance estimation accuracy scales inversely with the square of the signal-to-noise ratio, with a phase transition occurring at a critical noise level for a fixed $n$. We also compare our algorithm to the state-of-the-art RELION software \cite{scheres}, obtaining superior accuracy in shorter computation time without using an initial reference structure. The eigenvalue shrinkage variant outperforms the standard estimator, achieving the same accuracy for signal-to-noise ratios up to a factor of $1.4$ worse. Finally, we evaluate the algorithms on several experimental datasets, where we obtain state-of-the-art reconstruction results. GNU Octave/MATLAB code to reproduce the experiments and figures in this paper are provided by functions located in the \textsf{heterogeneity} folder of the ASPIRE toolbox available at \url{http://spr.math.princeton.edu/}.

The remainder of the paper is organized as follows. Section \ref{sec:setup} presents the heterogeneity problem in cryo-EM, while some background and existing approaches are described in Section \ref{sec:related}. Section \ref{sec:estimators} describes the least-squares estimators for the volume mean and covariance and proposes an eigenvalue shrinkage estimator to improve accuracy in the high-dimensional regime. The proposed algorithms for efficient calculation of these estimators are presented in Section \ref{sec:calc}. Once the mean and covariance have been estimated, we describe their use for image clustering in Section \ref{sec:recon}, while simulation and experimental results are provided in Sections \ref{sec:simulation} and \ref{sec:experimental}, respectively. Possible directions for future work are discussed in Section \ref{sec:future}.

\section{Image formation model with heterogeneity}
\label{sec:setup}

To model the cryo-EM imaging process, we equate molecular structure with its electric potentials in three dimensions. These potential maps, referred to as volumes, exist in a variety of states, characterizing the structural variability of the molecule. We consider the volume function $\Volfun: \Real^3 \rightarrow \Real$ to be a random field of unknown distribution such that $\Volfun \in \Leb^1(\Real^3)$. In other words, $\Volfun$ is a random variable in the form of a function. Each realization corresponds to a particular structural configuration of the molecule. The distribution can be discrete, where each realization of $\Volfun$ takes on one of some finite number $C$ of states, each of which is a function in $\Leb^1(\Real^3)$. Alternatively, the structures exist along some continuum, in which case the distribution is continuous, with each realization of $\Volfun$ being some function in $\Leb^1(\Real^3)$ randomly selected from this continuum.

The electron microscope sends a stream of electrons through the particle represented by $\Volfun$, which scatters the electrons. The result is a distorted tomographic projection of each volume, which can be modeled in the weak-phase approximation by an integral along a certain viewing angle followed by a convolution of the resulting image with a microscope-dependent point spread function \cite{frank}. The freezing process fixes each particle in a different orientation. We denote the rotation, or viewing direction, of the particle with respect to some reference frame by the $3$-by-$3$ rotation matrix $\rot$, which we assume is drawn from some distribution over the rotation group $\SO(3)$. We then define the projection of $\Volfun$ along $\rot$ to be the image
\begin{equation}
	\label{eq:projfun-def}
	\mathcal{Z}(\vu) \coloneqq \int_{\Real} \Volfun(\rot^\transp [\vu; z]) dz\mbox{,}
\end{equation}
where $\vu = \Real^2$ and $[\vu; z] \in \Real^3$ is the concatenation of $\vu$ with $z$. This mapping is also known as the X-ray transform of $\Volfun$ along $\rot$ \cite{natterer}. In addition to the tomographic projection, the configuration of the microscope induces a certain amount of optical aberration, which is modeled by a convolution with some point spread function $h \in \Leb^1(\Real^2)$ \cite{wade1992brief,erickson1971measurement}. Again, we assume that this is drawn from some (typically  discrete) distribution over $\Leb^1(\Real^2)$. The convolution is defined by
\begin{equation}
	\label{eq:tranfun-def}
	\Imfun(\vu) \coloneqq \int_{\Real^2} \mathcal{Z}(\vv-\vu) h(\vv) d\vv\mbox{,}
\end{equation}
for $\vu \in \Real^2$. Combining both operations, we have the projection mapping $\imagfun: \Leb^1(\Real^3) \rightarrow \Leb^1(\Real^2)$
\begin{equation}
    \label{eq:imagfun-def}
	\imagfun \Volfun(\vu) \coloneqq \int_{\Real^2} \left( \int_\Real \Volfun(\rot^\transp [\vu-\vv; z]) dz \right) h(\vv) d\vv\mbox{.}
\end{equation}

We can now state our forward model for the cryo-EM imaging process, which takes a volume $\Volfun$ and gives the image
\begin{equation}
    \label{eq:forward-cont}
	\Imfun = \imagfun \Volfun + \Noisefun\mbox{,}
\end{equation}
where $\Noisefun: \Real^2 \rightarrow \Real$ is a white Gaussian random field of variance $\sigma^2$. Since $\imagfun$, $\Volfun$ and $\Noisefun$ are random variables, $\Imfun$ is also a random variable. The noise $\Noisefun$ represents error introduced into the image due to non-interacting electrons, inelastic scattering, and quantum noise \cite{vulovic2013image,baxter2009determination,penczek2010chapter}. While these error sources follow a Poisson distribution, the counts are typically high enough for this to be well-approximated by a Gaussian distribution. Another potential problem is that the noise is rarely white but exhibits strong correlations which depend on the microscope configuration. Later in this section, we describe how to account for this discrepancy.

It is useful to consider the above mappings in the Fourier domain. Let us define the $d$-dimensional Fourier transform $\fourier{\mathcal{G}}$ of some function $\mathcal{G} \in \Leb^1(\Real^d)$ by
\begin{equation}
    \fourier{\mathcal{G}}(\vomega) \coloneqq \int_{\Real^d} \mathcal{G}(\vu) \euler^{-2\pi\ramuno \langle \vomega, \vu \rangle} d\vu\mbox{,}
\end{equation}
for any frequency $\vomega \in \Real^d$. In this case, the tomographic projection mapping $\imagfun$ satisfies
\begin{equation}
    \label{eq:forward-cont-fourier}
    \fourier{\projfun \Volfun}(\vomega) = \fourier{\Volfun}\left(\rot^\transp[\vomega; 0]\right) \cdot \hh(\vomega)\mbox{,}
\end{equation}
for any $\vomega \in \Real^2$, which is known as the Fourier Slice Theorem \cite{natterer}. In other words, projection in the spatial domain corresponds to restriction (or ``slicing'') to a plane in the Fourier domain and multiplication by a transfer function $\hh$.
The Fourier transform $\hh$ of the point spread function $h$ is known as the contrast transfer function (CTF). The CTF is an oscillatory function and is equal to zero for several frequencies. As a result, those frequencies are not available in that particular image. To mitigate this problem, cryo-EM datasets are collected for a number of different CTFs by varying certain microscope parameters.

The model presented above describes continuous images $\imfun$ obtained from continuous volume densities $\volfun$. While an accurate model of the physical process, it is not compatible with the output of an electron microscope, which is in the form of discrete images $\im$ with values on an $N$-by-$N$ pixel grid, where $N$ typically ranges from $100$ to $500$. The images are therefore limited in resolution, imposing limit on the resolution of the reconstructed volumes.

To discretize the images, we define the $N$-point grid $M_N$ as
\begin{equation}
    M_N \coloneqq \{-\lfloor N/2 \rfloor, \ldots, \lceil N/2-1 \rceil\}\mbox{.}
\end{equation}
An image $\imagfun$ is then represented by sampling evenly over the square $[-1, +1]^2$ at points $2M_N^2/N$, yielding a function $\im: M_N^2 \rightarrow \Real$. We treat this function as a vector in $\Real^\Nsq$.

There is more choice in representing the volumes. One popular approach is to consider the voxel samples of $\volfun$ at points $2M_N^3/N$ in the cube $[-1, +1]^3$, yielding a vector of dimension $\Ncu$ \cite{scheres-relion,vonesch,lanhui-firm}. While this basis has computational advantages, it is not always well suited to representing volumes of interest. As such, we will use a different basis described in Section \ref{sec:basis} and convert between this and the voxel basis. Let us denote by $\basis$ the matrix whose columns corresponding to the $p$ basis vectors of size $\Ncu$, each corresponding to an $N$-by-$N$-by-$N$ volume. Here $p = \bigO(\Ncu)$ and $\basis$ is of size $\Ncu$-by-$p$. We will assume that expansion and evaluation in this basis is fast, that is both $\basis$ and its transpose $\basis^\transp$ can be applied in $\bigO(\Ncu \log N)$. To simplify expressions, we introduce the notation $\volvox = \basis \vol$.

An important question is then how to properly discretize the forward model \eqref{eq:forward-cont}. One approach is embed the discretized volumes $\volvox \in \Real^\Ncu$ into $\Leb^1(\Real^3)$ using a $\sinc$ basis, apply $\imagfun$ and fit the result to a $\sinc$ basis expansion in $\Leb^1(\Real^2)$ using least-squares. While this provides a matrix representation of $\imagfun$ that is accurate in a least-squares sense and converges to $\imagfun$ as $N \rightarrow \infty$, the mapping is computationally inefficient, requiring a full $\Nsq$-by-$\Ncu$ matrix multiplication to apply.

Another approach is to mimic the Fourier slice structure of $\imagfun$ described in \eqref{eq:forward-cont-fourier}, enabling speedups associated with fast Fourier transforms (FFTs) \cite{cooley-tukey}. First, let us define the discrete Fourier transform of some function $g: M_N^d \rightarrow \Real$ in $d$ dimensions
\begin{equation}
    \label{eq:dft-def}
    \fourier{g}(\vk) \coloneqq \sum_{\vi \in M_N^d}
        g(\vi) \euler^{-2\pi \ramuno \langle \vi, \vk \rangle/N}\mbox{.}
\end{equation}
Here, we have abused notation slightly by having $\fourier$ signify both the continuous and discrete Fourier transforms. The nature of the mapping should be clear from context.
Although $\fourier{g}(\vk)$ is defined for any frequency vector $\vk \in \Real^d$, it is traditionally restricted to the grid $M_N^d$.

We now define the mapping $\summing$ transforming the voxel volume $\volvox$ into the image $\summing \volvox$ through
\begin{equation}
    \label{eq:summing-def}
    \summing \volvox(\vi) = \frac{1}{N^3} \sum_{\vj \in M_N^3} \volvox(\vj) \sum_{\vk \in M_{2\lceil N/2 \rceil-1}^2} \hh(\vk) \euler^{-2\pi \ramuno (\langle R^\transp [\vk; 0], \vj \rangle - \langle \vk, \vi \rangle)/N} \quad \mbox{for~all~} \vi \in M_N^2\mbox{.}
\end{equation}
Computing its discrete Fourier transform, we obtain
\begin{equation}
    \label{eq:summing-fourier-slice}
    \fourier{\summing \volvox}(\vk) =
        \left\{\begin{array}{ll}
            {\displaystyle \frac{1}{N} \fourier \volvox\left(R^\transp[\vk; 0]\right) \cdot \hh(\vk)\mbox{,}} &
            {\displaystyle \vk \in M_{2 \lceil N/2 \rceil-1}^2} \\
            {\displaystyle 0 \mbox{,}} & {\displaystyle \mbox{otherwise,}}
        \end{array}\right.
\end{equation}
for any $\vk \in M_N^2$. The operator $\summing$ therefore satisfies a discrete version of the Fourier Slice Theorem \eqref{eq:forward-cont-fourier}. Note that in the case of even $N$, the Nyquist frequencies at $-N/2$ are set to zero to ensure a real image $\summing \basis \vol$. Enforcing this one-to-one mapping of frequencies allows us to derive the convolutional formulations in Sections \ref{sec:mean-conv} and \ref{sec:conv}. The entire mapping, from $\vol$ to $\volvox = \basis \vol$ to $\summing \volvox = \summing \basis \vol$ is denoted by $\imag = \summing \basis$ and is called the volume imaging mapping. While $\imag$ describes the whole imaging process, that is both projection and convolution with the point spread function, we shall often refer to it as projection for simplicity. Similarly, its adjoint $\imag^\transp$ will be referred to as backprojection.

To project a volume $\vol \in \Real^\Ncu$, first evaluate it on the $2M_N^3/N$ voxel grid to obtain $\volvox = \basis \vol$, then calculate the discrete Fourier transform $\fourier \volvox$ using \eqref{eq:dft-def} on the grid defined by $R^\transp[\vk; 0]$. We then multiply the Fourier transform pointwise by the contrast transfer function $\hh(\vk)$, set Nyquist frequencies to zero, and apply the inverse discrete Fourier transform, which gives $\imag \vol \in \Real^\Nsq$. As mentioned earlier, the basis evaluation matrix $Q$ can be applied in $\bigO(\Ncu \log N)$ time. The first discrete Fourier transform $\fourier\volvox$ is computed using a non-uniform fast Fourier transform (NUFFT), which has a computational complexity of $\bigO(\Ncu \log N)$ \cite{nufft,greengard2004accelerating}, while pointwise multiplication and the 2D inverse FFT require $\bigO(\Nsq)$ and $\bigO(\Nsq \log N)$, respectively. The overall computational complexity is therefore $\bigO(\Ncu \log N)$, which is a significant improvement over the direct matrix multiplication approach, which has a complexity of $\bigO(N^5)$. Another important advantage is that calculating multiple projections of the same volume $\vol$, the overall complexity scales as $\bigO(\Ncu \log N + n\Nsq)$, where $n$ is the number of projection images.

With the projection mapping $\imag$, we can now formulate our discrete forward model as
\begin{equation}
    \label{eq:forward}
    \im \coloneqq \imag \vol + \noise\mbox{,}
\end{equation}
where $\noise \in \Real^\Nsq$ is a standard white Gaussian noise image. Note that again $\imag$ describes both projection and convolution with a point spread function $h$. To generate an image, a given volume density $\vol$ is rotated according to the viewing direction $\rot$, projected along the $z$-axis, convolved with a point spread function $h$, and finally a white Gaussian noise $\noise$ is added. We do not include translation as part of $\imag$ because we assume that translations have been previously estimated and subsequently removed from the images by translating them in the opposite direction. Apart from this and the white noise assumption, the above image formation model corresponds to those traditionally employed in single-particle cryo-EM \cite{frank,scheres,brubaker,tagare}.

As in the continuous case, we note that $\imag$, $\vol$, and $\noise$ are all random variables, so the same is true of $\im$. In particular, $\imag$ depends on $\rot$ and $h$, which are random variables with distributions over $\SO(3)$ and $\Leb^1(\Real^2)$, respectively (although in many cases we will condition $\im$ with respect to $\imag$, fixing that variable and by extension $\rot$ and $h$). The volume density $\vol$ is a random variable defined over $\Real^p$ as described above, and finally $\noise$ is a random variable with distribution over $\Real^\Nsq$. The distribution of $\im$ depends on those of $\rot$, $h$, $\vol$ and $\noise$ and drawing realizations from this distribution provides us with the experimental images. It is not necessary for us to know what these distributions are, but this probabilistic framework will make it easier to derive and reason about the estimators introduced in the following.

In a single-particle cryo-EM experiment, we have more than one image, with multiple copies of the same molecule being imaged separately, each with a different structural configuration, projected at a different viewing angle, subjected to convolution by a different point spread function, and degraded by a different realization of noise. As a result, we consider identically distributed, independent copies $\vol_1, \ldots, \vol_n$ of $\vol$. Similarly, we have rotations $\rot_1, \ldots, \rot_n$, point spread functions $h_1, \ldots, h_n$ and noise vectors $\noise_1, \ldots, \noise_n$ which are independent and identically distributed copies of $\rot$, $h$, and $\noise$, respectively. These yield the images
\begin{equation}
    \label{eq:forward-k}
	\im_s \coloneqq \imag_s \vol_s + \noise_s
\end{equation}
for $s = 1, \ldots, n$, where $\imag_s$ is the imaging operator corresponding to the viewing direction $\rot_s$ and contrast transfer function $\hh_s$. Each $\im_s$ is a different random variable representing a different projection image from an experiment. The contrast transfer functions $\hh_1, \ldots, \hh_n$ are not all distinct, but are typically shared between particles picked from the same micrograph.

We note that the $\imag_s$ mappings are not known for experimental data, but must be estimated. Several methods exist to estimate the CTFs $\hh_1, \ldots, \hh_n$ \cite{mindell2003accurate,zhang2016gctf}. Similarly, traditional methods for orientation estimation \cite{scheres,brubaker,frealign,lanhui-LUD} can be applied when $\Vol$ does not vary too much around its mean. Indeed, previous works have demonstrated that this is a feasible approach in a range of situations \cite{penczek-pca,liao-kaczmarz}. In this work, we assume such estimation of $\rot_1, \ldots \rot_n$ is possible and therefore $\imag_1, \ldots, \imag_n$ are known to a certain accuracy.\footnote{In the case of high variability in $\Vol$, the orientations must be estimated simultaneously with the clustering of the images, a more challenging problem studied in recent works by Lederman and Singer \cite{roy,lederman2017continuously}.}

In the process of estimating $\rot_1, \ldots, \rot_n$, these methods also estimate the translations of the individual images. This is what allows us to cancel their effect in our forward model \eqref{eq:forward}. Note that the effect of the viewing directions cannot be similarly removed since they define the tomographic projection from 3D to 2D. As such, it is necessary to include them in the forward model, unlike the translations.

As mentioned above, the noise component $\noise$ is typically not white. One approach to handle this is to estimate the power spectrum of the noise and include a non-white noise component in our forward model. Most of the results in the remainder of this work follow, but with a certain loss of simplicity. Instead, we choose to prewhiten the images, rendering the data more compatible with our proposed forward model \eqref{eq:forward} which specifies white Gaussian noise.

To achieve this, the noise power spectrum is first estimated for each image. Due to changing experimental conditions, the intensities of the various noise sources (background, inelastic scattering, quantum noise, etc.) vary from image to image. To account for this, we employ a noise estimation algorithm which exploits such low-rank variability \cite{sampta}. Alternatively, the software used to estimate the rotations $\rot_1, \ldots, \rot_n$ also estimates noise power spectra as part of the algorithm \cite{scheres-relion,brubaker}.

Let us denote the estimated noise power spectrum of $\noise_s$ to be $\fourier{m}_s$. To whiten $\im_s$, we filter it with the transfer function $\fourier{m}_s^{-1/2}$. The noise component is now white, but the signal component has been altered. This is remedied by including the whitening filter into the CTF $\hh_s$, replacing it with a new ``effective CTF`` $\fourier{m}_s^{-1/2} \cdot \hh_s$. The resulting images have an approximately white noise component, with a signal still equal to $\imag_s \vol_s$ since $\imag_s$ now includes the effect of the prewhitening filter. Consequently, the prewhitening filter is inverted whenever we use the new $\imag_1, \ldots, \imag_n$ to reconstruct $\vol$.

The heterogeneity problem can now be stated more formally. From projection images $\im_1, \ldots, \im_n$, we would like to characterize the distribution of $\Vol$, which is equivalent to reconstructing the energy landscape inhabited by the molecule. This problem is unfortunately ill-posed. Indeed, third- and higher-order moments of $\Vol$ are impossible to estimate from its projections. The ill-posedness may be removed by only estimating partial information on $\Vol$, such as its first- and second-order moments $\expect[\Vol]$ and $\var[\Vol]$, or by restricting the class of distributions, such as those supported on a low-dimensional subspace.  \ja{This is a little disingenuous. The low-dimensional assumption by itself does not make the distribution problem well-posed. For discrete variability, we should be able to capture the whole distribution by just estimating the mean and covariance and solving for the volumes and probabilities. However, for more general distributions supported on low-dimensional manifolds, this is not true.} This latter restriction includes discrete variability and continuous variability on a smooth, low-dimensional manifold. In this work we shall make use of both restrictions to render the heterogeneity problem more tractable. First, given images $\im_1, \ldots, \im_n$, we will first estimate $\expect[\Vol]$ and $\var[\Vol]$. Second, we will use the fact that the distribution of $\Vol$ is supported on a low-dimensional subspace to improve our estimate of $\var[\Vol]$.

A further goal is to reconstruct the underlying volumes $\vol_1, \ldots, \vol_n$ from their projection images $\im_1, \ldots, \im_n$. This problem is also ill-posed without any further assumptions. We therefore impose the same restriction on the class of distributions (namely, being supported on a low-dimensional subspace). Estimates of the mean and covariance lets us estimate the volumes themselves. However, this is only possible within limits dictated by the noise level. Indeed, if the noise is large enough, we may be able to estimate the mean and covariance accurately given enough images, but accurate reconstruction of individual volumes may not be feasible.

\section{Related work}
\label{sec:related}

Due to the importance of determining structural variability from cryo-EM projections, much work has been focused on resolving this heterogeneity problem. Although various methods have been introduced that have some degree of experimental success, they do not possess any accuracy guarantees, so it is sometimes difficult to validate the reconstructions. In addition, they also often rely on good initializations, which can significantly bias the final result. Finally, the computational complexity of these methods is typically quite high, requiring a large amount of computational resources.

\subsection{Maximum likelihood}
\label{sec:ml}

One popular method for solving the heterogeneity problem has been to set up a probabilistic model for image formation and maximizing the likelihood function with respect to the model parameters given the data. This was first considered for class averaging in the space of images by Sigworth \cite{fred-ml}, where the probability density of the images $\im$ was modeled as a mixture of Gaussians, with each component center constituting a distinct image class. These centers, along with other parameters such as shifts were estimated by maximizing the likelihood using an expectation-maximization algorithm \cite{dempster-laird}.

The maximum-likelihood method was subsequently extended by Scheres, who modeled the underlying volume vector $\vol$ as a mixture of Gaussians and regularized the likelihood function using Bayesian priors on the parameters, which includes the viewing directions $\rot_1, \ldots, \rot_n$ \cite{scheres}. The resulting algorithm, implemented in the RELION software package, has seen significant success and provides generally satisfactory volume estimates \cite{scheres-relion}. However, since the algorithm attempts to optimize a non-convex function, there is no guarantee that a globally optimal solution is obtained. The algorithm also needs to be initialized with single reference structure that is similar to the molecule being imaged, which can significantly bias the result if not chosen carefully. Similarly, the number of clusters is part of the model and needs to be specified in advance, limiting the method's flexibility. The performance also degrades when a large number of classes is specified, as more populated classes tend to absorb smaller ones, making rare conformational changes hard to characterize. Finally, the algorithm has a long running time, although a recent GPU-based implementation has mitigated this problem \cite{kimanius}.

\subsection{Common lines}
\label{sec:cline}

Another approach proposed by Shatsky et al. \cite{shatsky} clusters the projection images by defining a similarity measure between all pairs and applying a spectral clustering method. From the Fourier Slice Theorem \eqref{eq:forward-cont-fourier}, the Fourier transform $\imh$ of a clean projection image $\im$ corresponds to the restriction of the volume Fourier transform $\volh$ to a plane and multiplied by a contrast transfer function $\hh$. Two images $\im_s, \im_t$ thus occupy two central planes of the volume Fourier transform and intersect along a common line. The Fourier transforms $\imh_s, \imh_t$ of two noiseless projections of the same molecular structure should therefore coincide along this common line (up to differing contrast transfer functions), so their cross-correlation along this line provides a good similarity measure.

Using this common-lines affinity, the authors applied spectral clustering to group the images according to their underlying molecular structure. Unfortunately, the Fourier transforms of two projections of the same volume will not coincide exactly due to the image noise. In most cryo-EM experiments, the images are dominated by noise, making this particular approach unfeasible without some amount of denoising. Denoising images in cryo-EM is traditionally achieved by class averaging, where images that represent similar views are averaged together. However, for heterogeneous data this may break down since images belonging to different molecular structures could be averaged together.

\subsection{Covariance and low-rank approaches}
\label{sec:covar}

Instead of directly clustering the images, another line of work has focused on estimating the 3D covariance matrix $\var[\Vol]$ of the volume $\Vol$ as a random vector. The first of these, Liu \& Frank \cite{liu1995estimation}, introduced the notion of 3D covariance in the single-particle cryo-EM setting for the purposes of validation. The authors then proposed a method for estimating this covariance. Building on this, Penczek outlined a variant of the standard bootstrap algorithm for estimating the variance \cite{penczek-variance}. Here, multiple subsets of the images $\Im_1, \ldots, \Im_n$ are drawn, each yielding a different 3D reconstruction. The sample covariance of these 3D reconstructions then yields an estimate for the 3D covariance of $\Vol$. Since the distribution of molecular structures differs slightly between subsets, the idea is that this will capture the 3D variability of the volumes. Refinements of this method have successfully been applied to experimental data to estimate variance \cite{penczek-bootstrap} and perform 3D classification \cite{penczek-classification}. Related work by Doerschuk and others estimate the covariance by fitting Gaussian mixture models of the volumes \cite{zheng2012three,xu2018allosteric}.

A further refinement was proposed in Penczek et al. \cite{penczek-pca}, where the bootstrap method was used to perform principal component analysis of the reconstructed volumes. In this work, the top eigenvectors, or eigenvolumes, of the estimated 3D covariance matrix are used to reconstruct the volumes in the sample. Indeed, as discussed above, this covariance matrix is typically approximated by a low-rank matrix, so its top eigenvolumes together with the mean volume form an affine space containing most of the volumes in the dataset. Projecting the mean volume and the eigenvolumes, the authors find the coordinates of each image in this affine space through a least-squares fit. Clustering the images using these coordinates, each cluster is then used to reconstruct a different molecular structure. Unfortunately, the heuristic nature of the covariance estimation does not provide any accuracy guarantees. A maximum-likelihood approach has also been proposed to estimate the top eigenvolumes \cite{tagare}, but this suffers from the same initialization problems and lack of guarantees as other non-convex approaches (see Section \ref{sec:ml}).

To address these problems, Katsevich et al. \cite{gene} formulated the 3D covariance estimation problem as a linear inverse problem and proposed a least-squares estimator with proven consistency results. Unfortunately, direct calculation of the estimator proved computationally intractable, so the authors introduced an approximation which relied on uniform distribution of viewing directions and a single fixed contrast transfer function. These conditions are rarely satisfied in experimental data, so practical use of this method was limited. A more flexible approach has been proposed, based on calculating a related estimator using the block Kaczmarz method, but unfortunately this suffers from slow convergence and reduced accuracy \cite{liao-kaczmarz}.

An improved version of the method of Katsevich et al. \cite{gene} was introduced in And\'en et al. \cite{isbi15}, where the exact linear system was solved iteratively using the CG method. This had the advantage of allowing for non-uniform distributions of viewing directions and varying contrast transfer functions. However, the method required a pass through the dataset at each iteration and a large number of iterations was needed to reach convergence. As a result, this method proved unfeasible for large datasets.

\subsection{Other methods}
\label{sec:related-other}
Another successful approach has been to use an atomic reference structure to predict the possible motions of a molecule using normal mode analysis. These motions are then fit to the projection images to identify 3D variability in the sample \cite{jin2014iterative}. A significant drawback of this approach is its requirement for an atomic-resolution reference structure, which may not always be available for the imaged molecule.

To capture continuous variability, Dashti et al. \cite{dashti} group projection images by viewing direction and estimate the manifold structure in each group. The different manifolds are then assembled into a global manifold describing the variability of the entire molecule. Counting the number of projection images obtained from each point in that manifold, the authors derive an energy landscape for that molecule. The authors apply this method to a dataset of ribosome projections and obtain impressive results. However, the heuristic nature of this method and its lack of accuracy guarantees make it problematic to apply in a general setting.

For a survey methods related to the heterogeneity problem, see Joni\'{c} \cite{jonic2017computational}.

\section{Mean and covariance estimators}
\label{sec:estimators}

As discussed in the previous section, the 3D covariance is a powerful tool in characterizing variability for single particle cryo-EM \cite{liu1995estimation}. In particular, applying it to perform a principal component analysis of the volumes is especially useful \cite{penczek-pca,tagare}. Existing methods for covariance estimation, however, do not offer any accuracy guarantees \cite{zheng2012three,penczek-bootstrap,penczek-pca,tagare}. In the following, we describe the least-squares estimators for both volume mean and covariance previously introduced by Katsevich et al. \cite{gene}. The estimators take as input the projection images along with estimates of the viewing directions and CTFs and provide as output estimates of the mean and covariance of the volumes. These estimators have theoretical guarantees, ensuring that for a fixed noise level, the mean and covariance estimates converge to their population values as the number of images increases. We also introduce a modified variant of the original covariance estimator, where eigenvalue shrinkage is used to reduce error in the regime of high dimension.

\subsection{Mean estimator}
\label{sec:mean-est}

To estimate the mean $\expect[\Vol]$ of the volume density $\Vol$, we assume that the imaging operators $\imag_1, \ldots, \imag_n$ are known for all images (i.e., that the viewing directions $\rot_1, \ldots, \rot_n$ and CTFs $\hh_1, \ldots, \hh_n$ are known) and that the images are centered. As discussed in Section \ref{sec:setup}, the imaging operators can typically be estimated from the images with good accuracy. Likewise, their translations can be estimated and used to center the images.

We now consider the distribution of the images $\im_1, \ldots, \im_n$ with these mappings fixed. That is, we consider the distributions of $\im_s$ conditioned on $\imag_s$, where only the volume structure $\vol_s$ and noise $\noise_s$ are allowed to vary. Let us denote the expectation with respect to some variable $a$ conditioned on some other variable $b$ by $\expect_{a|b}$. From the forward model \eqref{eq:forward}, we obtain
\begin{equation}
	\label{eq:im-expect}
	\expect_{\vol_s,\noise_s|\imag_s}[\Im_s] = \Imag_s \expect[\Vol_s] = \imag_s \expect[\Vol]\mbox{,}
\end{equation}
for $s = 1, \ldots, n$, since $\Vol_1, \ldots, \Vol_n$ are all identically distributed and $\expect[\noise_1] = \cdots = \expect[\noise_n] = 0$. The above equation provides a constraint on $\expect[\Vol]$ for each $s = 1, \ldots, n$. We could therefore solve for $\expect[\Vol]$ if we were given the left-hand side expectations $\expect_{\vol_s,\noise_s|\imag_s}[\Im_s]$, but these are unavailable to us.

However, the image itself $\Im_s$ is an unbiased estimator of $\expect_{\vol_s,\noise_s|\Imag_s}[\Im_s]$, albeit one with significant variance. Substituting $\Im_s \approx \expect_{\vol_s,\noise_s|\Imag_s}[\Im_s]$ into \eqref{eq:im-expect} gives
\begin{equation}
    \label{eq:mean-constraint-s}
	\Im_s \approx \imag_s \expect[\Vol]\mbox{,}
\end{equation}
for all $s = 1, \ldots, n$. Combining these approximate constraints into a regularized least-squares objective, we obtain the following estimator $\Mean_n$ for $\expect[\Vol]$:
\begin{equation}
    \label{eq:mean-objective}
	\Mean_n \coloneqq \argmin[\mean\in\Real^p] \frac{1}{n} \sum_{s=1}^n \| \Im_s - \imag_s \mean \|^2 + \nu_n \|\mean\|^2 \mbox{,}
\end{equation}
where $\nu_n \ge 0$ is a regularization parameter which ensures the problem remains well-posed (mitigating ill-posedness due to CTFs and distribution of viewing directions, as discussed below). Since the problem is better conditioned for large $n$, $\nu_n$ typically decreases as $n$ grows. This estimator minimizes the average square distance of the $\imag_s \Mean_n$ (that is, $\Mean_n$ projected along $\rot_s$ and convolved with $h_s$) to each of the images $\Im_s$ subject to the regularization term $\nu_n \|\mean\|^2$. When there is no structural variability, that is, when $\var[x] = 0$, $\Mean_n$ is the regularized maximum-likelihood estimator for $\expect[x]$.

In order to calculate $\Mean_n$, we form its normal equations by differentiating the objective \eqref{eq:mean-objective} and setting the derivative to zero. We thus have
\begin{equation}
	\label{eq:mean-normal}
	A_n \Mean_n = b_n\mbox{,}
\end{equation}
where
\begin{align}
    \label{eq:An-def}
	A_n^{} &\coloneqq \frac{1}{n} \sum_{s=1}^n \imag_s^\transp \imag_s^{} + \nu_n \eye_p \mbox{,} \\
    \label{eq:bn-def}
	b_n^{} &\coloneqq \frac{1}{n} \sum_{s=1}^n \imag_s^\transp \Im_s^{}\mbox{.}
\end{align}
The right-hand side $b_n$ is the average of the backprojected images $\imag_s^\transp \Im_s^{}$, while $A_n$ is the projection-backprojection operator corresponding to the set of viewing directions $\rot_1, \ldots, \rot_n$ and CTFs $\hh_1, \ldots, \hh_n$ plus the regularization term $\nu_n \eye_p$.

This least-squares estimator is a good estimator of molecular structure in cryo-EM samples when heterogeneity is not present. Indeed, if $\vol$ has no variability, $\vol_1 = \cdots = \vol_n = \expect[\vol]$ then $\Mean_n$ estimates the single volume present in the sample. When there is heterogeneity, it is a consistent estimate of the average volume $\expect[\Vol]$. As a result, $\mu_n$ and closely related least-squares estimators have found widespread use in single-particle cryo-EM reconstruction \cite{harauz1986exact,vonesch,lanhui-firm,marina}.

The standard Tikhonov regularization term in \eqref{eq:mean-objective} can be replaced by more sophisticated regularizers to enforce smoothness and other properties. For example, RELION uses an adaptive weighting scheme where each radial frequency is assigned a different regularization parameter \cite{scheres-relion}. These are initialized using a reference structure and subsequently updated using the estimated structure at each iteration. The result is that higher frequencies are penalized more than lower frequencies, enforcing smoothness.

The constraints in \eqref{eq:mean-constraint-s} are by necessity loose due to the presence of noise in the data and the variability of $\Vol$. Still, aggregating these over a large number of images $n$ results in an estimator $\Mean_n$ that is consistent. Indeed, Katsevich et al. \cite{gene} showed that, for $\nu_n = 0$, we have
\begin{equation}
    \label{eq:mean-convergence}
	\| \Mean_n - \expect[\vol] \| = \bigO\left( \frac{1}{\sqrt{n}} \right) \mbox{,}
\end{equation}
with high probability as $n \rightarrow \infty$, provided that $\vol$ is bounded and $A_n$ is invertible when $n > n_0$ for some $n_0$. Informally, this invertibility condition is satisfied when the Fourier slices densely populate the Fourier domain and the zeros of the CTFs $\hh_1, \ldots, \hh_n$ do not overlap. We shall make this first requirement more specific below.

To understand the above result, it is helpful to consider a simpler toy example. Instead of a volume, we have a random vector $x = [a, b, c]$ containing three values. We have observations for every $s = 1, \ldots, n$, but we only observe two entries of $x$ for each $s$. That is, our observations $y_s$ are of the form $[a_s, b_s, ?]$, $[a_s, ?, c_s]$, and $[?, b_s, c_s]$, where $?$ denotes a missing value. From this data, we can estimate the mean of $x = [a, b, c]$.

First, to estimate $\expect[a]$, we collect all observations in which the first entry is present, that is, observations of the form $[a_s, b_s, ?]$ and $[a_s, ?, c_s]$. We then average all the first entries to obtain our estimate of $\expect[a]$. We can similarly estimate the means of the other entries. All that is required is that we know which entry is missing in each observation (i.e., the location of the $?$) and that all entries are represented among the observations.

The parallel with mean estimation in cryo-EM is established by considering the Fourier Slice Theorem \eqref{eq:forward-cont-fourier}. Let us first consider the case without CTF, that is for $\hh = 1$. In this case, $\imag$ is pure projection and acts as the restriction to a central slice in the Fourier domain. Its adjoint, the backprojection mapping $\imag^\transp$, therefore inserts a two-dimensional Fourier transform into that plane, with remaining frequencies set to zero. Just like in the toy example, the Fourier transform $\fourier\im$ of each observation is a ``subset'' of the entries of $\fourier\vol$. If we can place each observed value at its appropriate frequency in the Fourier domain and average across all observations, we could estimate the mean Fourier transform $\expect[\fourier\vol]$. Again, what is required is that we know the projection operators $\imag_1, \ldots, \imag_n$ (that is, we know the rotations $\rot_1, \ldots, \rot_n$) and that the collection of all central slices adequately covers the Fourier domain. The requirement on the viewing directions is not very strict. For example, a set of Fourier slices forming a fan-like pattern (where their normals are contained in a single plane) or a tilt series with no missing wedge are enough to guarantee accurate estimation of the mean.

This is exactly what it done by the least-squares estimator $\mean_n$ as defined by \eqref{eq:mean-normal}--\eqref{eq:bn-def} but in a more formal way. The right-hand side $b_n$ takes the Fourier transform $\fourier{\Im_s}$ of each image, places it onto the proper plane in three-dimensional Fourier space as defined by $\rot_s$, and averages across all images for $s = 1, \ldots, n$. If the central slices cover enough of the three-dimensional Fourier domain, this will ``reconstruct'' the average volume $\expect[\vol]$ up to a frequency-dependent reweighting that depends on the distribution of viewing directions. Indeed, even in the case of uniform distribution of viewing directions $\rot$ over $\SO(3)$, frequencies close to the origin will be oversampled compared to frequencies farther away, and this must be compensated for. This reweighting is encoded in $A_n$ through the average of the projection-backprojection operators $\imag_s^\transp \imag_s^{}$. Again, \eqref{eq:forward-cont-fourier} tells us that $\imag_s^\transp \imag_s^{}$ in the Fourier domain is restriction followed by insertion, which is equivalent to multiplication by the indicator function of the plane corresponding to $\rot_s$. Adding all of these indicator functions together yields the reweighting $A_n$ relating $\mean_n$ to $b_n$. Here, having an adequate coverage of the Fourier domain by the central slices implies that $A_n$ is invertible.

While $A_n$ may be invertible, it may still have a high condition number. This can happen, for example, if certain viewing directions are more common than others. In a given direction, higher frequencies are also undersampled with respect to lower frequencies. Reconstruction at a higher resolution $N$ is therefore less well-conditioned compared to at lower $N$. These conditioning problems can be partially remedied by choosing an appropriate regularization parameter $\nu_n$ at the cost of some bias in the estimation.

We note here that if the viewing directions $\rot_1, \ldots, \rot_n$ are sampled from the uniform distribution over $\SO(3)$, the inverse of $A = \lim_{n\rightarrow\infty} A_n$ is a ramp filter $\|\vomega\|$ so inverting it is particularly straightforward. When $n$ is large, we can therefore calculate $A^{-1}(b_n)$, which is known as filtered backprojection \cite{herman2009fundamentals,natterer,radon1917uber}, a popular estimator for reconstruction in computerized tomography (CT) and related fields. As $n \rightarrow \infty$, this estimate will converge to $\expect[\Vol]$. However, in cryo-EM, the distribution of viewing directions is typically non-uniform, so this idealized ramp filter is not appropriate and the exact $A_n$ must be used \cite{harauz1986exact}.

In the case of non-trivial CTFs $\hh_1, \ldots, \hh_n$, the same ideas hold, except that backprojection $\imag_s^\transp$ includes multiplication by the CTF $\hh_s$ and projection-backprojection $\imag_s^\transp\imag_s^{}$ involves multiplication by $|\hh_s|^2$. As a result, for each $s$, certain frequencies are zeroed out in the corresponding term of $\sum_{s=1}^n \imag_s^\transp \imag_s^{}$ due to the CTF since $\im_s$ does not contain any information at those frequencies. However, this is mitigated by the fact that we have different CTFs for different images, and therefore different sets of zeros. As long as these do not all overlap, the matrix $A_n$ is invertible. In addition, the CTF is small at low and high frequencies, acting as a bandpass filter. This would not make $A_n$ invertible, but it does make it ill-conditioned. As before, increasing the regularization parameter $\nu_n$ partly mitigates this.

\subsection{Covariance estimator}
\label{sec:covar-est}

To capture the variability of the volume density $\Vol$ as a random vector in $\Real^p$, we consider its covariance $\var[\Vol]$. The same construction outlined in the previous section to estimate the mean can be used to estimate the covariance. Specifically, for each $s = 1, \ldots, n$, computing the covariance of \eqref{eq:forward} conditioned on $\imag_s$ gives
\begin{equation}
	\label{eq:im-covar}
	\var_{\vol_s,\noise_s|\Imag_s}[\Im_s] = \imag_s^{} \var[\Vol] \imag_s^\transp + \sigma^2 \eye_\Nsq\mbox{,}
\end{equation}
where $\var[\Noise] = \sigma^2 \eye_\Nsq$. The left-hand side is the covariance of the image $\Im_s$ where $\imag_s$ is fixed, but $\vol_s$ and $\noise_s$ are allowed to vary. To compute it, we would need an infinite number of realizations of $\im_s$ for a fixed viewing direction and CTF. However, we do not have an infinite number of images. We are only guaranteed to have one, but we can use this image to estimate the conditional covariance as in
\begin{equation}
	(\Im_s - \imag_s \Mean_n)(\Im_s - \imag_s \Mean_n)^\transp \approx \var_{\vol_s,\noise_s|\Imag_s}[\Im_s]\mbox{.}
\end{equation}
The left-hand side is available to us since we have already estimated $\Mean_n$ and $\imag_s$ is known. In expectation it equals the right-hand side, so it forms an unbiased estimator, albeit again with high variance. Plugging this into \eqref{eq:im-covar}, we obtain
\begin{equation}
    \label{eq:covar-constraint-s}
	(\Im_s - \imag_s \Mean_n)(\Im_s - \imag_s \Mean_n)^\transp \approx \imag_s^{} \var[\Vol] \imag_s^\transp + \sigma^2 \eye_\Nsq\mbox{.}
\end{equation}
The high variance in the left-hand side makes this a loose constraint and we do not expect it to hold exactly. Instead, we compute the mean squared error between the two sides and attempt to minimize it over all images. This yields the regularized least-squares estimator $\Covar_n$ of $\var[\Vol]$ defined by
\begin{equation}
    \label{eq:covar-objective}
	\Covar_n \coloneqq \argmin[\covar] \frac{1}{n} \sum_{s=1}^n \left\| (\Im_s - \imag_s \Mean_n)(\Im_s - \imag_s \Mean_n)^\transp - (\imag_s \covar \imag_s^\transp + \sigma^2 \eye_\Nsq) \right\|^2_\frob + \xi_n \|\covar\|^2_\frob \mbox{,}
\end{equation}
where $\xi_n \ge 0$ is a regularization parameter. As with estimating the mean, any potential ill-posedness can be mitigated by the regularization term $\xi_n \|\covar\|^2_\frob$. Typically, the problem is less ill-posed at large $n$, so $\xi_n$ decreases with growing $n$.

Similar to $\Mean_n$, this estimator finds the covariance matrix that, when projected along $\rot_s$ and convolved by $h_s$ according to $\Sigma \mapsto \imag_s^{} \Sigma \imag_s^\transp$, minimizes the square Frobenius distance to the outer product of the mean-subtracted images. Note that $\Sigma_n$ is not the regularized maximum-likelihood estimator for $\var[\vol]$. However, as we shall see later in this section, it converges to $\var[\vol]$ as $n \rightarrow \infty$ with high probability under a wide range of conditions.

To solve the least-squares optimization problem in \eqref{eq:covar-objective}, we again differentiate and set the derivative to zero, obtaining
\begin{equation}
	\label{eq:covar-normal}
	L_n(\Covar_n) = B_n\mbox{,}
\end{equation}
where
\begin{align}
	\label{eq:Ln-def}
	L_n(\covar) &\coloneqq \frac{1}{n} \sum_{s=1}^n \imag_s^\transp \imag_s^{} \covar \imag_s^\transp \imag_s^{} + \xi_n \eye_{p^2} \mbox{,} \\
    \label{eq:Bn-def}
	B_n &\coloneqq \frac{1}{n} \sum_{s=1}^n \imag_s^\transp (\Im_s-\imag_s\Mean_n)(\Im_s-\imag_s\Mean_n)^\transp\imag_s^{} - \sigma^2 \imag_s^\transp \imag_s^{}\mbox{.}
\end{align}
Calculating least-squares estimator defined in \eqref{eq:covar-objective} is therefore equivalent to solving the linear system \eqref{eq:covar-normal}. In the same spirit as $b_n$, the right-hand side matrix $B_n$ averages the backprojected outer product covariance estimators for each image with a noise term correction. The covariance projection-backprojection operator $L_n$ plays the same role as $A_n$ by describing the reweighting of the backprojected covariance matrix estimators.

As for the mean estimator, the Tikhonov regularization term in \eqref{eq:covar-objective} may be replaced by other regularization terms. We also note that $\Sigma_n$ is not constrained to be positive semi-definite and so may not qualify as a covariance matrix. Again, however, we are only interested in the dominant eigenvectors, so imposing this condition would not appreciably alter our results at the cost of significant computational expense.

Each of the constraints in \eqref{eq:covar-constraint-s} are loose because of the noise and the potentially large amount of variability in $\Vol$. Consequently, it may appear that their derived least-squares estimator \eqref{eq:covar-objective} would be a poor one. However, it has been shown that, given enough images, $\Covar_n$ does in fact provide a reasonable estimate of $\var[\vol]$. Indeed, Katsevich et al. \cite{gene} have shown that, for $\xi_n = 0$,
\begin{align}
    \label{eq:covar-convergence}
	\| \Covar_n - \var[\vol] \|_\frob = \bigO\left( \frac{\log n}{\sqrt{n}} \right) \mbox{,}
\end{align}
with high probability as $n \rightarrow \infty$, as long as that $\vol$ is bounded and $L_n$ is invertible when $n > n_0$ for some $n_0 > 0$. This invertibility condition is satisfied when the central planes defined by the viewing directions $\rot_1, \ldots, \rot_n$ contain enough frequency pairs in the Fourier domain and the zeros of the CTFs $\hh_1, \ldots, \hh_n$ do not overlap. Note that this requirement on the viewing directions is more strict than the corresponding requirement for the invertibility of $A_n$. Indeed, since the Fourier transform of the covariance matrix $\var[\vol]$ describes the correlation between any pairs of frequencies in the 3D Fourier domain, each pair must be represented in the data.

To clarify this, we return to the toy example introduced in the previous section. As before, we have a random vector $x = [a, b, c]$ containing three values and observations $y_s$ of the form $[a_s, b_s, ?]$, $[a_s, ?, c_s]$, and $[?, b_s, c_s]$ for $s = 1, \ldots, n$. It is not possible to reconstruct the joint probability density of $x = [a, b, c]$, but we can estimate its covariance.
Indeed, to estimate the variance of $a$, we collect all observations of the form $[a_s, b_s, ?]$ and $[a_s, ?, c_s]$, extract the first entry, and compute its variance. Similarly, for the covariance between $a$ and $b$, we take all observations of the form $[a_s, b_s, ?]$ and compute the covariance between the first two entries. Proceeding like this, we can ``fill up'' an entire matrix, which is a consistent estimator of the population covariance (that is, it converges to the latter as $n \rightarrow \infty$). Note that at no point do we need to draw samples of the complete vector $x = [a, b, c]$ or characterize its full distribution in order to estimate the covariance. All that is necessary is that we know which observation has what entries missing (which entry has the $?$ in place of a value) and that all pairs of entries are observed. However, it is not possible to estimate all third-order moments in this scenario since this would require observing all three entries at once.

We can draw parallels to the cryo-EM covariance estimation problem by again making use of the Fourier Slice Theorem, which tells us that the imaging operator $\imag$ in the Fourier domain corresponds to extracting a slice of the Fourier transform of a volume. The entries of $\fourier{x}$ lying on the plane defined by $\rot$ are kept, while others are discarded. The remaining entries are then multiplied by the CTF $\hh$. If we are to successfully estimate the covariance, we have to make sure that all pairs of 3D frequencies appear in our observed projections. For any given pair, we could then find the projection images whose Fourier slices contain that pairs and use these to compute the covariance. A given pair of frequencies, together with the origin, uniquely define a plane in 3D. Consequently, all such central planes must be present in our sample. In other words, the set of rotations $\rot_1, \ldots, \rot_n$ must cover all of $\SO(3)$. This does not mean that they have to be uniformly distributed, but that any given rotation in $\SO(3)$ is sampled with non-zero probability.

Again, this is much stricter than the requirement for accurate reconstruction of the mean $\mean_n$, where a great circle of viewing directions is sufficient. This requirement was first observed by Liu \& Frank \cite{liu1995estimation}, who argue that its stringency precludes accurate estimation of the covariance, which they refer to as ``type-II variance.'' In our work, this problem is mitigated by several factors. The first is that while we attempt to estimate the whole covariance matrix $\var[\vol]$, our ultimate goal is its top eigenvectors. It follows from the Davis-Kahan $\sin(\theta)$ theorem that it is possible to accurately estimate the leading eigenvectors of a matrix provided its eigenvalues are well-separated from the remaining eigenvalues (that is, there is a large eigengap) \cite{davis-kahan}. As will be described in the Section \ref{sec:high-dim}, random matrix theory suggests that such a separation in the eigenvalues of $\Covar_n$ does indeed appear as $n$ grows. As a result, we can expect to obtain good estimates for the top eigenvectors of $\Sigma_n$ even though $\Sigma_n$ is not very accurate overall. The second point is that our proposed algorithm is typically applied at low resolution $N$. As a result, we only need the viewing directions to cover the sphere up to this low resolution (i.e., gaps smaller than $1/N$ in the distribution of rotations are acceptable). Finally, adjusting the regularization parameter $\xi_n$ allows us to regularize the entries of $\Sigma_n$ whose corresponding pairs of frequencies are missing from the data.

To see how $\Sigma_n$ performs this estimation, we apply the Fourier Slice Theorem to the continuous covariance matrix $\mathcal{C}: \Real^3 \times \Real^3 \longrightarrow \Real$ satisfying $\mathcal{C} \in \Leb^1(\Real^3 \times \Real^3)$. This gives
\begin{equation}
    \label{eq:forward-covar-cont-fourier}
    (\fourier \times \fourier) (\projfun \times \projfun) \mathcal{C} (\vomega_1, \vomega_2) = (\fourier \times \fourier) \mathcal{C} (\rot^\transp[\vomega_1; 0], \rot^\transp[\vomega_2; 0]) \hh(\vomega_1) \hh(\vomega_2) \mbox{,}
\end{equation}
where $(\mathcal{A} \times \mathcal{B})$ denotes the mapping that applies $\mathcal{A}$ along the first variable and $\mathcal{B}$ along the second variable \cite{tagare}. As a result, projecting the Fourier transform of a covariance matrix along its columns and rows therefore corresponds to restriction to frequency pairs belonging to a certain plane defined by $\rot$ followed by multiplication by the CTF. 

The dual formulation of \eqref{eq:forward-covar-cont-fourier} says that backprojecting a two-dimensional covariance matrix (i.e., applying $(\projfun^\transp \times \projfun^\transp)$) corresponds to inserting its Fourier transform into a three-dimensional Fourier transform along pairs of frequencies both belonging to a certain plane. Each term in the sum \eqref{eq:Bn-def} defining $B_n$ therefore takes the two-dimensional matrix estimate $(\Im_s-\imag_s\Mean_s)(\Im_s-\imag_s\Mean_s)^\transp - \sigma^2 \eye_\Nsq$, places its Fourier transform along the correct plane in three-dimensional covariance Fourier space and multiplies by the appropriate CTF. These are then averaged across all images. Slice by slice, this provides a ``reconstruction'' of the three-dimensional covariance matrix.

Much like the case for mean estimation, however, this reconstruction by backprojection needs to be reweighted in order to obtain an accurate covariance estimate. The weighting is encoded by the covariance projection-backprojection operator $L_n$ and depends on the distribution of viewing directions and CTFs. In the case of uniform distribution of viewing angles, the same consideration of $A_n$ applies, where frequencies closer to the origin are weighted higher with respect to frequencies farther away. However, for the covariance we also need to take into account the relationship between pairs of frequencies. Indeed, for a given pair of frequencies, its weight depends on how many central planes, that is images, pass through both frequencies. Since frequencies that are nearly co-linear have more central planes passing through them, this results in higher weights compared to other pairs. By inverting $L_n$, we renormalize the backprojected covariance estimate $B_n$ with respect to this density. For a more detailed discussion of this phenomenon, see Katsevich et al. \cite{gene}.

As discussed above, the invertibility of $L_n$ depends on the viewing directions $\rot_1, \ldots, \rot_n$ sufficiently covering $\SO(3)$ and the zeros of the CTFs not completely overlapping. However, as with $A_n$, $L_n$ may still be highly ill-conditioned if certain viewing directions are more common than others. Higher frequencies are also less well-conditioned, so a high resolution $N$ gives a worse conditioning for $L_n$. Finally, the bandpass effect of the CTF increases its condition number. These can again be mitigated by an appropriate choice of the regularization parameter $\xi_n$, ensuring that $L_n$ is well-conditioned without introducing too much bias.

\subsection{Resolution limits}
\label{sec:resolution}

While their construction is similar, the least-squares mean $\Mean_n$ and covariance $\Covar_n$ estimators differ in their well-posedness and conditioning properties, as seen in the previous section. For a fixed number of images $n$, this results in the achievable resolution $N$ for the covariance estimator being significantly lower compared to that of the mean.

To illustrate, we estimate the mean $\expect[\Vol]$ at resolution $N$. Using the Fourier Slice Theorem, Section \ref{sec:mean-est} shows that this is achieved by ``filling up'' the 3D Fourier domain with $n$ central slices (each corresponding to a projection image) to obtain $b_n$ and applying $A_n^{-1}$. Each central slice has $\bigO(\Nsq)$ points, yielding a total of $\bigO(n\Nsq)$ points. Since the 3D Fourier domain contains $\bigO(\Ncu)$ points, we require that $n N^2 \gg N^3$, so $n$ must be of order of at least $N$.

If we have clean data, $n = \bigO(N)$ images would suffice. However, adding noise of variance $\sigma^2$, more images are needed for an accurate estimate. Specifically, for each of the $\bigO(\Ncu)$ points in the 3D Fourier domain, we need $\bigO(\sigma^2)$ samples to reduce the noise in $b_n$ to order $1$. The total number of required samples is therefore $\bigO(\Ncu \sigma^2)$, so $n$ has to be of order of at least $N \sigma^2$. This does not account for non-uniform distributions of viewing directions (which increases the constant of proportionality) or the effect of the CTF, the power spectrum of the volumes, and non-white noise (which increase the number of samples necessary to estimate the higher frequencies). However, this provides a lower bound, requiring the number of images to be at least proportional to the desired resolution times the noise variance.

For the covariance, on the other hand, each image contributes $\bigO(N^4)$ entries in the Fourier domain 3D covariance matrix as described in Section \ref{sec:covar-est}. The number of entries to estimate in the covariance matrix is $\bigO(N^6)$. Consequently, we need $n$ to be at least of order $N^2$ to adequately estimate $\var[\Vol]$ from clean data. From another perspective, the viewing directions need to cover the unit sphere with separation $1/N$, so an order of $N^2$ images is required.

Each term of the sum in $B_n$ concerns the outer product of a (mean-centered) image with itself. As a result, adding noise of variance $\sigma^2$ to the images results in noise of variance $\sigma^4$ in each term. To reduce the total variance of the noise in $B_n$ to $\bigO(1)$, we therefore need $\bigO(N^6 \sigma^4)$ samples. Consequently, $n$ must be of order at least $N^2 \sigma^4$. Again, this is an idealized setting, but this relationship provides a reasonable lower bound for $n$.

The difference in the required number of images for $\Mean_n$ and $\Covar_n$ is quite stark. Instead fixing the number of images $n$, we see that the best resolution $N$ that we can achieve for $\Mean_n$ is $\bigO(n \sigma^{-2})$. The resolution limit for $\Covar_n$, on the other hand, is $\bigO(\sqrt{n} \sigma^{-2})$. In other words, to increase the resolution by a factor of two, we need four times as many images. Estimating the covariance is therefore not only more computationally demanding (in terms of running time and memory usage), but is also more demanding in terms of data.

More concretely, let us suppose we have $n = 10000$ images with $\sigma^2 = 10$. Assuming the clean projection images have a mean square intensity of order $1$, this implies a signal-to-noise ratio of $0.1$. The highest achievable resolution $N$ for the $\Mean_n$ is then $1000$. Again, this is an upper bound. The achieved resolution is much lower in practice due to the effects of pixel size, CTF, and lower signal-to-noise ratio at high frequencies. In contrast, the maximum achievable resolution $N$ for $\Covar_n$ in our idealized setting is only $10$.

One way to overcome these limitations is to impose strong assumptions on the covariance. For example, that it is low-rank, satisfies certain smoothness or sparsity constraints, or that it is generated by certain types of deformations. In this work, we make use of the low-rank property since we ultimately extract the leading eigenvectors of our covariance estimate $\Covar_n$. This may not be optimal, however, as the low-rank constraint is not imposed when estimating the covariance matrix. We will explore these directions in future work.

\subsection{High-dimensional PCA}
\label{sec:high-dim}

The consistency result \eqref{eq:covar-convergence} shows that $\Covar_n$ converges to $\var[\Vol]$ as $n \rightarrow \infty$. However, in many applications, while $n$ may be large, it is not necessarily large with respect to the size $\Ncu$ of the volume vectors.

In this case, a more appropriate setting is to consider the behavior of $\Covar_n$ as $n$ and $N$ both tend to infinity, but at potentially different rates. Indeed, Section \ref{sec:resolution} suggests that $N$ should grow no faster than $\sqrt{n}$ to ensure estimation is well-posed. To better understand the behavior of $\Covar_n$ in this high-dimensional regime, we first review related results from the literature on sample covariance. Let us consider the sample covariance of a set of independently sampled Gaussian noise vectors $w_1, \ldots, w_n$
\begin{equation}
    \label{eq:sample-cov}
	\frac{1}{n} \sum_{s=1}^n w_s^{} w_s^\transp\mbox{.}
\end{equation}
where $\var[w_1] = \ldots = \var[w_n] = \sigma^2 \eye_p$ for some dimension $p > 1$.

In the low-dimensional regime where $n \gg p$, all eigenvalues of this sample covariance are concentrated around the single population eigenvalue $\sigma^2$. However, for $n, p \rightarrow \infty$ where $p/n \rightarrow \gamma < 1$, the spectrum will instead spread between $\sigma^2(1-\sqrt{\gamma})^2$ and $\sigma^2(1+\sqrt{\gamma})^2$, following the Mar\v{c}enko-Pastur distribution \cite{marcenko-pastur}.

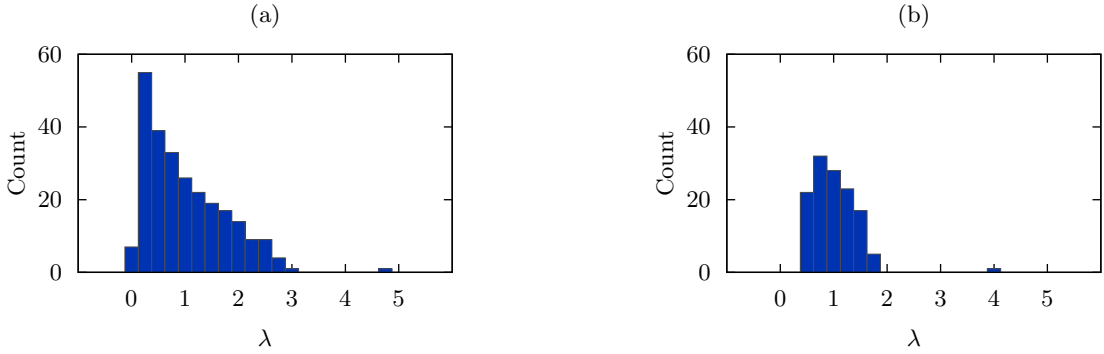
\begin{figure}
\begin{minipage}[t]{7cm}
\begin{center}
\input{mp_gplt}
\end{center}
\end{minipage}
\hfill
\begin{minipage}[t]{7cm}
\begin{center}
\input{spiked_cov_gplt}
\end{center}
\end{minipage}
\caption{
\label{fig:spiked-cov-sample}
The eigenvalue distribution of the sample covariance matrix for (a) $p = 256$, $n = 512$ and (b) $p = 128$, $n = 1024$ with $\sigma = 1$ and $\ell = 3$ in both regimes. For (a), we have $\gamma = 1/2$ and the spiked covariance model predicts a maximum noise eigenvalue at $(1+\sqrt{1/2})^2 \approx 2.91$ and a signal eigenvalue at $\lambda(3, 1/2) \approx 4.67$, while for (b), $\gamma = 1/8$ gives $(1+\sqrt{1/8})^2 \approx 1.83$ and $\lambda(3, 1/8) \approx 4.17$.
}
\end{figure}

In the spiked covariance model \cite{johnstone}, we have a clean signal $a_s = \sqrt{\ell} v z_s$ for $s = 1, \ldots, n$, where $v$ is a unit vector, $z_1, \ldots, z_n$ are i.i.d., zero-mean and unit variance random variables, while $\ell$ is the signal strength. The covariance of $a_s$ is then equal to $\ell v v^\transp$ and has rank one. Adding noise then gives the measurements
\begin{equation}
	\label{eq:spiked-cov}
	d_s = a_s + w_s\mbox{,}
\end{equation}
for all $s = 1, \ldots, n$. As before, the sample covariance is
\begin{equation}
    \label{eq:spiked-sample-cov}
	\frac{1}{n} \sum_{s=1}^n d_s^{} d_s^\transp\mbox{.}
\end{equation}
When $n \gg p$, its spectrum converges to $\{\ell + \sigma^2, \sigma^2, \ldots, \sigma^2\}$, with the dominant eigenvector equal to $v$. However, when $n, p \rightarrow \infty$ and $p/n \rightarrow \gamma < 1$, there are two possible scenarios. If $\ell/\sigma^2 < \sqrt{\gamma}$, the spectrum will be the same as the pure noise case---the signal is lost in the noise. If instead $\ell/\sigma^2 \ge \sqrt{\gamma}$, all eigenvalues will follow the Mar\v{c}enko-Pastur distribution except one \cite{paul2007asymptotics}. This signal eigenvalue will ``pop out'' at
\begin{equation}
    \label{eq:mp-lambda}
	\lambda(\ell, \gamma) = (\sigma^2+\ell)(1+\gamma\sigma^2/\ell)\mbox{.}
\end{equation}
These distributions are illustrated for two regimes $\gamma = 1/2$ and $\gamma = 1/8$ in Figure \ref{fig:spiked-cov-sample}. As $\ell/(\sigma^{2}\sqrt{\gamma})$ increases, the dominant eigenvector converges to $v$ \cite{paul2007asymptotics}. Specifically, the square correlation $|\langle v, u \rangle|^2$ between $v$ and the dominant eigenvector $u$ of \eqref{eq:spiked-sample-cov} tends to
\begin{equation}
    \label{eq:mp-corr}
    c(\ell, \gamma) = \frac{1-\gamma\sigma^4/\ell^2}{1+\gamma\sigma^2/\ell}\mbox{.}
\end{equation}

The spiked covariance model suggests a solution for estimating $v$ from the noisy observations $d_1, \ldots, d_n$. For signal covariance estimation, the eigenvalues below $\sigma^2(1+\sqrt{\gamma})^2$ are set to zero while those above are shrunk by an appropriate amount. A first approach would be to invert \eqref{eq:mp-lambda} to obtain
\begin{equation}
    \ell(\lambda, \gamma) = \frac{1}{2} \left (\lambda + \sigma^2(1-\gamma)+\sqrt{(\lambda+\sigma^2(1-\gamma))^2 - 4\sigma^2 \lambda} \right) - \sigma^2
\end{equation}
and replace a given eigenvalue $\lambda$ by $\ell(\lambda, \gamma)$. This ``shrinks'' the eigenvalues of the covariance matrix to provide better approximations of the population eigenvalues and consequently a better approximation of the covariance. Other functions can similarly be used to improve the estimation of the eigenvalues and are known as ``shrinkers.'' This leads to the question of which shrinker is optimal given a certain loss function on the covariance matrix. Such shrinkers have been derived by Donoho et al. \cite{donoho-gavish} for $26$ different loss functions.

This more general approach succeeds quite well in recovering covariance matrices for high-dimensional data. Given a sample covariance matrix, we calculate its eigendecomposition
\begin{equation}
	\frac{1}{n} \sum_{s=1}^n d_s^{} d_s^\transp = \sum_{m=1}^p \lambda_m^{} u_m^{} u_m^\transp\mbox{,}
\end{equation}
where $\{v_1, \ldots, v_p\}$ form an orthonormal basis. The eigenvalues $\{\lambda_1,\ldots,\lambda_p\}$ are transformed using a shrinker function $\rho$ into $\{\rho(\lambda_1),\ldots,\rho(\lambda_p)\}$.
Putting everything back together gives us
\begin{equation}
	\label{eq:shrinkage}
	\sum_{m=1}^p \rho(\lambda_m^{}) u_m^{} u_m^\transp\mbox{,}
\end{equation}
which is an estimator for the population covariance $\var[a]$. 

Donoho et al. showed how to choose the shrinker $\rho$ to optimize the error with respect to some loss function on the covariance estimator \cite{donoho-gavish}. The shrinker which achieves the lowest expected loss in the Frobenius norm as $n, p \rightarrow \infty$ is given by
\begin{equation}
	\label{eq:frob-norm-shrinker}
    \rho(\lambda) = \ell(\lambda, \gamma) c(\ell(\lambda, \gamma), \gamma)\mbox{.}
\end{equation}
Among all shrinkage estimators of the form \eqref{eq:shrinkage}, the shrinker $\rho$ given by \eqref{eq:frob-norm-shrinker} provides the lowest expected loss in the Frobenius norm \cite{donoho-gavish} as $n, p \rightarrow \infty$. The expected loss with respect to the operator norm is minimized by $\rho(\lambda) = \ell(\lambda, \gamma)$. The authors consider a variety of norms, each of which is assigned a corresponding optimal shrinker \cite{donoho-gavish}. Since our least-squares objective is formulated with respect to the Frobenius norm, we shall use the corresponding shrinker \eqref{eq:frob-norm-shrinker} in the following. Note that this estimator is not restricted to rank-one signals $a$ but is optimal for arbitrary fixed finite rank.

By slight abuse of notation, we extend the action of $\rho$ from scalars to symmetric matrices by its action on the eigenvalues, so that
\begin{equation}
	\rho(A) \coloneqq \rho\left( \sum_{m=1}^p \lambda_m^{} v_m^{} v_m^\transp \right) = \sum_{i=1}^p \rho(\lambda_m^{}) v_m^{} v_m^\transp\mbox{,}
\end{equation}
provided $\sum_{m=1}^p \lambda_m^{} v_m^{} v_m^\transp$ is an eigendecomposition of $A$.

\subsection{Shrinkage of $B_n$}
\label{sec:shrinkage}

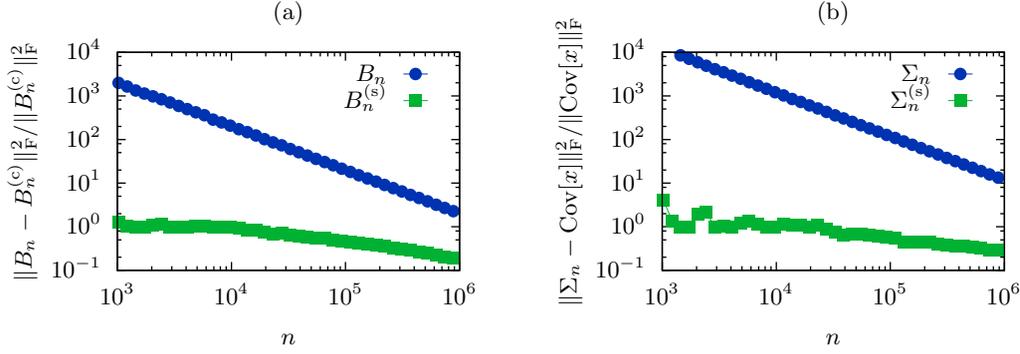
\begin{figure}
\input{shrinkage_Bn_gplt.tex}
\input{shrinkage_Sigman_gplt.tex}
\caption{\label{fig:shrinkage-Bn-Sigman} Effect of shrinkage on the relative error of (a) $B_n$ and (b) $\Sigma_n$ for simulated data with $N = 16$, $C = 2$ and $7$ distinct CTFs. The signal-to-noise ratio (see \eqref{eq:snr-def}) is $0.001$.}
\end{figure}

We can apply the above ideas to obtain a better estimate for the right-hand side $B_n$. This is the first major improvement over previous work on least-squares estimators for 3D covariance matrices in cryo-EM \cite{gene,isbi15}. Specifically, we will replace $B_n$ by
\begin{equation}
    B_n^\shrink = A_n^{1/2} \rho(A_n^{-1/2} B_n A_n^{-1/2} + \sigma^2 \eye_p) A_n^{1/2}\mbox{,}
\end{equation}
where $A_n$ is given by \eqref{eq:An-def}.

For clean data, $\sigma = 0$ so $\im_s = \imag_s \vol_s$. In this case, the definition of $B_n$ \eqref{eq:Bn-def} gives
\begin{equation}
	B_n^\clean = \frac{1}{n} \sum_{s=1}^n \imag_s^\transp (\Im_s-\imag_s\Mean_n)(\Im_s-\imag_s\Mean_n)^\transp\imag_s^{}\mbox{,}
\end{equation}
Plugging in our forward model \eqref{eq:forward-k} for $\Im_s$, we get
\begin{equation}
    \label{eq:clean-rhs}
	B_n^\clean = \frac{1}{n} \sum_{s=1}^n \imag_s^\transp \imag_s^{} (\Vol_s-\Mean_n)(\Vol_s-\Mean_n)^\transp\imag_s^\transp\imag_s^{}\mbox{,}
\end{equation}
which is the sample covariance of the mean-subtracted and projected-backprojected volumes $\imag_s^\transp \imag_s^{} \Vol_s$ for $s = 1, \ldots, n$.

We would now like to obtain something close to \eqref{eq:clean-rhs} also in the noisy case. Mean-subtracting and backprojecting noisy images $\im_s$, we have
\begin{equation}
	\imag_s^\transp (\Im_s-\imag_s\Mean_n) = \imag_s^\transp \imag_s^{} (\Vol_s-\Mean_n) + \imag_s^\transp \Noise_s^{}\mbox{.}
\end{equation}
This is similar to the spiked covariance model \eqref{eq:spiked-cov}, except the noise term is not white but has covariance $\expect[\Imag^\transp \Imag]$. We can approximate the noise covariance using the regularized estimator
\begin{equation}
	A_n = \frac{1}{n} \sum_{s=1}^n \Imag_s^\transp \Imag_s^{} + \nu_n\eye_p \mbox{,}
\end{equation}
since $\expect[A_n] = \expect[\Imag^\transp \Imag] + \nu_n\eye_p$ and the law of large numbers guarantees that $A_n$ converges to its expectation as $n \rightarrow \infty$. The regularization term $\nu_n\eye_p$ ensures that the inverse of $A_n$ will be bounded. Note that the high-dimensional phenomena studied in the previous section do not appear here since $\Imag$ is sampled from a low-dimensional space of fixed dimension. Indeed, the viewing direction $\rot$ is sampled from $\SO(3)$ (which has dimension three) while the CTF $\hh$ depends on two defocus values and an azimuth angle, yielding a total of $6$ dimensions.

Multiplying the backprojected images by $A_n^{-1/2}$ whitens the noise, and we define
\begin{equation}
	\Z_s = A_n^{-1/2} \Imag_s^\transp (\Im_s^{}-\Imag_s^{} \Mean_n^{}) = A_n^{-1/2} \Imag_s^\transp \Imag_s^{} (\Vol_s^{}-\Mean_n^{}) + A_n^{-1/2} \Imag_s^\transp \Noise_s^{}\mbox{.}
\end{equation}
We now apply the standard shrinkage operator $\rho$ defined in \eqref{eq:frob-norm-shrinker} and obtain
\begin{equation}
	\rho\left(\frac{1}{n} \sum_{s=1}^n \Z_s^{} \Z_s^\transp\right) = \rho\left( A_n^{-1/2} B_n A_n^{-1/2} + \sigma^2 \eye_p\right)\mbox{.}
\end{equation}
Conjugating the shrunken covariance by $A_n^{1/2}$, we obtain a shrinkage variant $B_n^\shrink$ of $B_n$ as
\begin{equation}
    \label{eq:Bn-shrink-def}
	B_n^\shrink \coloneqq A_n^{1/2} \rho\left( A_n^{-1/2} B_n A_n^{-1/2} + \sigma^2 \eye_p \right) A_n^{1/2}\mbox{,}
\end{equation}
providing an estimator of the clean right-hand side $B_n^\clean$.
Note that the loss is minimized with respect to the Frobenius norm on $A_n^{-1/2} B_n^\shrink A_n^{-1/2}$, not $B_n^\shrink$, so there still might be some room for improvement. This is the subject of future work.

Replacing $B_n$ by $B_n^\shrink$ in \eqref{eq:covar-normal} yields a more accurate estimator $\Sigma_n^\shrink = L_n^{-1}(B_n^\shrink)$ since $B_n^\shrink$ is a more accurate estimate of $B_n^\clean$ compared to $B_n$. We will refer to $\Sigma_n^\shrink$ as the shrinkage covariance estimator, in contrast to the standard $\Sigma_n$ least-squares covariance estimator. To evaluate the effect of the shrinkage on estimation accuracy, we plot the error of $B_n$ and $B_n^\shrink$ with respect to the clean $B_n^\clean$ as a function of $n$ in Figure \ref{fig:shrinkage-Bn-Sigman}(a). For the values of $n$ considered, shrinkage introduces a significant reduction in error. The same effect is found for the resulting estimators $\Sigma_n$ and $\Sigma_n^\shrink$ when compared with $\var[\vol]$ in Figure \ref{fig:shrinkage-Bn-Sigman}(b).

\section{Efficient computation}
\label{sec:calc}

The covariance estimators as formulated in the previous section are computed by solving the corresponding normal equations. However, direct matrix inversion is intractable for typical problem sizes. We therefore consider an iterative solution based on the conjugate gradient method applied to a convolutional formulation of the normal equations. We first illustrate this for the mean least-squares estimator $\Mean_n$ and then generalize this technique to the covariance estimator $\Covar_n$. To speed up convergence, we employ circulant preconditioners for both $A_n$ and $L_n$. The use of the conjugate gradient method to estimate the covariance was previously considered in And\'en et al. \cite{isbi15}, but this work lacked the convolutional formulation and appropriate preconditioners necessary for rapid convergence.

\subsection{Mean deconvolution}
\label{sec:mean-conv}

The normal equations \eqref{eq:mean-normal} for the mean estimator $\Mean_n$ can be solved by calculating the matrix $A_n$, the right-hand side $b_n$, and solving for $\Mean_n$ in $A_n \Mean_n = b_n$. This is a linear system in $\Ncu$ variables, so solving it directly has complexity $\bigO(N^9)$. A more sophisticated approach is therefore needed. Here, we shall exploit the convolutional structure of $A_n$. This approach has been successful in several related applications \cite{wajer-pruessmann,fessler-toeplitz,guerquin-kern,vonesch}, but we shall focus on its use in the Fourier-based iterative reconstruction method (FIRM) introduced by Wang et al. \cite{lanhui-firm}. This section will rederive that method with the goal of applying these ideas to the computation of $\Covar_n$.

We first note that the projection-backprojection operator $\imag_s^\transp \imag_s^{}$ is factored into $\basis^\transp \summing_s^\transp \summing_s^{} \basis$, where $\summing_s: \Real^\Ncu \rightarrow \Real^\Nsq$ is the voxel projection mapping corresponding to $\imag_s$. In the voxel domain, $\summing_s^\transp \summing_s^{}$ is a convolution. Indeed, in the continuous case, projection integrates along a certain viewing direction and convolves with a point spread function, while backprojection ``fills up'' a volume along a certain viewing angle using an image convolved with that point spread function. The resulting volume is then constant along that viewing direction. The projection-backprojection operator is therefore a low-pass filter.

In the frequency domain, the Fourier slice theorem tells the same story. Projection is restriction of the volume Fourier transform to a plane followed by multiplication by the CTF. Backprojection multiplies a two-dimensional Fourier transform by a CTF and inserts it into a plane in a three-dimensional Fourier transform, filling the rest with zeros. The combined projection-backprojection operator is therefore a multiplication by a Dirac delta function along the projection direction times the squared CTF along the transverse direction.

Having engineered our voxel discretization $\summing_s$ of the projection operator to satisfy a discrete Fourier slice theorem \eqref{eq:summing-fourier-slice}, these properties carry over to the discrete case. Consequently,
\begin{equation}
    \label{eq:proj-backproj-conv-s}
    \summing_s^\transp \summing_s^{} \vol(\vi) = \vol \conv k_s(\vi) = \sum_{\vj \in M_{2N-1}^3} \vol(\vi-\vj) k_s(\vj)\mbox{,}
\end{equation}
where $\conv$ denotes convolution and
\begin{equation}
    \label{eq:proj-backproj-kernel-s}
    k_s(\vi) \coloneqq \frac{1}{N^4} \sum_{\vl \in M_{N-1}^2} |\hh_s(\vl)|^2 \euler^{\frac{2\pi \ramuno}{N} \langle \vi, \rot_s^\transp [\vl; 0] \rangle}\mbox{,}
\end{equation}
for all $\vi \in M_{2N-1}^3$. This follows from calculating the dual $\summing_s^\transp$ of the voxel projection matrix \eqref{eq:summing-def} and applying it to the matrix $\summing_s$ itself. In other words, $\summing_s^\transp \summing_s^{}$ is a $3$-Toeplitz matrix. The projection-backprojection operator $\imag_s^\transp \imag_s^{}$ is thus factored into a basis evaluation $\basis$, a convolution $\summing_s^\transp \summing_s^{}$, and a basis expansion $\basis^\transp$.

In the definition \eqref{eq:An-def} of $A_n$, plugging in \eqref{eq:proj-backproj-conv-s} now gives
\begin{align}
    \nonumber
    A_n \vol &= \frac{1}{n} \sum_{s=1}^n \imag_s^\transp \imag_s^{} \vol + \nu_n \vol
    =\frac{1}{n} \sum_{s=1}^n \basis^\transp (\basis\vol \conv k_s) + \nu_n\vol
    = \basis^\transp \left(\basis\vol \conv \frac{1}{n} \sum_{s=1}^n k_s\right) + \nu_n\vol \\
    &= \basis^\transp(\basis\vol \conv f_n) + \nu_n\vol \mbox{,}
    \label{eq:proj-backproj-conv}
\end{align}
where
\begin{equation}
    \label{eq:proj-backproj-kernel}
    f_n \coloneqq \frac{1}{n} \sum_{s=1}^n k_s\mbox{.}
\end{equation}
The sum over $n$ is therefore factorized as basis evaluation $\basis$, followed by application of a $3$-Toeplitz matrix, then a basis expansion $\basis^\transp$.

The convolution kernel $f_n$ can be calculated in one pass over the dataset using NUFFTs with complexity $\bigO(N^3 \log N + n N^2)$. Once this has been calculated, however, each application of $A_n$ using the convolutional formulation of \eqref{eq:proj-backproj-conv} is achieved in $\bigO(N^3 \log N)$ time using FFTs, independent of the number of images.

We can exploit this fast application of $A_n$ to solve the system $A_n \Mean_n = b_n$. Specifically, we apply the conjugate gradient (CG) method, which is an iterative algorithm for solving linear systems \cite{hestenes}. It computes an approximate solution at each iteration through a single application of $A_n$, so its efficiency depends on being able to apply this operator fast, which is the case for Toeplitz operators, as seen above \cite{chan1996conjugate}. To reach a given accuracy $\bigO(\sqrt{\kappa(A_n)})$ iterations are needed, where $\kappa(A_n)$ is the condition number of $A_n$. As we shall see, $\kappa(A_n)$ can be reduced by the circulant preconditioner described in Section \ref{sec:renorm}.

The above algorithm therefore consists of two steps. First, we precalculate the kernel $f_n$ and the right-hand side $b_n$. These are both achieved in $\bigO(N^3 \log N + n N^2)$ using NUFFTs. Second, $\bigO(\sqrt{\kappa(A_n)})$ iterations CG are performed, each at a cost of $\bigO(N^3 \log N)$. The overall complexity is then $\bigO(\sqrt{\kappa(A_n)} N^3 \log N + n N^2)$. Note that this method is nearly optimal in the sense that simply reading the images requires $\bigO(n N^2)$ operations, while the reconstructed volume takes up $\bigO(N^3)$ in memory.

\subsection{Covariance deconvolution}
\label{sec:conv}

As discussed in the previous section, directly solving the normal equations for $\Mean_n$ can be computationally expensive. This is also the case for the covariance estimator $\Covar_n$, which scales worse in $N$. Indeed, volume vectors $\Vol$ are of size $\Ncu$ so the covariance estimate $\Covar_n$ is of size $\Ncu$-by-$\Ncu$ and thus contains $N^6$ elements. Since $L_n$ maps covariance matrices to covariance matrices, the matrix representation of $L_n$ requires $(N^6)^2 = N^{12}$ elements, which stored at single precision occupies $256~\mathrm{GB}$ for $N = 8$. Direct inversion of this matrix would have computational complexity $\bigO(N^{18})$.

Previously, Katsevich et al. \cite{gene} defined a volume basis based on spherical harmonics in which $L = \lim_{n\rightarrow\infty} L_n$ is a block diagonal matrix with sparse blocks under certain conditions. Unfortunately, the approximation is only valid if $\rot$ is uniformly distributed on $\SO(3)$ and the CTF is fixed. These conditions typically do not hold for experimental data. In addition, the approximation of $L_n$ by $L$ only holds as $n \rightarrow \infty$ and is not appropriate for smaller datasets. These problems are mitigated by solving the exact system $L_n(\Covar_n) = B_n$ using the CG method \cite{isbi15}, but this approach passes through the entire dataset at each iteration and converges slowly.

A more practical approach is to apply the ideas from the mean estimation algorithm described in the previous section. According to \eqref{eq:Ln-def}, $L_n$ is a sum of linear matrix operators
\begin{equation}
	\covar \mapsto \imag_s^\transp \imag_s^{} \covar \imag_s^\transp \imag_s^{}
\end{equation}
plus a regularization term $\xi_n \eye_{p^2}$. Since $\imag_s^\transp\imag_s^{}$ can be factored into basis evaluation/expansion and convolution in three dimensions, this mapping enjoys a similar factorization
\begin{equation}
    \imag_s^\transp \imag_s^{} \covar \imag_s^\transp \imag_s^{} = \basis^\transp \left( \summing_s^\transp \summing_s^{} \left(\basis \covar \basis^\transp \right) \summing_s^\transp \summing_s^{} \right) \basis\mbox{.}
\end{equation}
The conjugation by $\summing_s^\transp \summing_s^{}$ convolves both the rows and the columns of the matrix by $k_s$, which is a convolution in six dimensions by the outer product of $k_s$ with itself. Specifically, we have
\begin{equation}
    \summing_s^\transp \summing_s^{} Z \summing_s^\transp \summing_s^{} = Z \conv K_s\mbox{,}
\end{equation}
for any matrix $Z \in \Real^{\Ncu \times \Ncu}$, where
\begin{equation}
    K_s[\vi_1,\vi_2] \coloneqq k_s[\vi_1]k_s[\vi_2]\mbox{,}
\end{equation}
for all $(\vi_1, \vi_2) \in M_{2N-1}^6$. Consequently
\begin{equation}
    \imag_s^\transp \imag_s^{} \covar \imag_s^\transp \imag_s^{} = \basis^\transp \left( \basis \covar \basis^\transp \conv K_s \right) \basis\mbox{.}
\end{equation}

One advantage of this formulation is that we average the convolutional kernels over the whole dataset to obtain a convolutional representation for $L_n$. This gives
\begin{equation}
    \label{eq:covar-proj-backproj-conv}
    L_n(\covar) = \basis^\transp \left( \basis \covar \basis^\transp \conv F_n \right) \basis + \xi_n \covar \mbox{,}
\end{equation}
where
\begin{equation}
    \label{eq:covar-proj-backproj-kernel}
    F_n \coloneqq \frac{1}{n} \sum_{s=1}^n K_s\mbox{.}
\end{equation}
Similar to $A_n$, the sum over $s$ in $L_n$ is factored into basis evaluations/expansions and a $6$-Toeplitz matrix operator, allowing for rapid calculation of $L_n(\Sigma)$.

The kernel $F_n$ is calculated using an NUFFT at a computational cost of $\bigO(N^6 \log N + n N^4)$ by rewriting \eqref{eq:covar-proj-backproj-kernel} as the six-dimensional non-uniform discrete Fourier transform of the $\Nsq$-by-$\Nsq$-by-$n$ array formed by the outer products of $\hh_s$ sampled on a two-dimensional $M_{N-1}^2$ grid. Once this kernel is computed, applying $L_n$ using the convolution formulation \eqref{eq:covar-proj-backproj-conv} costs $\bigO(N^6 \log N)$. This follows because the basis evaluations/expansions each have complexity $\bigO(N^3 \log N)$ and there are $\bigO(N^3)$ rows and columns in the matrix, while the six-dimensional convolution achieves complexity $\bigO(N^6 \log N)$ using FFTs.

The right-hand side matrix $B_n$ can also be computed as an NUFFT with complexity $\bigO(N^6 \log N + n N^4)$. As a result, the overall complexity for solving the normal equations \eqref{eq:covar-normal} using CG is $\bigO(\sqrt{\kappa(L_n)} N^6 \log N + n N^4)$. We again note that this complexity is nearly optimal, since with respect to storing the $\bigO(N^6)$ size covariance matrix, we only lose a condition number and logarithmic factor, while we only require a $N^2$ factor increase with respect to storing the input images of size $\bigO(n N^2)$.

\subsection{Circulant preconditioners for $A_n$ and $L_n$}
\label{sec:renorm}

The number of iterations required for CG to converge scales with the square root of the condition number of the linear system \cite{saad2003iterative,axelsson1996iterative,trefethen-bau}. As we shall see, both $A_n$ and $L_n$ are badly conditioned due to geometric considerations and the influence of the CTFs. One way to solve this is to increase the regularization parameters $\nu_n$ and $\xi_n$. However, this increases the bias of the estimator and may not be desirable in all situations. Fortunately, the number of iterations in CG can be reduced without regularizing the original problem by introducing an appropriate preconditioner. In the following, we describe how this can be achieved for $A_n$ and $L_n$.

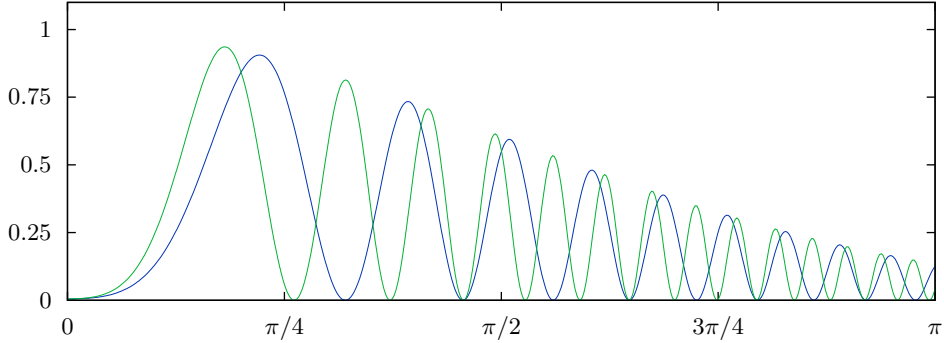
\begin{figure}
\begin{center}
\input{ctf2_gplt.tex}
\end{center}
\caption{\label{fig:ctf-envelope} The square magnitudes $|\hh(\vomega)|^2$ of two sample CTFs. Their sum forms a bandpass filter, worsening the conditioning of the least-squares estimators.}
\end{figure}

For a uniform distribution of viewing directions $\rot$ over $\SO(3)$ and with no microscope aberration, that is $\hh(\omega) = 1$, the unregularized (i.e., $\nu_n = 0$), continuous version of $A_n$ is approximated in the Fourier domain by the filter $2/\|\vomega\|$ as $n \rightarrow \infty$ (e.g., see \cite{natterer,gene}). Qualitatively, this is the limit of the Fourier transform $\fourier f_n$ of $f_n$ as $N, n \rightarrow \infty$. As $\fourier f_n$ approaches this limit, the $1/\|\vomega\|$ decay results in worse conditioning of $A_n$. The influence of the CTFs do not improve this situation. Indeed, the square magnitudes of the CTFs $|\hh_s(\vomega)|^2$ form a bandpass filter (see Figure \ref{fig:ctf-envelope}), attenuating low and high frequencies, which is replicated in the Fourier transform of $f_n$, further worsening the conditioning of $A_n$.

Similar results hold for $L_n$. Indeed, for uniform distribution of viewing directions and $\hh(\omega) = 1$ with no regularization (i.e., $\xi_n = 0$), it was shown by Katsevich et al. \cite{gene} that its continuous version acts as multiplication in the Fourier domain by $2/\|\vomega_1 \times \vomega_2\|$ as $n \rightarrow \infty$. Not only does this decay as $1/\|\vomega_1\|\|\vomega_2\|$, but the kernel is singular for $\vomega_1$ parallel to $\vomega_2$. Again, adding CTFs results in attenuation for low and high frequencies $\|\vomega_1\|$ and $\|\vomega_2\|$. As a result, the Fourier transform $\fourier F_n$ of the kernel $F_n$ representing $L_n$ is close to singular and decays rapidly, resulting in poor conditioning of $L_n$.

If the distribution of viewing directions is non-uniform, the condition number will be even larger. As a consequence of the above phenomena, a large number of iterations will be required in order to achieve convergence, which is not a desirable situation.

To remedy this, we precondition $A_n$ and $L_n$. In other words, we define operators $C_n$ and $D_n$ that can be easily inverted and such that $C_n^{-1} A_n^{}$ and $D_n^{-1} L_n^{}$ are close to the identity. As a result, $\kappa(C_n^{-1} A_n^{})$ and $\kappa(D_n^{-1} L_n^{})$ are small, allowing us to reformulate the linear systems to achieve better conditioning. This is the idea of the preconditioned CG method, which converges in $\bigO(\sqrt{\kappa(C_n^{-1} A_n^{})})$ and $\bigO(\sqrt{\kappa(D_n^{-1} L_n^{})})$ steps for the mean and covariance estimators, respectively \cite{saad2003iterative,axelsson1996iterative,trefethen-bau}. Note that the overall conditioning of the problem is still the same, it is only the CG convergence rate that changes.

A variety of preconditioners have been developed to improve convergence of the CG method. Popular alternatives include diagonal, or Jacobi, preconditioners and incomplete Cholesky or LU factorizations \cite{saad2003iterative,axelsson1996iterative}. As mentioned above, a good preconditioner is one whose inverse closely approximates the operator itself. This is commonly achieved by exploiting its structure. For example, a Toeplitz operator is well approximated by a circulant operator. Inverting a circulant operator is in turn achieved efficiently using FFTs \cite{strang1986proposal}.

In the current work, we therefore consider circulant approximations to the convolution factors of $A_n$ and $L_n$ as preconditioners. A circulant operator $C$ operating on some $d$-dimensional vector $w$ defined on $M_N^d$ is given by a circular convolution
\begin{equation}
    Cw(\vi) = \sum_{\vj \in M_N^d} w^\per(\vi-\vj) g(\vj)\mbox{,}
\end{equation}
where $w^\per$ is the periodized extension of $w$
\begin{equation}
    w^\per(\vi) = w(\vi+N\vm)\quad\mbox{for~}\vm\mbox{~such~that~}\vi+N\vm \in M_N^d\mbox{,}
\end{equation}
and $g$ is a convolution kernel defined on $M_N^d$. The difference with the standard convolution encountered before in $A_n$ and $L_n$ (see \eqref{eq:proj-backproj-conv} and \eqref{eq:covar-proj-backproj-conv}) is that the functions are periodized at the boundary instead of padded with zeros.

An important advantage of circulant operators is that they are diagonalized by the Fourier basis, and we can therefore write the action of $C$ in the Fourier domain as
\begin{equation}
    \fourier{Cw}(\vk) = \fourier{w}(\vk)\fourier{g}(\vk)\mbox{.}
\end{equation}
This makes circulant operators fast to apply, but also fast to invert, since
\begin{equation}
    \fourier{C^{-1}w}(\vk) = \fourier{w}(\vk)\fourier{g}(\vk)^{-1}\mbox{,}
\end{equation}
provided $\fourier{g}(\vk) \neq 0$ for all $\vk \in M_N^d$. Since standard and circular convolutions differ principally at the boundary, they are similar when both $w$ and $g$ concentrate around the origin.

Circulant operators are therefore good preconditioners to standard convolutions, provided we can calculate them efficiently. Let $C_n$ be a circulant operator with kernel $g_n$ such that
\begin{equation}
    \widetilde{C}_n \coloneqq \argmin[\widetilde{C}_n~\mathrm{circ.}] ~ \|\widetilde{A}_n - \widetilde{C}_n\|_\frob\mbox{,}
\end{equation}
and let
\begin{equation}
    \widetilde{A}_n = \frac{1}{n} \sum_{s=1}^n \summing_s^\transp \summing_s^{} + \nu_n\eye_{\Ncu} \mbox{,}
\end{equation}
be the voxel version of $A_n$ such that $\widetilde{A}_n \volvox = \volvox \conv (f_n + \nu_n \delta_0)$, where $\delta_0$ is the three-dimensional Dirac delta function with value $1$ at $\vi = 0$ and zero elsewhere. Such approximations have been previously studied by Tyrtyshnikov \cite{circulant}, from whose results we derive the formula
\begin{equation}
    \label{eq:gn-def}
    g_n(\vi) \coloneqq \frac{1}{N^3} \sum_{\substack{\vm \in M_{2N-1}^3 \\ \vm = \vi~(\mathrm{mod}~N)}} (N-|m_1|)(N-|m_2|)(N-|m_3|) f_n(\vm) + \nu_n\delta_0(\vi)\mbox{,}
\end{equation}
for all $\vi \in M_N^3$. This periodizes the original kernel $f_n + \nu_n\delta_0$ with periodicity $N$ and weights by a multiplier that attenuates points far from the origin.

This particular circulant approximation $\widetilde{C}_n$ of the Toeplitz operator $\widetilde{A}_n$ has the advantage of being computed with low computational complexity. Indeed, calculating $g_n$ using \eqref{eq:gn-def} has complexity $\bigO(N^3)$. Furthermore, it minimizes the distance in Frobenius norm to $\widetilde{A}_n$ and preserves the positive semidefiniteness and invertibility of $\widetilde{A}_n$ \cite{circulant}, which is not true for other circulant preconditioners, such as those proposed by Strang and Chan \cite{strang1986proposal,chan1988optimal}.

Since the circulant approximation $\widetilde{C}_n$ can be inverted easily using three-dimensional FFTs, we use $C_n^{-1} \coloneqq \basis^\transp \widetilde{C}_n^{-1} \basis$ as a preconditioner in the CG method when solving $A_n\Mean_n = b_n$. In this case, the effective condition number of the preconditioned linear system is equal to $\kappa(C_n^{-1} A_n^{})$, which is small if the approximation is accurate. Numerical experiments in Section \ref{sec:sim-cond} indicate that this preconditioner brings the condition number down to $1$--$50$.

The same type of circulant approximation can also be found for the convolution factor in $L_n$, with the circulant operator
\begin{equation}
    \widetilde{D}_n \coloneqq \argmin[\widetilde{D}_n~\mathrm{circ.}] ~ \|\widetilde{L}_n-\widetilde{D}_n\|_\frob\mbox{,}
\end{equation}
where
\begin{equation}
    \widetilde{L}_n(Z) = \frac{1}{n} \sum_{s=1}^n \summing_s^\transp \summing_s^{} Z \summing_s^\transp \summing_s^{} + \xi_n Z = Z \conv (F_n + \xi_n \delta_0)
\end{equation}
is the voxel version of $L_n$ where $\delta_0$ is now the six-dimensional Dirac delta function. The kernel $G_n$ of $\widetilde{D}_n$ can be found to equal
\begin{equation}
    G_n(\vi) \coloneqq \frac{1}{N^6} \sum_{\substack{\vm \in M_{2N-1}^6 \\ \vm = \vi~(\mathrm{mod}~N)}} (N-|m_1|) \cdots (N-|m_6|) F_n(\vm) + \xi_n \delta_0(\vi) \mbox{,}
\end{equation}
for all $\vi \in M_N^6$. Again, $G_n$ is a weighted periodization of $F_n + \xi_n\delta_0$. The computational complexity of calculating $G_n$ is $\bigO(N^6)$.

The circulant operator $\widetilde{D}_n$ can also be inverted quickly using six-dimensional FFTs, so we use $\Sigma \mapsto D_n^{-1}(\Sigma) \coloneqq \basis^\transp \widetilde{D}_n^{-1}(\basis \Sigma \basis^\transp) \basis$ to precondition the normal equations $L_n(\Covar_n) = B_n$ of the least-squares covariance estimator or $L_n(\Covar_n) = B_n^{\mathrm{(s)}}$ for the shrinkage variant. For the same reasons as in the mean estimation case, the condition number $\kappa(D_n^{-1} L_n^{})$ is expected to be small. Again, numerical simulations in Section \ref{sec:sim-cond} indicate that the condition number of this operator stays in the regime $1$--$200$.

\subsection{Conjugate gradient \& thresholding}
\label{sec:cg}

\begin{algorithm}[t]
\begin{algorithmic}
\Function{MeanEstimation}{$\{R_s\}_{s=1}^n$, $\{h_s\}_{s=1}^n$, $\{y_s\}_{s=1}^n$, $\basis$, $\nu_n$}
%\State Calculate $f_n$ using \eqref{eq:proj-backproj-kernel} and \eqref{eq:proj-backproj-kernel-s}
\State Set $f_n[\vi] \gets \frac{1}{nN^4} \sum_{s=1}^n \sum_{\vl \in M_{N-1}^2} |\hh_s(\vl)|^2 \euler^{\frac{2\pi \ramuno}{N} \langle \vi, \rot_s^\transp [\vl; 0] \rangle}$
%\State Obtain $g_n$ from $f_n$ using \eqref{eq:gn-def}
\State Set $g_n[\vi] \gets \frac{1}{N^3} \sum_{\substack{\vm \in M_{2N-1}^3 \\ \vm = \vi~(\mathrm{mod}~N)}} (N-|m_1|)(N-|m_2|)(N-|m_3|) f_n(\vm) + \nu_n\delta_0(\vi) \quad \forall \vi \in M_N^3$
%\State Calculate $g_n^{-1}$ by inverting its Fourier transform
\State Calculate $\overline{g}_n$ such that $\fourier{\overline{g}_n} = (\fourier{g_n})^{-1}$
%\State Calculate $b_n$ using \eqref{eq:bn-def}
\State Set $b_n \gets \frac{1}{n} \sum_{s=1}^n \imag_s^\transp \Im_s^{}$
\State Apply CG to $\basis^\transp (\basis \Mean_n \conv f_n) + \nu_n\Mean_n = b_n$ with preconditioner $x \mapsto \basis^\transp(\basis x \conv \overline{g}_n)$
\State \Return $\Mean_n$
\EndFunction
\end{algorithmic}
\caption{
\label{algo:mean-estimation}
The least-squares mean estimator $\Mean_n$
}
\end{algorithm}

\begin{algorithm}[t]
\begin{algorithmic}
\Function{CovarianceEstimation}{$\{R_s\}_{s=1}^n$, $\{h_s\}_{s=1}^n$, $\{y_s\}_{s=1}^n$, $\Mean_n$, $\sigma$, $\basis$, $\xi_n$, \textit{do\_shrink}}
\State Set $f_n[\vi_1,\vi_2] \gets \frac{1}{nN^8} \sum_{s=1}^n \sum_{\vl_1, \vl_2 \in M_{N-1}^2} |\hh_s(\vl_1)|^2 |\hh_s(\vl_2)|^2 \euler^{\frac{2\pi \ramuno}{N} \left( \langle \vi_1, \rot_s^\transp [\vl_1; 0] \rangle - \langle \vi_2, \rot_s^\transp [\vl_2; 0] \rangle \right)}$
\State Set $G_n[\vi] \gets \frac{1}{N^6} \sum_{\substack{\vm \in M_{2N-1}^6 \\ \vm = \vi~(\mathrm{mod}~N)}} (N-|m_1|) \cdots (N-|m_6|) F_n(\vm) + \xi_n \delta_0(\vi) \quad \forall \vi \in M_N^6$
\State Calculate $\overline{G}_n$ such that $\fourier{\overline{G}_n} = (\fourier{G_n})^{-1}$
\State Set $B_n \gets \frac{1}{n} \sum_{s=1}^n \imag_s^\transp (\Im_s-\imag_s\Mean_n)(\Im_s-\imag_s\Mean_n)^\transp\imag_s^{} - \sigma^2 \imag_s^\transp \imag_s^{}$
\If{\textit{do\_shrink}}
\State Set $B_n \gets B_n^\shrink = A_n^{1/2} \rho(A_n^{-1/2} B_n^{} A_n^{-1/2} + \sigma^2\eye_p) A_n^{1/2}$
\EndIf
\State Apply CG to $\basis^\transp (\basis \Covar_n \basis^\transp \conv F_n) + \xi_n \Covar_n = B_n$ with preconditioner $X \mapsto \basis^\transp(\basis X \basis^\transp \conv \overline{G}_n) \basis^\transp$
\State \Return $\Covar_n$
\EndFunction
\end{algorithmic}
\caption{
\label{algo:covar-estimation}
The covariance estimators $\Covar_n$ ($\mbox{\textit{do\_shrink}} = \mathrm{false}$) and $\Covar^\shrink_n$ ($\mbox{\textit{do\_shrink}} = \mathrm{true}$)
}
\end{algorithm}

We are now ready to formulate the algorithms for estimating $\Mean_n$ and $\Covar_n$ given the input images $\im_1, \ldots, \im_n$ and projection mappings $\imag_1, \ldots, \imag_n$ (or equivalently, the viewing directions $\rot_1, \ldots, \rot_n$ and CTFs $\hh_1, \ldots \hh_n$).

The mean estimation algorithm is given in Algorithm \ref{algo:mean-estimation}. First, the convolutional kernel $f_n$ associated with $A_n$ and the right-hand side $b_n$ are computed at a cost of $\bigO(N^3 \log N + n N^2)$. The circulant approximation kernel $g_n$ is then calculated from $f_n$, which takes $\bigO(N^3)$. We then apply CG to \eqref{eq:mean-normal}, with each iteration requiring application of $A_n$ and $C_n^{-1}$ which are obtained by multiplications by $\basis$, $\basis^\transp$ and convolutions by $f_n$ and $\overline{g}_n$, all of which have computational complexity of $\bigO(N^3 \log N)$. After $\bigO(\sqrt{\kappa(C_n^{-1} A_n)})$ iterations, we have $\mean_n$. The overall computational complexity of Algorithm \ref{algo:mean-estimation} is then
\begin{equation}
    O\left(\sqrt{\kappa(C_n^{-1} A_n)} N^3 \log N + n N^2\right)\mbox{,}
\end{equation}
where $\kappa(C_n^{-1} A_n^{})$ is typically in the range $1$--$50$.

The covariance estimation method listed in Algorithm \ref{algo:covar-estimation} is qualitatively similar. The convolutional kernel $F_n$ associated with $L_n$ and the right-hand side matrix $B_n$ are computed with complexity $\bigO(N^6 \log N + n N^4)$. The circulant kernel $G_n$ is obtained at cost $\bigO(N^6)$. Applying $L_n$ and $D_n$ now involves multiplication by $\basis$ and $\basis^\transp$ as well as convolution with $F_n$ and $G_n^{-1}$, each of which has computational complexity $\bigO(N^6 \log N)$. Now $\bigO(\sqrt{\kappa(D_n^{-1} L_n)})$ iterations are needed to obtain $\covar_n$. The overall complexity of Algorithm \ref{algo:covar-estimation} is then
\begin{equation}
    \label{eq:covar-estimation-complexity}
    O\left(\sqrt{\kappa(D_n^{-1} L_n)} N^6 \log N + n N^4 \right)\mbox{,}
\end{equation}
where $\kappa(D_n^{-1} L_n^{})$ is in the range $1$--$200$.

To obtain the shrinkage variant of the estimator, $\Covar_n^\shrink$, the additional step of calculating $B_n^\shrink$ from $B_n$ is added before the CG step in Algorithm \ref{algo:covar-estimation}. This is done using \eqref{eq:Bn-shrink-def}, where $B_n$ is whitened by conjugation with $A_n^{-1/2}$, the whitened matrix is shrunk using $\rho$, and the result is unwhitened by conjugation with $A_n^{1/2}$. The number of top eigenvalues of $A_n^{-1/2} B_n A_n^{-1/2}$ which exceed the Mar\v{c}enko-Pastur threshold is typically small, so we can exploit Lanczos method for finding the top eigenvalues and eigenvectors. For this, we need to apply $A_n^{-1/2} B_n A_n^{-1/2}$ fast. Since $A_n$ can be applied fast using its convolution kernel, applying its inverse square root $A_n^{-1/2}$ to a volume can be approximated iteratively using Krylov subspace methods \cite{druskin1989two,saad1992analysis,higham2008functions}. The number of iterations needed is typically small, so we take its complexity to be $\bigO(N^3 \log N)$. As a result, applying $A_n^{-1/2} B_n A_n^{-1/2}$ has complexity $O(N^6)$, since matrix multiplication by $B_n$ takes $\bigO(N^6)$. The overall eigendecomposition calculation therefore has complexity $\bigO(N^6)$, where we have assumed that the number of non-trivial eigenvectors is taken to be constant.

We note that an alternative to Krylov subspace methods for approximating $A_n^{-1/2}$ is to exploit the Toeplitz structure in $A_n$ and use this to calculate its Cholesky factors, which have similar properties to the matrix square root but can be inverted efficiently. In one dimension, this can be done in $\bigO(N^2)$ using the Schur algorithm \cite{schur1917potenzreihen,musicus1984levinson,ammar1987generalized}. While this has been generalized to matrices of block Toeplitz structure \cite{ammar1987generalized}, these do not take into account 2-Toeplitz structure, also known as Toeplitz-block-Toeplitz, and so have complexity $O(N^5)$ instead of the desired $O(N^4)$. Designing an appropriate generalization of the Schur algorithm for d-Toeplitz operators where $d \ge 2$ is the subject of future work.

Both in the case of the standard estimator and the shrinkage variant, the estimated covariance matrix $\Covar_n$ will contain a considerable amount of error in the form of a bulk noise distribution similar to that observed in the spiked covariance model. A final step of selecting the dominant eigenvectors is therefore necessary to extract the relevant part of the covariance matrix structure. Since we expect the population covariance matrix to be of low rank, it must have a small number of non-zero eigenvectors. This number can be estimated by looking for a ``knee'' in the spectrum of $\Covar_n$ where the dominant eigenvalues separate from the bulk noise distribution. In the case of a discrete distribution of molecular structures, this is at most one minus the number of resolved structures in the dataset. However, this determination has to be done manually. A heuristic method for validating this choice could be to inspect the corresponding eigenvectors and determine how ``noise-like'' they appear, using a suitable prior. Future work will focus on enabling the algorithm to perform this selection automatically. The computational complexity of calculating the leading $r = O(1)$ eigenvectors of $\Sigma_n$ is $O(rN^6)$.

An important feature of \eqref{eq:covar-estimation-complexity} is that the algorithm scales as $N^6$ in the resolution $N$ of the images. Since we are estimating the entire covariance matrix $\var[\vol]$, this is unavoidable since that matrix has $N^6$ entries. However, it has the unfortunate consequence of limiting the attainable resolution of covariance estimation using the proposed algorithm. For example, at $N = 16$, the covariance estimate $\Covar_n$ requires $128~\mathrm{MB}$ to store in double precision. The kernel $f_n$ is of size $2N$ and therefore requires $2^6 = 64$ times as much space, or $8~\mathrm{GB}$. Increasing $N$ beyond this becomes impractical for a typical workstation.

That being said, a large amount of useful information can be obtained at these resolutions. Indeed, since our goal is to classify rather than reconstruct, all we need is for the features that discriminate between various conformations to be present at low resolution. This is not an unreasonable assumption. Indeed, if one subunit of a molecule moves with respect to another, we can capture that movement at low resolution as long as that subunit is large enough. Similarly, the binding of external complexes to larger molecule is visible provided that those complexes are large enough. Once we can distinguish such differences, the dataset can be partitioned and higher-resolution reconstructions can then be produced from each subset. We have observed this in experimental data, suggesting that the restriction to $N = 16$ is not as debilitating as it may first seem to be. In our experiments in this paper, we shall therefore restrict ourselves to $N = 16$.

\subsection{Choice of basis}
\label{sec:basis}

To represent a volume $x$, we can store its values on the voxel grid $M_N^3/N$. We will call this a decomposition in the voxel basis. The problem with this basis is that the electric potential of a molecule is supported in the central ball $\{\|\vu\| < 1\}$, with no energy in the ``corners`` of the cube $[-1, +1]^3$. Indeed, any energy in this region will be captured by projections along a subset of viewing angles and will not be well reconstructed. We can therefore safely assume that the support is contained in the central ball. The same holds in the frequency domain, where frequency samples outside the Nyquist ball $\{\|\vk\| < N/4\}$ are expected to be negligible. In addition, the low sampling density of these frequencies leads to ill-conditioning of the reconstruction problem, which we would like to avoid.

To solve this, we will use different bases which are concentrated on $\{\|\vu\| < 1\}$ in space and within $\{\|\vk\| < N/4\}$ in frequency. One solution to this spectral concentration is given by the 3D Slepian functions \cite{slepian}, but their implementation is quite complicated. Instead, we will focus on an alternative basis with similar properties, the spherical Fourier-Bessel basis. It consists of functions given in spherical coordinates $(r, \theta, \phi)$ by
\begin{equation}
    \phi_{\ell, k, m}(r, \theta, \phi) =
    \left\{
    \begin{array}{ll}
        C_{\ell,k} j_\ell(r z_{\ell, k}) Y_{\ell, m}(\theta, \phi) & 0 \le r < 1 \\
        0 & 1 \le r
    \end{array}
    \right.
\end{equation}
where $j_\ell$ is the spherical Bessel function of order $\ell$, $z_{\ell, k}$ is the $k$th zero of $j_\ell$, and $Y_{\ell, m}$ is the spherical harmonic function of order $\ell$ and degree $m$, and $C_{\ell,k} = \sqrt{2}|j_{\ell+1}(z_{\ell,k})|^{-1}$. The indices $m$ and $k$ satisfy $|m| \le \ell$ and $k \le k_\max(\ell)$, where $k_\max(\ell)$ is the largest integer such that  $z_{\ell, k_\max(\ell)+1} < N \pi/4$. This is the same sampling criterion used in Bhamre et al. \cite{twicing} and Cheng \cite{xiuyuan}, which generalizes similar criteria for the 2D Fourier-Bessel basis \cite{klug,zhao}. This condition on $k$ ensures that $\fourier \phi$ is concentrated within the Nyquist ball since this function is concentrated around a ring centered at $\|\vk\| = z_{\ell, k}/\pi$. Finally, the constant $C_{\ell, k}$ ensures that the basis functions have unit norm. For $\ell$ up to some $\ell_\max$, we therefore have the basis $\{ \phi_{\ell, k, m} \}_{\ell \le \ell_\max, k \le k_\max(\ell), |m| \le \ell}$ which we use to decompose $x$.

However, as discussed in Section \ref{sec:setup}, the standard voxel basis allows for fast projection through using NUFFTs. To take advantage of this, we need a fast change-of-basis mapping between the voxel basis and the spherical Fourier-Bessel basis. For this, we can use NUFFTs and separation of variables to evaluate the basis at voxel grid points in $\bigO(N^4)$ complexity. Using fast spherical harmonic transforms \cite{tygert-sph-3,kunis-potts-sph} and fast Fourier-Bessel transforms \cite{michielssen1996multilevel}, we can reduce this further to $\bigO(N^3 \log N)$.

A simpler alternative is provided by the truncated Fourier basis
\begin{equation}
    \phi_{\vk}(x) =
    \left\{
    \begin{array}{ll}
        C \euler^{2\pi i \langle x, \vk \rangle} & x \in M_{N-2}^3/N \\
        0 & \mbox{otherwise}
    \end{array}
    \right.
\end{equation}
for $\vk \in M_{N-2}^3 \cap \{\|\vomega\| < (N-2)/2\}$. Again, $C$ is chosen so that $\phi_\vk$ has unit norm for all $\vk$. The functions are zero outside a central box $M_{N-2}$, providing a padding of one voxel in each direction, and only consists of frequencies inside the Nyquist ball. While less concentrated compared to the spherical Fourier-Bessel basis, it has the advantage of providing efficient change-of-basis mappings through standard FFTs.

In the following, we will use the spherical Fourier-Bessel basis since it enjoys better concentration properties. We note, however, that for large values of $N$, it may be more computationally efficient to use the truncated Fourier basis since the constant associated with FFTs is much smaller than those of the fast spherical harmonic and Fourier-Bessel transforms.

\section{Reconstruction of states}
\label{sec:recon}

Having estimates $\Mean_n$ and $\Covar_n$ of the mean $\expect[\Vol]$ and covariance $\var[\Vol]$ provides us with partial information on the distribution of the volumes $\Vol_1, \ldots, \Vol_n$. However, unless the distribution of $\Vol$ is Gaussian, this is not enough to fully characterize it. To do so, more information has to be extracted from the images $\Im_1, \ldots, \Im_n$. We shall consider two types of singular distributions: those supported on a finite number of points and those supported on a continuous low-dimensional manifold.

\subsection{Wiener filter}
\label{sec:wiener}

For a fixed viewing direction $\rot$, the variability in the random density $x$ encoded by the 3D covariance $\var[\Vol]$ induces variability in the clean images $\imag x$ through the 2D covariance $\imag \var[\Vol] \imag^\transp$. Classical Wiener filtering leverages this covariance to denoise images or estimate the underlying volume corresponding to each image.

Recall that we have the image formation model \eqref{eq:forward-k}, restated here as
\begin{equation}
	\Im_s = \imag_s \Vol_s + \Noise_s\mbox{,}
\end{equation}
where, as before, $\imag_s$ defines projection along viewing direction $\rot_s$ and convolution with $h_s$. As we saw earlier, this induces the relation
\begin{equation}
    \label{eq:proj-sig-covar}
	\var_{\Vol_s|\imag_s}[\imag_s \Vol_s] = \imag_s^{} \var[\Vol] \imag_s^\transp\mbox{,}
\end{equation}
for the signal term, which allows us to estimate the image covariance as
\begin{equation}
    \var_{\vol_s|\imag_s}[\imag_s \vol_s] \approx \imag_s^{} \covar_n \imag_s^\transp\mbox{.}
\end{equation}
The noise covariance $\var[\Noise_s]$ is assumed to be $\sigma^2 \eye_\Nsq$.

We now use the estimated mean and covariance to define a Wiener filter estimator \cite{stephane-book}
\begin{equation}
	\volest_s \coloneqq H_s(\Im_s - \imag_s \Mean_n) + \Mean_n\mbox{,}
\end{equation}
of $\vol_s$ where
\begin{equation}
    \label{eq:wiener-matrix-def}
	H_s \coloneqq \Covar_{n,r} \imag_s^\transp(\imag_s^{} \Covar_{n,r} \imag_s^\transp+\sigma^2 \eye_\Nsq^{})^{-1}\mbox{.}
\end{equation}
If $\Covar_{n,r} = \var[x]$ and $\mean_n = \expect[\vol]$, this linear filter minimizes the expected mean-squared error
\begin{equation}
	\expect_{\Vol_s,\Noise_s|\Imag_s}\left\| \volest_s - \vol_s\right\|^2\mbox{.}
\end{equation}
In the finite-sample case, this no longer holds, but we can expect the Wiener filter to perform better than the pseudo-inverse for reasonably accurate mean and covariance estimates.

As discussed in the previous section, $\Covar_n$ is often of low rank $r$ following the final thresholding step, giving $\Covar_{n,r}$. We can use this to significantly reduce the complexity of calculating $\volest_s$, which through direct evaluation of \eqref{eq:wiener-matrix-def} takes $\bigO(N^6 \log N)$ operations since it involves calculating $\imag_s^{} Q \Covar_n^{} Q^\transp \imag_s^\transp$. Let
\begin{equation}
    O_s U_s = \imag_s V_{n,r}
\end{equation}
be the ``thin'' QR decomposition of $\imag_s V_{n,r}$, where $O_s$ is an $\Nsq$-by-$r$ orthonormal matrix and $U_s$ is an $r$-by-$r$ upper triangular matrix \cite{trefethen-bau,golub-vanloan}. Using these matrices, we rewrite $\volest_s$ as
\begin{equation}
    \label{eq:fast-wiener}
    \volest_s = V_{n,r} \Lambda_{n,r} U_s^\transp (U_s \Lambda_{n,r} U_s^\transp + \sigma^2 \eye_r)^{-1} O_s^\transp (\im_s - \Imag_s \mean_n) + \mean_n\mbox{,}
\end{equation}
for $s = 1, \ldots, n$. This shows that $\volest_s$ equals $\mean_n$ plus a linear combination of the vectors $V_{n,r}$. A more compact, but isometrically equivalent, representation is therefore given by
\begin{equation}
    \coordest_s \coloneqq V_{n,r}^\transp (\volest_s - \mean_n) = \Lambda_{n,r} U_s^\transp (U_s \Lambda_{n,r} U_s^\transp + \sigma^2 \eye_r)^{-1} O_s^\transp (\im_s - \Imag_s \mean_n)\mbox{.}
\end{equation}
These coordinates can be calculated in $\bigO(r N^3 \log N + n r^2 N^2)$, since the images $\Imag_s V_{n,r}$ and $\Imag_s \mean_n$ require $\bigO(r N^3 \log N + n r N^2)$, the QR decompositions have computational complexity $\bigO(n r^2 N^2)$ and inverting n $r$-by-$r$ matrices takes $\bigO(n r^3)$, where we assume that $N^2 \gg r$. We note that this is close to optimal, since $r$ eigenvolumes require $\bigO(r N^3)$ in storage, while the images are stored in $\bigO(n N^2)$, so we only lose a factor of $\log N$ and $r^2$, respectively.

The Wiener filter estimate of the volumes is now
\begin{equation}
    \volest_s = V_{n,r} \coordest_s + \mean_n\mbox{.}
\end{equation}
The traditional denoising Wiener filter of the 2D images is obtained by projecting these volume estimates. Specifically, we define
\begin{equation}
    \imest_s \coloneqq \imag_s \volest_s = \imag_s (V_{n,r} \coordest_s + \mean_n)\mbox{.}
\end{equation}
This is the same estimator obtained by minimizing the expected loss $\expect_{\vol_s,\noise_s|\imag_s}[\|\imest_s - \im_s\|^2]$ of a linear estimator $\imest_s$ and substituting our estimates for the volume mean and covariance.

\subsection{Volume distance measures}
\label{sec:distances}

Given the images $\im_1, \ldots, \im_n$ together with our mean and covariance estimates $\mean_n$ and $\covar_n$, we can also define distance measures on the underlying volumes. This will allow us to cluster them using methods described in Section \ref{sec:clustering} or to describe their manifold structure using the manifold learning techniques in Section \ref{sec:manifold}.

The simplest distance is the Euclidean norm on the volume estimates $\volest_1, \ldots, \volest_n$ given by
\begin{equation}
    d_{st}^{\mathrm{(eucl)}} = \| \volest_s - \volest_t \|\mbox{,}
\end{equation}
for $s, t = 1, \ldots, n$.

Unfortunately, this distance measure weights all directions equally regardless of their accuracy for a given pair. To see why this is a problem, consider the fact that the columns of $\imag_s V_{n,r}$ have different norms which depend on $s$. For example, a volume which is highly oscillatory along one axis will project to almost zero for viewing directions along that axis. Since these vectors are used to estimate $\volest_s$, this means that the power of the noise is different for each coordinate. A distance measure that takes this into account would therefore be more robust than $d_{st}^{\mathrm{(eucl)}}$.

One way to do this is to instead consider distances on the denoised images $\imest_1, \ldots, \imest_n$. While we still have the problem of low-energy basis vectors, these do not have a large energy once reprojected, so the situation is better. We then use the common-lines distance between the images \cite{shatsky}. From the Fourier Slice Theorem \eqref{eq:forward-cont-fourier}, we see that the Fourier transforms of two images $\im_s$ and $\im_t$ occupy the planes orthogonal to $R_s^{(3)}$ and $R_t^{(3)}$, respectively, where $R_s^{(i)}$ denotes the $i$th row of $R_s$. As such, they intersect along the line defined by the unit vector
\begin{equation}
    \frac{R_s^{(3)} \times R_t^{(3)}}{\|R_s^{(3)} \times R_t^{(3)}\|}\mbox{.}
\end{equation}
We therefore define the common-lines vector $\vc_{st} \in \Real^2$ for image $s$ with respect to image $t$ by rotating this vector into the image coordinates
\begin{equation}
    \vc_{st} = \left[ R_s^{(1)}; R_s^{(2)} \right] \left( \frac{R_s^{(3)} \times R_t^{(3)}}{\|R_s^{(3)} \times R_t^{(3)}\|} \right)\mbox{.}
\end{equation}
The common lines of $\imest_s$ and $\imest_t$ with respect to one another are then $\fourier \imest_s (k \vc_{st})$ and $\fourier \imest_t (k \vc_{ts})$, respectively, where $k \in M_N$. If $\im_s$ and $\im_t$ are projections of the same molecular structure, that is $\vol_s = \vol_t$, we expect that these common lines should be close since they are restrictions of the same volume Fourier transform along the same line. A useful distance is therefore
\begin{equation}
    d_{st}^{\mathrm{(cl)}} = \sum_{k \in M_N} | \fourier \imest_s (k \vc_{st}) - \fourier\imest_t (k \vc_{ts}) |^2\mbox{,}
\end{equation}
for $s, t = 1, \ldots, n$, which we call the common-lines distance. Note that this does not take into account the different CTFs of $\imest_s$ and $\imest_t$, which puts it at a disadvantage compared to the Euclidean distance $d_{st}^{\mathrm{(eucl)}}$.

\subsection{Clustering}
\label{sec:clustering}

In the previous sections, we estimated the 3D covariance matrix and used it to calculate estimates $\volest_1, \ldots, \volest_n$ of the volumes $\vol_1, \ldots, \vol_n$, or more specifically, their low-dimensional coordinate vectors $\coordest_1, \ldots, \coordest_n$. These were then used to define distances $d^{\mathrm{(eucl)}}$ and $d^{\mathrm{(cl)}}$ between the volumes. Without any additional assumptions, we cannot extract more information. For this, we need prior information on the distribution of the volume $\vol$.

For example, if $\Vol$ is a discrete random variable, we can fit a discrete distribution by clustering the volume estimates $\volest_1, \ldots, \volest_n$. Such a model is realistic for many molecules, where the majority of their time is spent in a small number of states.

Instead of clustering the volume estimates $\volest_1, \ldots, \volest_n$ themselves, we work on the coordinates $\coordest_1, \ldots, \coordest_n$, since these are lower-dimensional but isometrically parametrize the volumes. One clustering approach is the $k$-means vector quantization algorithm \cite{lloyd}. While widely used for clustering, $k$-means has several problems, one of which is that it favors partitions that distribute points uniformly between clusters. This can be partly mitigated by modeling $\Vol$ using a Gaussian mixture model (GMM) and fitting its parameters using the expectation-maximization algorithm \cite{dempster-laird}.

Given the distances $d^{\mathrm{(eucl)}}$ or $d^{\mathrm{(cl)}}$ instead of coordinates, we can use standard graph clustering algorithms such as normalized cut \cite{shi-malik}. These algorithms partition the points into subsets that optimize certain criteria, the goal being to minimize the distances within clusters while maximizing distances between clusters.

Regardless of the clustering mechanism, we obtain a cluster assignment associated with each image. We then average the corresponding volume estimates to obtain an reconstruction of that class. However, as we shall see, the algorithm will typically be applied to downsampled images, so this reconstruction is by necessity of low resolution. A more accurate reconstruction is obtained by partitioning the dataset according to the cluster assignments and reconstructing each subset separately at full resolution using tools such as RELION \cite{scheres-relion}, cryoSPARC \cite{brubaker}, FREALIGN \cite{frealign}, or ASPIRE \cite{zhao2014rotationally,shkolnisky2012viewing,singer2010detecting}.

We note that performing full-resolution reconstruction for each subset provides refined estimates of the viewing directions  $\rot_1, \ldots, \rot_n$. While these are assumed given for our algorithm, they are not necessarily very accurate, since they must be estimated from the average molecular structure as discussed in Section \ref{sec:setup}. Since these subsets of particles given by clustering should be more homogeneous, we expect the estimates to be more accurate. The covariance estimation and clustering steps can then be repeated using the refined estimates to achieve better results. This approach is known as iterative refinement, and has proved useful for other cryo-EM problems \cite{van2000single,frank}.

\subsection{Continuous variability and diffusion maps}
\label{sec:manifold}

Certain molecules do not primarily exist in a discrete set of states, but exhibit continuous variability. In this case, the clustering approach outlined above fails. However, due to physical constraints on the molecular dynamics, this continuum of states can often be described by a small number of dominant flexible motions. In this section, we describe a method for analyzing this low-dimensional manifold using diffusion maps \cite{lafon}.

These tools have previously been applied to study the continuous variability of molecular structure by calculating diffusion maps for images in each viewing direction and ``patching'' these together to yield a diffusion map for the whole volume \cite{dashti}. Unfortunately, this is a heuristic method which is not guaranteed to yield an accurate description of the low-dimensional conformation manifold. We propose to use the mean and covariance estimates, together with the derived distance measures $d^{\mathrm{(eucl)}}$ or $d^{\mathrm{(cl)}}$ to calculate a diffusion map embedding that more closely captures the underlying structural variability.

The first step is to form a similarity matrix $W$ whose entries are
\begin{equation}
	W_{st} \coloneqq \exp\left(-\frac{d_{st}^2}{\epsilon}\right)\mbox{,}
\end{equation}
for $s, t = 1, \ldots, n$, where $d_{st}$ is a distance between volumes $s$ and $t$ while $\epsilon$ is a scale parameter that depends on the smoothness of the manifold. The distance $d_{st}$ can be one of $d^{\mathrm{(eucl)}}$ or $d^{\mathrm{(cl)}}$. In the following, we shall use $d^{\mathrm{(eucl)}}$.

We now sum the entries along each row and define the diagonal matrix $D$ with entries
\begin{equation}
    D_{ss} \coloneqq \sum_{t=1}^n W_{st}\mbox{,}
\end{equation}
for $s = 1, \ldots, n$. Renormalizing $W$ by $D$, we obtain the row-stochastic Markov transition matrix $A \coloneqq D^{-1} W$. This defines a random walk on the graph of volume estimates, with the transition probability between two points proportional to their similarity. Let us calculate the eigenvalues and eigenvectors of $A$. We then have
\begin{equation}
    A\phi_i = \lambda_i \phi_i\mbox{,}
\end{equation}
for $i = 1, \ldots, n$. These eigenvalues satisfy $|\lambda_i| \le 1$ for all $i$ and there is at least one eigenvalue equal to $1$ with the corresponding eigenvector parallel to the all-ones vector (this follows from the fact that $A$ is row-stochastic). We thus order the eigenvalues as $1 = \lambda_1 \ge |\lambda_2| \ge \ldots \ge |\lambda_n|$.

The diffusion coordinates at diffusion time $\tau$ for the $s$th volume are now
\begin{equation}
    \dmapest_s^{(\tau)} \coloneqq [\lambda_2^\tau\phi_{2,s}^{}, \ldots, \lambda_n^\tau\phi_{n,s}^{}]\mbox{,}
\end{equation}
where $\phi_{i,s}$ is the $s$th element of the $i$th eigenvector $\phi_i$ \cite{lafon}. The diffusion time $\tau$ specifies the scale of the diffusion embedding. Specifically, the distance $\|\dmapest_s^{(\tau)}-\dmapest_t^{(\tau)}\|$ between two diffusion map coordinates $\dmapest_s^{(\tau)}$ and $\dmapest_t^{(\tau)}$ approximates the distance between the probability distributions obtained from random walks on the volume graph starting at $\volest_s$ and $\volest_t$ after $\tau$ steps. As $\tau$ increases, these distributions start to overlap for points close together on the manifold. By increasing $\tau$, we obtain a set of coordinates $\dmapest_1^{\mathrm{(\tau)}}, \ldots, \dmapest_n^{\mathrm{(\tau)}}$ that are more robust to noise compared to $\coordest_1, \ldots, \coordest_n$ at the expense of smoothing out the fine-scale manifold structure.

Restricting $\dmapest_s$ to the first two or three coordinates, we obtain a two- or three-dimensional embedding of the estimated volumes, which provides a helpful visualization to determine the global geometric structure of the continuous manifold of conformations. We shall see some examples of this in Section \ref{sec:sim-manifold}.

\begin{figure}[t]
\begin{center}
\begin{minipage}{1cm}
\rotatebox[origin=c]{90}{$N = 130$}
\end{minipage}
\begin{minipage}{4.2cm}
\begin{center}
{\footnotesize (a)}

\includegraphics[height=3.0cm]{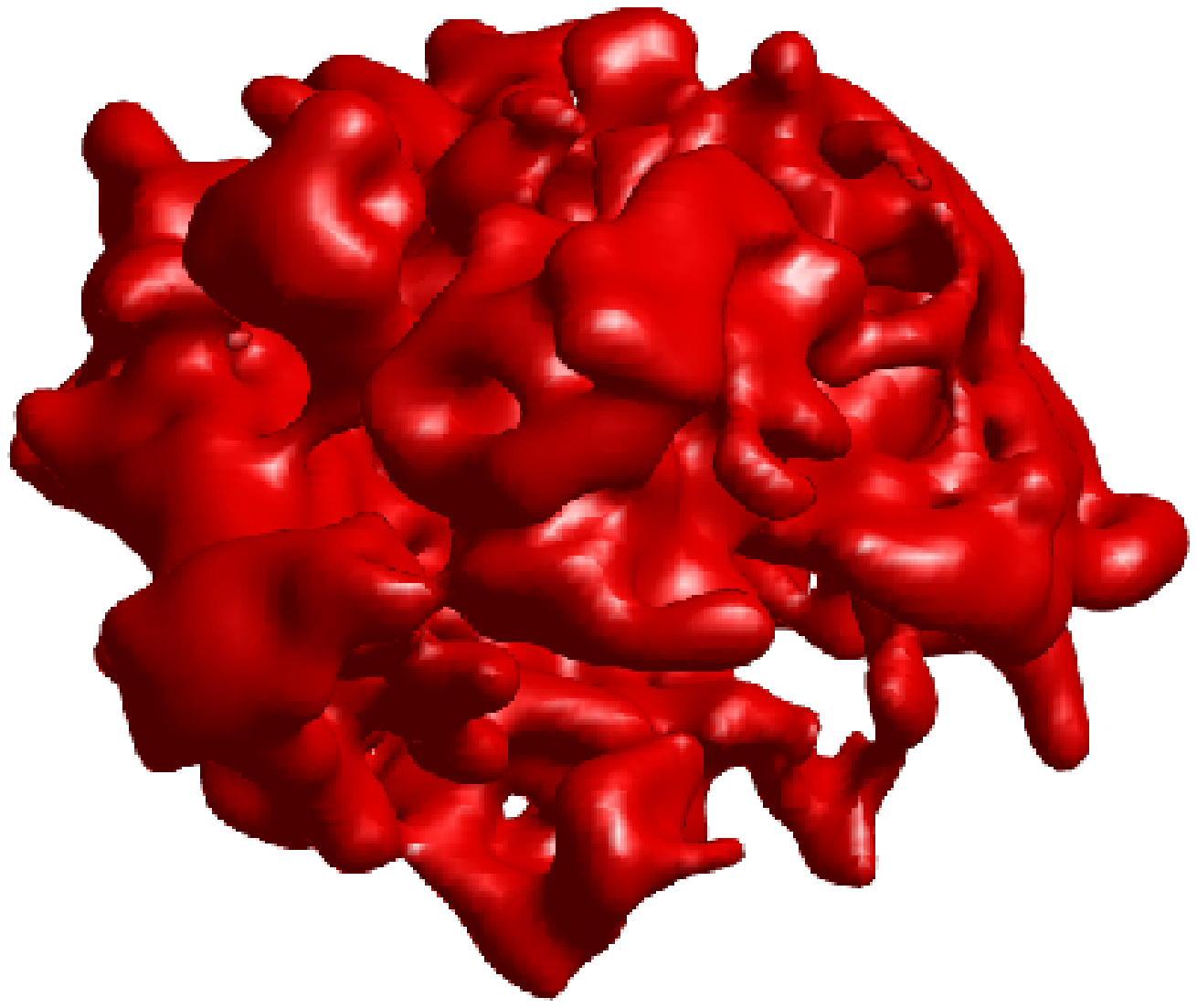}
\end{center}
\end{minipage}
\begin{minipage}{4.2cm}
\begin{center}
{\footnotesize (b)}

\includegraphics[height=3.0cm]{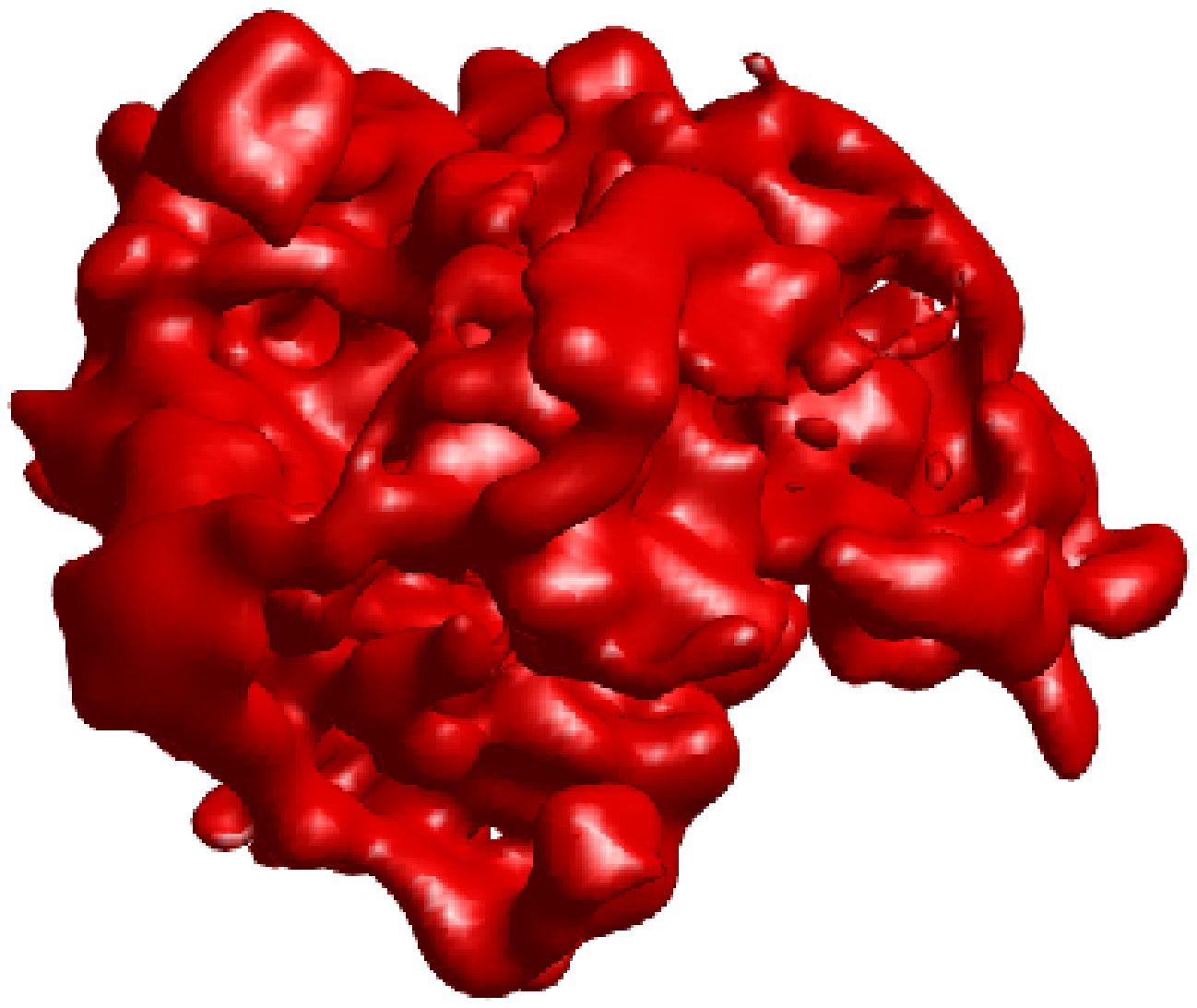}
\end{center}
\end{minipage}
\begin{minipage}{4.2cm}
\begin{center}
{\footnotesize (c)}

\includegraphics[height=3.0cm]{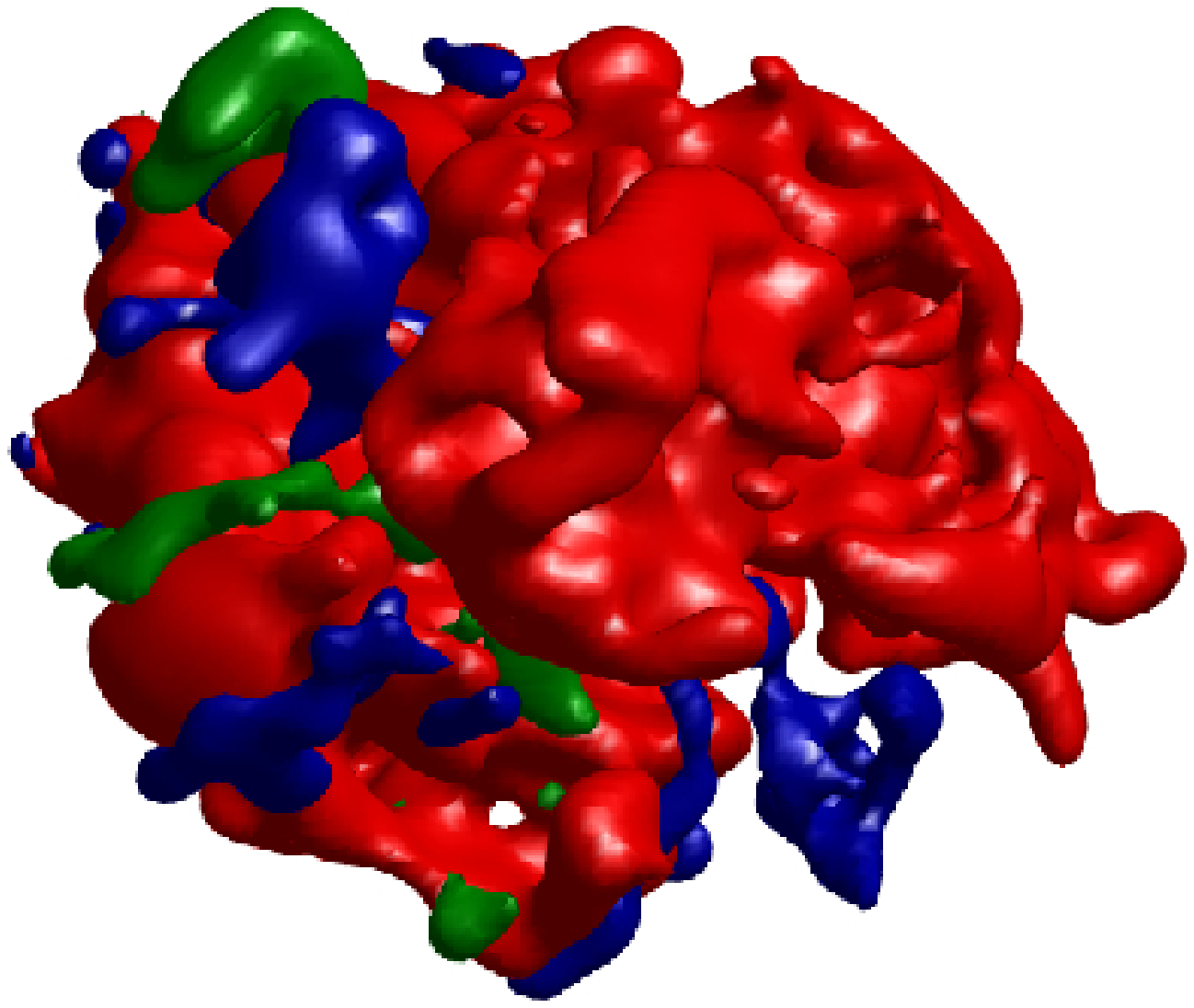}
\end{center}
\end{minipage}
\end{center}

\begin{center}
\begin{minipage}{1cm}
\rotatebox[origin=c]{90}{$N = 16$}
\end{minipage}
\begin{minipage}{4.2cm}
\begin{center}

\includegraphics[height=2.3cm]{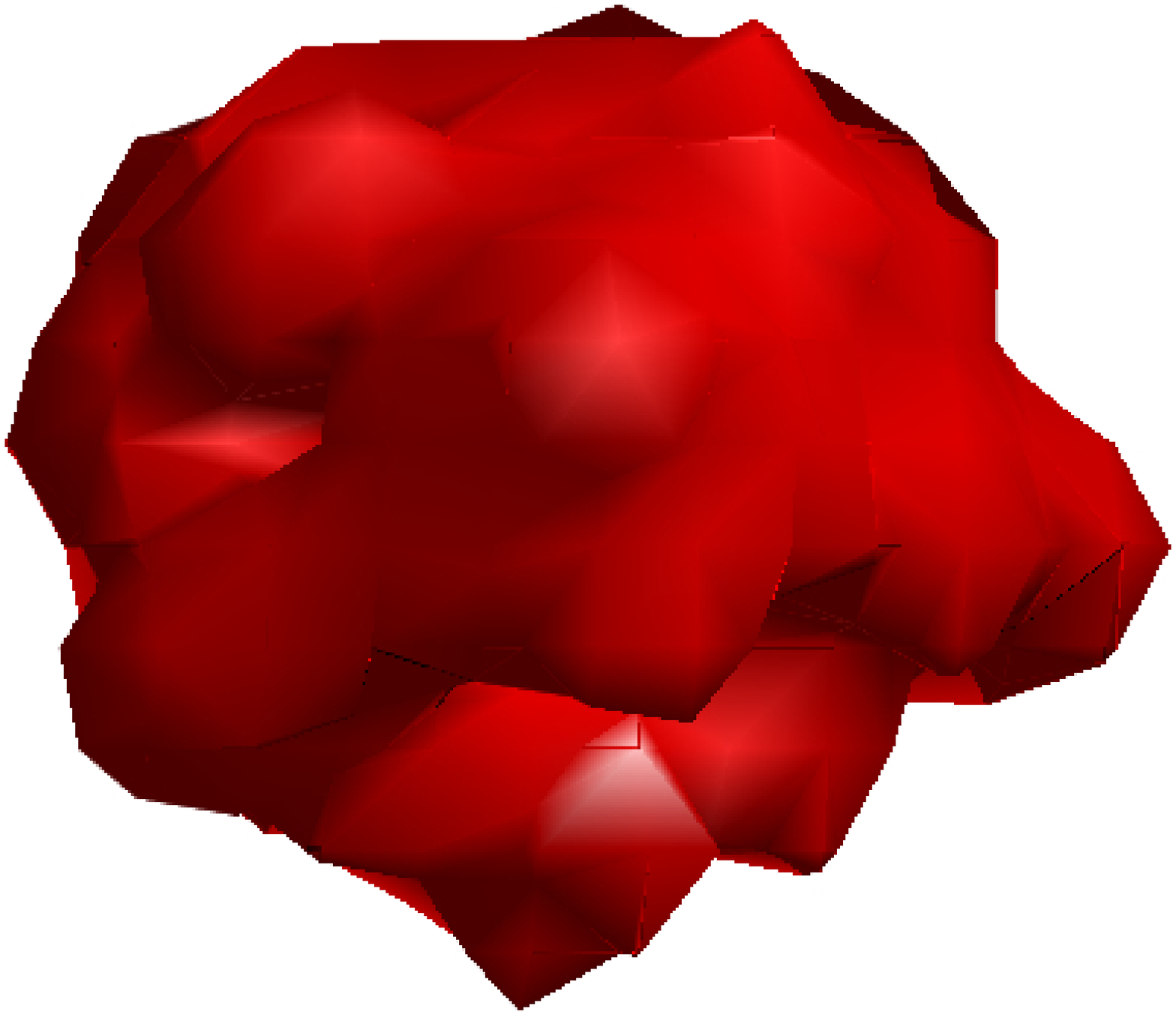}
\end{center}
\end{minipage}
\begin{minipage}{4.2cm}
\begin{center}

\includegraphics[height=2.3cm]{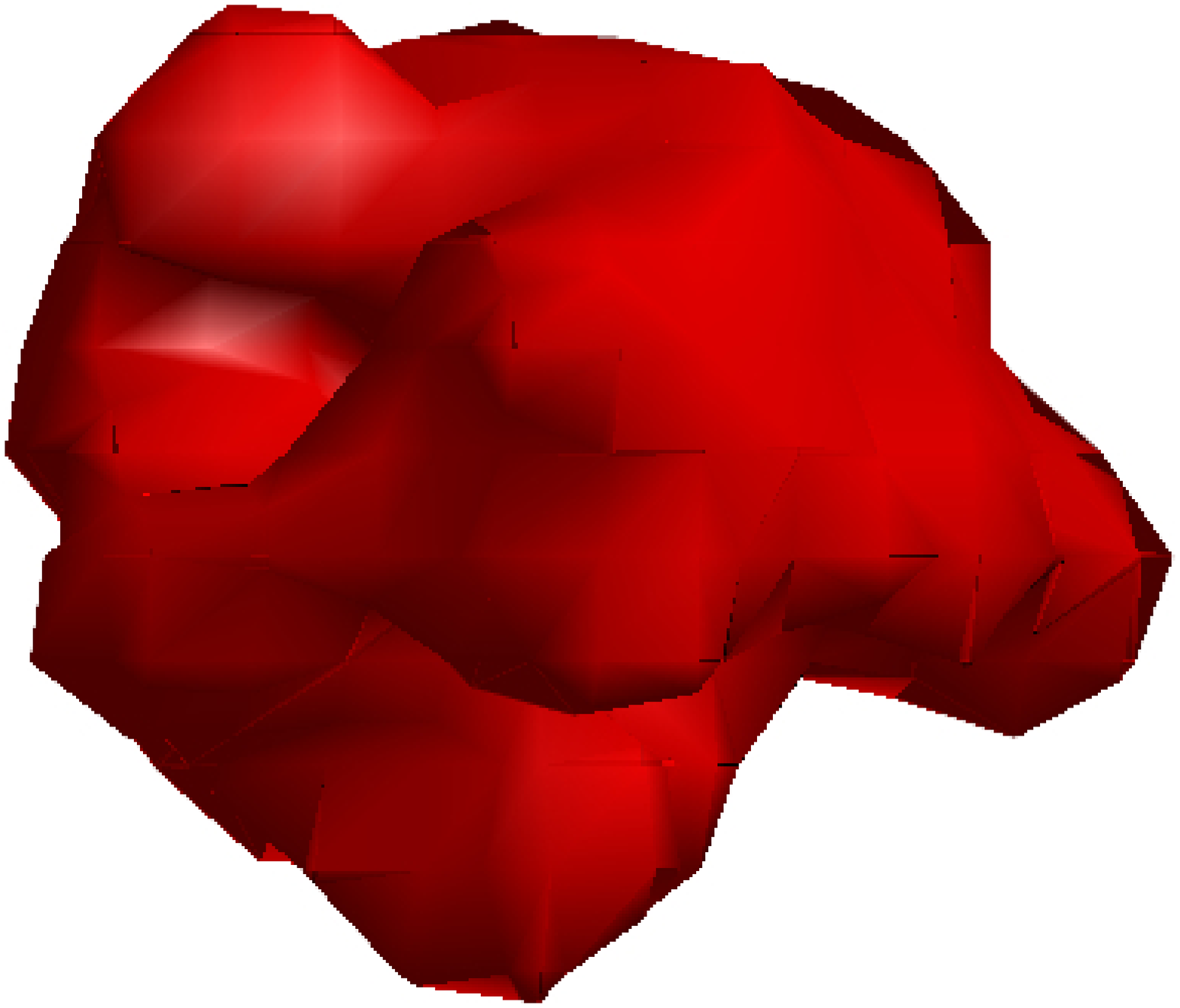}
\end{center}
\end{minipage}
\begin{minipage}{4.2cm}
\begin{center}
\includegraphics[height=2.3cm]{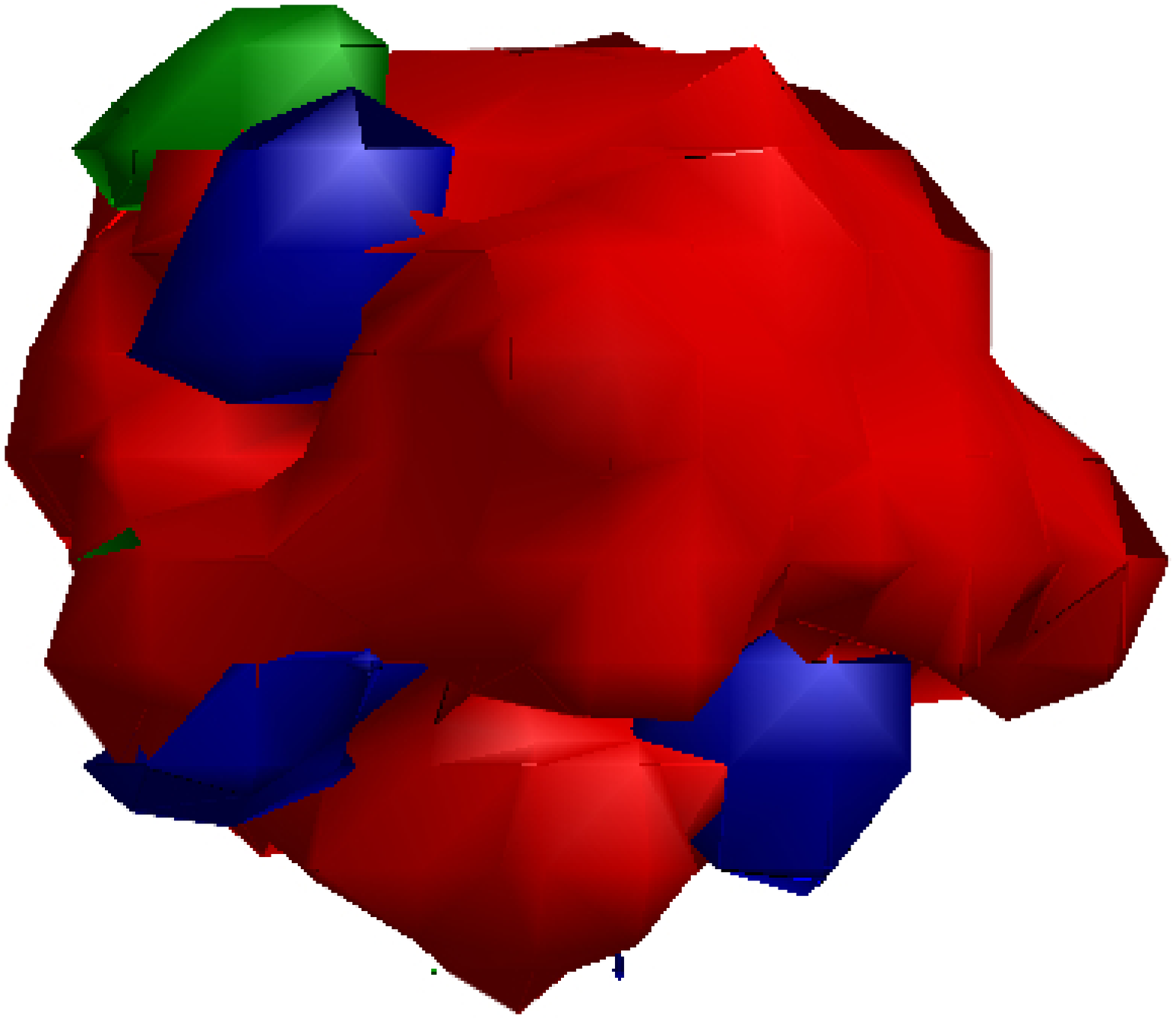}
\end{center}
\end{minipage}
\end{center}

\caption{\label{fig:sim-model} The simulation ground truth at $N = 130$ (top) and $N = 16$ (bottom). (a,b) Two conformations of the 70S ribosome. (c) Their mean volume (red) and difference map (positive in blue, negative in green).}
\end{figure}

\section{Simulation results}
\label{sec:simulation}

We evaluate the performance of the covariance estimation algorithm using simulated data. These are obtained by applying the forward model \eqref{eq:forward} to different configurations of volumes, projection mappings, and noise sources. The resulting images are then given as input to the covariance estimation algorithms outlined in Sections \ref{sec:estimators}--\ref{sec:recon} to study their computational efficiency and accuracy.

\subsection{Covariance estimation results}
\label{sec:sim-covar}

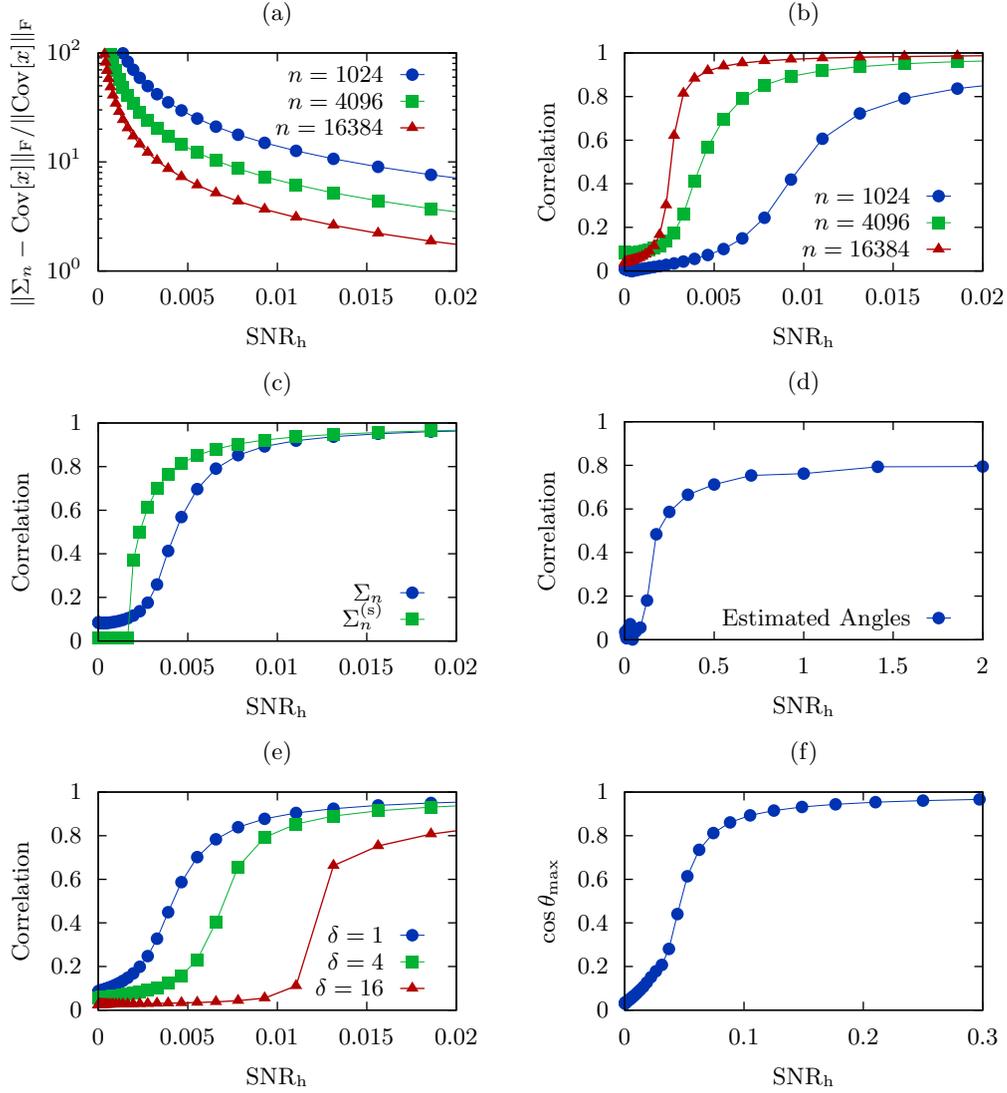
\begin{figure}[h!]
\begin{center}
\begin{minipage}{7cm}
\begin{center}
\include{varying_n_err_gplt}
\end{center}
\end{minipage}
\begin{minipage}{7cm}
\begin{center}
\include{varying_n_corr_gplt}
\end{center}
\end{minipage}

\vspace{-1.2cm}

\begin{minipage}{7cm}
\begin{center}
\include{shrinkage_gplt}
\end{center}
\end{minipage}
\begin{minipage}{7cm}
\begin{center}
\input{est_angles_gplt}
\end{center}
\end{minipage}

\vspace{-0.7cm}

\begin{minipage}{7cm}
\begin{center}
\input{nonuniform_gplt}
\end{center}
\end{minipage}
\begin{minipage}{7cm}
\begin{center}
\input{moreclasses_gplt}
\end{center}
\end{minipage}
\end{center}

\caption{\label{fig:baseline-sim} Covariance estimation results for different simulations with $N = 16$, $C = 2$, seven distinct CTFs, and unless otherwise noted, $n = 4096$ and uniform distribution of orientations. (a) The relative error in $\Covar_n$ as a function of $\SNRh$ for $n = 1024$, $n = 4096$, and $n = 16384$. (b) The correlation of the top eigenvector of $\Sigma_n$ with that of $\var[\vol]$. (c) Top eigenvector correlations for $\Covar_n$ and $\Covar_n^\shrink$. (d) Top eigenvector correlations for $\Covar_n$ with orientations estimated using the ASPIRE toolbox. (e) Top eigenvector correlations for $\Covar_n$ with different orientation distributions over $\SO(3)$ described by \eqref{eq:nonuniform-dist}. (f) The cosine of the maximum principal angle between the top three eigenvectors of $\Covar_n$ and those of $\var[\vol]$ for a simulation with $C = 4$ classes.}
\end{figure}

We first evaluate the performance of the covariance matrix in the absence of noise. We create a synthetic dataset from two 70S ribosome maps downsampled to $N = 16$ shown in the bottom row of Figure \ref{fig:sim-model}. These are projected in viewing directions drawn from a uniform distribution over $\SO(3)$ and convolved with one of seven distinct CTFs. From this we obtain $n = 1024$ simulated images of size $N = 16$. Applying the mean estimation algorithm (Algorithm \ref{algo:mean-estimation}) followed by the covariance estimation algorithm (Algorithm \ref{algo:covar-estimation}), we obtain a covariance matrix estimate $\Covar_n$. Here and throughout this section, we use the unregularized variants of $\mean_n$ and $\covar_n$, setting $\nu_n = 0$ and $\xi_n = 0$.

The relative error of $\Covar_n$ compared to $\var[\vol]$ is $\|\Covar_n-\var[\vol]\|_\frob/\|\var[\vol]\|_\frob \approx 4.4 \cdot 10^{-2}$. Note that the error is not zero due to the finite size of the dataset. Since we have two configurations, $C = 2$ and $\var[\vol]$ has rank one. Consequently, we are only interested in the top eigenvector of $\Covar_n$. If it is well-correlated with the single non-trivial eigenvector $\var[\vol]$, this is another important performance measure. In this case, that correlation is $1-8 \cdot 10^{-5}$. If we redo the experiment for $n = 16384$, the relative error of $\Covar_n$ drops to $2.3 \cdot 10^{-3}$ while the correlation of the top eigenvector increases to $1-2 \cdot 10^{-7}$. For clean data, the proposed method accurately estimates the covariance matrix as $n$ increases.

We now add noise to the above simulation and consider the performance of our algorithm with respect to the signal-to-noise ratio ($\SNR$) of the images. Since the task is to extract the heterogeneity structure of the data, the standard $\SNR$ comparing the average signal power to that of the noise is insufficient. We instead consider the heterogeneous signal-to-noise ratio
\begin{equation}
\label{eq:snr-def}
    \SNRh = \frac{\sum_{s=1}^n \| \imag_s (x_s - \expect[x]) \|^2}{n N^2 \sigma^2}\mathrm{.}
\end{equation}
That is, we center the clean images $\imag_s x_s$ by subtracting the projection of the mean volume $\expect[x]$ and compute the square norm of these coefficients before dividing by the noise power $\sigma^2$. As a result, for a fixed $\sigma^2$, a dataset with low variability will yield a lower $\SNRh$ compared to a dataset with higher variability. This is the same definition used by Katsevich et al. \cite{gene}.

We now consider the simulation described previously at different noise levels with $n = 1024$ images and $N = 16$. The relative error in the Frobenius norm $\|\Sigma_n-\var[\vol]\|_\frob/\|\var[\vol]\|_\frob$ is shown in Figure \ref{fig:baseline-sim}(a) as a function of $\SNRh$ for $n = 1024$, $n = 4096$, and $n = 16384$. This agrees with the guarantee provided by \eqref{eq:covar-convergence} in that for larger $n$, the error goes to zero. Note that this is independent of the noise level. Higher noise simply requires a larger number of images $n$ to achieve a given accuracy.

Although decreasing, the error in $\Covar_n$ is still high for $\SNRh$ range shown in Figure \ref{fig:baseline-sim}(a). As we saw above, however, we are only concerned with the top eigenvector. Figure \ref{fig:baseline-sim}(b) therefore plots the correlation of the top eigenvector of $\covar_n$ with that of $\var[\vol]$ for different values of $\SNRh$ and $n$. Below a certain $\SNRh$, the correlation drops to zero while above a critical threshold the eigenvector correlation approaches one. This behavior is typical of the high-dimensional PCA model discussed in Section \ref{sec:high-dim} and the critical $\SNRh$ threshold can similarly be observed to vary proportionally to the inverse square root of the number of images $n^{-1/2}$. Indeed, this behavior is present in Figure \ref{fig:baseline-sim} as the threshold $\SNRh$ decreases with increasing $n$. Note that this is consistent with the derivations of Section \ref{sec:resolution}, where we found that the achievable resolution $N$ was proportional to $n^{1/2} \sigma^{-2}$ which is proportional to $n^{1/2} \SNRh$. From this, we expect the critical $\SNRh$ to grow faster than $n^{-1/2} N$.

Having established a baseline performance, we study the effect of replacing $B_n$ with its shrinkage variant $B_n^{\mathrm{(s)}}$ in the estimator. Figure \ref{fig:baseline-sim}(c) plots the eigenvector correlation as a function of $\SNRh$ for both $\covar_n$ and $\covar_n^\shrink$. The shrinkage makes a difference when $\SNRh$ is between $0.001$ and $0.01$. Above $0.01$, while the shrinkage provides a good estimate of the covariance, the top eigenvector is already well-correlated even without shrinkage, so there is little difference in performance. Below $0.001$, the signal eigenvalues are absorbed by the noise bulk, so there is no possibility of extracting accurate eigenvectors and both variants perform badly. Between these values, however, shrinkage makes a difference, obtaining an eigenvector correlation of $0.8$ for an $\SNRh$ about $1.4$ times lower than the standard estimator.

We also study the robustness of the estimation algorithm with respect to errors in viewing angle estimation. Running the standard orientation estimation algorithms in the ASPIRE toolbox, which relies on class averaging \cite{zhao2014rotationally} followed by a common-lines based synchronization \cite{shkolnisky2012viewing,singer2010detecting}, we apply the covariance estimation method using the estimated viewing angles. The results are shown in Figure \ref{fig:baseline-sim}(d). We see that for this particular set of molecular structures, we recover the viewing angles with enough accuracy to allow accurate covariance estimation.

So far, the distribution of viewing angles has been uniform over $\SO(3)$, but this is not necessary. To demonstrate robustness of $\Sigma_n$ to non-uniform distributions, we draw the rotations $\rot_s$ from a family of distributions on $\SO(3)$ indexed by a parameter $\delta$ which determines the skew of the distribution towards the identity rotation $\eye_3$. Representing the rotation matrices using Euler angles $(\alpha, \beta, \gamma)$ in the relative z-y-z convention, we consider the distributions
\begin{equation}
\label{eq:nonuniform-dist}
\begin{array}{l}
\alpha \sim U[0, 2\pi] \\
\beta \sim \cos^{-1}(2 U[0, 1]^{\delta}-1) \\
\gamma \sim U[0, 2\pi] \\
\end{array}\mbox{,}
\end{equation}
where $\delta = 1$ is a uniform distribution over $\SO(3)$ and higher values of $\delta$ concentrate the distribution closer to $\eye_3$. The resulting eigenvector correlations are shown in Figure \ref{fig:baseline-sim}(e) as a function of $\SNRh$ for different values of $\delta$. As long as the distribution is not too skewed, we are able to recover the covariance structure accurately at low signal-to-noise ratios.

Increasing the number of classes to $C = 4$, we see that the method handles this type of variability just as well as for two classes. Since $C = 4$, the population covariance is of rank $3$, so instead of evaluating the top eigenvector of $\Covar_n$, we need to consider the top $3$ eigenvectors. A simple correlation will not do, and we instead consider the maximum principal angles between the subspace spanned by the top three eigenvectors of $\Covar_n$ and those of $\var[\vol]$. The cosine $\cos \theta_\max$ of the maximum principal angle $\theta_\max$ between two subspaces $U$ and $V$ is given by
\begin{equation}
    \cos \theta_\max = \mathop{\min}_{u \in U, v \in V} \frac{|\langle u, v \rangle|}{\|u\|\|v\|}
\end{equation}
If $U$ and $V$ are the ranges of two orthogonal matrices $Q_U$ and $Q_V$, $\cos \theta_\max$ is smallest non-zero singular value of $Q_U^\transp Q_V^{}$. This value is plotted as a function of $\SNRh$ in Figure \ref{fig:baseline-sim}(f).

\subsection{Clustering results}
\label{sec:sim-cluster}

\begin{figure}
\begin{center}
\begin{minipage}{7cm}
\begin{center}
\input{clustering_spectrum_gplt.tex}
\end{center}
\end{minipage}
\begin{minipage}{7cm}
\begin{center}
\input{clustering_coords_gplt.tex}
\end{center}
\end{minipage}

\vspace{-0.1cm}

\begin{minipage}{7cm}
\begin{center}
\input{clustering_accuracies_gplt.tex}
\end{center}
\end{minipage}
\begin{minipage}{7cm}
\begin{center}
\input{clustering_errors_gplt.tex}
\end{center}
\end{minipage}
\end{center}
\caption{\label{fig:sim-clustering} Clustering results for discrete variability with $C = 2$ classes imaged using $n = 4096$ images with resolution $N = 16$ for uniform distribution of viewing angles and seven distinct CTFs. (a) The top $32$ eigenvalues of $\Covar_n$ obtained at $\SNRh = 0.01$. (b) A histogram of the coordinates $\coordest_s^{(1)}$ corresponding to the images $\im_s$ for $s = 1, \ldots, n$ subject to the same $\SNRh$. (c) The fraction of images classified correctly as a function of $\SNRh$. (d) The normalized root mean square error (NRMSE) of the reconstructed volumes.}
\end{figure}
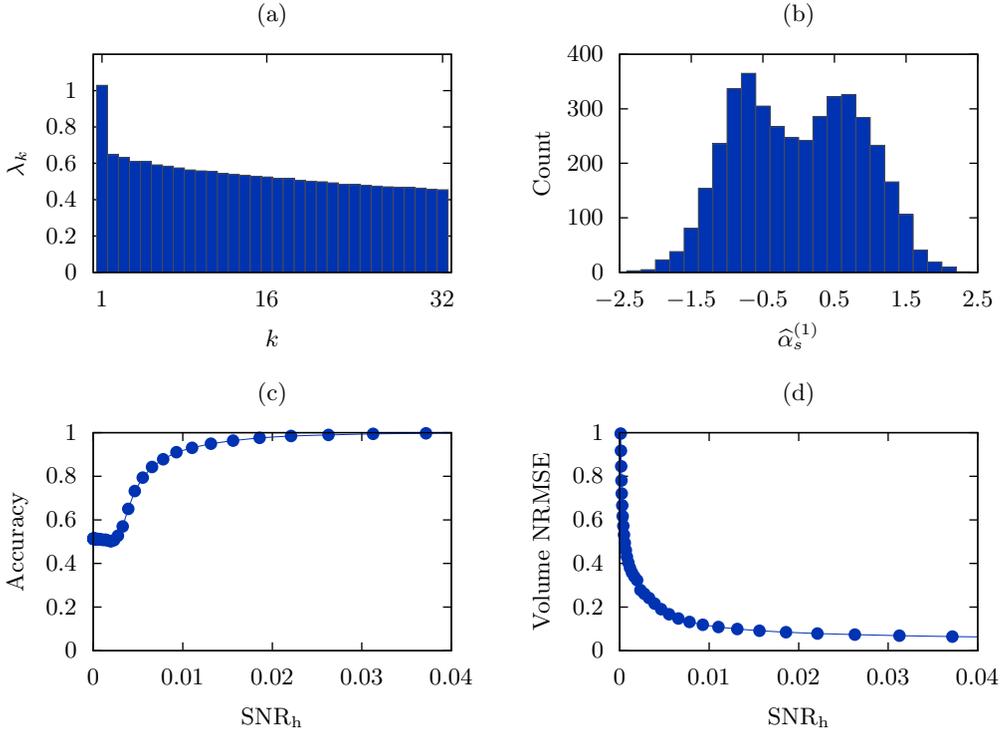

We now consider the clustering and reconstruction steps for the baseline estimator at $\SNRh = 0.01$, where the top eigenvector correlation equals $0.91$. Figure \ref{fig:sim-clustering}(a) shows the spectrum of the covariance estimate $\covar_n$. Since $C = 2$, we expect there to be one dominant eigenvalue since the population covariance is of rank one. Indeed, there is one eigenvalue that stands out from the bulk noise distribution, so we form $\covar_{n,1}$ by extracting the dominant eigenvector and eigenvalue. The Wiener filter described in Section \ref{sec:wiener} gives us a set of scalar coordinate estimates $\{\coordest_1^{(1)}, \ldots, \coordest_n^{(1)}\}$. Their histogram is shown in Figure \ref{fig:sim-clustering}(b). A clear bimodal distribution suggests that we do indeed have two molecular structures present in the data. Clustering the coordinates using $k$-means as described in Section \ref{sec:clustering} and comparing with the ground truth assignments, we achieve $91.8\%$ accuracy.

We now plot the clustering accuracy with respect to $\SNRh$ in Figure \ref{fig:sim-clustering}(c). Similarly, the reconstruction error with respect to $\SNRh$ is plotted in Figure \ref{fig:sim-clustering}(d). As expected, we observe a phase transition phenomenon similar to that of the top eigenvector correlations. Once a certain threshold is passed, we classify well and obtain high-quality reconstructions. Below the threshold, however, the estimated eigenvectors correlate badly with the population eigenvectors so we do not identify the important directions of variability in the molecules. As a result, the subsequent clustering and reconstruction steps fail.

\subsection{Manifold learning results}
\label{sec:sim-manifold}

\begin{figure}[h!]
\begin{center}
{\footnotesize (a)}

\newcommand{\rotatearrow}{%
    \tikz [x=0.25cm,y=0.60cm,line width=0.3ex,-stealth] \draw (0,0) arc (-150:150:1 and 1);%
}

\begin{tikzpicture}
\useasboundingbox (0, 0) rectangle (10, 3.45);

\node at (5.0, 1.4) {\includegraphics[angle=90,origin=c,height=4cm,trim=5cm 2cm 4.5cm 1.5cm,clip]{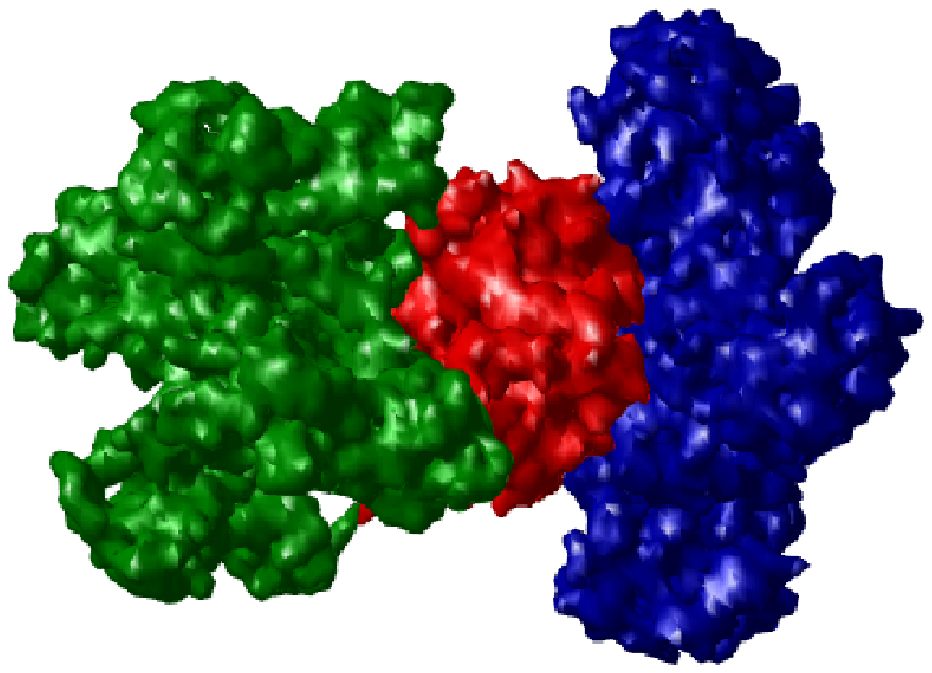}};

\node at (1.5, 1.7) {\rotatearrow};
\node at (2.2, 1.7) {$\theta_1$};
\node at (8.0, 1.7) {\rotatearrow};
\node at (8.7, 1.7) {$\theta_2$};
\end{tikzpicture}
\end{center}

\vspace{-0.6cm}

\begin{center}
\begin{minipage}{7cm}
\input{manifold_corrs_gplt.tex}
\end{minipage}
\begin{minipage}{7cm}
\begin{tikzpicture}
\useasboundingbox (0, 0) rectangle (7, 6);

\node at (3.5, 5.0) {\footnotesize (c)};

\node at (1.75, 3) {\includegraphics[width=3.5cm]{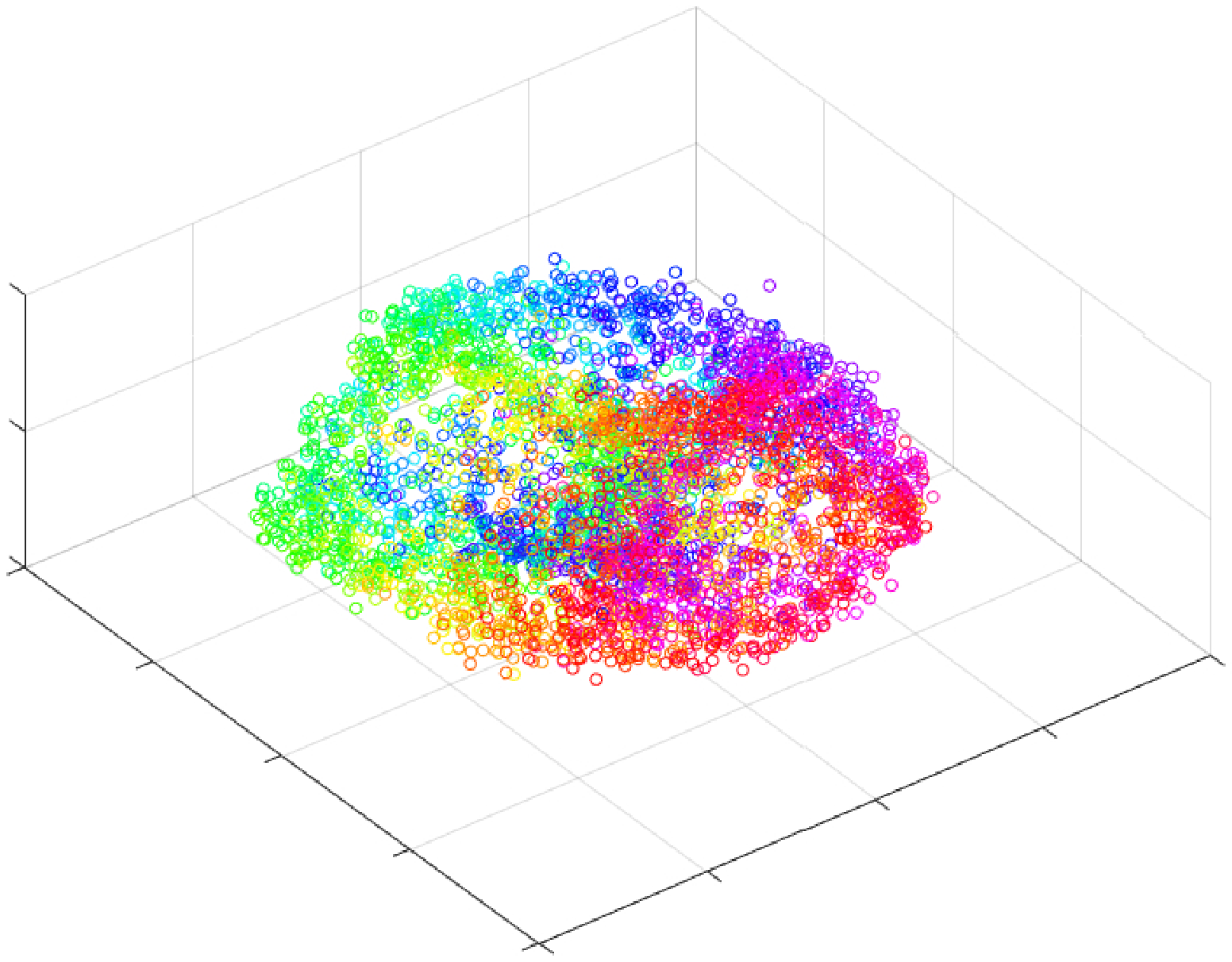}};
\node at (5.25, 3) {\includegraphics[width=3.5cm]{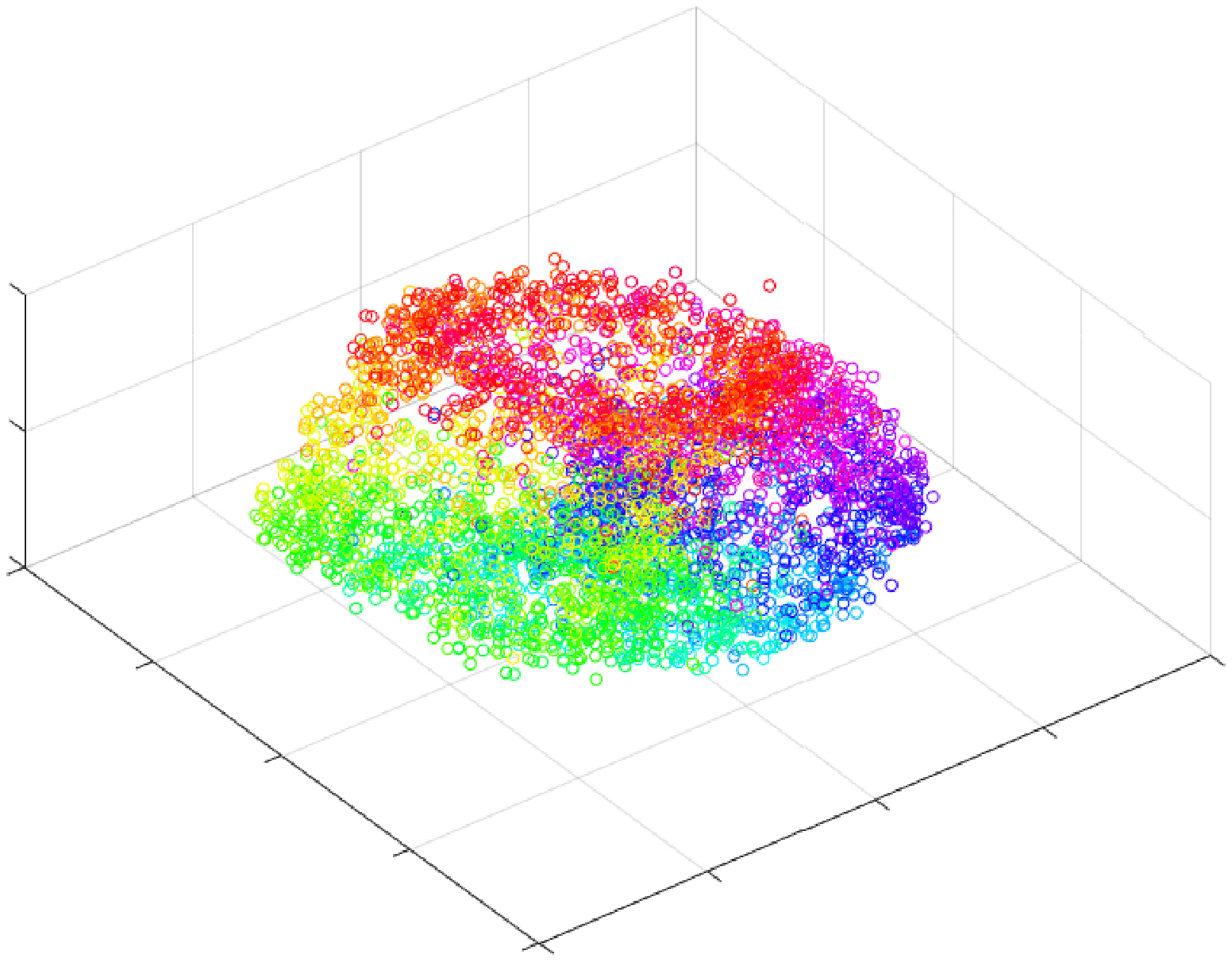}};

\node at (3.50, 1) {\includegraphics[width=3.5cm]{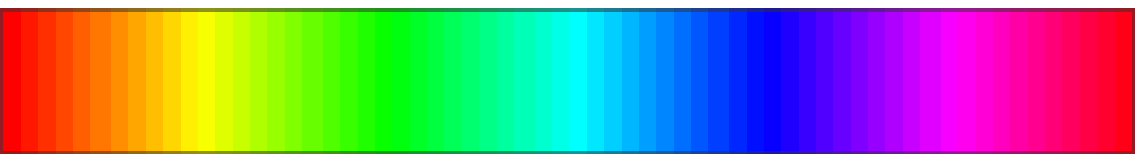}};
\node at (1.25, 1) {\footnotesize $0$};
\node at (6.00, 1) {\footnotesize $2\pi$};
\end{tikzpicture}
\end{minipage}

\vspace{-1cm}

\begin{minipage}{7cm}
\begin{tikzpicture}
\useasboundingbox (0, 0) rectangle (7, 6);

\node at (3.5, 5.0) {\footnotesize (d)};

\node at (3.50, 3) {\includegraphics[width=5cm]{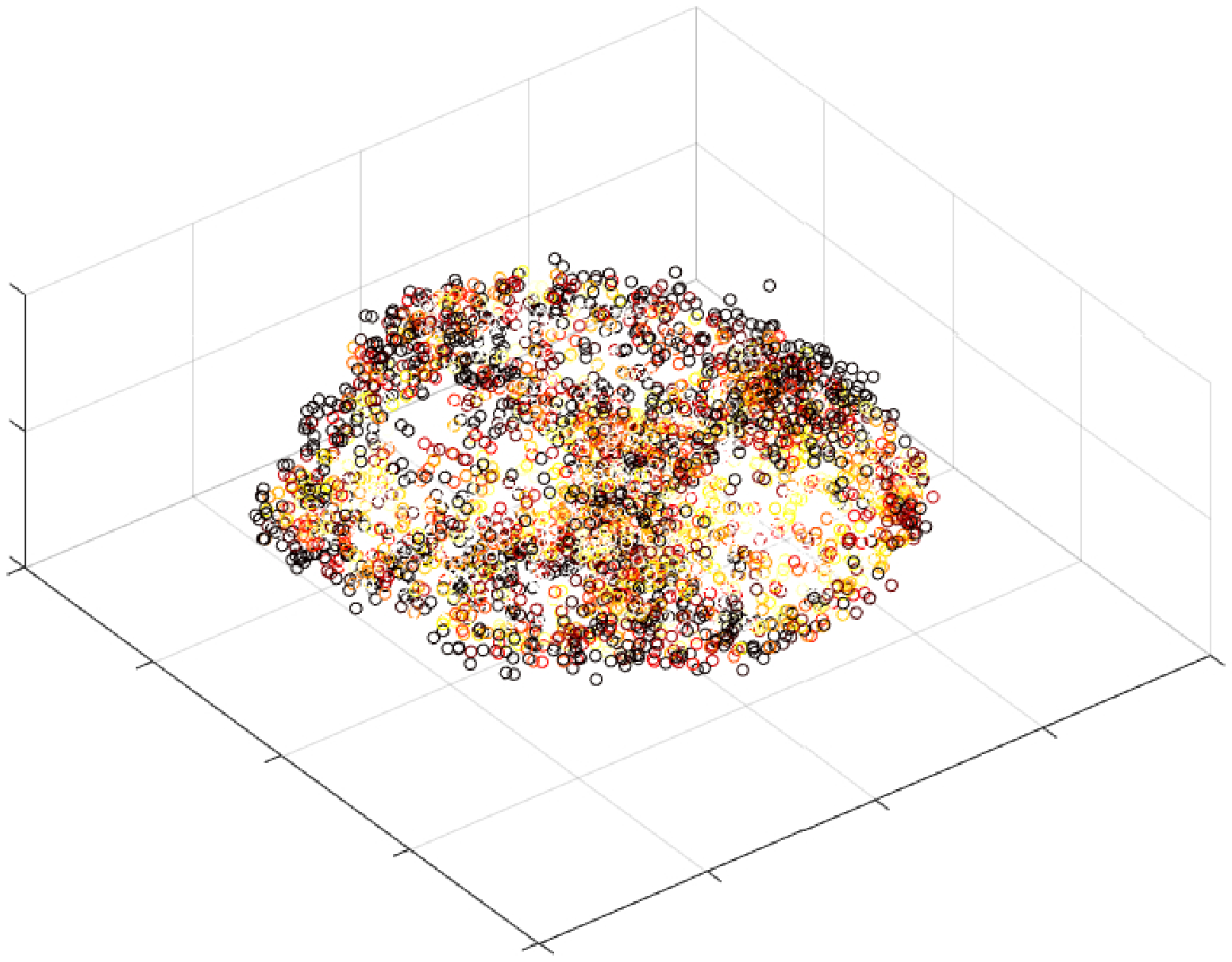}};
\node at (3.50, 1) {\includegraphics[width=3.5cm]{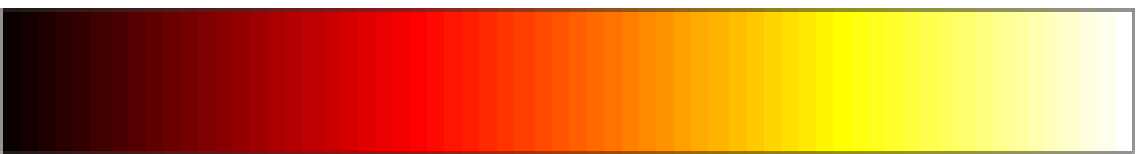}};
\node at (1.25, 1) {\footnotesize 0};
\node at (6.00, 1) {\footnotesize 1};
\end{tikzpicture}
\end{minipage}
\begin{minipage}{7cm}
\input{manifold_coord_errs_gplt.tex}
\end{minipage}
\end{center}
\caption{\label{fig:continuous-sim} Manifold learning results for continuous variability. (a) Volumes are generated by independently rotating two parts (green and blue) by angles $\theta_1$ and $\theta_2$ while keeping the remainder (red) fixed.  (b) The cosine of the maximum principal angle between the top four population eigenvectors and those of $\Covar_n$. (c) Three-dimensional diffusion map embedding coordinates of the volume coordinates $\{\coordest_1, \ldots, \coordest_n\}$, colored according to the first and second rotation angles. (d) The NRMSE of each volume estimate as a function of its diffusion map coordinate. (e) The NRMSE of the reconstructed volumes as a function of $\SNRh$.}
\end{figure}

To simulate continuous variability, we deform a potassium channel molecule by independently rotating two parts by angles $\theta_1$ and $\theta_2$ as shown in Figure \ref{fig:continuous-sim}(a). This yields a two-parameter family of molecular structures. The manifold described by these volumes is the two-dimensional torus, which can be embedded in three dimensions.

The population covariance of the volumes is not strictly low-rank, but $83\%$ of the variance is concentrated in the leading $4$ eigenvectors. If we can recover this eigenspace using our covariance estimation method, we should be able to estimate the manifold structure of the continuous variability. The cosine of the maximum principal angle between the top four population eigenvectors and those of the estimated covariance $\Covar_n$ (again with $\nu_n = \xi_n = 0$) is plotted in Figure \ref{fig:continuous-sim}(b) as a function of $\SNRh$. For an $\SNRh$ above $0.05$ these top eigenvectors are well-estimated, with the cosine of the maximum principal angle in excess of $0.90$.

Fixing the $\SNRh$ at $0.125$, we calculate the coordinates $\{\coordest_1, \ldots, \coordest_n\}$ of the images using the mean estimate $\Mean_n$ and the top four eigenvectors of $\Covar_n$. From these we compute a diffusion map embedding. Figure \ref{fig:continuous-sim}(c) plots first three embedding coordinates: first colored according to one rotation angle, second colored according to the other. The embedding successfully reproduces these angles, indicating that the procedure captures the two-parameter structure quite well. Comparing the estimated volumes $\volest_s$ with the ground truth volumes $\vol_s$ for $s = 1, \ldots, 4096$, we obtain an NRMSE of $0.31$. From this, the recovered range of molecular structure appears to be quite accurate. Plotting the NRMSE as a function of the diffusion map coordinates in Figure \ref{fig:continuous-sim}(d), we see that in areas with high sampling density, the reconstruction is more accurate compared to areas with more sparse sampling. Since the neighborhoods in the sparsely sampled regions have fewer images, this loss of accuracy is expected.

The accuracy of manifold learning reconstruction as a function of $\SNRh$ is shown in Figure \ref{fig:continuous-sim}(e). As in the discrete case, higher $\SNRh$ yields more accurate reconstructions for fixed $n$.

\subsection{Conditioning and convergence results}
\label{sec:sim-cond}

\begin{table}
\caption{\label{tab:cond} The condition numbers obtained for $A_n$, $C_n^{-1} A_n^{}$, $L_n$, and $D_n^{-1} L_n^{}$ defined from projection mappings $\imag_1, \ldots, \imag_n$ where $n = 16384$ or $n = 1024$ and $N = 16$. The projection mappings are obtained from uniform or non-uniform viewing direction distributions with or without CTFs.}
\begin{center}
{\renewcommand{\arraystretch}{1.1}%
\begin{tabular}{|c||c|c|c|c|c|c|}
\hline
$n$ & Distribution of $\rot$ & CTF & $\kappa(A_n)$ & $\kappa(C_n^{-1} A_n^{})$ & $\kappa(L_n)$ & $\kappa(D_n^{-1} L_n^{})$ \\
\hline
\multirow{3}{*}{\rotatebox[origin=c]{90}{$16384$}} & uniform & no & $23$ & $6.6$ & $650$ & $170$ \\
& uniform & yes & $16$ & $12$ & $1400$ & $230$ \\
& non-uniform & yes & $45$ & $17$ & $4800$ & $720$ \\
\hline
\multirow{3}{*}{\rotatebox[origin=c]{90}{$1024$}} & uniform & no & $24$ & $6.5$ & $800$ & $180$ \\
& uniform & yes & $18$ & $12$ & $4600$ & $400$ \\
& non-uniform & yes & $53$ & $17$ & $24000$ & $1800$ \\
\hline
\end{tabular}
}
\end{center}
\end{table}

The conjugate gradient method requires $O(\sqrt{\kappa(Z)})$ iterations to invert the operator $Z$ up to a fixed accuracy, where $\kappa(Z)$ is its condition number \cite{saad2003iterative,axelsson1996iterative,trefethen-bau}. As such, we would like to compare this quantity for the operators $A_n$ and $C_n^{-1} A_n$, as well as $L_n$ and $D_n^{-1} L_n^{}$ in order to evaluate the effect of the strategy outlined in Section \ref{sec:renorm}. In addition, we examine its effect on the convergence rate of the CG method.

To estimate the condition numbers, we generate a dataset with $n = 16384$ projection mappings $\imag_1, \ldots, \imag_n$ with $N = 16$. Setting the regularization parameters $\nu_n$ and $\xi_n$ to zero, we estimate condition numbers in three scenarios. First, we have no CTF (that is $\hh_s = 1$ for all $s = 1, \ldots, n$) and uniform distribution of viewing angles over $\SO(3)$. The second scenario includes three distinct CTFs, but with uniform distribution of viewing directions. Finally, we generate viewing directions using a non-uniform distribution and combine with CTFs. The resulting condition numbers for $A_n$, $C_n^{-1} A_n^{}$, $L_n$, and $D_n^{-1} L_n^{}$ are shown in Table \ref{tab:cond}.

For all operators, adding CTFs and making the viewing direction distribution non-uniform generally worsens the condition number. The exception is adding the CTF for $A_n$, which improves its conditioning slightly. Since the CTF may boost certain high frequencies, such an improvement is not unreasonable. Another feature of these results is that covariance estimation is inherently more ill-conditioned compared to mean estimation for a given resolution $N$ and sample size $n$. This confirms our analysis of Section \ref{sec:resolution} showing that the larger number of unknowns in covariance estimation renders it fundamentally harder than mean estimation.

From the same analysis, we see that $A_n$ requires $n$ to scale as $N$ to maintain well-posedness, while $L_n$ requires $n$ to scale as $N^2$. Consequently, reducing $n$ for fixed $N$ should have a greater impact on the conditioning of $L_n$ compared to $A_n$. This is indeed what we see in Table \ref{tab:cond} when we redo the experiments for $n = 1024$. While the condition numbers of $A_n$ and $C_n^{-1} A_n^{}$ remain unchanged, those of $L_n$ and $D_n^{-1} L_n^{}$ deteriorate significantly.

Finally, we see that the preconditioning gives a decent improvement in the condition number for all operators, even for small $n$. Notably, it reduces $\kappa$ by a factor of $6$ for $L_n$ in the case of non-uniform distribution of viewing directions and with CTF included when $n = 16384$. Since the number of iterations required for convergence of CG scales with $\sqrt{\kappa}$, this should reduce the number of iterations by a factor of at least $2.5$.

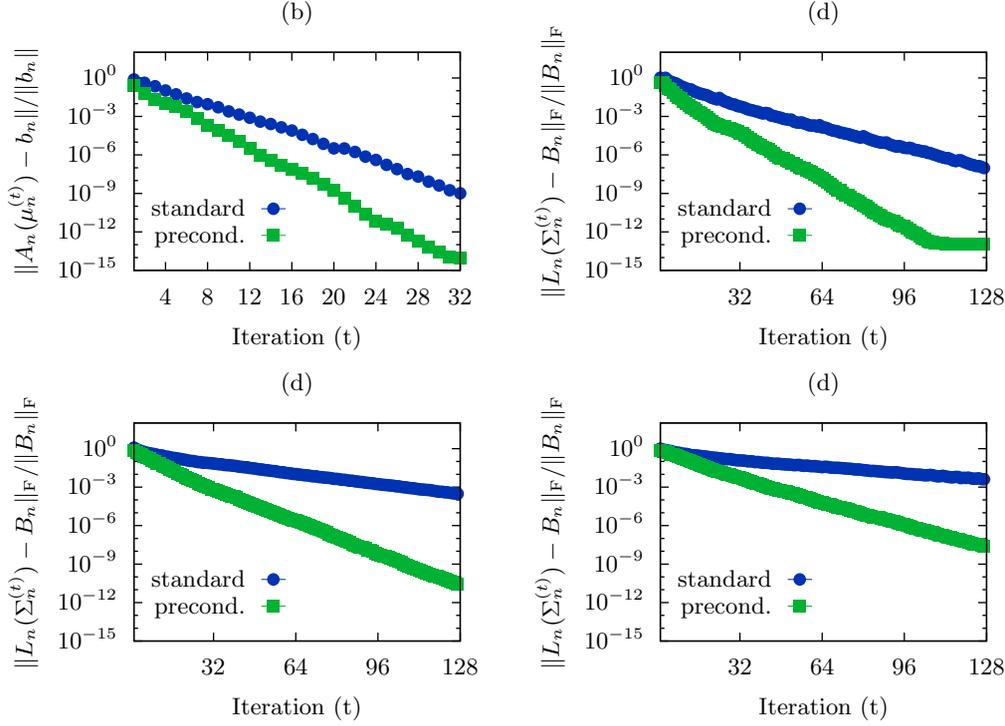
\begin{figure}
\begin{minipage}[t]{7cm}
\begin{center}
\input{An_conv_baseline_gplt}
\end{center}
\end{minipage}
\begin{minipage}[t]{7cm}
\begin{center}
\input{Ln_conv_baseline_gplt}
\end{center}
\end{minipage}

\vspace{-0.2cm}

\begin{minipage}[t]{7cm}
\begin{center}
\input{Ln_conv_ctf_gplt}
\end{center}
\end{minipage}
\begin{minipage}[t]{7cm}
\begin{center}
\input{Ln_conv_ctf_nu_gplt}
\end{center}
\end{minipage}
\caption{\label{fig:sim-conv} (a) The relative residuals for each iteration of CG applied to $A_n \mean_n = b_n$, denoted by $\mean_n^{(t)}$, with no CTF and uniform distribution of viewing angles. (b-d) The relative residuals for each CG iterate $\Covar_n^{(t)}$ of  $L_n(\Covar_n) = B_n$ with (b) no CTF and uniform distribution of viewing angles, (c) three distinct CTFs and uniform distribution of viewing angles, and (d) three distinct CTFs and non-uniform distribution of viewing angles. For all plots, the residuals of the standard (unpreconditioned) CG method is compared with using a circulant preconditioner. All methods were applied to $n = 16384$ images with size $N = 16$ and $\sigma^2 = 1$.}
\end{figure}

We can observe the effects of preconditioning on the convergence rate of the CG method. Here we generate a dataset with $C = 2$ classes by applying the forward model specified by $\imag_1, \ldots, \imag_n$ as generated previously and adding a noise of variance $\sigma^2 = 1$. \ja{Note that the noise variance is kept the same here, so the $\SNR$ changes for each simulation. This shouldn't matter too much in terms of convergence, but we may want to rerun this holding $\SNR$ constant.}

In Figure \ref{fig:sim-conv}(a) we have the CG residuals at different iterations for $A_n \mean_n = b_n$, with and without preconditioning, for the baseline case of no CTF and uniform viewing angles. Preconditioning allows CG to converge much faster, reaching a relative residual of $10^{-3}$ in $8$ iterations instead of $13$, a reduction by a factor of $1.6$. This is close to the factor $1.9$ predicted by examining the condition numbers $\kappa(A_n)$ and $\kappa(C_n^{-1} A_n^{})$.

Similar plots are obtained for CG applied to $L_n(\Covar_n) = B_n$ in Figure \ref{fig:sim-conv}(b), which similarly has no CTF and uniform distribution of viewing angles. Reaching relative residual $10^{-3}$ takes $20$ iterations compared to $47$ for the non-preconditioned algorithm. In Figure \ref{fig:sim-conv}(c), we see that adding CTFs slows convergence for both methods but with the preconditioned algorithm showing a similar improvement. It reduces the number of iterations required to attain a relative residual of $10^{-3}$ from $107$ to $31$, resulting in a speedup of at least $3$ times. Finally, making the distribution of viewing angles non-uniform gives an even worse convergence rate as shown in Figure \ref{fig:sim-conv}(d). However, the preconditioned algorithm only requires $44$ iterations to converge while without preconditioning, more than $128$ iterations are required. These reduction in the number of iterations are consistent with the condition numbers in Table \ref{tab:cond}.

We note that applying $D_n^{-1} L_n^{}$ requires more time than $L_n$. However, $D_n^{-1}$ only involves 6D FFTs of size $N$, while the 6D FFTs in $L_n$ have size $2N$, that is $64$ times larger. The additional time required is therefore expected to be small. Indeed, one preconditioned iteration of CG requires, on average, $15\%$ more time compared to the non-preconditioned iteration.

\subsection{Running time results}
\label{sec:runtime}

\begin{figure}
\begin{center}
\input{running_times_gplt}
\end{center}
\caption{\label{fig:running-time} Running times for the whole covariance estimation algorithm applied to a dataset of size $n = 16384$ with varying image size $N$. Three scenarios are considered: no CTF with uniform distribution of viewing angles, three distinct CTFs with uniform distribution of viewing angles, and three CTFs with non-uniform distribution of viewing angles.}
\end{figure}
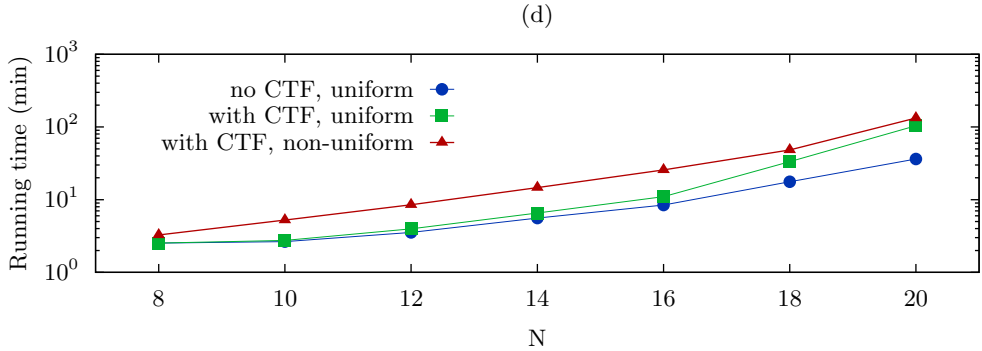

To evaluate the efficiency of our method, we measure its running time on the datasets outlined in the previous section. That is, we have $n = 16384$ images in three different configurations. The first has no CTF and uniform distribution of viewing angles, whereas the second introduces three distinct CTFs into the imaging model. Finally, the third set of experiments adds a non-uniform distribution of viewing directions. The running times on dual $14$-core $2.4~\mathrm{GHz}$ Intel Xeon CPUs are reported in Figure \ref{fig:running-time}.

We see that in the relatively well-conditioned case of uniform distribution of viewing angles for $N = 16$, the algorithm terminates quite quickly. It takes around $5$ minutes when no CTFs are present and $6$ minutes with CTFs. When we no longer have a uniform distribution of viewing angles, we have a running time of $8$ minutes. In all scenarios, the dominant step is the conjugate gradient method, since each iteration requires a 6D FFT of size $2N$ and a large number of iterations can be required when the system is ill-conditioned. That being said, the results compare favorably with other methods for CPU-based heterogeneity, such as RELION, which can take up to $800$ minutes to run on the same dataset.

\section{Experimental results}
\label{sec:experimental}

\begin{figure}
\begin{center}
\begin{minipage}{7cm}
\begin{center}
\input{frank70s_spectrum_gplt.tex}
\end{center}
\end{minipage}
\begin{minipage}{7cm}
\begin{center}
{\footnotesize (b)}

\vspace{0.4cm}

\includegraphics[width=4cm]{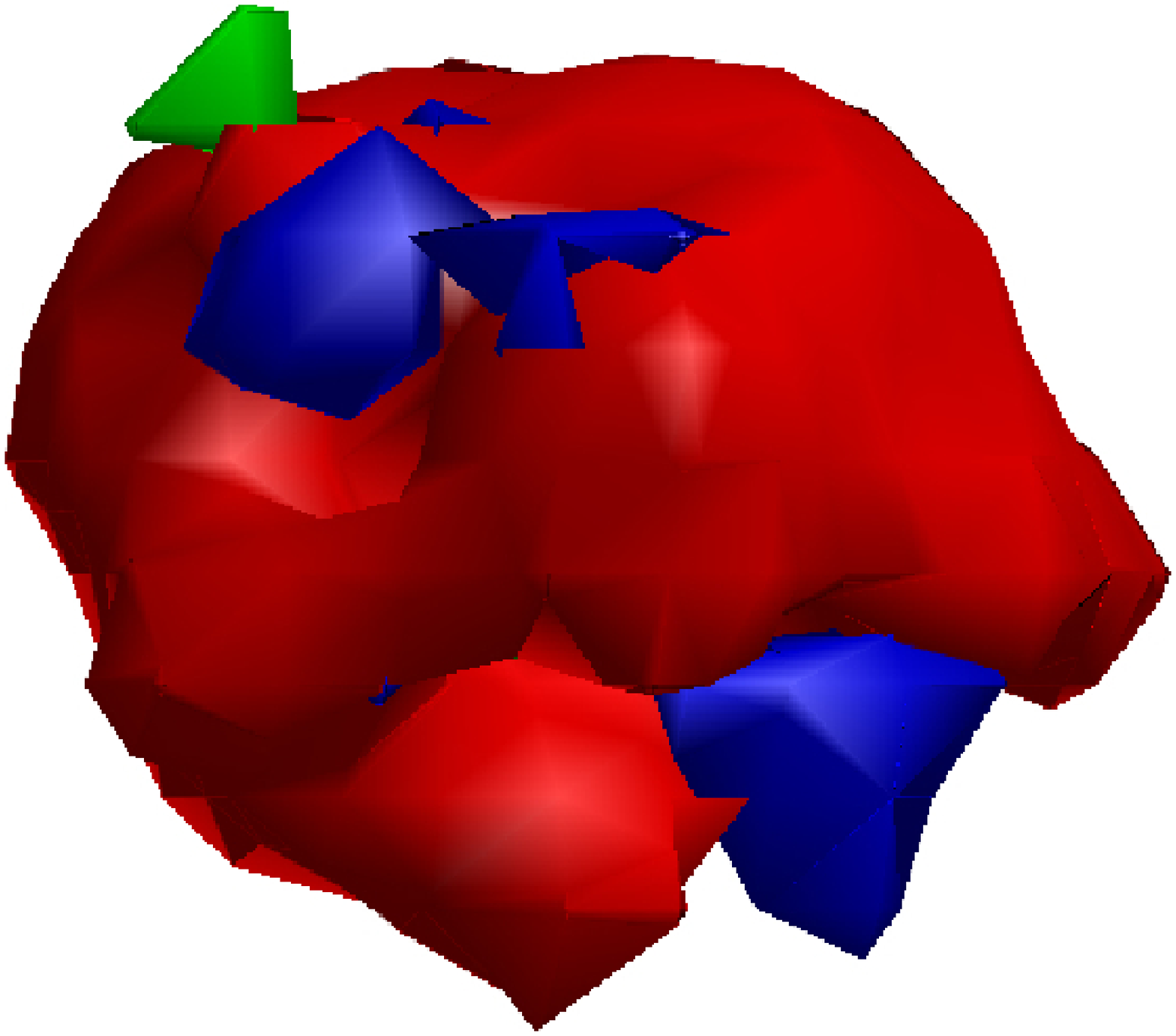}
\end{center}
\end{minipage}

\vspace{-0.2cm}

\begin{minipage}{7cm}
\begin{center}
\input{frank70s_coords_gplt.tex}
\end{center}
\end{minipage}
\begin{minipage}{7cm}
\begin{center}
\input{frank70s_coords_2d_gplt.tex}
\end{center}
\end{minipage}

\vspace{-0.2cm}

\begin{minipage}{7cm}
\begin{center}
\vspace{0.4cm}

{\footnotesize (e)}

\includegraphics[width=5cm,trim=4cm 6cm 4cm 6cm,clip]{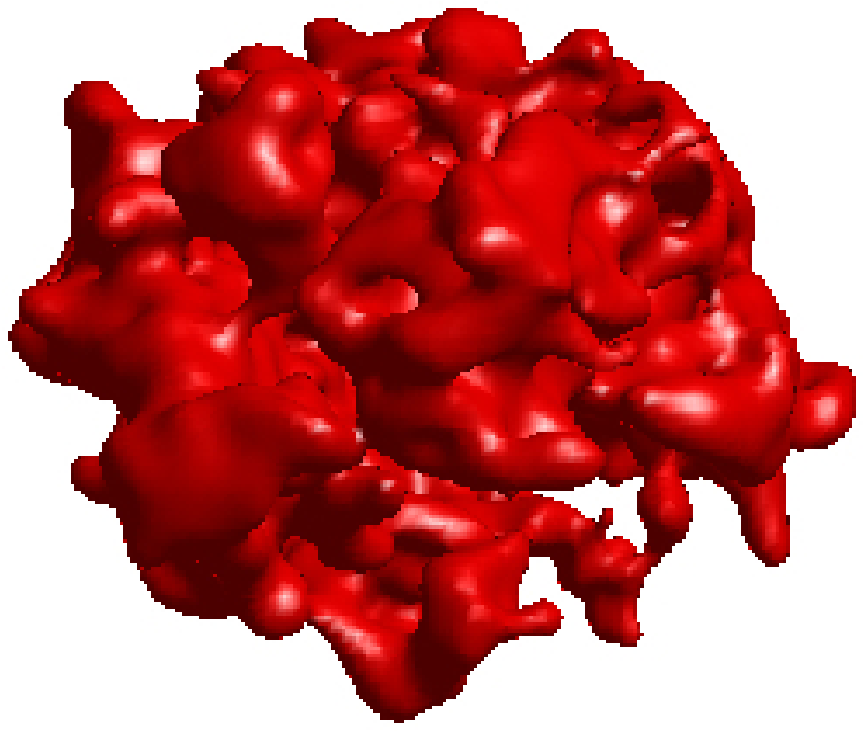}
\end{center}
\end{minipage}
\begin{minipage}{7cm}
\begin{center}
\vspace{0.4cm}

{\footnotesize (f)}

\includegraphics[width=5cm,trim=4cm 6cm 4cm 6cm,clip]{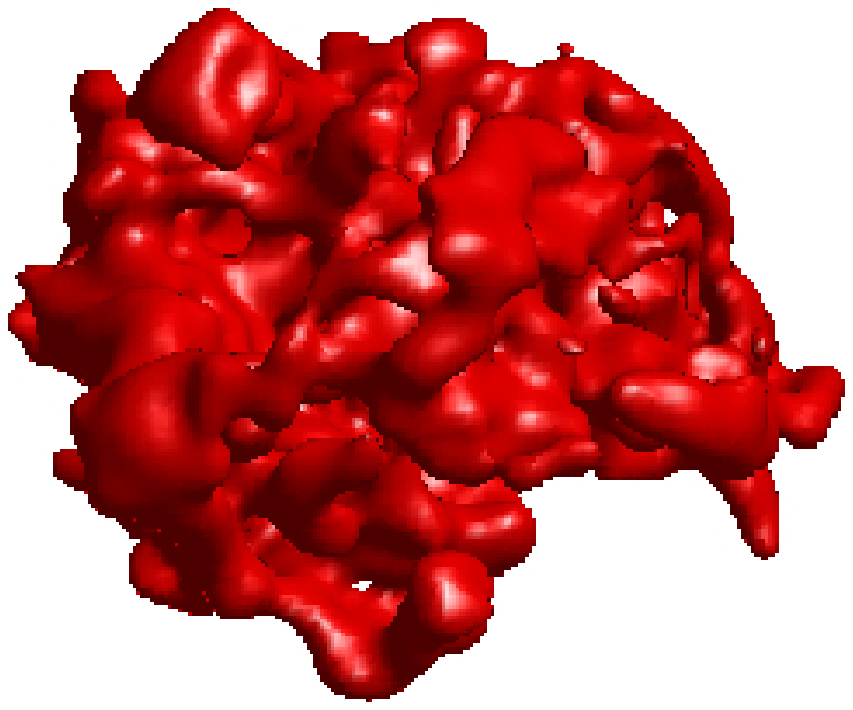}
\end{center}
\end{minipage}
\end{center}
\caption{\label{fig:frank70s-results} Covariance estimation on the 70S ribosome dataset. (a) Largest eigenvalues of the estimated covariance matrix $\Covar_n$. (b) The estimated mean volume (red), together with the positive (blue) and negative (green) components of the top eigenvector. (c) Histogram of coordinates $\{\coordest_1, \ldots, \coordest_n\}$ from the Wiener filter estimator. (d) Two-dimensional histogram of coordinates $\{(\coordest_1^{(1)}, \coordest_1^{(2)}), \ldots, (\coordest_n^{(1)}, \coordest_n^{(2)})\}$ from the Wiener filter estimator. (e,f) Full-resolution reconstructions obtained using RELION applied to the clusters identified in (c).}
\end{figure}

Although simulations are useful to understand the workings of an algorithm, they do not suffice to demonstrate the practical usefulness of a tool. We therefore investigate our covariance estimation approach on an experimental dataset obtained from real cryo-EM samples. This is a standard dataset published by Joachim Frank's lab \cite{frank70s_10k} consisting of projections of a 70S ribosome in two distinct conformations. The dataset comes with a labeling of the images as coming from one of the two states. This labeling was obtained using supervised classification.

For the 70S ribosome dataset, we have $10000$ images of size $130$-by-$130$. We first run the RELION software \cite{scheres-relion} with the number of classes set to one. This provides an estimate of the viewing angles and translations associated with each projection image which are then given as input to our algorithms. To speed up computations, we downsample the images to $16$-by-$16$. Since our goal is to cluster the images, it is not necessary to do this at full resolution if the discriminant features are already present at low resolution. The images are then whitened so that the noise is approximately white with variance $\sigma^2 = 1$. Given the images and the estimates of viewing angles and translations, we apply our mean estimation algorithm (see Algorithm \ref{algo:mean-estimation}). Using the mean estimate $\mean_n$ (for $\nu_n = 0$), we then apply our covariance estimation method (see Algorithm \ref{algo:covar-estimation}) with $\xi_n = 2^{-10}$.

The top eigenvalues of the covariance matrix estimate $\Covar_n$ are shown in Figure \ref{fig:frank70s-results}(a). There is a large eigenvalue well-separated from the rest, suggesting that the variability in molecular structure is at least one-dimensional, which is consistent with a two-class configuration. However, there is also a second eigenvalue of significant amplitude, possibly indicating the presence of a small third class. Figure \ref{fig:frank70s-results}(b) shows the estimated mean volume (red) with the leading eigenvector superimposed (positive part in blue, negative part in green). In the top part of the molecule, we observe the rotation of a small subunit indicated by the negative values on one side and positive values on the other side of that subunit. In the bottom, there is a subunit that attaches and detaches depending on the class.

To investigate this further, we plot the one-dimensional histogram obtained from the first coordinates $\coordest_{s}^{(1)}$ for $s = 1, \ldots, n$, and a two-dimensional heatmap obtained from $(\coordest_{s}^{(1)}, \coordest_{s}^{(2)})$ for $s = 1, \ldots, n$ in Figures \ref{fig:frank70s-results}(c) and \ref{fig:frank70s-results}(d), respectively. The heatmap is obtained by dividing the plane into square boxes and counting the number of points in each box. There does seem to be at least two structures while a third structure is hard to discern. It is therefore likely that this second dimension is due to some continuous variability between the two states.

Clustering the coordinates $\coordest^{(1)}_1, \ldots, \coordest^{(1)}_n$ into two classes and reconstructing from these subsets at full resolution, we obtain the two molecular structures shown in Figures \ref{fig:frank70s-results}(e,f). The two are very similar, with two differences consisting of the rotating subunit at the top and the appearance of the bottom subunit, which agrees the changes observed in the leading eigenvector (see Figure \ref{fig:frank70s-results}(b)). This is in agreement with the presumed structures of the dataset, which consists of a ribosome with and without EF-G. If we compare our clustering with the given annotation of the dataset, we obtain $88.7\%$ accuracy. To compare, the accuracy achieved by RELION on the same dataset with number of classes set to two is $84.6\%$.

The total running time of Algorithms \ref{algo:mean-estimation} and \ref{algo:covar-estimation} was $9$ minutes on a quad-core 3.4 GHz Intel Core i7 CPU. The initial estimation of viewing angles and translations using RELION took $356$ minutes. This compares with running RELION configured with two classes, which had a running time of $520$ minutes. Enabling support for GPU with a GeForce GTX 980 Ti, these running times dropped to $22$ and $28$ minutes for one and two classes, respectively.

\section{Future Work}
\label{sec:future}

The proposed algorithm scales as $N^6 \log N$ in the image size $N$. As a result, doubling the size of the image results in more than a $64$-fold increase in running time and memory usage. This makes estimation for $N$ larger than $20$ prohibitive in most cases. In addition, the large number of unknowns severely limits the achievable resolution as described in Section \ref{sec:resolution}. To resolve these problems, we need to incorporate further structure into the estimation problem. One approach is to better leverage the approximate low rank of the population covariance. Since the number of unknowns drops from $\bigO(N^6)$ to $\bigO(N^3)$, this is a more well-posed problem which could be solved faster and using less memory. This could be done by explicitly fixing the rank of $\Sigma_n$ during estimation. We will also explore related approaches for low-rank matrix recovery such as low-rank matrix sensing via alternating minimization \cite{jain2013low} and direct shrinkage of singular values \cite{dobriban}.

Another drawback of the above approach is that it requires the viewing directions and translations to be known in advance. While this may be possible for small, localized heterogeneity, where homogeneous reconstruction methods can provide reasonable estimates, this is not always feasible. In that case, methods which combine heterogeneous reconstruction and parameter estimation enjoy a significant advantage \cite{scheres,brubaker,frealign,roy,lederman2017continuously}. Extending the proposed method for this more general setting provides another avenue for future research.

Finally, more work is needed to process the coordinate estimates $\coordest_1, \ldots, \coordest_n$. While standard clustering and manifold learning approaches outlined in this paper provide reasonable results, they do not perform well at high noise levels. A subject of further investigation is therefore how to incorporate informative priors on the space of volumes, such as variability caused by deformation, which should prove useful in this regime.

\section{Conclusion}
\label{sec:conclusion}

We have introduced a computationally efficient method for least-squares estimation of the 3D covariance matrix of the molecular structure from noisy 2D projection images. Given $n$ images of size $N$-by-$N$, it has computational complexity $\bigO(\sqrt{\kappa} N^6\log N + n N^4)$, where $\kappa$ is in the range $1$--$200$ for typical problems. This is achieved by reformulating the normal equations as a deconvolution problem solved using the preconditioned conjugate gradient method. We also introduced a shrinkage variant which improves accuracy at low signal-to-noise ratios, decreasing by a factor of $1.4$ the necessary signal power for accurate estimation. The estimated covariance matrices are then used to reconstruct the three-dimensional structures a Wiener filter, which are then clustered into a number of discrete states or fitted to a continuous manifold structure. The accurate performance of both methods is confirmed through experiments on simulated and experimental datasets.

\section{Acknowledgments}
\label{sec:acknowledgments}

The authors would like to thank Fred Sigworth and Joachim Frank for invaluable discussions regarding the heterogeneity problem and single-particle reconstruction more generally. They are also very grateful to the reviewers for their valuable comments. Initial results on manifold learning for continuous heterogeneity were obtained by Hugh Wilson. This work was performed while the first author was a postdoctoral research associate in the Program in Applied and Computational Mathematics at Princeton University.

\bibliographystyle{siamplain}
\bibliography{refs}

\end{document}

%% file: mp_gplt.tex
% GNUPLOT: LaTeX picture with Postscript
\begingroup
\newcommand{\fnt}[0]{\footnotesize}
  \makeatletter
  \providecommand\color[2][]{%
    \GenericError{(gnuplot) \space\space\space\@spaces}{%
      Package color not loaded in conjunction with
      terminal option `colourtext'%
    }{See the gnuplot documentation for explanation.%
    }{Either use 'blacktext' in gnuplot or load the package
      color.sty in LaTeX.}%
    \renewcommand\color[2][]{}%
  }%
  \providecommand\includegraphics[2][]{%
    \GenericError{(gnuplot) \space\space\space\@spaces}{%
      Package graphicx or graphics not loaded%
    }{See the gnuplot documentation for explanation.%
    }{The gnuplot epslatex terminal needs graphicx.sty or graphics.sty.}%
    \renewcommand\includegraphics[2][]{}%
  }%
  \providecommand\rotatebox[2]{#2}%
  \@ifundefined{ifGPcolor}{%
    \newif\ifGPcolor
    \GPcolortrue
  }{}%
  \@ifundefined{ifGPblacktext}{%
    \newif\ifGPblacktext
    \GPblacktextfalse
  }{}%
  % define a \g@addto@macro without @ in the name:
  \let\gplgaddtomacro\g@addto@macro
  % define empty templates for all commands taking text:
  \gdef\gplbacktext{}%
  \gdef\gplfronttext{}%
  \makeatother
  \ifGPblacktext
    % no textcolor at all
    \def\colorrgb#1{}%
    \def\colorgray#1{}%
  \else
    % gray or color?
    \ifGPcolor
      \def\colorrgb#1{\color[rgb]{#1}}%
      \def\colorgray#1{\color[gray]{#1}}%
      \expandafter\def\csname LTw\endcsname{\color{white}}%
      \expandafter\def\csname LTb\endcsname{\color{black}}%
      \expandafter\def\csname LTa\endcsname{\color{black}}%
      \expandafter\def\csname LT0\endcsname{\color[rgb]{1,0,0}}%
      \expandafter\def\csname LT1\endcsname{\color[rgb]{0,1,0}}%
      \expandafter\def\csname LT2\endcsname{\color[rgb]{0,0,1}}%
      \expandafter\def\csname LT3\endcsname{\color[rgb]{1,0,1}}%
      \expandafter\def\csname LT4\endcsname{\color[rgb]{0,1,1}}%
      \expandafter\def\csname LT5\endcsname{\color[rgb]{1,1,0}}%
      \expandafter\def\csname LT6\endcsname{\color[rgb]{0,0,0}}%
      \expandafter\def\csname LT7\endcsname{\color[rgb]{1,0.3,0}}%
      \expandafter\def\csname LT8\endcsname{\color[rgb]{0.5,0.5,0.5}}%
    \else
      % gray
      \def\colorrgb#1{\color{black}}%
      \def\colorgray#1{\color[gray]{#1}}%
      \expandafter\def\csname LTw\endcsname{\color{white}}%
      \expandafter\def\csname LTb\endcsname{\color{black}}%
      \expandafter\def\csname LTa\endcsname{\color{black}}%
      \expandafter\def\csname LT0\endcsname{\color{black}}%
      \expandafter\def\csname LT1\endcsname{\color{black}}%
      \expandafter\def\csname LT2\endcsname{\color{black}}%
      \expandafter\def\csname LT3\endcsname{\color{black}}%
      \expandafter\def\csname LT4\endcsname{\color{black}}%
      \expandafter\def\csname LT5\endcsname{\color{black}}%
      \expandafter\def\csname LT6\endcsname{\color{black}}%
      \expandafter\def\csname LT7\endcsname{\color{black}}%
      \expandafter\def\csname LT8\endcsname{\color{black}}%
    \fi
  \fi
  \setlength{\unitlength}{0.0500bp}%
  \begin{picture}(3960.00,2880.00)%
    \gplgaddtomacro\gplbacktext{%
      \csname LTb\endcsname%
      \put(660,640){\makebox(0,0)[r]{\strut{}\fnt $0$}}%
      \put(660,1186){\makebox(0,0)[r]{\strut{}\fnt $20$}}%
      \put(660,1733){\makebox(0,0)[r]{\strut{}\fnt $40$}}%
      \put(660,2279){\makebox(0,0)[r]{\strut{}\fnt $60$}}%
      \put(1183,440){\makebox(0,0){\strut{}\fnt $0$}}%
      \put(1585,440){\makebox(0,0){\strut{}\fnt $1$}}%
      \put(1988,440){\makebox(0,0){\strut{}\fnt $2$}}%
      \put(2391,440){\makebox(0,0){\strut{}\fnt $3$}}%
      \put(2794,440){\makebox(0,0){\strut{}\fnt $4$}}%
      \put(3196,440){\makebox(0,0){\strut{}\fnt $5$}}%
      \put(320,1459){\rotatebox{-270}{\makebox(0,0){\strut{}\fnt Count}}}%
      \put(2189,140){\makebox(0,0){\strut{}\fnt $\lambda$}}%
      \put(2189,2579){\makebox(0,0){\strut{}\fnt (a)}}%
    }%
    \gplgaddtomacro\gplfronttext{%
    }%
    \gplbacktext
    \put(0,0){\includegraphics{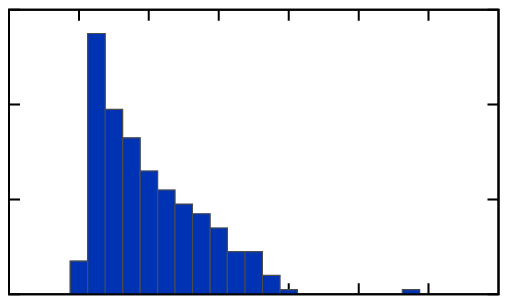}}%
    \gplfronttext
  \end{picture}%
\endgroup

%% file: spiked_cov_gplt.tex
% GNUPLOT: LaTeX picture with Postscript
\begingroup
\newcommand{\fnt}[0]{\footnotesize}
  \makeatletter
  \providecommand\color[2][]{%
    \GenericError{(gnuplot) \space\space\space\@spaces}{%
      Package color not loaded in conjunction with
      terminal option `colourtext'%
    }{See the gnuplot documentation for explanation.%
    }{Either use 'blacktext' in gnuplot or load the package
      color.sty in LaTeX.}%
    \renewcommand\color[2][]{}%
  }%
  \providecommand\includegraphics[2][]{%
    \GenericError{(gnuplot) \space\space\space\@spaces}{%
      Package graphicx or graphics not loaded%
    }{See the gnuplot documentation for explanation.%
    }{The gnuplot epslatex terminal needs graphicx.sty or graphics.sty.}%
    \renewcommand\includegraphics[2][]{}%
  }%
  \providecommand\rotatebox[2]{#2}%
  \@ifundefined{ifGPcolor}{%
    \newif\ifGPcolor
    \GPcolortrue
  }{}%
  \@ifundefined{ifGPblacktext}{%
    \newif\ifGPblacktext
    \GPblacktextfalse
  }{}%
  % define a \g@addto@macro without @ in the name:
  \let\gplgaddtomacro\g@addto@macro
  % define empty templates for all commands taking text:
  \gdef\gplbacktext{}%
  \gdef\gplfronttext{}%
  \makeatother
  \ifGPblacktext
    % no textcolor at all
    \def\colorrgb#1{}%
    \def\colorgray#1{}%
  \else
    % gray or color?
    \ifGPcolor
      \def\colorrgb#1{\color[rgb]{#1}}%
      \def\colorgray#1{\color[gray]{#1}}%
      \expandafter\def\csname LTw\endcsname{\color{white}}%
      \expandafter\def\csname LTb\endcsname{\color{black}}%
      \expandafter\def\csname LTa\endcsname{\color{black}}%
      \expandafter\def\csname LT0\endcsname{\color[rgb]{1,0,0}}%
      \expandafter\def\csname LT1\endcsname{\color[rgb]{0,1,0}}%
      \expandafter\def\csname LT2\endcsname{\color[rgb]{0,0,1}}%
      \expandafter\def\csname LT3\endcsname{\color[rgb]{1,0,1}}%
      \expandafter\def\csname LT4\endcsname{\color[rgb]{0,1,1}}%
      \expandafter\def\csname LT5\endcsname{\color[rgb]{1,1,0}}%
      \expandafter\def\csname LT6\endcsname{\color[rgb]{0,0,0}}%
      \expandafter\def\csname LT7\endcsname{\color[rgb]{1,0.3,0}}%
      \expandafter\def\csname LT8\endcsname{\color[rgb]{0.5,0.5,0.5}}%
    \else
      % gray
      \def\colorrgb#1{\color{black}}%
      \def\colorgray#1{\color[gray]{#1}}%
      \expandafter\def\csname LTw\endcsname{\color{white}}%
      \expandafter\def\csname LTb\endcsname{\color{black}}%
      \expandafter\def\csname LTa\endcsname{\color{black}}%
      \expandafter\def\csname LT0\endcsname{\color{black}}%
      \expandafter\def\csname LT1\endcsname{\color{black}}%
      \expandafter\def\csname LT2\endcsname{\color{black}}%
      \expandafter\def\csname LT3\endcsname{\color{black}}%
      \expandafter\def\csname LT4\endcsname{\color{black}}%
      \expandafter\def\csname LT5\endcsname{\color{black}}%
      \expandafter\def\csname LT6\endcsname{\color{black}}%
      \expandafter\def\csname LT7\endcsname{\color{black}}%
      \expandafter\def\csname LT8\endcsname{\color{black}}%
    \fi
  \fi
  \setlength{\unitlength}{0.0500bp}%
  \begin{picture}(3960.00,2880.00)%
    \gplgaddtomacro\gplbacktext{%
      \csname LTb\endcsname%
      \put(660,640){\makebox(0,0)[r]{\strut{}\fnt $0$}}%
      \put(660,1186){\makebox(0,0)[r]{\strut{}\fnt $20$}}%
      \put(660,1733){\makebox(0,0)[r]{\strut{}\fnt $40$}}%
      \put(660,2279){\makebox(0,0)[r]{\strut{}\fnt $60$}}%
      \put(1183,440){\makebox(0,0){\strut{}\fnt $0$}}%
      \put(1585,440){\makebox(0,0){\strut{}\fnt $1$}}%
      \put(1988,440){\makebox(0,0){\strut{}\fnt $2$}}%
      \put(2391,440){\makebox(0,0){\strut{}\fnt $3$}}%
      \put(2794,440){\makebox(0,0){\strut{}\fnt $4$}}%
      \put(3196,440){\makebox(0,0){\strut{}\fnt $5$}}%
      \put(320,1459){\rotatebox{-270}{\makebox(0,0){\strut{}\fnt Count}}}%
      \put(2189,140){\makebox(0,0){\strut{}\fnt $\lambda$}}%
      \put(2189,2579){\makebox(0,0){\strut{}\fnt (b)}}%
    }%
    \gplgaddtomacro\gplfronttext{%
    }%
    \gplbacktext
    \put(0,0){\includegraphics{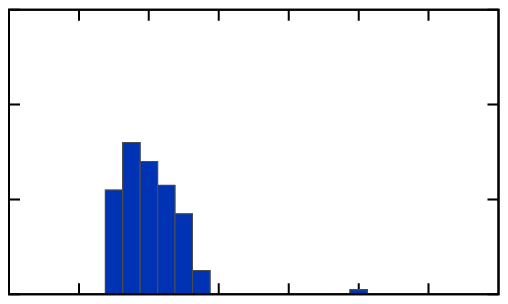}}%
    \gplfronttext
  \end{picture}%
\endgroup

%% file: shrinkage_Bn_gplt.tex
% GNUPLOT: LaTeX picture with Postscript
\begingroup
\newcommand{\fnt}[0]{\footnotesize}
  \makeatletter
  \providecommand\color[2][]{%
    \GenericError{(gnuplot) \space\space\space\@spaces}{%
      Package color not loaded in conjunction with
      terminal option `colourtext'%
    }{See the gnuplot documentation for explanation.%
    }{Either use 'blacktext' in gnuplot or load the package
      color.sty in LaTeX.}%
    \renewcommand\color[2][]{}%
  }%
  \providecommand\includegraphics[2][]{%
    \GenericError{(gnuplot) \space\space\space\@spaces}{%
      Package graphicx or graphics not loaded%
    }{See the gnuplot documentation for explanation.%
    }{The gnuplot epslatex terminal needs graphicx.sty or graphics.sty.}%
    \renewcommand\includegraphics[2][]{}%
  }%
  \providecommand\rotatebox[2]{#2}%
  \@ifundefined{ifGPcolor}{%
    \newif\ifGPcolor
    \GPcolortrue
  }{}%
  \@ifundefined{ifGPblacktext}{%
    \newif\ifGPblacktext
    \GPblacktextfalse
  }{}%
  % define a \g@addto@macro without @ in the name:
  \let\gplgaddtomacro\g@addto@macro
  % define empty templates for all commands taking text:
  \gdef\gplbacktext{}%
  \gdef\gplfronttext{}%
  \makeatother
  \ifGPblacktext
    % no textcolor at all
    \def\colorrgb#1{}%
    \def\colorgray#1{}%
  \else
    % gray or color?
    \ifGPcolor
      \def\colorrgb#1{\color[rgb]{#1}}%
      \def\colorgray#1{\color[gray]{#1}}%
      \expandafter\def\csname LTw\endcsname{\color{white}}%
      \expandafter\def\csname LTb\endcsname{\color{black}}%
      \expandafter\def\csname LTa\endcsname{\color{black}}%
      \expandafter\def\csname LT0\endcsname{\color[rgb]{1,0,0}}%
      \expandafter\def\csname LT1\endcsname{\color[rgb]{0,1,0}}%
      \expandafter\def\csname LT2\endcsname{\color[rgb]{0,0,1}}%
      \expandafter\def\csname LT3\endcsname{\color[rgb]{1,0,1}}%
      \expandafter\def\csname LT4\endcsname{\color[rgb]{0,1,1}}%
      \expandafter\def\csname LT5\endcsname{\color[rgb]{1,1,0}}%
      \expandafter\def\csname LT6\endcsname{\color[rgb]{0,0,0}}%
      \expandafter\def\csname LT7\endcsname{\color[rgb]{1,0.3,0}}%
      \expandafter\def\csname LT8\endcsname{\color[rgb]{0.5,0.5,0.5}}%
    \else
      % gray
      \def\colorrgb#1{\color{black}}%
      \def\colorgray#1{\color[gray]{#1}}%
      \expandafter\def\csname LTw\endcsname{\color{white}}%
      \expandafter\def\csname LTb\endcsname{\color{black}}%
      \expandafter\def\csname LTa\endcsname{\color{black}}%
      \expandafter\def\csname LT0\endcsname{\color{black}}%
      \expandafter\def\csname LT1\endcsname{\color{black}}%
      \expandafter\def\csname LT2\endcsname{\color{black}}%
      \expandafter\def\csname LT3\endcsname{\color{black}}%
      \expandafter\def\csname LT4\endcsname{\color{black}}%
      \expandafter\def\csname LT5\endcsname{\color{black}}%
      \expandafter\def\csname LT6\endcsname{\color{black}}%
      \expandafter\def\csname LT7\endcsname{\color{black}}%
      \expandafter\def\csname LT8\endcsname{\color{black}}%
    \fi
  \fi
  \setlength{\unitlength}{0.0500bp}%
  \begin{picture}(3960.00,2880.00)%
    \gplgaddtomacro\gplbacktext{%
      \csname LTb\endcsname%
      \put(900,640){\makebox(0,0)[r]{\strut{}\fnt $10^{-1}$}}%
      \put(900,968){\makebox(0,0)[r]{\strut{}\fnt $10^{0}$}}%
      \put(900,1296){\makebox(0,0)[r]{\strut{}\fnt $10^{1}$}}%
      \put(900,1623){\makebox(0,0)[r]{\strut{}\fnt $10^{2}$}}%
      \put(900,1951){\makebox(0,0)[r]{\strut{}\fnt $10^{3}$}}%
      \put(900,2279){\makebox(0,0)[r]{\strut{}\fnt $10^{4}$}}%
      \put(1020,440){\makebox(0,0){\strut{}\fnt $10^{3}$}}%
      \put(1880,440){\makebox(0,0){\strut{}\fnt $10^{4}$}}%
      \put(2739,440){\makebox(0,0){\strut{}\fnt $10^{5}$}}%
      \put(3599,440){\makebox(0,0){\strut{}\fnt $10^{6}$}}%
      \put(320,1459){\rotatebox{-270}{\makebox(0,0){\strut{}\fnt $\|B_n-B_n^{\mathrm{(c)}}\|_\frob^2/\|B_n^{\mathrm{(c)}}\|_\frob^2$}}}%
      \put(2309,140){\makebox(0,0){\strut{}\fnt $n$}}%
      \put(2309,2579){\makebox(0,0){\strut{}\fnt (a)}}%
    }%
    \gplgaddtomacro\gplfronttext{%
      \csname LTb\endcsname%
      \put(3056,2116){\makebox(0,0)[r]{\strut{}\fnt $B_n$}}%
      \csname LTb\endcsname%
      \put(3056,1916){\makebox(0,0)[r]{\strut{}\fnt $B_n^{\mathrm{(s)}}$}}%
    }%
    \gplbacktext
    \put(0,0){\includegraphics{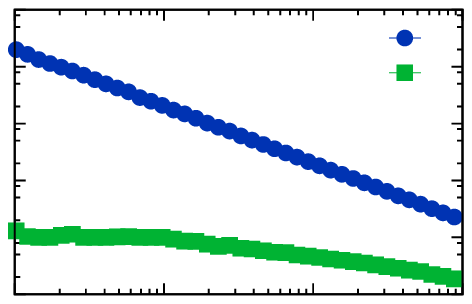}}%
    \gplfronttext
  \end{picture}%
\endgroup

%% file: shrinkage_Sigman_gplt.tex
% GNUPLOT: LaTeX picture with Postscript
\begingroup
\newcommand{\fnt}[0]{\footnotesize}
  \makeatletter
  \providecommand\color[2][]{%
    \GenericError{(gnuplot) \space\space\space\@spaces}{%
      Package color not loaded in conjunction with
      terminal option `colourtext'%
    }{See the gnuplot documentation for explanation.%
    }{Either use 'blacktext' in gnuplot or load the package
      color.sty in LaTeX.}%
    \renewcommand\color[2][]{}%
  }%
  \providecommand\includegraphics[2][]{%
    \GenericError{(gnuplot) \space\space\space\@spaces}{%
      Package graphicx or graphics not loaded%
    }{See the gnuplot documentation for explanation.%
    }{The gnuplot epslatex terminal needs graphicx.sty or graphics.sty.}%
    \renewcommand\includegraphics[2][]{}%
  }%
  \providecommand\rotatebox[2]{#2}%
  \@ifundefined{ifGPcolor}{%
    \newif\ifGPcolor
    \GPcolortrue
  }{}%
  \@ifundefined{ifGPblacktext}{%
    \newif\ifGPblacktext
    \GPblacktextfalse
  }{}%
  % define a \g@addto@macro without @ in the name:
  \let\gplgaddtomacro\g@addto@macro
  % define empty templates for all commands taking text:
  \gdef\gplbacktext{}%
  \gdef\gplfronttext{}%
  \makeatother
  \ifGPblacktext
    % no textcolor at all
    \def\colorrgb#1{}%
    \def\colorgray#1{}%
  \else
    % gray or color?
    \ifGPcolor
      \def\colorrgb#1{\color[rgb]{#1}}%
      \def\colorgray#1{\color[gray]{#1}}%
      \expandafter\def\csname LTw\endcsname{\color{white}}%
      \expandafter\def\csname LTb\endcsname{\color{black}}%
      \expandafter\def\csname LTa\endcsname{\color{black}}%
      \expandafter\def\csname LT0\endcsname{\color[rgb]{1,0,0}}%
      \expandafter\def\csname LT1\endcsname{\color[rgb]{0,1,0}}%
      \expandafter\def\csname LT2\endcsname{\color[rgb]{0,0,1}}%
      \expandafter\def\csname LT3\endcsname{\color[rgb]{1,0,1}}%
      \expandafter\def\csname LT4\endcsname{\color[rgb]{0,1,1}}%
      \expandafter\def\csname LT5\endcsname{\color[rgb]{1,1,0}}%
      \expandafter\def\csname LT6\endcsname{\color[rgb]{0,0,0}}%
      \expandafter\def\csname LT7\endcsname{\color[rgb]{1,0.3,0}}%
      \expandafter\def\csname LT8\endcsname{\color[rgb]{0.5,0.5,0.5}}%
    \else
      % gray
      \def\colorrgb#1{\color{black}}%
      \def\colorgray#1{\color[gray]{#1}}%
      \expandafter\def\csname LTw\endcsname{\color{white}}%
      \expandafter\def\csname LTb\endcsname{\color{black}}%
      \expandafter\def\csname LTa\endcsname{\color{black}}%
      \expandafter\def\csname LT0\endcsname{\color{black}}%
      \expandafter\def\csname LT1\endcsname{\color{black}}%
      \expandafter\def\csname LT2\endcsname{\color{black}}%
      \expandafter\def\csname LT3\endcsname{\color{black}}%
      \expandafter\def\csname LT4\endcsname{\color{black}}%
      \expandafter\def\csname LT5\endcsname{\color{black}}%
      \expandafter\def\csname LT6\endcsname{\color{black}}%
      \expandafter\def\csname LT7\endcsname{\color{black}}%
      \expandafter\def\csname LT8\endcsname{\color{black}}%
    \fi
  \fi
  \setlength{\unitlength}{0.0500bp}%
  \begin{picture}(3960.00,2880.00)%
    \gplgaddtomacro\gplbacktext{%
      \csname LTb\endcsname%
      \put(900,640){\makebox(0,0)[r]{\strut{}\fnt $10^{-1}$}}%
      \put(900,968){\makebox(0,0)[r]{\strut{}\fnt $10^{0}$}}%
      \put(900,1296){\makebox(0,0)[r]{\strut{}\fnt $10^{1}$}}%
      \put(900,1623){\makebox(0,0)[r]{\strut{}\fnt $10^{2}$}}%
      \put(900,1951){\makebox(0,0)[r]{\strut{}\fnt $10^{3}$}}%
      \put(900,2279){\makebox(0,0)[r]{\strut{}\fnt $10^{4}$}}%
      \put(1020,440){\makebox(0,0){\strut{}\fnt $10^{3}$}}%
      \put(1880,440){\makebox(0,0){\strut{}\fnt $10^{4}$}}%
      \put(2739,440){\makebox(0,0){\strut{}\fnt $10^{5}$}}%
      \put(3599,440){\makebox(0,0){\strut{}\fnt $10^{6}$}}%
      \put(320,1459){\rotatebox{-270}{\makebox(0,0){\strut{}\fnt $\|\Sigma_n-\var[\vol]\|_\frob^2/\|\var[\vol]\|_\frob^2$}}}%
      \put(2309,140){\makebox(0,0){\strut{}\fnt $n$}}%
      \put(2309,2579){\makebox(0,0){\strut{}\fnt (b)}}%
    }%
    \gplgaddtomacro\gplfronttext{%
      \csname LTb\endcsname%
      \put(3056,2116){\makebox(0,0)[r]{\strut{}\fnt $\Sigma_n$}}%
      \csname LTb\endcsname%
      \put(3056,1916){\makebox(0,0)[r]{\strut{}\fnt $\Sigma_n^{\mathrm{(s)}}$}}%
    }%
    \gplbacktext
    \put(0,0){\includegraphics{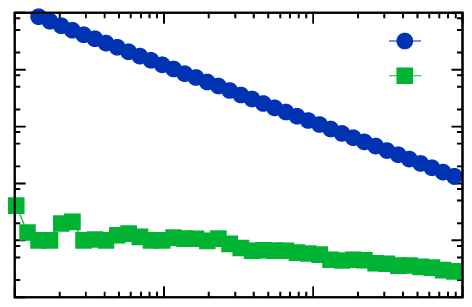}}%
    \gplfronttext
  \end{picture}%
\endgroup

%% file: ctf2_gplt.tex
% GNUPLOT: LaTeX picture with Postscript
\begingroup
\newcommand{\fnt}[0]{\footnotesize}
  \makeatletter
  \providecommand\color[2][]{%
    \GenericError{(gnuplot) \space\space\space\@spaces}{%
      Package color not loaded in conjunction with
      terminal option `colourtext'%
    }{See the gnuplot documentation for explanation.%
    }{Either use 'blacktext' in gnuplot or load the package
      color.sty in LaTeX.}%
    \renewcommand\color[2][]{}%
  }%
  \providecommand\includegraphics[2][]{%
    \GenericError{(gnuplot) \space\space\space\@spaces}{%
      Package graphicx or graphics not loaded%
    }{See the gnuplot documentation for explanation.%
    }{The gnuplot epslatex terminal needs graphicx.sty or graphics.sty.}%
    \renewcommand\includegraphics[2][]{}%
  }%
  \providecommand\rotatebox[2]{#2}%
  \@ifundefined{ifGPcolor}{%
    \newif\ifGPcolor
    \GPcolortrue
  }{}%
  \@ifundefined{ifGPblacktext}{%
    \newif\ifGPblacktext
    \GPblacktextfalse
  }{}%
  % define a \g@addto@macro without @ in the name:
  \let\gplgaddtomacro\g@addto@macro
  % define empty templates for all commands taking text:
  \gdef\gplbacktext{}%
  \gdef\gplfronttext{}%
  \makeatother
  \ifGPblacktext
    % no textcolor at all
    \def\colorrgb#1{}%
    \def\colorgray#1{}%
  \else
    % gray or color?
    \ifGPcolor
      \def\colorrgb#1{\color[rgb]{#1}}%
      \def\colorgray#1{\color[gray]{#1}}%
      \expandafter\def\csname LTw\endcsname{\color{white}}%
      \expandafter\def\csname LTb\endcsname{\color{black}}%
      \expandafter\def\csname LTa\endcsname{\color{black}}%
      \expandafter\def\csname LT0\endcsname{\color[rgb]{1,0,0}}%
      \expandafter\def\csname LT1\endcsname{\color[rgb]{0,1,0}}%
      \expandafter\def\csname LT2\endcsname{\color[rgb]{0,0,1}}%
      \expandafter\def\csname LT3\endcsname{\color[rgb]{1,0,1}}%
      \expandafter\def\csname LT4\endcsname{\color[rgb]{0,1,1}}%
      \expandafter\def\csname LT5\endcsname{\color[rgb]{1,1,0}}%
      \expandafter\def\csname LT6\endcsname{\color[rgb]{0,0,0}}%
      \expandafter\def\csname LT7\endcsname{\color[rgb]{1,0.3,0}}%
      \expandafter\def\csname LT8\endcsname{\color[rgb]{0.5,0.5,0.5}}%
    \else
      % gray
      \def\colorrgb#1{\color{black}}%
      \def\colorgray#1{\color[gray]{#1}}%
      \expandafter\def\csname LTw\endcsname{\color{white}}%
      \expandafter\def\csname LTb\endcsname{\color{black}}%
      \expandafter\def\csname LTa\endcsname{\color{black}}%
      \expandafter\def\csname LT0\endcsname{\color{black}}%
      \expandafter\def\csname LT1\endcsname{\color{black}}%
      \expandafter\def\csname LT2\endcsname{\color{black}}%
      \expandafter\def\csname LT3\endcsname{\color{black}}%
      \expandafter\def\csname LT4\endcsname{\color{black}}%
      \expandafter\def\csname LT5\endcsname{\color{black}}%
      \expandafter\def\csname LT6\endcsname{\color{black}}%
      \expandafter\def\csname LT7\endcsname{\color{black}}%
      \expandafter\def\csname LT8\endcsname{\color{black}}%
    \fi
  \fi
  \setlength{\unitlength}{0.0500bp}%
  \begin{picture}(7920.00,2880.00)%
    \gplgaddtomacro\gplbacktext{%
      \csname LTb\endcsname%
      \put(900,400){\makebox(0,0)[r]{\strut{}\fnt $0$}}%
      \put(900,909){\makebox(0,0)[r]{\strut{}\fnt $0.25$}}%
      \put(900,1418){\makebox(0,0)[r]{\strut{}\fnt $0.5$}}%
      \put(900,1927){\makebox(0,0)[r]{\strut{}\fnt $0.75$}}%
      \put(900,2435){\makebox(0,0)[r]{\strut{}\fnt $1$}}%
      \put(1020,200){\makebox(0,0){\strut{}\fnt 0}}%
      \put(2655,200){\makebox(0,0){\strut{}\fnt $\pi/4$}}%
      \put(4290,200){\makebox(0,0){\strut{}\fnt $\pi/2$}}%
      \put(5924,200){\makebox(0,0){\strut{}\fnt $3\pi/4$}}%
      \put(7559,200){\makebox(0,0){\strut{}\fnt $\pi$}}%
    }%
    \gplgaddtomacro\gplfronttext{%
    }%
    \gplbacktext
    \put(0,0){\includegraphics{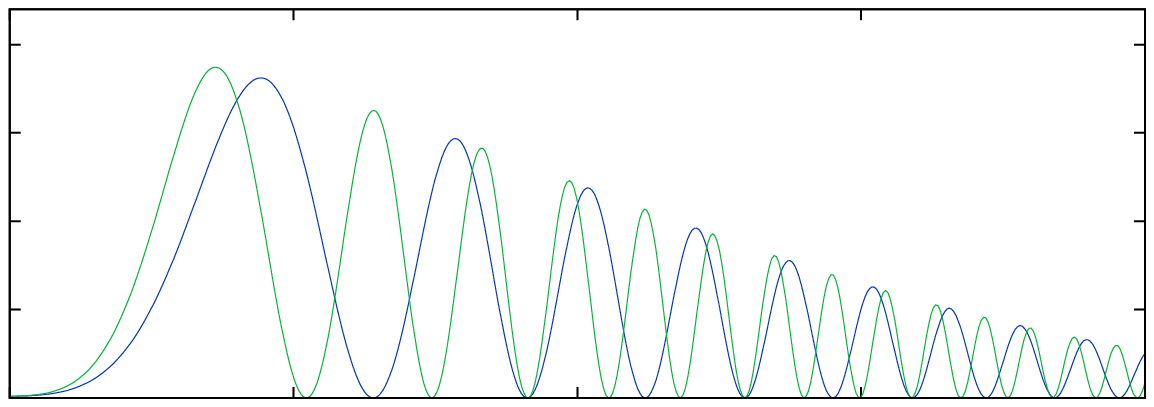}}%
    \gplfronttext
  \end{picture}%
\endgroup

%% file: varying_n_err_gplt.tex
% GNUPLOT: LaTeX picture with Postscript
\begingroup
\newcommand{\fnt}[0]{\footnotesize}
  \makeatletter
  \providecommand\color[2][]{%
    \GenericError{(gnuplot) \space\space\space\@spaces}{%
      Package color not loaded in conjunction with
      terminal option `colourtext'%
    }{See the gnuplot documentation for explanation.%
    }{Either use 'blacktext' in gnuplot or load the package
      color.sty in LaTeX.}%
    \renewcommand\color[2][]{}%
  }%
  \providecommand\includegraphics[2][]{%
    \GenericError{(gnuplot) \space\space\space\@spaces}{%
      Package graphicx or graphics not loaded%
    }{See the gnuplot documentation for explanation.%
    }{The gnuplot epslatex terminal needs graphicx.sty or graphics.sty.}%
    \renewcommand\includegraphics[2][]{}%
  }%
  \providecommand\rotatebox[2]{#2}%
  \@ifundefined{ifGPcolor}{%
    \newif\ifGPcolor
    \GPcolortrue
  }{}%
  \@ifundefined{ifGPblacktext}{%
    \newif\ifGPblacktext
    \GPblacktextfalse
  }{}%
  % define a \g@addto@macro without @ in the name:
  \let\gplgaddtomacro\g@addto@macro
  % define empty templates for all commands taking text:
  \gdef\gplbacktext{}%
  \gdef\gplfronttext{}%
  \makeatother
  \ifGPblacktext
    % no textcolor at all
    \def\colorrgb#1{}%
    \def\colorgray#1{}%
  \else
    % gray or color?
    \ifGPcolor
      \def\colorrgb#1{\color[rgb]{#1}}%
      \def\colorgray#1{\color[gray]{#1}}%
      \expandafter\def\csname LTw\endcsname{\color{white}}%
      \expandafter\def\csname LTb\endcsname{\color{black}}%
      \expandafter\def\csname LTa\endcsname{\color{black}}%
      \expandafter\def\csname LT0\endcsname{\color[rgb]{1,0,0}}%
      \expandafter\def\csname LT1\endcsname{\color[rgb]{0,1,0}}%
      \expandafter\def\csname LT2\endcsname{\color[rgb]{0,0,1}}%
      \expandafter\def\csname LT3\endcsname{\color[rgb]{1,0,1}}%
      \expandafter\def\csname LT4\endcsname{\color[rgb]{0,1,1}}%
      \expandafter\def\csname LT5\endcsname{\color[rgb]{1,1,0}}%
      \expandafter\def\csname LT6\endcsname{\color[rgb]{0,0,0}}%
      \expandafter\def\csname LT7\endcsname{\color[rgb]{1,0.3,0}}%
      \expandafter\def\csname LT8\endcsname{\color[rgb]{0.5,0.5,0.5}}%
    \else
      % gray
      \def\colorrgb#1{\color{black}}%
      \def\colorgray#1{\color[gray]{#1}}%
      \expandafter\def\csname LTw\endcsname{\color{white}}%
      \expandafter\def\csname LTb\endcsname{\color{black}}%
      \expandafter\def\csname LTa\endcsname{\color{black}}%
      \expandafter\def\csname LT0\endcsname{\color{black}}%
      \expandafter\def\csname LT1\endcsname{\color{black}}%
      \expandafter\def\csname LT2\endcsname{\color{black}}%
      \expandafter\def\csname LT3\endcsname{\color{black}}%
      \expandafter\def\csname LT4\endcsname{\color{black}}%
      \expandafter\def\csname LT5\endcsname{\color{black}}%
      \expandafter\def\csname LT6\endcsname{\color{black}}%
      \expandafter\def\csname LT7\endcsname{\color{black}}%
      \expandafter\def\csname LT8\endcsname{\color{black}}%
    \fi
  \fi
  \setlength{\unitlength}{0.0500bp}%
  \begin{picture}(3960.00,2880.00)%
    \gplgaddtomacro\gplbacktext{%
      \csname LTb\endcsname%
      \put(780,640){\makebox(0,0)[r]{\strut{}\fnt $10^{0}$}}%
      \put(780,1460){\makebox(0,0)[r]{\strut{}\fnt $10^{1}$}}%
      \put(780,2279){\makebox(0,0)[r]{\strut{}\fnt $10^{2}$}}%
      \put(900,440){\makebox(0,0){\strut{}\fnt $0$}}%
      \put(1575,440){\makebox(0,0){\strut{}\fnt $0.005$}}%
      \put(2250,440){\makebox(0,0){\strut{}\fnt $0.01$}}%
      \put(2924,440){\makebox(0,0){\strut{}\fnt $0.015$}}%
      \put(3599,440){\makebox(0,0){\strut{}\fnt $0.02$}}%
      \put(320,1459){\rotatebox{-270}{\makebox(0,0){\strut{}\fnt $\|\Covar_n-\var[\vol]\|_\frob/\|\var[\vol]\|_\frob$}}}%
      \put(2249,140){\makebox(0,0){\strut{}\fnt $\SNRh$}}%
      \put(2249,2579){\makebox(0,0){\strut{}\fnt (a)}}%
    }%
    \gplgaddtomacro\gplfronttext{%
      \csname LTb\endcsname%
      \put(3056,2116){\makebox(0,0)[r]{\strut{}\fnt $n = 1024$}}%
      \csname LTb\endcsname%
      \put(3056,1916){\makebox(0,0)[r]{\strut{}\fnt $n = 4096$}}%
      \csname LTb\endcsname%
      \put(3056,1716){\makebox(0,0)[r]{\strut{}\fnt $n = 16384$}}%
    }%
    \gplbacktext
    \put(0,0){\includegraphics{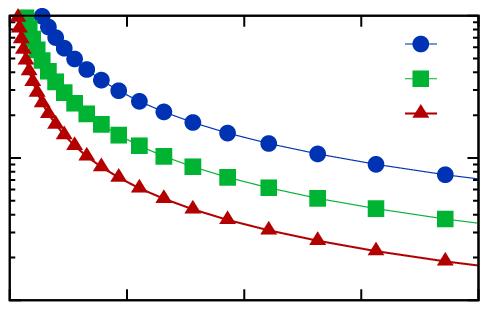}}%
    \gplfronttext
  \end{picture}%
\endgroup

%% file: varying_n_corr_gplt.tex
% GNUPLOT: LaTeX picture with Postscript
\begingroup
\newcommand{\fnt}[0]{\footnotesize}
  \makeatletter
  \providecommand\color[2][]{%
    \GenericError{(gnuplot) \space\space\space\@spaces}{%
      Package color not loaded in conjunction with
      terminal option `colourtext'%
    }{See the gnuplot documentation for explanation.%
    }{Either use 'blacktext' in gnuplot or load the package
      color.sty in LaTeX.}%
    \renewcommand\color[2][]{}%
  }%
  \providecommand\includegraphics[2][]{%
    \GenericError{(gnuplot) \space\space\space\@spaces}{%
      Package graphicx or graphics not loaded%
    }{See the gnuplot documentation for explanation.%
    }{The gnuplot epslatex terminal needs graphicx.sty or graphics.sty.}%
    \renewcommand\includegraphics[2][]{}%
  }%
  \providecommand\rotatebox[2]{#2}%
  \@ifundefined{ifGPcolor}{%
    \newif\ifGPcolor
    \GPcolortrue
  }{}%
  \@ifundefined{ifGPblacktext}{%
    \newif\ifGPblacktext
    \GPblacktextfalse
  }{}%
  % define a \g@addto@macro without @ in the name:
  \let\gplgaddtomacro\g@addto@macro
  % define empty templates for all commands taking text:
  \gdef\gplbacktext{}%
  \gdef\gplfronttext{}%
  \makeatother
  \ifGPblacktext
    % no textcolor at all
    \def\colorrgb#1{}%
    \def\colorgray#1{}%
  \else
    % gray or color?
    \ifGPcolor
      \def\colorrgb#1{\color[rgb]{#1}}%
      \def\colorgray#1{\color[gray]{#1}}%
      \expandafter\def\csname LTw\endcsname{\color{white}}%
      \expandafter\def\csname LTb\endcsname{\color{black}}%
      \expandafter\def\csname LTa\endcsname{\color{black}}%
      \expandafter\def\csname LT0\endcsname{\color[rgb]{1,0,0}}%
      \expandafter\def\csname LT1\endcsname{\color[rgb]{0,1,0}}%
      \expandafter\def\csname LT2\endcsname{\color[rgb]{0,0,1}}%
      \expandafter\def\csname LT3\endcsname{\color[rgb]{1,0,1}}%
      \expandafter\def\csname LT4\endcsname{\color[rgb]{0,1,1}}%
      \expandafter\def\csname LT5\endcsname{\color[rgb]{1,1,0}}%
      \expandafter\def\csname LT6\endcsname{\color[rgb]{0,0,0}}%
      \expandafter\def\csname LT7\endcsname{\color[rgb]{1,0.3,0}}%
      \expandafter\def\csname LT8\endcsname{\color[rgb]{0.5,0.5,0.5}}%
    \else
      % gray
      \def\colorrgb#1{\color{black}}%
      \def\colorgray#1{\color[gray]{#1}}%
      \expandafter\def\csname LTw\endcsname{\color{white}}%
      \expandafter\def\csname LTb\endcsname{\color{black}}%
      \expandafter\def\csname LTa\endcsname{\color{black}}%
      \expandafter\def\csname LT0\endcsname{\color{black}}%
      \expandafter\def\csname LT1\endcsname{\color{black}}%
      \expandafter\def\csname LT2\endcsname{\color{black}}%
      \expandafter\def\csname LT3\endcsname{\color{black}}%
      \expandafter\def\csname LT4\endcsname{\color{black}}%
      \expandafter\def\csname LT5\endcsname{\color{black}}%
      \expandafter\def\csname LT6\endcsname{\color{black}}%
      \expandafter\def\csname LT7\endcsname{\color{black}}%
      \expandafter\def\csname LT8\endcsname{\color{black}}%
    \fi
  \fi
  \setlength{\unitlength}{0.0500bp}%
  \begin{picture}(3960.00,2880.00)%
    \gplgaddtomacro\gplbacktext{%
      \csname LTb\endcsname%
      \put(780,640){\makebox(0,0)[r]{\strut{}\fnt $0$}}%
      \put(780,968){\makebox(0,0)[r]{\strut{}\fnt $0.2$}}%
      \put(780,1296){\makebox(0,0)[r]{\strut{}\fnt $0.4$}}%
      \put(780,1623){\makebox(0,0)[r]{\strut{}\fnt $0.6$}}%
      \put(780,1951){\makebox(0,0)[r]{\strut{}\fnt $0.8$}}%
      \put(780,2279){\makebox(0,0)[r]{\strut{}\fnt $1$}}%
      \put(900,440){\makebox(0,0){\strut{}\fnt $0$}}%
      \put(1575,440){\makebox(0,0){\strut{}\fnt $0.005$}}%
      \put(2250,440){\makebox(0,0){\strut{}\fnt $0.01$}}%
      \put(2924,440){\makebox(0,0){\strut{}\fnt $0.015$}}%
      \put(3599,440){\makebox(0,0){\strut{}\fnt $0.02$}}%
      \put(320,1459){\rotatebox{-270}{\makebox(0,0){\strut{}\fnt Correlation}}}%
      \put(2249,140){\makebox(0,0){\strut{}\fnt $\SNRh$}}%
      \put(2249,2579){\makebox(0,0){\strut{}\fnt (b)}}%
    }%
    \gplgaddtomacro\gplfronttext{%
      \csname LTb\endcsname%
      \put(3056,1203){\makebox(0,0)[r]{\strut{}\fnt $n = 1024$}}%
      \csname LTb\endcsname%
      \put(3056,1003){\makebox(0,0)[r]{\strut{}\fnt $n = 4096$}}%
      \csname LTb\endcsname%
      \put(3056,803){\makebox(0,0)[r]{\strut{}\fnt $n = 16384$}}%
    }%
    \gplbacktext
    \put(0,0){\includegraphics{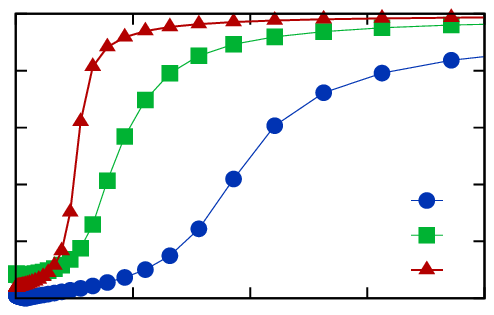}}%
    \gplfronttext
  \end{picture}%
\endgroup

%% file: shrinkage_gplt.tex
% GNUPLOT: LaTeX picture with Postscript
\begingroup
\newcommand{\fnt}[0]{\footnotesize}
  \makeatletter
  \providecommand\color[2][]{%
    \GenericError{(gnuplot) \space\space\space\@spaces}{%
      Package color not loaded in conjunction with
      terminal option `colourtext'%
    }{See the gnuplot documentation for explanation.%
    }{Either use 'blacktext' in gnuplot or load the package
      color.sty in LaTeX.}%
    \renewcommand\color[2][]{}%
  }%
  \providecommand\includegraphics[2][]{%
    \GenericError{(gnuplot) \space\space\space\@spaces}{%
      Package graphicx or graphics not loaded%
    }{See the gnuplot documentation for explanation.%
    }{The gnuplot epslatex terminal needs graphicx.sty or graphics.sty.}%
    \renewcommand\includegraphics[2][]{}%
  }%
  \providecommand\rotatebox[2]{#2}%
  \@ifundefined{ifGPcolor}{%
    \newif\ifGPcolor
    \GPcolortrue
  }{}%
  \@ifundefined{ifGPblacktext}{%
    \newif\ifGPblacktext
    \GPblacktextfalse
  }{}%
  % define a \g@addto@macro without @ in the name:
  \let\gplgaddtomacro\g@addto@macro
  % define empty templates for all commands taking text:
  \gdef\gplbacktext{}%
  \gdef\gplfronttext{}%
  \makeatother
  \ifGPblacktext
    % no textcolor at all
    \def\colorrgb#1{}%
    \def\colorgray#1{}%
  \else
    % gray or color?
    \ifGPcolor
      \def\colorrgb#1{\color[rgb]{#1}}%
      \def\colorgray#1{\color[gray]{#1}}%
      \expandafter\def\csname LTw\endcsname{\color{white}}%
      \expandafter\def\csname LTb\endcsname{\color{black}}%
      \expandafter\def\csname LTa\endcsname{\color{black}}%
      \expandafter\def\csname LT0\endcsname{\color[rgb]{1,0,0}}%
      \expandafter\def\csname LT1\endcsname{\color[rgb]{0,1,0}}%
      \expandafter\def\csname LT2\endcsname{\color[rgb]{0,0,1}}%
      \expandafter\def\csname LT3\endcsname{\color[rgb]{1,0,1}}%
      \expandafter\def\csname LT4\endcsname{\color[rgb]{0,1,1}}%
      \expandafter\def\csname LT5\endcsname{\color[rgb]{1,1,0}}%
      \expandafter\def\csname LT6\endcsname{\color[rgb]{0,0,0}}%
      \expandafter\def\csname LT7\endcsname{\color[rgb]{1,0.3,0}}%
      \expandafter\def\csname LT8\endcsname{\color[rgb]{0.5,0.5,0.5}}%
    \else
      % gray
      \def\colorrgb#1{\color{black}}%
      \def\colorgray#1{\color[gray]{#1}}%
      \expandafter\def\csname LTw\endcsname{\color{white}}%
      \expandafter\def\csname LTb\endcsname{\color{black}}%
      \expandafter\def\csname LTa\endcsname{\color{black}}%
      \expandafter\def\csname LT0\endcsname{\color{black}}%
      \expandafter\def\csname LT1\endcsname{\color{black}}%
      \expandafter\def\csname LT2\endcsname{\color{black}}%
      \expandafter\def\csname LT3\endcsname{\color{black}}%
      \expandafter\def\csname LT4\endcsname{\color{black}}%
      \expandafter\def\csname LT5\endcsname{\color{black}}%
      \expandafter\def\csname LT6\endcsname{\color{black}}%
      \expandafter\def\csname LT7\endcsname{\color{black}}%
      \expandafter\def\csname LT8\endcsname{\color{black}}%
    \fi
  \fi
  \setlength{\unitlength}{0.0500bp}%
  \begin{picture}(3960.00,2880.00)%
    \gplgaddtomacro\gplbacktext{%
      \csname LTb\endcsname%
      \put(780,640){\makebox(0,0)[r]{\strut{}\fnt $0$}}%
      \put(780,968){\makebox(0,0)[r]{\strut{}\fnt $0.2$}}%
      \put(780,1296){\makebox(0,0)[r]{\strut{}\fnt $0.4$}}%
      \put(780,1623){\makebox(0,0)[r]{\strut{}\fnt $0.6$}}%
      \put(780,1951){\makebox(0,0)[r]{\strut{}\fnt $0.8$}}%
      \put(780,2279){\makebox(0,0)[r]{\strut{}\fnt $1$}}%
      \put(900,440){\makebox(0,0){\strut{}\fnt $0$}}%
      \put(1575,440){\makebox(0,0){\strut{}\fnt $0.005$}}%
      \put(2250,440){\makebox(0,0){\strut{}\fnt $0.01$}}%
      \put(2924,440){\makebox(0,0){\strut{}\fnt $0.015$}}%
      \put(3599,440){\makebox(0,0){\strut{}\fnt $0.02$}}%
      \put(320,1459){\rotatebox{-270}{\makebox(0,0){\strut{}\fnt Correlation}}}%
      \put(2249,140){\makebox(0,0){\strut{}\fnt $\SNRh$}}%
      \put(2249,2579){\makebox(0,0){\strut{}\fnt (c)}}%
    }%
    \gplgaddtomacro\gplfronttext{%
      \csname LTb\endcsname%
      \put(3056,1003){\makebox(0,0)[r]{\strut{}\fnt $\Sigma_n$}}%
      \csname LTb\endcsname%
      \put(3056,803){\makebox(0,0)[r]{\strut{}\fnt $\Sigma_n^\shrink$}}%
    }%
    \gplbacktext
    \put(0,0){\includegraphics{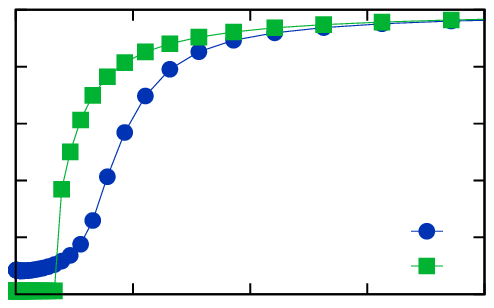}}%
    \gplfronttext
  \end{picture}%
\endgroup

%% file: est_angles_gplt.tex
% GNUPLOT: LaTeX picture with Postscript
\begingroup
\newcommand{\fnt}[0]{\footnotesize}
  \makeatletter
  \providecommand\color[2][]{%
    \GenericError{(gnuplot) \space\space\space\@spaces}{%
      Package color not loaded in conjunction with
      terminal option `colourtext'%
    }{See the gnuplot documentation for explanation.%
    }{Either use 'blacktext' in gnuplot or load the package
      color.sty in LaTeX.}%
    \renewcommand\color[2][]{}%
  }%
  \providecommand\includegraphics[2][]{%
    \GenericError{(gnuplot) \space\space\space\@spaces}{%
      Package graphicx or graphics not loaded%
    }{See the gnuplot documentation for explanation.%
    }{The gnuplot epslatex terminal needs graphicx.sty or graphics.sty.}%
    \renewcommand\includegraphics[2][]{}%
  }%
  \providecommand\rotatebox[2]{#2}%
  \@ifundefined{ifGPcolor}{%
    \newif\ifGPcolor
    \GPcolortrue
  }{}%
  \@ifundefined{ifGPblacktext}{%
    \newif\ifGPblacktext
    \GPblacktextfalse
  }{}%
  % define a \g@addto@macro without @ in the name:
  \let\gplgaddtomacro\g@addto@macro
  % define empty templates for all commands taking text:
  \gdef\gplbacktext{}%
  \gdef\gplfronttext{}%
  \makeatother
  \ifGPblacktext
    % no textcolor at all
    \def\colorrgb#1{}%
    \def\colorgray#1{}%
  \else
    % gray or color?
    \ifGPcolor
      \def\colorrgb#1{\color[rgb]{#1}}%
      \def\colorgray#1{\color[gray]{#1}}%
      \expandafter\def\csname LTw\endcsname{\color{white}}%
      \expandafter\def\csname LTb\endcsname{\color{black}}%
      \expandafter\def\csname LTa\endcsname{\color{black}}%
      \expandafter\def\csname LT0\endcsname{\color[rgb]{1,0,0}}%
      \expandafter\def\csname LT1\endcsname{\color[rgb]{0,1,0}}%
      \expandafter\def\csname LT2\endcsname{\color[rgb]{0,0,1}}%
      \expandafter\def\csname LT3\endcsname{\color[rgb]{1,0,1}}%
      \expandafter\def\csname LT4\endcsname{\color[rgb]{0,1,1}}%
      \expandafter\def\csname LT5\endcsname{\color[rgb]{1,1,0}}%
      \expandafter\def\csname LT6\endcsname{\color[rgb]{0,0,0}}%
      \expandafter\def\csname LT7\endcsname{\color[rgb]{1,0.3,0}}%
      \expandafter\def\csname LT8\endcsname{\color[rgb]{0.5,0.5,0.5}}%
    \else
      % gray
      \def\colorrgb#1{\color{black}}%
      \def\colorgray#1{\color[gray]{#1}}%
      \expandafter\def\csname LTw\endcsname{\color{white}}%
      \expandafter\def\csname LTb\endcsname{\color{black}}%
      \expandafter\def\csname LTa\endcsname{\color{black}}%
      \expandafter\def\csname LT0\endcsname{\color{black}}%
      \expandafter\def\csname LT1\endcsname{\color{black}}%
      \expandafter\def\csname LT2\endcsname{\color{black}}%
      \expandafter\def\csname LT3\endcsname{\color{black}}%
      \expandafter\def\csname LT4\endcsname{\color{black}}%
      \expandafter\def\csname LT5\endcsname{\color{black}}%
      \expandafter\def\csname LT6\endcsname{\color{black}}%
      \expandafter\def\csname LT7\endcsname{\color{black}}%
      \expandafter\def\csname LT8\endcsname{\color{black}}%
    \fi
  \fi
  \setlength{\unitlength}{0.0500bp}%
  \begin{picture}(3960.00,2880.00)%
    \gplgaddtomacro\gplbacktext{%
      \csname LTb\endcsname%
      \put(780,640){\makebox(0,0)[r]{\strut{}\fnt $0$}}%
      \put(780,968){\makebox(0,0)[r]{\strut{}\fnt $0.2$}}%
      \put(780,1296){\makebox(0,0)[r]{\strut{}\fnt $0.4$}}%
      \put(780,1623){\makebox(0,0)[r]{\strut{}\fnt $0.6$}}%
      \put(780,1951){\makebox(0,0)[r]{\strut{}\fnt $0.8$}}%
      \put(780,2279){\makebox(0,0)[r]{\strut{}\fnt $1$}}%
      \put(900,440){\makebox(0,0){\strut{}\fnt $0$}}%
      \put(1575,440){\makebox(0,0){\strut{}\fnt $0.5$}}%
      \put(2250,440){\makebox(0,0){\strut{}\fnt $1$}}%
      \put(2924,440){\makebox(0,0){\strut{}\fnt $1.5$}}%
      \put(3599,440){\makebox(0,0){\strut{}\fnt $2$}}%
      \put(320,1459){\rotatebox{-270}{\makebox(0,0){\strut{}\fnt Correlation}}}%
      \put(2249,140){\makebox(0,0){\strut{}\fnt $\SNRh$}}%
      \put(2249,2579){\makebox(0,0){\strut{}\fnt (d)}}%
    }%
    \gplgaddtomacro\gplfronttext{%
      \csname LTb\endcsname%
      \put(3056,803){\makebox(0,0)[r]{\strut{}\fnt Estimated Angles}}%
    }%
    \gplbacktext
    \put(0,0){\includegraphics{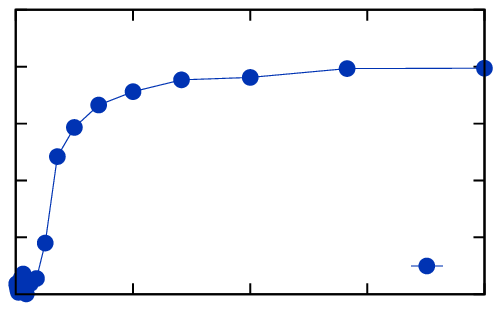}}%
    \gplfronttext
  \end{picture}%
\endgroup

%% file: nonuniform_gplt.tex
% GNUPLOT: LaTeX picture with Postscript
\begingroup
\newcommand{\fnt}[0]{\footnotesize}
  \makeatletter
  \providecommand\color[2][]{%
    \GenericError{(gnuplot) \space\space\space\@spaces}{%
      Package color not loaded in conjunction with
      terminal option `colourtext'%
    }{See the gnuplot documentation for explanation.%
    }{Either use 'blacktext' in gnuplot or load the package
      color.sty in LaTeX.}%
    \renewcommand\color[2][]{}%
  }%
  \providecommand\includegraphics[2][]{%
    \GenericError{(gnuplot) \space\space\space\@spaces}{%
      Package graphicx or graphics not loaded%
    }{See the gnuplot documentation for explanation.%
    }{The gnuplot epslatex terminal needs graphicx.sty or graphics.sty.}%
    \renewcommand\includegraphics[2][]{}%
  }%
  \providecommand\rotatebox[2]{#2}%
  \@ifundefined{ifGPcolor}{%
    \newif\ifGPcolor
    \GPcolortrue
  }{}%
  \@ifundefined{ifGPblacktext}{%
    \newif\ifGPblacktext
    \GPblacktextfalse
  }{}%
  % define a \g@addto@macro without @ in the name:
  \let\gplgaddtomacro\g@addto@macro
  % define empty templates for all commands taking text:
  \gdef\gplbacktext{}%
  \gdef\gplfronttext{}%
  \makeatother
  \ifGPblacktext
    % no textcolor at all
    \def\colorrgb#1{}%
    \def\colorgray#1{}%
  \else
    % gray or color?
    \ifGPcolor
      \def\colorrgb#1{\color[rgb]{#1}}%
      \def\colorgray#1{\color[gray]{#1}}%
      \expandafter\def\csname LTw\endcsname{\color{white}}%
      \expandafter\def\csname LTb\endcsname{\color{black}}%
      \expandafter\def\csname LTa\endcsname{\color{black}}%
      \expandafter\def\csname LT0\endcsname{\color[rgb]{1,0,0}}%
      \expandafter\def\csname LT1\endcsname{\color[rgb]{0,1,0}}%
      \expandafter\def\csname LT2\endcsname{\color[rgb]{0,0,1}}%
      \expandafter\def\csname LT3\endcsname{\color[rgb]{1,0,1}}%
      \expandafter\def\csname LT4\endcsname{\color[rgb]{0,1,1}}%
      \expandafter\def\csname LT5\endcsname{\color[rgb]{1,1,0}}%
      \expandafter\def\csname LT6\endcsname{\color[rgb]{0,0,0}}%
      \expandafter\def\csname LT7\endcsname{\color[rgb]{1,0.3,0}}%
      \expandafter\def\csname LT8\endcsname{\color[rgb]{0.5,0.5,0.5}}%
    \else
      % gray
      \def\colorrgb#1{\color{black}}%
      \def\colorgray#1{\color[gray]{#1}}%
      \expandafter\def\csname LTw\endcsname{\color{white}}%
      \expandafter\def\csname LTb\endcsname{\color{black}}%
      \expandafter\def\csname LTa\endcsname{\color{black}}%
      \expandafter\def\csname LT0\endcsname{\color{black}}%
      \expandafter\def\csname LT1\endcsname{\color{black}}%
      \expandafter\def\csname LT2\endcsname{\color{black}}%
      \expandafter\def\csname LT3\endcsname{\color{black}}%
      \expandafter\def\csname LT4\endcsname{\color{black}}%
      \expandafter\def\csname LT5\endcsname{\color{black}}%
      \expandafter\def\csname LT6\endcsname{\color{black}}%
      \expandafter\def\csname LT7\endcsname{\color{black}}%
      \expandafter\def\csname LT8\endcsname{\color{black}}%
    \fi
  \fi
  \setlength{\unitlength}{0.0500bp}%
  \begin{picture}(3960.00,2880.00)%
    \gplgaddtomacro\gplbacktext{%
      \csname LTb\endcsname%
      \put(780,640){\makebox(0,0)[r]{\strut{}\fnt $0$}}%
      \put(780,968){\makebox(0,0)[r]{\strut{}\fnt $0.2$}}%
      \put(780,1296){\makebox(0,0)[r]{\strut{}\fnt $0.4$}}%
      \put(780,1623){\makebox(0,0)[r]{\strut{}\fnt $0.6$}}%
      \put(780,1951){\makebox(0,0)[r]{\strut{}\fnt $0.8$}}%
      \put(780,2279){\makebox(0,0)[r]{\strut{}\fnt $1$}}%
      \put(900,440){\makebox(0,0){\strut{}\fnt $0$}}%
      \put(1575,440){\makebox(0,0){\strut{}\fnt $0.005$}}%
      \put(2250,440){\makebox(0,0){\strut{}\fnt $0.01$}}%
      \put(2924,440){\makebox(0,0){\strut{}\fnt $0.015$}}%
      \put(3599,440){\makebox(0,0){\strut{}\fnt $0.02$}}%
      \put(320,1459){\rotatebox{-270}{\makebox(0,0){\strut{}\fnt Correlation}}}%
      \put(2249,140){\makebox(0,0){\strut{}\fnt $\SNRh$}}%
      \put(2249,2579){\makebox(0,0){\strut{}\fnt (e)}}%
    }%
    \gplgaddtomacro\gplfronttext{%
      \csname LTb\endcsname%
      \put(3056,1203){\makebox(0,0)[r]{\strut{}\fnt $\delta = 1$}}%
      \csname LTb\endcsname%
      \put(3056,1003){\makebox(0,0)[r]{\strut{}\fnt $\delta = 4$}}%
      \csname LTb\endcsname%
      \put(3056,803){\makebox(0,0)[r]{\strut{}\fnt $\delta = 16$}}%
    }%
    \gplbacktext
    \put(0,0){\includegraphics{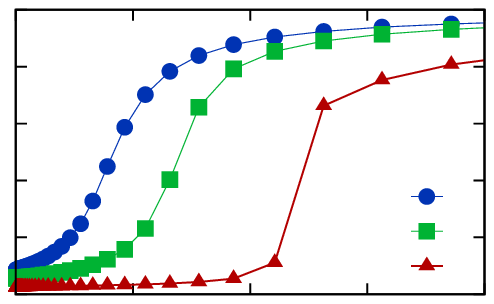}}%
    \gplfronttext
  \end{picture}%
\endgroup

%% file: moreclasses_gplt.tex
% GNUPLOT: LaTeX picture with Postscript
\begingroup
\newcommand{\fnt}[0]{\footnotesize}
  \makeatletter
  \providecommand\color[2][]{%
    \GenericError{(gnuplot) \space\space\space\@spaces}{%
      Package color not loaded in conjunction with
      terminal option `colourtext'%
    }{See the gnuplot documentation for explanation.%
    }{Either use 'blacktext' in gnuplot or load the package
      color.sty in LaTeX.}%
    \renewcommand\color[2][]{}%
  }%
  \providecommand\includegraphics[2][]{%
    \GenericError{(gnuplot) \space\space\space\@spaces}{%
      Package graphicx or graphics not loaded%
    }{See the gnuplot documentation for explanation.%
    }{The gnuplot epslatex terminal needs graphicx.sty or graphics.sty.}%
    \renewcommand\includegraphics[2][]{}%
  }%
  \providecommand\rotatebox[2]{#2}%
  \@ifundefined{ifGPcolor}{%
    \newif\ifGPcolor
    \GPcolortrue
  }{}%
  \@ifundefined{ifGPblacktext}{%
    \newif\ifGPblacktext
    \GPblacktextfalse
  }{}%
  % define a \g@addto@macro without @ in the name:
  \let\gplgaddtomacro\g@addto@macro
  % define empty templates for all commands taking text:
  \gdef\gplbacktext{}%
  \gdef\gplfronttext{}%
  \makeatother
  \ifGPblacktext
    % no textcolor at all
    \def\colorrgb#1{}%
    \def\colorgray#1{}%
  \else
    % gray or color?
    \ifGPcolor
      \def\colorrgb#1{\color[rgb]{#1}}%
      \def\colorgray#1{\color[gray]{#1}}%
      \expandafter\def\csname LTw\endcsname{\color{white}}%
      \expandafter\def\csname LTb\endcsname{\color{black}}%
      \expandafter\def\csname LTa\endcsname{\color{black}}%
      \expandafter\def\csname LT0\endcsname{\color[rgb]{1,0,0}}%
      \expandafter\def\csname LT1\endcsname{\color[rgb]{0,1,0}}%
      \expandafter\def\csname LT2\endcsname{\color[rgb]{0,0,1}}%
      \expandafter\def\csname LT3\endcsname{\color[rgb]{1,0,1}}%
      \expandafter\def\csname LT4\endcsname{\color[rgb]{0,1,1}}%
      \expandafter\def\csname LT5\endcsname{\color[rgb]{1,1,0}}%
      \expandafter\def\csname LT6\endcsname{\color[rgb]{0,0,0}}%
      \expandafter\def\csname LT7\endcsname{\color[rgb]{1,0.3,0}}%
      \expandafter\def\csname LT8\endcsname{\color[rgb]{0.5,0.5,0.5}}%
    \else
      % gray
      \def\colorrgb#1{\color{black}}%
      \def\colorgray#1{\color[gray]{#1}}%
      \expandafter\def\csname LTw\endcsname{\color{white}}%
      \expandafter\def\csname LTb\endcsname{\color{black}}%
      \expandafter\def\csname LTa\endcsname{\color{black}}%
      \expandafter\def\csname LT0\endcsname{\color{black}}%
      \expandafter\def\csname LT1\endcsname{\color{black}}%
      \expandafter\def\csname LT2\endcsname{\color{black}}%
      \expandafter\def\csname LT3\endcsname{\color{black}}%
      \expandafter\def\csname LT4\endcsname{\color{black}}%
      \expandafter\def\csname LT5\endcsname{\color{black}}%
      \expandafter\def\csname LT6\endcsname{\color{black}}%
      \expandafter\def\csname LT7\endcsname{\color{black}}%
      \expandafter\def\csname LT8\endcsname{\color{black}}%
    \fi
  \fi
  \setlength{\unitlength}{0.0500bp}%
  \begin{picture}(3960.00,2880.00)%
    \gplgaddtomacro\gplbacktext{%
      \csname LTb\endcsname%
      \put(780,640){\makebox(0,0)[r]{\strut{}\fnt $0$}}%
      \put(780,968){\makebox(0,0)[r]{\strut{}\fnt $0.2$}}%
      \put(780,1296){\makebox(0,0)[r]{\strut{}\fnt $0.4$}}%
      \put(780,1623){\makebox(0,0)[r]{\strut{}\fnt $0.6$}}%
      \put(780,1951){\makebox(0,0)[r]{\strut{}\fnt $0.8$}}%
      \put(780,2279){\makebox(0,0)[r]{\strut{}\fnt $1$}}%
      \put(900,440){\makebox(0,0){\strut{}\fnt $0$}}%
      \put(1800,440){\makebox(0,0){\strut{}\fnt $0.1$}}%
      \put(2699,440){\makebox(0,0){\strut{}\fnt $0.2$}}%
      \put(3599,440){\makebox(0,0){\strut{}\fnt $0.3$}}%
      \put(320,1459){\rotatebox{-270}{\makebox(0,0){\strut{}\fnt $\cos \theta_\max$}}}%
      \put(2249,140){\makebox(0,0){\strut{}\fnt $\SNRh$}}%
      \put(2249,2579){\makebox(0,0){\strut{}\fnt (f)}}%
    }%
    \gplgaddtomacro\gplfronttext{%
    }%
    \gplbacktext
    \put(0,0){\includegraphics{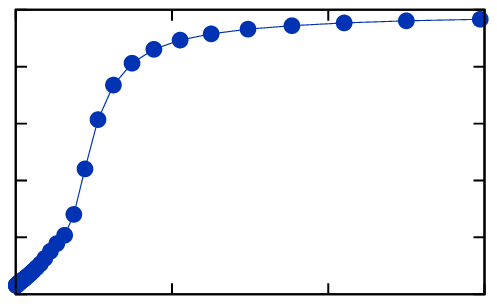}}%
    \gplfronttext
  \end{picture}%
\endgroup

%% file: clustering_spectrum_gplt.tex
% GNUPLOT: LaTeX picture with Postscript
\begingroup
\newcommand{\fnt}[0]{\footnotesize}
  \makeatletter
  \providecommand\color[2][]{%
    \GenericError{(gnuplot) \space\space\space\@spaces}{%
      Package color not loaded in conjunction with
      terminal option `colourtext'%
    }{See the gnuplot documentation for explanation.%
    }{Either use 'blacktext' in gnuplot or load the package
      color.sty in LaTeX.}%
    \renewcommand\color[2][]{}%
  }%
  \providecommand\includegraphics[2][]{%
    \GenericError{(gnuplot) \space\space\space\@spaces}{%
      Package graphicx or graphics not loaded%
    }{See the gnuplot documentation for explanation.%
    }{The gnuplot epslatex terminal needs graphicx.sty or graphics.sty.}%
    \renewcommand\includegraphics[2][]{}%
  }%
  \providecommand\rotatebox[2]{#2}%
  \@ifundefined{ifGPcolor}{%
    \newif\ifGPcolor
    \GPcolortrue
  }{}%
  \@ifundefined{ifGPblacktext}{%
    \newif\ifGPblacktext
    \GPblacktextfalse
  }{}%
  % define a \g@addto@macro without @ in the name:
  \let\gplgaddtomacro\g@addto@macro
  % define empty templates for all commands taking text:
  \gdef\gplbacktext{}%
  \gdef\gplfronttext{}%
  \makeatother
  \ifGPblacktext
    % no textcolor at all
    \def\colorrgb#1{}%
    \def\colorgray#1{}%
  \else
    % gray or color?
    \ifGPcolor
      \def\colorrgb#1{\color[rgb]{#1}}%
      \def\colorgray#1{\color[gray]{#1}}%
      \expandafter\def\csname LTw\endcsname{\color{white}}%
      \expandafter\def\csname LTb\endcsname{\color{black}}%
      \expandafter\def\csname LTa\endcsname{\color{black}}%
      \expandafter\def\csname LT0\endcsname{\color[rgb]{1,0,0}}%
      \expandafter\def\csname LT1\endcsname{\color[rgb]{0,1,0}}%
      \expandafter\def\csname LT2\endcsname{\color[rgb]{0,0,1}}%
      \expandafter\def\csname LT3\endcsname{\color[rgb]{1,0,1}}%
      \expandafter\def\csname LT4\endcsname{\color[rgb]{0,1,1}}%
      \expandafter\def\csname LT5\endcsname{\color[rgb]{1,1,0}}%
      \expandafter\def\csname LT6\endcsname{\color[rgb]{0,0,0}}%
      \expandafter\def\csname LT7\endcsname{\color[rgb]{1,0.3,0}}%
      \expandafter\def\csname LT8\endcsname{\color[rgb]{0.5,0.5,0.5}}%
    \else
      % gray
      \def\colorrgb#1{\color{black}}%
      \def\colorgray#1{\color[gray]{#1}}%
      \expandafter\def\csname LTw\endcsname{\color{white}}%
      \expandafter\def\csname LTb\endcsname{\color{black}}%
      \expandafter\def\csname LTa\endcsname{\color{black}}%
      \expandafter\def\csname LT0\endcsname{\color{black}}%
      \expandafter\def\csname LT1\endcsname{\color{black}}%
      \expandafter\def\csname LT2\endcsname{\color{black}}%
      \expandafter\def\csname LT3\endcsname{\color{black}}%
      \expandafter\def\csname LT4\endcsname{\color{black}}%
      \expandafter\def\csname LT5\endcsname{\color{black}}%
      \expandafter\def\csname LT6\endcsname{\color{black}}%
      \expandafter\def\csname LT7\endcsname{\color{black}}%
      \expandafter\def\csname LT8\endcsname{\color{black}}%
    \fi
  \fi
  \setlength{\unitlength}{0.0500bp}%
  \begin{picture}(3960.00,2880.00)%
    \gplgaddtomacro\gplbacktext{%
      \csname LTb\endcsname%
      \put(780,640){\makebox(0,0)[r]{\strut{}\fnt $0$}}%
      \put(780,913){\makebox(0,0)[r]{\strut{}\fnt $0.2$}}%
      \put(780,1186){\makebox(0,0)[r]{\strut{}\fnt $0.4$}}%
      \put(780,1460){\makebox(0,0)[r]{\strut{}\fnt $0.6$}}%
      \put(780,1733){\makebox(0,0)[r]{\strut{}\fnt $0.8$}}%
      \put(780,2006){\makebox(0,0)[r]{\strut{}\fnt $1$}}%
      \put(966,440){\makebox(0,0){\strut{}\fnt $1$}}%
      \put(2208,440){\makebox(0,0){\strut{}\fnt $16$}}%
      \put(3533,440){\makebox(0,0){\strut{}\fnt $32$}}%
      \put(320,1459){\rotatebox{-270}{\makebox(0,0){\strut{}\fnt $\lambda_k$}}}%
      \put(2249,140){\makebox(0,0){\strut{}\fnt $k$}}%
      \put(2249,2579){\makebox(0,0){\strut{}\fnt (a)}}%
    }%
    \gplgaddtomacro\gplfronttext{%
    }%
    \gplbacktext
    \put(0,0){\includegraphics{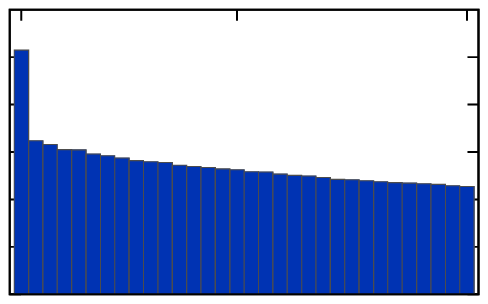}}%
    \gplfronttext
  \end{picture}%
\endgroup

%% file: clustering_coords_gplt.tex
% GNUPLOT: LaTeX picture with Postscript
\begingroup
\newcommand{\fnt}[0]{\footnotesize}
  \makeatletter
  \providecommand\color[2][]{%
    \GenericError{(gnuplot) \space\space\space\@spaces}{%
      Package color not loaded in conjunction with
      terminal option `colourtext'%
    }{See the gnuplot documentation for explanation.%
    }{Either use 'blacktext' in gnuplot or load the package
      color.sty in LaTeX.}%
    \renewcommand\color[2][]{}%
  }%
  \providecommand\includegraphics[2][]{%
    \GenericError{(gnuplot) \space\space\space\@spaces}{%
      Package graphicx or graphics not loaded%
    }{See the gnuplot documentation for explanation.%
    }{The gnuplot epslatex terminal needs graphicx.sty or graphics.sty.}%
    \renewcommand\includegraphics[2][]{}%
  }%
  \providecommand\rotatebox[2]{#2}%
  \@ifundefined{ifGPcolor}{%
    \newif\ifGPcolor
    \GPcolortrue
  }{}%
  \@ifundefined{ifGPblacktext}{%
    \newif\ifGPblacktext
    \GPblacktextfalse
  }{}%
  % define a \g@addto@macro without @ in the name:
  \let\gplgaddtomacro\g@addto@macro
  % define empty templates for all commands taking text:
  \gdef\gplbacktext{}%
  \gdef\gplfronttext{}%
  \makeatother
  \ifGPblacktext
    % no textcolor at all
    \def\colorrgb#1{}%
    \def\colorgray#1{}%
  \else
    % gray or color?
    \ifGPcolor
      \def\colorrgb#1{\color[rgb]{#1}}%
      \def\colorgray#1{\color[gray]{#1}}%
      \expandafter\def\csname LTw\endcsname{\color{white}}%
      \expandafter\def\csname LTb\endcsname{\color{black}}%
      \expandafter\def\csname LTa\endcsname{\color{black}}%
      \expandafter\def\csname LT0\endcsname{\color[rgb]{1,0,0}}%
      \expandafter\def\csname LT1\endcsname{\color[rgb]{0,1,0}}%
      \expandafter\def\csname LT2\endcsname{\color[rgb]{0,0,1}}%
      \expandafter\def\csname LT3\endcsname{\color[rgb]{1,0,1}}%
      \expandafter\def\csname LT4\endcsname{\color[rgb]{0,1,1}}%
      \expandafter\def\csname LT5\endcsname{\color[rgb]{1,1,0}}%
      \expandafter\def\csname LT6\endcsname{\color[rgb]{0,0,0}}%
      \expandafter\def\csname LT7\endcsname{\color[rgb]{1,0.3,0}}%
      \expandafter\def\csname LT8\endcsname{\color[rgb]{0.5,0.5,0.5}}%
    \else
      % gray
      \def\colorrgb#1{\color{black}}%
      \def\colorgray#1{\color[gray]{#1}}%
      \expandafter\def\csname LTw\endcsname{\color{white}}%
      \expandafter\def\csname LTb\endcsname{\color{black}}%
      \expandafter\def\csname LTa\endcsname{\color{black}}%
      \expandafter\def\csname LT0\endcsname{\color{black}}%
      \expandafter\def\csname LT1\endcsname{\color{black}}%
      \expandafter\def\csname LT2\endcsname{\color{black}}%
      \expandafter\def\csname LT3\endcsname{\color{black}}%
      \expandafter\def\csname LT4\endcsname{\color{black}}%
      \expandafter\def\csname LT5\endcsname{\color{black}}%
      \expandafter\def\csname LT6\endcsname{\color{black}}%
      \expandafter\def\csname LT7\endcsname{\color{black}}%
      \expandafter\def\csname LT8\endcsname{\color{black}}%
    \fi
  \fi
  \setlength{\unitlength}{0.0500bp}%
  \begin{picture}(3960.00,2880.00)%
    \gplgaddtomacro\gplbacktext{%
      \csname LTb\endcsname%
      \put(780,640){\makebox(0,0)[r]{\strut{}\fnt $0$}}%
      \put(780,1050){\makebox(0,0)[r]{\strut{}\fnt $100$}}%
      \put(780,1460){\makebox(0,0)[r]{\strut{}\fnt $200$}}%
      \put(780,1869){\makebox(0,0)[r]{\strut{}\fnt $300$}}%
      \put(780,2279){\makebox(0,0)[r]{\strut{}\fnt $400$}}%
      \put(900,440){\makebox(0,0){\strut{}\fnt $-2.5$}}%
      \put(1440,440){\makebox(0,0){\strut{}\fnt $-1.5$}}%
      \put(1980,440){\makebox(0,0){\strut{}\fnt $-0.5$}}%
      \put(2519,440){\makebox(0,0){\strut{}\fnt $0.5$}}%
      \put(3059,440){\makebox(0,0){\strut{}\fnt $1.5$}}%
      \put(3599,440){\makebox(0,0){\strut{}\fnt $2.5$}}%
      \put(320,1459){\rotatebox{-270}{\makebox(0,0){\strut{}\fnt Count}}}%
      \put(2249,140){\makebox(0,0){\strut{}\fnt $\coordest_s^{(1)}$}}%
      \put(2249,2579){\makebox(0,0){\strut{}\fnt (b)}}%
    }%
    \gplgaddtomacro\gplfronttext{%
    }%
    \gplbacktext
    \put(0,0){\includegraphics{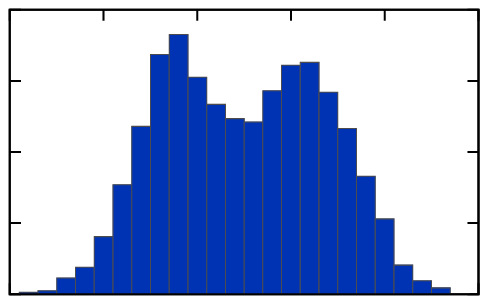}}%
    \gplfronttext
  \end{picture}%
\endgroup

%% file: clustering_accuracies_gplt.tex
% GNUPLOT: LaTeX picture with Postscript
\begingroup
\newcommand{\fnt}[0]{\footnotesize}
  \makeatletter
  \providecommand\color[2][]{%
    \GenericError{(gnuplot) \space\space\space\@spaces}{%
      Package color not loaded in conjunction with
      terminal option `colourtext'%
    }{See the gnuplot documentation for explanation.%
    }{Either use 'blacktext' in gnuplot or load the package
      color.sty in LaTeX.}%
    \renewcommand\color[2][]{}%
  }%
  \providecommand\includegraphics[2][]{%
    \GenericError{(gnuplot) \space\space\space\@spaces}{%
      Package graphicx or graphics not loaded%
    }{See the gnuplot documentation for explanation.%
    }{The gnuplot epslatex terminal needs graphicx.sty or graphics.sty.}%
    \renewcommand\includegraphics[2][]{}%
  }%
  \providecommand\rotatebox[2]{#2}%
  \@ifundefined{ifGPcolor}{%
    \newif\ifGPcolor
    \GPcolortrue
  }{}%
  \@ifundefined{ifGPblacktext}{%
    \newif\ifGPblacktext
    \GPblacktextfalse
  }{}%
  % define a \g@addto@macro without @ in the name:
  \let\gplgaddtomacro\g@addto@macro
  % define empty templates for all commands taking text:
  \gdef\gplbacktext{}%
  \gdef\gplfronttext{}%
  \makeatother
  \ifGPblacktext
    % no textcolor at all
    \def\colorrgb#1{}%
    \def\colorgray#1{}%
  \else
    % gray or color?
    \ifGPcolor
      \def\colorrgb#1{\color[rgb]{#1}}%
      \def\colorgray#1{\color[gray]{#1}}%
      \expandafter\def\csname LTw\endcsname{\color{white}}%
      \expandafter\def\csname LTb\endcsname{\color{black}}%
      \expandafter\def\csname LTa\endcsname{\color{black}}%
      \expandafter\def\csname LT0\endcsname{\color[rgb]{1,0,0}}%
      \expandafter\def\csname LT1\endcsname{\color[rgb]{0,1,0}}%
      \expandafter\def\csname LT2\endcsname{\color[rgb]{0,0,1}}%
      \expandafter\def\csname LT3\endcsname{\color[rgb]{1,0,1}}%
      \expandafter\def\csname LT4\endcsname{\color[rgb]{0,1,1}}%
      \expandafter\def\csname LT5\endcsname{\color[rgb]{1,1,0}}%
      \expandafter\def\csname LT6\endcsname{\color[rgb]{0,0,0}}%
      \expandafter\def\csname LT7\endcsname{\color[rgb]{1,0.3,0}}%
      \expandafter\def\csname LT8\endcsname{\color[rgb]{0.5,0.5,0.5}}%
    \else
      % gray
      \def\colorrgb#1{\color{black}}%
      \def\colorgray#1{\color[gray]{#1}}%
      \expandafter\def\csname LTw\endcsname{\color{white}}%
      \expandafter\def\csname LTb\endcsname{\color{black}}%
      \expandafter\def\csname LTa\endcsname{\color{black}}%
      \expandafter\def\csname LT0\endcsname{\color{black}}%
      \expandafter\def\csname LT1\endcsname{\color{black}}%
      \expandafter\def\csname LT2\endcsname{\color{black}}%
      \expandafter\def\csname LT3\endcsname{\color{black}}%
      \expandafter\def\csname LT4\endcsname{\color{black}}%
      \expandafter\def\csname LT5\endcsname{\color{black}}%
      \expandafter\def\csname LT6\endcsname{\color{black}}%
      \expandafter\def\csname LT7\endcsname{\color{black}}%
      \expandafter\def\csname LT8\endcsname{\color{black}}%
    \fi
  \fi
  \setlength{\unitlength}{0.0500bp}%
  \begin{picture}(3960.00,2880.00)%
    \gplgaddtomacro\gplbacktext{%
      \csname LTb\endcsname%
      \put(780,640){\makebox(0,0)[r]{\strut{}\fnt $0$}}%
      \put(780,968){\makebox(0,0)[r]{\strut{}\fnt $0.2$}}%
      \put(780,1296){\makebox(0,0)[r]{\strut{}\fnt $0.4$}}%
      \put(780,1623){\makebox(0,0)[r]{\strut{}\fnt $0.6$}}%
      \put(780,1951){\makebox(0,0)[r]{\strut{}\fnt $0.8$}}%
      \put(780,2279){\makebox(0,0)[r]{\strut{}\fnt $1$}}%
      \put(900,440){\makebox(0,0){\strut{}\fnt $0$}}%
      \put(1575,440){\makebox(0,0){\strut{}\fnt $0.01$}}%
      \put(2250,440){\makebox(0,0){\strut{}\fnt $0.02$}}%
      \put(2924,440){\makebox(0,0){\strut{}\fnt $0.03$}}%
      \put(3599,440){\makebox(0,0){\strut{}\fnt $0.04$}}%
      \put(320,1459){\rotatebox{-270}{\makebox(0,0){\strut{}\fnt Accuracy}}}%
      \put(2249,140){\makebox(0,0){\strut{}\fnt $\SNRh$}}%
      \put(2249,2579){\makebox(0,0){\strut{}\fnt (c)}}%
    }%
    \gplgaddtomacro\gplfronttext{%
    }%
    \gplbacktext
    \put(0,0){\includegraphics{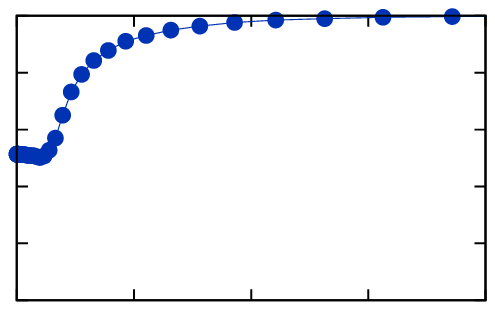}}%
    \gplfronttext
  \end{picture}%
\endgroup

%% file: clustering_errors_gplt.tex
% GNUPLOT: LaTeX picture with Postscript
\begingroup
\newcommand{\fnt}[0]{\footnotesize}
  \makeatletter
  \providecommand\color[2][]{%
    \GenericError{(gnuplot) \space\space\space\@spaces}{%
      Package color not loaded in conjunction with
      terminal option `colourtext'%
    }{See the gnuplot documentation for explanation.%
    }{Either use 'blacktext' in gnuplot or load the package
      color.sty in LaTeX.}%
    \renewcommand\color[2][]{}%
  }%
  \providecommand\includegraphics[2][]{%
    \GenericError{(gnuplot) \space\space\space\@spaces}{%
      Package graphicx or graphics not loaded%
    }{See the gnuplot documentation for explanation.%
    }{The gnuplot epslatex terminal needs graphicx.sty or graphics.sty.}%
    \renewcommand\includegraphics[2][]{}%
  }%
  \providecommand\rotatebox[2]{#2}%
  \@ifundefined{ifGPcolor}{%
    \newif\ifGPcolor
    \GPcolortrue
  }{}%
  \@ifundefined{ifGPblacktext}{%
    \newif\ifGPblacktext
    \GPblacktextfalse
  }{}%
  % define a \g@addto@macro without @ in the name:
  \let\gplgaddtomacro\g@addto@macro
  % define empty templates for all commands taking text:
  \gdef\gplbacktext{}%
  \gdef\gplfronttext{}%
  \makeatother
  \ifGPblacktext
    % no textcolor at all
    \def\colorrgb#1{}%
    \def\colorgray#1{}%
  \else
    % gray or color?
    \ifGPcolor
      \def\colorrgb#1{\color[rgb]{#1}}%
      \def\colorgray#1{\color[gray]{#1}}%
      \expandafter\def\csname LTw\endcsname{\color{white}}%
      \expandafter\def\csname LTb\endcsname{\color{black}}%
      \expandafter\def\csname LTa\endcsname{\color{black}}%
      \expandafter\def\csname LT0\endcsname{\color[rgb]{1,0,0}}%
      \expandafter\def\csname LT1\endcsname{\color[rgb]{0,1,0}}%
      \expandafter\def\csname LT2\endcsname{\color[rgb]{0,0,1}}%
      \expandafter\def\csname LT3\endcsname{\color[rgb]{1,0,1}}%
      \expandafter\def\csname LT4\endcsname{\color[rgb]{0,1,1}}%
      \expandafter\def\csname LT5\endcsname{\color[rgb]{1,1,0}}%
      \expandafter\def\csname LT6\endcsname{\color[rgb]{0,0,0}}%
      \expandafter\def\csname LT7\endcsname{\color[rgb]{1,0.3,0}}%
      \expandafter\def\csname LT8\endcsname{\color[rgb]{0.5,0.5,0.5}}%
    \else
      % gray
      \def\colorrgb#1{\color{black}}%
      \def\colorgray#1{\color[gray]{#1}}%
      \expandafter\def\csname LTw\endcsname{\color{white}}%
      \expandafter\def\csname LTb\endcsname{\color{black}}%
      \expandafter\def\csname LTa\endcsname{\color{black}}%
      \expandafter\def\csname LT0\endcsname{\color{black}}%
      \expandafter\def\csname LT1\endcsname{\color{black}}%
      \expandafter\def\csname LT2\endcsname{\color{black}}%
      \expandafter\def\csname LT3\endcsname{\color{black}}%
      \expandafter\def\csname LT4\endcsname{\color{black}}%
      \expandafter\def\csname LT5\endcsname{\color{black}}%
      \expandafter\def\csname LT6\endcsname{\color{black}}%
      \expandafter\def\csname LT7\endcsname{\color{black}}%
      \expandafter\def\csname LT8\endcsname{\color{black}}%
    \fi
  \fi
  \setlength{\unitlength}{0.0500bp}%
  \begin{picture}(3960.00,2880.00)%
    \gplgaddtomacro\gplbacktext{%
      \csname LTb\endcsname%
      \put(780,640){\makebox(0,0)[r]{\strut{}\fnt $0$}}%
      \put(780,968){\makebox(0,0)[r]{\strut{}\fnt $0.2$}}%
      \put(780,1296){\makebox(0,0)[r]{\strut{}\fnt $0.4$}}%
      \put(780,1623){\makebox(0,0)[r]{\strut{}\fnt $0.6$}}%
      \put(780,1951){\makebox(0,0)[r]{\strut{}\fnt $0.8$}}%
      \put(780,2279){\makebox(0,0)[r]{\strut{}\fnt $1$}}%
      \put(900,440){\makebox(0,0){\strut{}\fnt $0$}}%
      \put(1575,440){\makebox(0,0){\strut{}\fnt $0.01$}}%
      \put(2250,440){\makebox(0,0){\strut{}\fnt $0.02$}}%
      \put(2924,440){\makebox(0,0){\strut{}\fnt $0.03$}}%
      \put(3599,440){\makebox(0,0){\strut{}\fnt $0.04$}}%
      \put(320,1459){\rotatebox{-270}{\makebox(0,0){\strut{}\fnt Volume NRMSE}}}%
      \put(2249,140){\makebox(0,0){\strut{}\fnt $\SNRh$}}%
      \put(2249,2579){\makebox(0,0){\strut{}\fnt (d)}}%
    }%
    \gplgaddtomacro\gplfronttext{%
    }%
    \gplbacktext
    \put(0,0){\includegraphics{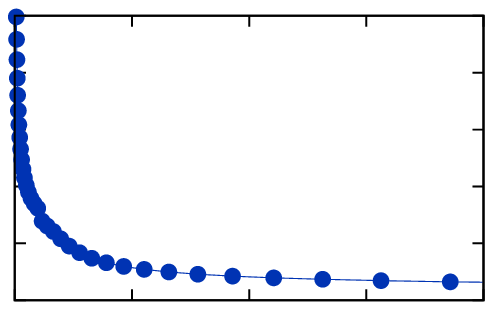}}%
    \gplfronttext
  \end{picture}%
\endgroup

%% file: manifold_corrs_gplt.tex
% GNUPLOT: LaTeX picture with Postscript
\begingroup
\newcommand{\fnt}[0]{\footnotesize}
  \makeatletter
  \providecommand\color[2][]{%
    \GenericError{(gnuplot) \space\space\space\@spaces}{%
      Package color not loaded in conjunction with
      terminal option `colourtext'%
    }{See the gnuplot documentation for explanation.%
    }{Either use 'blacktext' in gnuplot or load the package
      color.sty in LaTeX.}%
    \renewcommand\color[2][]{}%
  }%
  \providecommand\includegraphics[2][]{%
    \GenericError{(gnuplot) \space\space\space\@spaces}{%
      Package graphicx or graphics not loaded%
    }{See the gnuplot documentation for explanation.%
    }{The gnuplot epslatex terminal needs graphicx.sty or graphics.sty.}%
    \renewcommand\includegraphics[2][]{}%
  }%
  \providecommand\rotatebox[2]{#2}%
  \@ifundefined{ifGPcolor}{%
    \newif\ifGPcolor
    \GPcolortrue
  }{}%
  \@ifundefined{ifGPblacktext}{%
    \newif\ifGPblacktext
    \GPblacktextfalse
  }{}%
  % define a \g@addto@macro without @ in the name:
  \let\gplgaddtomacro\g@addto@macro
  % define empty templates for all commands taking text:
  \gdef\gplbacktext{}%
  \gdef\gplfronttext{}%
  \makeatother
  \ifGPblacktext
    % no textcolor at all
    \def\colorrgb#1{}%
    \def\colorgray#1{}%
  \else
    % gray or color?
    \ifGPcolor
      \def\colorrgb#1{\color[rgb]{#1}}%
      \def\colorgray#1{\color[gray]{#1}}%
      \expandafter\def\csname LTw\endcsname{\color{white}}%
      \expandafter\def\csname LTb\endcsname{\color{black}}%
      \expandafter\def\csname LTa\endcsname{\color{black}}%
      \expandafter\def\csname LT0\endcsname{\color[rgb]{1,0,0}}%
      \expandafter\def\csname LT1\endcsname{\color[rgb]{0,1,0}}%
      \expandafter\def\csname LT2\endcsname{\color[rgb]{0,0,1}}%
      \expandafter\def\csname LT3\endcsname{\color[rgb]{1,0,1}}%
      \expandafter\def\csname LT4\endcsname{\color[rgb]{0,1,1}}%
      \expandafter\def\csname LT5\endcsname{\color[rgb]{1,1,0}}%
      \expandafter\def\csname LT6\endcsname{\color[rgb]{0,0,0}}%
      \expandafter\def\csname LT7\endcsname{\color[rgb]{1,0.3,0}}%
      \expandafter\def\csname LT8\endcsname{\color[rgb]{0.5,0.5,0.5}}%
    \else
      % gray
      \def\colorrgb#1{\color{black}}%
      \def\colorgray#1{\color[gray]{#1}}%
      \expandafter\def\csname LTw\endcsname{\color{white}}%
      \expandafter\def\csname LTb\endcsname{\color{black}}%
      \expandafter\def\csname LTa\endcsname{\color{black}}%
      \expandafter\def\csname LT0\endcsname{\color{black}}%
      \expandafter\def\csname LT1\endcsname{\color{black}}%
      \expandafter\def\csname LT2\endcsname{\color{black}}%
      \expandafter\def\csname LT3\endcsname{\color{black}}%
      \expandafter\def\csname LT4\endcsname{\color{black}}%
      \expandafter\def\csname LT5\endcsname{\color{black}}%
      \expandafter\def\csname LT6\endcsname{\color{black}}%
      \expandafter\def\csname LT7\endcsname{\color{black}}%
      \expandafter\def\csname LT8\endcsname{\color{black}}%
    \fi
  \fi
  \setlength{\unitlength}{0.0500bp}%
  \begin{picture}(3960.00,2880.00)%
    \gplgaddtomacro\gplbacktext{%
      \csname LTb\endcsname%
      \put(780,640){\makebox(0,0)[r]{\strut{}\fnt $0$}}%
      \put(780,968){\makebox(0,0)[r]{\strut{}\fnt $0.2$}}%
      \put(780,1296){\makebox(0,0)[r]{\strut{}\fnt $0.4$}}%
      \put(780,1623){\makebox(0,0)[r]{\strut{}\fnt $0.6$}}%
      \put(780,1951){\makebox(0,0)[r]{\strut{}\fnt $0.8$}}%
      \put(780,2279){\makebox(0,0)[r]{\strut{}\fnt $1$}}%
      \put(900,440){\makebox(0,0){\strut{}\fnt $0$}}%
      \put(1800,440){\makebox(0,0){\strut{}\fnt $0.1$}}%
      \put(2699,440){\makebox(0,0){\strut{}\fnt $0.2$}}%
      \put(3599,440){\makebox(0,0){\strut{}\fnt $0.3$}}%
      \put(320,1459){\rotatebox{-270}{\makebox(0,0){\strut{}\fnt $\cos \theta_\max$}}}%
      \put(2249,140){\makebox(0,0){\strut{}\fnt $\SNRh$}}%
      \put(2249,2579){\makebox(0,0){\strut{}\fnt (b)}}%
    }%
    \gplgaddtomacro\gplfronttext{%
    }%
    \gplbacktext
    \put(0,0){\includegraphics{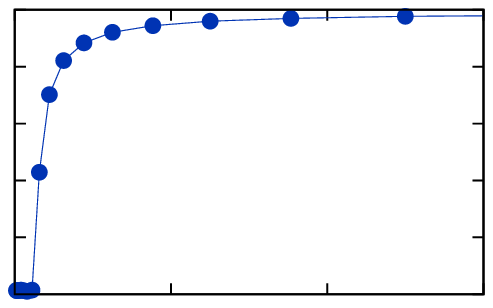}}%
    \gplfronttext
  \end{picture}%
\endgroup

%% file: manifold_coord_errs_gplt.tex
% GNUPLOT: LaTeX picture with Postscript
\begingroup
\newcommand{\fnt}[0]{\footnotesize}
  \makeatletter
  \providecommand\color[2][]{%
    \GenericError{(gnuplot) \space\space\space\@spaces}{%
      Package color not loaded in conjunction with
      terminal option `colourtext'%
    }{See the gnuplot documentation for explanation.%
    }{Either use 'blacktext' in gnuplot or load the package
      color.sty in LaTeX.}%
    \renewcommand\color[2][]{}%
  }%
  \providecommand\includegraphics[2][]{%
    \GenericError{(gnuplot) \space\space\space\@spaces}{%
      Package graphicx or graphics not loaded%
    }{See the gnuplot documentation for explanation.%
    }{The gnuplot epslatex terminal needs graphicx.sty or graphics.sty.}%
    \renewcommand\includegraphics[2][]{}%
  }%
  \providecommand\rotatebox[2]{#2}%
  \@ifundefined{ifGPcolor}{%
    \newif\ifGPcolor
    \GPcolortrue
  }{}%
  \@ifundefined{ifGPblacktext}{%
    \newif\ifGPblacktext
    \GPblacktextfalse
  }{}%
  % define a \g@addto@macro without @ in the name:
  \let\gplgaddtomacro\g@addto@macro
  % define empty templates for all commands taking text:
  \gdef\gplbacktext{}%
  \gdef\gplfronttext{}%
  \makeatother
  \ifGPblacktext
    % no textcolor at all
    \def\colorrgb#1{}%
    \def\colorgray#1{}%
  \else
    % gray or color?
    \ifGPcolor
      \def\colorrgb#1{\color[rgb]{#1}}%
      \def\colorgray#1{\color[gray]{#1}}%
      \expandafter\def\csname LTw\endcsname{\color{white}}%
      \expandafter\def\csname LTb\endcsname{\color{black}}%
      \expandafter\def\csname LTa\endcsname{\color{black}}%
      \expandafter\def\csname LT0\endcsname{\color[rgb]{1,0,0}}%
      \expandafter\def\csname LT1\endcsname{\color[rgb]{0,1,0}}%
      \expandafter\def\csname LT2\endcsname{\color[rgb]{0,0,1}}%
      \expandafter\def\csname LT3\endcsname{\color[rgb]{1,0,1}}%
      \expandafter\def\csname LT4\endcsname{\color[rgb]{0,1,1}}%
      \expandafter\def\csname LT5\endcsname{\color[rgb]{1,1,0}}%
      \expandafter\def\csname LT6\endcsname{\color[rgb]{0,0,0}}%
      \expandafter\def\csname LT7\endcsname{\color[rgb]{1,0.3,0}}%
      \expandafter\def\csname LT8\endcsname{\color[rgb]{0.5,0.5,0.5}}%
    \else
      % gray
      \def\colorrgb#1{\color{black}}%
      \def\colorgray#1{\color[gray]{#1}}%
      \expandafter\def\csname LTw\endcsname{\color{white}}%
      \expandafter\def\csname LTb\endcsname{\color{black}}%
      \expandafter\def\csname LTa\endcsname{\color{black}}%
      \expandafter\def\csname LT0\endcsname{\color{black}}%
      \expandafter\def\csname LT1\endcsname{\color{black}}%
      \expandafter\def\csname LT2\endcsname{\color{black}}%
      \expandafter\def\csname LT3\endcsname{\color{black}}%
      \expandafter\def\csname LT4\endcsname{\color{black}}%
      \expandafter\def\csname LT5\endcsname{\color{black}}%
      \expandafter\def\csname LT6\endcsname{\color{black}}%
      \expandafter\def\csname LT7\endcsname{\color{black}}%
      \expandafter\def\csname LT8\endcsname{\color{black}}%
    \fi
  \fi
  \setlength{\unitlength}{0.0500bp}%
  \begin{picture}(3960.00,2880.00)%
    \gplgaddtomacro\gplbacktext{%
      \csname LTb\endcsname%
      \put(780,640){\makebox(0,0)[r]{\strut{}\fnt $0$}}%
      \put(780,1050){\makebox(0,0)[r]{\strut{}\fnt $0.5$}}%
      \put(780,1460){\makebox(0,0)[r]{\strut{}\fnt $1$}}%
      \put(780,1869){\makebox(0,0)[r]{\strut{}\fnt $1.5$}}%
      \put(780,2279){\makebox(0,0)[r]{\strut{}\fnt $2$}}%
      \put(900,440){\makebox(0,0){\strut{}\fnt $0$}}%
      \put(1575,440){\makebox(0,0){\strut{}\fnt $0.05$}}%
      \put(2250,440){\makebox(0,0){\strut{}\fnt $0.1$}}%
      \put(2924,440){\makebox(0,0){\strut{}\fnt $0.15$}}%
      \put(3599,440){\makebox(0,0){\strut{}\fnt $0.2$}}%
      \put(320,1459){\rotatebox{-270}{\makebox(0,0){\strut{}\fnt $\mathrm{NRMSE}$}}}%
      \put(2249,140){\makebox(0,0){\strut{}\fnt $\SNRh$}}%
      \put(2249,2579){\makebox(0,0){\strut{}\fnt (e)}}%
    }%
    \gplgaddtomacro\gplfronttext{%
    }%
    \gplbacktext
    \put(0,0){\includegraphics{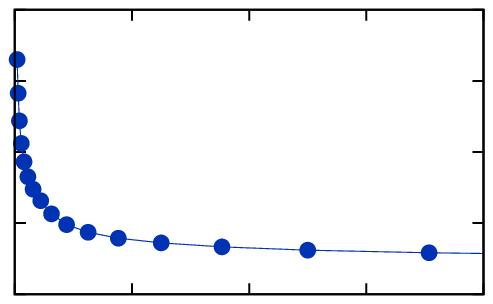}}%
    \gplfronttext
  \end{picture}%
\endgroup

%% file: An_conv_baseline_gplt.tex
% GNUPLOT: LaTeX picture with Postscript
\begingroup
\newcommand{\fnt}[0]{\footnotesize}
  \makeatletter
  \providecommand\color[2][]{%
    \GenericError{(gnuplot) \space\space\space\@spaces}{%
      Package color not loaded in conjunction with
      terminal option `colourtext'%
    }{See the gnuplot documentation for explanation.%
    }{Either use 'blacktext' in gnuplot or load the package
      color.sty in LaTeX.}%
    \renewcommand\color[2][]{}%
  }%
  \providecommand\includegraphics[2][]{%
    \GenericError{(gnuplot) \space\space\space\@spaces}{%
      Package graphicx or graphics not loaded%
    }{See the gnuplot documentation for explanation.%
    }{The gnuplot epslatex terminal needs graphicx.sty or graphics.sty.}%
    \renewcommand\includegraphics[2][]{}%
  }%
  \providecommand\rotatebox[2]{#2}%
  \@ifundefined{ifGPcolor}{%
    \newif\ifGPcolor
    \GPcolortrue
  }{}%
  \@ifundefined{ifGPblacktext}{%
    \newif\ifGPblacktext
    \GPblacktextfalse
  }{}%
  % define a \g@addto@macro without @ in the name:
  \let\gplgaddtomacro\g@addto@macro
  % define empty templates for all commands taking text:
  \gdef\gplbacktext{}%
  \gdef\gplfronttext{}%
  \makeatother
  \ifGPblacktext
    % no textcolor at all
    \def\colorrgb#1{}%
    \def\colorgray#1{}%
  \else
    % gray or color?
    \ifGPcolor
      \def\colorrgb#1{\color[rgb]{#1}}%
      \def\colorgray#1{\color[gray]{#1}}%
      \expandafter\def\csname LTw\endcsname{\color{white}}%
      \expandafter\def\csname LTb\endcsname{\color{black}}%
      \expandafter\def\csname LTa\endcsname{\color{black}}%
      \expandafter\def\csname LT0\endcsname{\color[rgb]{1,0,0}}%
      \expandafter\def\csname LT1\endcsname{\color[rgb]{0,1,0}}%
      \expandafter\def\csname LT2\endcsname{\color[rgb]{0,0,1}}%
      \expandafter\def\csname LT3\endcsname{\color[rgb]{1,0,1}}%
      \expandafter\def\csname LT4\endcsname{\color[rgb]{0,1,1}}%
      \expandafter\def\csname LT5\endcsname{\color[rgb]{1,1,0}}%
      \expandafter\def\csname LT6\endcsname{\color[rgb]{0,0,0}}%
      \expandafter\def\csname LT7\endcsname{\color[rgb]{1,0.3,0}}%
      \expandafter\def\csname LT8\endcsname{\color[rgb]{0.5,0.5,0.5}}%
    \else
      % gray
      \def\colorrgb#1{\color{black}}%
      \def\colorgray#1{\color[gray]{#1}}%
      \expandafter\def\csname LTw\endcsname{\color{white}}%
      \expandafter\def\csname LTb\endcsname{\color{black}}%
      \expandafter\def\csname LTa\endcsname{\color{black}}%
      \expandafter\def\csname LT0\endcsname{\color{black}}%
      \expandafter\def\csname LT1\endcsname{\color{black}}%
      \expandafter\def\csname LT2\endcsname{\color{black}}%
      \expandafter\def\csname LT3\endcsname{\color{black}}%
      \expandafter\def\csname LT4\endcsname{\color{black}}%
      \expandafter\def\csname LT5\endcsname{\color{black}}%
      \expandafter\def\csname LT6\endcsname{\color{black}}%
      \expandafter\def\csname LT7\endcsname{\color{black}}%
      \expandafter\def\csname LT8\endcsname{\color{black}}%
    \fi
  \fi
  \setlength{\unitlength}{0.0500bp}%
  \begin{picture}(3960.00,2880.00)%
    \gplgaddtomacro\gplbacktext{%
      \csname LTb\endcsname%
      \put(1020,640){\makebox(0,0)[r]{\strut{}\fnt $10^{-15}$}}%
      \put(1020,929){\makebox(0,0)[r]{\strut{}\fnt $10^{-12}$}}%
      \put(1020,1218){\makebox(0,0)[r]{\strut{}\fnt $10^{-9}$}}%
      \put(1020,1508){\makebox(0,0)[r]{\strut{}\fnt $10^{-6}$}}%
      \put(1020,1797){\makebox(0,0)[r]{\strut{}\fnt $10^{-3}$}}%
      \put(1020,2086){\makebox(0,0)[r]{\strut{}\fnt $10^{0}$}}%
      \put(1378,440){\makebox(0,0){\strut{}\fnt $4$}}%
      \put(1695,440){\makebox(0,0){\strut{}\fnt $8$}}%
      \put(2013,440){\makebox(0,0){\strut{}\fnt $12$}}%
      \put(2330,440){\makebox(0,0){\strut{}\fnt $16$}}%
      \put(2647,440){\makebox(0,0){\strut{}\fnt $20$}}%
      \put(2964,440){\makebox(0,0){\strut{}\fnt $24$}}%
      \put(3282,440){\makebox(0,0){\strut{}\fnt $28$}}%
      \put(3599,440){\makebox(0,0){\strut{}\fnt $32$}}%
      \put(320,1459){\rotatebox{-270}{\makebox(0,0){\strut{}\fnt $\|A_n(\mean_n^{(t)})-b_n\|/\|b_n\|$}}}%
      \put(2369,140){\makebox(0,0){\strut{}\fnt Iteration (t)}}%
      \put(2369,2579){\makebox(0,0){\strut{}\fnt (b)}}%
    }%
    \gplgaddtomacro\gplfronttext{%
      \csname LTb\endcsname%
      \put(1980,1103){\makebox(0,0)[r]{\strut{}\fnt standard}}%
      \csname LTb\endcsname%
      \put(1980,903){\makebox(0,0)[r]{\strut{}\fnt precond.}}%
    }%
    \gplbacktext
    \put(0,0){\includegraphics{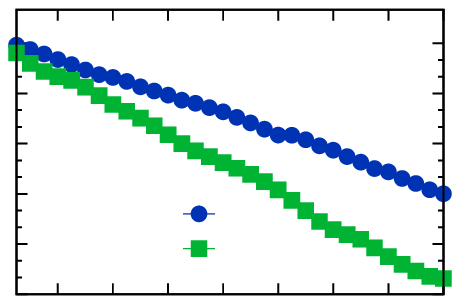}}%
    \gplfronttext
  \end{picture}%
\endgroup

%% file: Ln_conv_baseline_gplt.tex
% GNUPLOT: LaTeX picture with Postscript
\begingroup
\newcommand{\fnt}[0]{\footnotesize}
  \makeatletter
  \providecommand\color[2][]{%
    \GenericError{(gnuplot) \space\space\space\@spaces}{%
      Package color not loaded in conjunction with
      terminal option `colourtext'%
    }{See the gnuplot documentation for explanation.%
    }{Either use 'blacktext' in gnuplot or load the package
      color.sty in LaTeX.}%
    \renewcommand\color[2][]{}%
  }%
  \providecommand\includegraphics[2][]{%
    \GenericError{(gnuplot) \space\space\space\@spaces}{%
      Package graphicx or graphics not loaded%
    }{See the gnuplot documentation for explanation.%
    }{The gnuplot epslatex terminal needs graphicx.sty or graphics.sty.}%
    \renewcommand\includegraphics[2][]{}%
  }%
  \providecommand\rotatebox[2]{#2}%
  \@ifundefined{ifGPcolor}{%
    \newif\ifGPcolor
    \GPcolortrue
  }{}%
  \@ifundefined{ifGPblacktext}{%
    \newif\ifGPblacktext
    \GPblacktextfalse
  }{}%
  % define a \g@addto@macro without @ in the name:
  \let\gplgaddtomacro\g@addto@macro
  % define empty templates for all commands taking text:
  \gdef\gplbacktext{}%
  \gdef\gplfronttext{}%
  \makeatother
  \ifGPblacktext
    % no textcolor at all
    \def\colorrgb#1{}%
    \def\colorgray#1{}%
  \else
    % gray or color?
    \ifGPcolor
      \def\colorrgb#1{\color[rgb]{#1}}%
      \def\colorgray#1{\color[gray]{#1}}%
      \expandafter\def\csname LTw\endcsname{\color{white}}%
      \expandafter\def\csname LTb\endcsname{\color{black}}%
      \expandafter\def\csname LTa\endcsname{\color{black}}%
      \expandafter\def\csname LT0\endcsname{\color[rgb]{1,0,0}}%
      \expandafter\def\csname LT1\endcsname{\color[rgb]{0,1,0}}%
      \expandafter\def\csname LT2\endcsname{\color[rgb]{0,0,1}}%
      \expandafter\def\csname LT3\endcsname{\color[rgb]{1,0,1}}%
      \expandafter\def\csname LT4\endcsname{\color[rgb]{0,1,1}}%
      \expandafter\def\csname LT5\endcsname{\color[rgb]{1,1,0}}%
      \expandafter\def\csname LT6\endcsname{\color[rgb]{0,0,0}}%
      \expandafter\def\csname LT7\endcsname{\color[rgb]{1,0.3,0}}%
      \expandafter\def\csname LT8\endcsname{\color[rgb]{0.5,0.5,0.5}}%
    \else
      % gray
      \def\colorrgb#1{\color{black}}%
      \def\colorgray#1{\color[gray]{#1}}%
      \expandafter\def\csname LTw\endcsname{\color{white}}%
      \expandafter\def\csname LTb\endcsname{\color{black}}%
      \expandafter\def\csname LTa\endcsname{\color{black}}%
      \expandafter\def\csname LT0\endcsname{\color{black}}%
      \expandafter\def\csname LT1\endcsname{\color{black}}%
      \expandafter\def\csname LT2\endcsname{\color{black}}%
      \expandafter\def\csname LT3\endcsname{\color{black}}%
      \expandafter\def\csname LT4\endcsname{\color{black}}%
      \expandafter\def\csname LT5\endcsname{\color{black}}%
      \expandafter\def\csname LT6\endcsname{\color{black}}%
      \expandafter\def\csname LT7\endcsname{\color{black}}%
      \expandafter\def\csname LT8\endcsname{\color{black}}%
    \fi
  \fi
  \setlength{\unitlength}{0.0500bp}%
  \begin{picture}(3960.00,2880.00)%
    \gplgaddtomacro\gplbacktext{%
      \csname LTb\endcsname%
      \put(1020,640){\makebox(0,0)[r]{\strut{}\fnt $10^{-15}$}}%
      \put(1020,929){\makebox(0,0)[r]{\strut{}\fnt $10^{-12}$}}%
      \put(1020,1218){\makebox(0,0)[r]{\strut{}\fnt $10^{-9}$}}%
      \put(1020,1508){\makebox(0,0)[r]{\strut{}\fnt $10^{-6}$}}%
      \put(1020,1797){\makebox(0,0)[r]{\strut{}\fnt $10^{-3}$}}%
      \put(1020,2086){\makebox(0,0)[r]{\strut{}\fnt $10^{0}$}}%
      \put(1740,440){\makebox(0,0){\strut{}\fnt $32$}}%
      \put(2360,440){\makebox(0,0){\strut{}\fnt $64$}}%
      \put(2979,440){\makebox(0,0){\strut{}\fnt $96$}}%
      \put(3599,440){\makebox(0,0){\strut{}\fnt $128$}}%
      \put(320,1459){\rotatebox{-270}{\makebox(0,0){\strut{}\fnt $\|L_n(\Sigma_n^{(t)})-B_n\|_\frob/\|B_n\|_\frob$}}}%
      \put(2369,140){\makebox(0,0){\strut{}\fnt Iteration (t)}}%
      \put(2369,2579){\makebox(0,0){\strut{}\fnt (d)}}%
    }%
    \gplgaddtomacro\gplfronttext{%
      \csname LTb\endcsname%
      \put(1980,1103){\makebox(0,0)[r]{\strut{}\fnt standard}}%
      \csname LTb\endcsname%
      \put(1980,903){\makebox(0,0)[r]{\strut{}\fnt precond.}}%
    }%
    \gplbacktext
    \put(0,0){\includegraphics{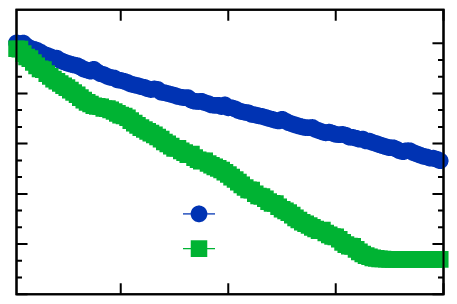}}%
    \gplfronttext
  \end{picture}%
\endgroup

%% file: Ln_conv_ctf_gplt.tex
% GNUPLOT: LaTeX picture with Postscript
\begingroup
\newcommand{\fnt}[0]{\footnotesize}
  \makeatletter
  \providecommand\color[2][]{%
    \GenericError{(gnuplot) \space\space\space\@spaces}{%
      Package color not loaded in conjunction with
      terminal option `colourtext'%
    }{See the gnuplot documentation for explanation.%
    }{Either use 'blacktext' in gnuplot or load the package
      color.sty in LaTeX.}%
    \renewcommand\color[2][]{}%
  }%
  \providecommand\includegraphics[2][]{%
    \GenericError{(gnuplot) \space\space\space\@spaces}{%
      Package graphicx or graphics not loaded%
    }{See the gnuplot documentation for explanation.%
    }{The gnuplot epslatex terminal needs graphicx.sty or graphics.sty.}%
    \renewcommand\includegraphics[2][]{}%
  }%
  \providecommand\rotatebox[2]{#2}%
  \@ifundefined{ifGPcolor}{%
    \newif\ifGPcolor
    \GPcolortrue
  }{}%
  \@ifundefined{ifGPblacktext}{%
    \newif\ifGPblacktext
    \GPblacktextfalse
  }{}%
  % define a \g@addto@macro without @ in the name:
  \let\gplgaddtomacro\g@addto@macro
  % define empty templates for all commands taking text:
  \gdef\gplbacktext{}%
  \gdef\gplfronttext{}%
  \makeatother
  \ifGPblacktext
    % no textcolor at all
    \def\colorrgb#1{}%
    \def\colorgray#1{}%
  \else
    % gray or color?
    \ifGPcolor
      \def\colorrgb#1{\color[rgb]{#1}}%
      \def\colorgray#1{\color[gray]{#1}}%
      \expandafter\def\csname LTw\endcsname{\color{white}}%
      \expandafter\def\csname LTb\endcsname{\color{black}}%
      \expandafter\def\csname LTa\endcsname{\color{black}}%
      \expandafter\def\csname LT0\endcsname{\color[rgb]{1,0,0}}%
      \expandafter\def\csname LT1\endcsname{\color[rgb]{0,1,0}}%
      \expandafter\def\csname LT2\endcsname{\color[rgb]{0,0,1}}%
      \expandafter\def\csname LT3\endcsname{\color[rgb]{1,0,1}}%
      \expandafter\def\csname LT4\endcsname{\color[rgb]{0,1,1}}%
      \expandafter\def\csname LT5\endcsname{\color[rgb]{1,1,0}}%
      \expandafter\def\csname LT6\endcsname{\color[rgb]{0,0,0}}%
      \expandafter\def\csname LT7\endcsname{\color[rgb]{1,0.3,0}}%
      \expandafter\def\csname LT8\endcsname{\color[rgb]{0.5,0.5,0.5}}%
    \else
      % gray
      \def\colorrgb#1{\color{black}}%
      \def\colorgray#1{\color[gray]{#1}}%
      \expandafter\def\csname LTw\endcsname{\color{white}}%
      \expandafter\def\csname LTb\endcsname{\color{black}}%
      \expandafter\def\csname LTa\endcsname{\color{black}}%
      \expandafter\def\csname LT0\endcsname{\color{black}}%
      \expandafter\def\csname LT1\endcsname{\color{black}}%
      \expandafter\def\csname LT2\endcsname{\color{black}}%
      \expandafter\def\csname LT3\endcsname{\color{black}}%
      \expandafter\def\csname LT4\endcsname{\color{black}}%
      \expandafter\def\csname LT5\endcsname{\color{black}}%
      \expandafter\def\csname LT6\endcsname{\color{black}}%
      \expandafter\def\csname LT7\endcsname{\color{black}}%
      \expandafter\def\csname LT8\endcsname{\color{black}}%
    \fi
  \fi
  \setlength{\unitlength}{0.0500bp}%
  \begin{picture}(3960.00,2880.00)%
    \gplgaddtomacro\gplbacktext{%
      \csname LTb\endcsname%
      \put(1020,640){\makebox(0,0)[r]{\strut{}\fnt $10^{-15}$}}%
      \put(1020,929){\makebox(0,0)[r]{\strut{}\fnt $10^{-12}$}}%
      \put(1020,1218){\makebox(0,0)[r]{\strut{}\fnt $10^{-9}$}}%
      \put(1020,1508){\makebox(0,0)[r]{\strut{}\fnt $10^{-6}$}}%
      \put(1020,1797){\makebox(0,0)[r]{\strut{}\fnt $10^{-3}$}}%
      \put(1020,2086){\makebox(0,0)[r]{\strut{}\fnt $10^{0}$}}%
      \put(1740,440){\makebox(0,0){\strut{}\fnt $32$}}%
      \put(2360,440){\makebox(0,0){\strut{}\fnt $64$}}%
      \put(2979,440){\makebox(0,0){\strut{}\fnt $96$}}%
      \put(3599,440){\makebox(0,0){\strut{}\fnt $128$}}%
      \put(320,1459){\rotatebox{-270}{\makebox(0,0){\strut{}\fnt $\|L_n(\Sigma_n^{(t)})-B_n\|_\frob/\|B_n\|_\frob$}}}%
      \put(2369,140){\makebox(0,0){\strut{}\fnt Iteration (t)}}%
      \put(2369,2579){\makebox(0,0){\strut{}\fnt (d)}}%
    }%
    \gplgaddtomacro\gplfronttext{%
      \csname LTb\endcsname%
      \put(1980,1103){\makebox(0,0)[r]{\strut{}\fnt standard}}%
      \csname LTb\endcsname%
      \put(1980,903){\makebox(0,0)[r]{\strut{}\fnt precond.}}%
    }%
    \gplbacktext
    \put(0,0){\includegraphics{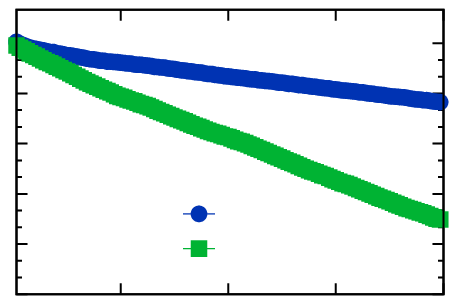}}%
    \gplfronttext
  \end{picture}%
\endgroup

%% file: Ln_conv_ctf_nu_gplt.tex
% GNUPLOT: LaTeX picture with Postscript
\begingroup
\newcommand{\fnt}[0]{\footnotesize}
  \makeatletter
  \providecommand\color[2][]{%
    \GenericError{(gnuplot) \space\space\space\@spaces}{%
      Package color not loaded in conjunction with
      terminal option `colourtext'%
    }{See the gnuplot documentation for explanation.%
    }{Either use 'blacktext' in gnuplot or load the package
      color.sty in LaTeX.}%
    \renewcommand\color[2][]{}%
  }%
  \providecommand\includegraphics[2][]{%
    \GenericError{(gnuplot) \space\space\space\@spaces}{%
      Package graphicx or graphics not loaded%
    }{See the gnuplot documentation for explanation.%
    }{The gnuplot epslatex terminal needs graphicx.sty or graphics.sty.}%
    \renewcommand\includegraphics[2][]{}%
  }%
  \providecommand\rotatebox[2]{#2}%
  \@ifundefined{ifGPcolor}{%
    \newif\ifGPcolor
    \GPcolortrue
  }{}%
  \@ifundefined{ifGPblacktext}{%
    \newif\ifGPblacktext
    \GPblacktextfalse
  }{}%
  % define a \g@addto@macro without @ in the name:
  \let\gplgaddtomacro\g@addto@macro
  % define empty templates for all commands taking text:
  \gdef\gplbacktext{}%
  \gdef\gplfronttext{}%
  \makeatother
  \ifGPblacktext
    % no textcolor at all
    \def\colorrgb#1{}%
    \def\colorgray#1{}%
  \else
    % gray or color?
    \ifGPcolor
      \def\colorrgb#1{\color[rgb]{#1}}%
      \def\colorgray#1{\color[gray]{#1}}%
      \expandafter\def\csname LTw\endcsname{\color{white}}%
      \expandafter\def\csname LTb\endcsname{\color{black}}%
      \expandafter\def\csname LTa\endcsname{\color{black}}%
      \expandafter\def\csname LT0\endcsname{\color[rgb]{1,0,0}}%
      \expandafter\def\csname LT1\endcsname{\color[rgb]{0,1,0}}%
      \expandafter\def\csname LT2\endcsname{\color[rgb]{0,0,1}}%
      \expandafter\def\csname LT3\endcsname{\color[rgb]{1,0,1}}%
      \expandafter\def\csname LT4\endcsname{\color[rgb]{0,1,1}}%
      \expandafter\def\csname LT5\endcsname{\color[rgb]{1,1,0}}%
      \expandafter\def\csname LT6\endcsname{\color[rgb]{0,0,0}}%
      \expandafter\def\csname LT7\endcsname{\color[rgb]{1,0.3,0}}%
      \expandafter\def\csname LT8\endcsname{\color[rgb]{0.5,0.5,0.5}}%
    \else
      % gray
      \def\colorrgb#1{\color{black}}%
      \def\colorgray#1{\color[gray]{#1}}%
      \expandafter\def\csname LTw\endcsname{\color{white}}%
      \expandafter\def\csname LTb\endcsname{\color{black}}%
      \expandafter\def\csname LTa\endcsname{\color{black}}%
      \expandafter\def\csname LT0\endcsname{\color{black}}%
      \expandafter\def\csname LT1\endcsname{\color{black}}%
      \expandafter\def\csname LT2\endcsname{\color{black}}%
      \expandafter\def\csname LT3\endcsname{\color{black}}%
      \expandafter\def\csname LT4\endcsname{\color{black}}%
      \expandafter\def\csname LT5\endcsname{\color{black}}%
      \expandafter\def\csname LT6\endcsname{\color{black}}%
      \expandafter\def\csname LT7\endcsname{\color{black}}%
      \expandafter\def\csname LT8\endcsname{\color{black}}%
    \fi
  \fi
  \setlength{\unitlength}{0.0500bp}%
  \begin{picture}(3960.00,2880.00)%
    \gplgaddtomacro\gplbacktext{%
      \csname LTb\endcsname%
      \put(1020,640){\makebox(0,0)[r]{\strut{}\fnt $10^{-15}$}}%
      \put(1020,929){\makebox(0,0)[r]{\strut{}\fnt $10^{-12}$}}%
      \put(1020,1218){\makebox(0,0)[r]{\strut{}\fnt $10^{-9}$}}%
      \put(1020,1508){\makebox(0,0)[r]{\strut{}\fnt $10^{-6}$}}%
      \put(1020,1797){\makebox(0,0)[r]{\strut{}\fnt $10^{-3}$}}%
      \put(1020,2086){\makebox(0,0)[r]{\strut{}\fnt $10^{0}$}}%
      \put(1740,440){\makebox(0,0){\strut{}\fnt $32$}}%
      \put(2360,440){\makebox(0,0){\strut{}\fnt $64$}}%
      \put(2979,440){\makebox(0,0){\strut{}\fnt $96$}}%
      \put(3599,440){\makebox(0,0){\strut{}\fnt $128$}}%
      \put(320,1459){\rotatebox{-270}{\makebox(0,0){\strut{}\fnt $\|L_n(\Sigma_n^{(t)})-B_n\|_\frob/\|B_n\|_\frob$}}}%
      \put(2369,140){\makebox(0,0){\strut{}\fnt Iteration (t)}}%
      \put(2369,2579){\makebox(0,0){\strut{}\fnt (d)}}%
    }%
    \gplgaddtomacro\gplfronttext{%
      \csname LTb\endcsname%
      \put(1980,1103){\makebox(0,0)[r]{\strut{}\fnt standard}}%
      \csname LTb\endcsname%
      \put(1980,903){\makebox(0,0)[r]{\strut{}\fnt precond.}}%
    }%
    \gplbacktext
    \put(0,0){\includegraphics{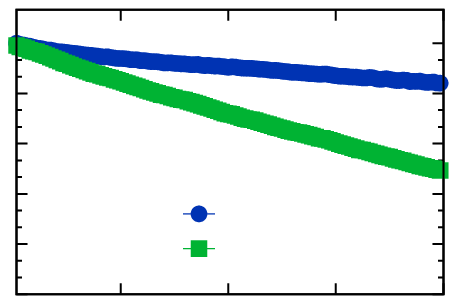}}%
    \gplfronttext
  \end{picture}%
\endgroup

%% file: running_times_gplt.tex
% GNUPLOT: LaTeX picture with Postscript
\begingroup
\newcommand{\fnt}[0]{\footnotesize}
  \makeatletter
  \providecommand\color[2][]{%
    \GenericError{(gnuplot) \space\space\space\@spaces}{%
      Package color not loaded in conjunction with
      terminal option `colourtext'%
    }{See the gnuplot documentation for explanation.%
    }{Either use 'blacktext' in gnuplot or load the package
      color.sty in LaTeX.}%
    \renewcommand\color[2][]{}%
  }%
  \providecommand\includegraphics[2][]{%
    \GenericError{(gnuplot) \space\space\space\@spaces}{%
      Package graphicx or graphics not loaded%
    }{See the gnuplot documentation for explanation.%
    }{The gnuplot epslatex terminal needs graphicx.sty or graphics.sty.}%
    \renewcommand\includegraphics[2][]{}%
  }%
  \providecommand\rotatebox[2]{#2}%
  \@ifundefined{ifGPcolor}{%
    \newif\ifGPcolor
    \GPcolortrue
  }{}%
  \@ifundefined{ifGPblacktext}{%
    \newif\ifGPblacktext
    \GPblacktextfalse
  }{}%
  % define a \g@addto@macro without @ in the name:
  \let\gplgaddtomacro\g@addto@macro
  % define empty templates for all commands taking text:
  \gdef\gplbacktext{}%
  \gdef\gplfronttext{}%
  \makeatother
  \ifGPblacktext
    % no textcolor at all
    \def\colorrgb#1{}%
    \def\colorgray#1{}%
  \else
    % gray or color?
    \ifGPcolor
      \def\colorrgb#1{\color[rgb]{#1}}%
      \def\colorgray#1{\color[gray]{#1}}%
      \expandafter\def\csname LTw\endcsname{\color{white}}%
      \expandafter\def\csname LTb\endcsname{\color{black}}%
      \expandafter\def\csname LTa\endcsname{\color{black}}%
      \expandafter\def\csname LT0\endcsname{\color[rgb]{1,0,0}}%
      \expandafter\def\csname LT1\endcsname{\color[rgb]{0,1,0}}%
      \expandafter\def\csname LT2\endcsname{\color[rgb]{0,0,1}}%
      \expandafter\def\csname LT3\endcsname{\color[rgb]{1,0,1}}%
      \expandafter\def\csname LT4\endcsname{\color[rgb]{0,1,1}}%
      \expandafter\def\csname LT5\endcsname{\color[rgb]{1,1,0}}%
      \expandafter\def\csname LT6\endcsname{\color[rgb]{0,0,0}}%
      \expandafter\def\csname LT7\endcsname{\color[rgb]{1,0.3,0}}%
      \expandafter\def\csname LT8\endcsname{\color[rgb]{0.5,0.5,0.5}}%
    \else
      % gray
      \def\colorrgb#1{\color{black}}%
      \def\colorgray#1{\color[gray]{#1}}%
      \expandafter\def\csname LTw\endcsname{\color{white}}%
      \expandafter\def\csname LTb\endcsname{\color{black}}%
      \expandafter\def\csname LTa\endcsname{\color{black}}%
      \expandafter\def\csname LT0\endcsname{\color{black}}%
      \expandafter\def\csname LT1\endcsname{\color{black}}%
      \expandafter\def\csname LT2\endcsname{\color{black}}%
      \expandafter\def\csname LT3\endcsname{\color{black}}%
      \expandafter\def\csname LT4\endcsname{\color{black}}%
      \expandafter\def\csname LT5\endcsname{\color{black}}%
      \expandafter\def\csname LT6\endcsname{\color{black}}%
      \expandafter\def\csname LT7\endcsname{\color{black}}%
      \expandafter\def\csname LT8\endcsname{\color{black}}%
    \fi
  \fi
  \setlength{\unitlength}{0.0500bp}%
  \begin{picture}(7920.00,2880.00)%
    \gplgaddtomacro\gplbacktext{%
      \csname LTb\endcsname%
      \put(780,640){\makebox(0,0)[r]{\strut{}\fnt $10^{0}$}}%
      \put(780,1186){\makebox(0,0)[r]{\strut{}\fnt $10^{1}$}}%
      \put(780,1733){\makebox(0,0)[r]{\strut{}\fnt $10^{2}$}}%
      \put(780,2279){\makebox(0,0)[r]{\strut{}\fnt $10^{3}$}}%
      \put(1376,440){\makebox(0,0){\strut{}\fnt $8$}}%
      \put(2327,440){\makebox(0,0){\strut{}\fnt $10$}}%
      \put(3278,440){\makebox(0,0){\strut{}\fnt $12$}}%
      \put(4230,440){\makebox(0,0){\strut{}\fnt $14$}}%
      \put(5181,440){\makebox(0,0){\strut{}\fnt $16$}}%
      \put(6132,440){\makebox(0,0){\strut{}\fnt $18$}}%
      \put(7083,440){\makebox(0,0){\strut{}\fnt $20$}}%
      \put(320,1459){\rotatebox{-270}{\makebox(0,0){\strut{}\fnt Running time (min)}}}%
      \put(4229,140){\makebox(0,0){\strut{}\fnt N}}%
      \put(4229,2579){\makebox(0,0){\strut{}\fnt (d)}}%
    }%
    \gplgaddtomacro\gplfronttext{%
      \csname LTb\endcsname%
      \put(3300,2016){\makebox(0,0)[r]{\strut{}\fnt no CTF, uniform}}%
      \csname LTb\endcsname%
      \put(3300,1816){\makebox(0,0)[r]{\strut{}\fnt with CTF, uniform}}%
      \csname LTb\endcsname%
      \put(3300,1616){\makebox(0,0)[r]{\strut{}\fnt with CTF, non-uniform}}%
    }%
    \gplbacktext
    \put(0,0){\includegraphics{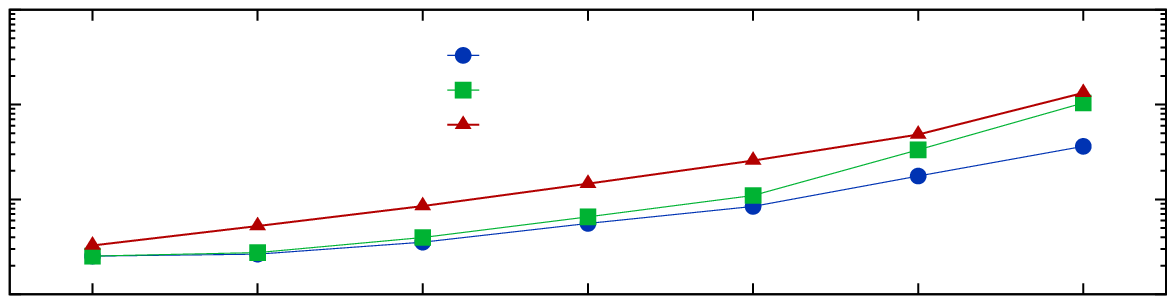}}%
    \gplfronttext
  \end{picture}%
\endgroup

%% file: frank70s_spectrum_gplt.tex
% GNUPLOT: LaTeX picture with Postscript
\begingroup
\newcommand{\fnt}[0]{\footnotesize}
  \makeatletter
  \providecommand\color[2][]{%
    \GenericError{(gnuplot) \space\space\space\@spaces}{%
      Package color not loaded in conjunction with
      terminal option `colourtext'%
    }{See the gnuplot documentation for explanation.%
    }{Either use 'blacktext' in gnuplot or load the package
      color.sty in LaTeX.}%
    \renewcommand\color[2][]{}%
  }%
  \providecommand\includegraphics[2][]{%
    \GenericError{(gnuplot) \space\space\space\@spaces}{%
      Package graphicx or graphics not loaded%
    }{See the gnuplot documentation for explanation.%
    }{The gnuplot epslatex terminal needs graphicx.sty or graphics.sty.}%
    \renewcommand\includegraphics[2][]{}%
  }%
  \providecommand\rotatebox[2]{#2}%
  \@ifundefined{ifGPcolor}{%
    \newif\ifGPcolor
    \GPcolortrue
  }{}%
  \@ifundefined{ifGPblacktext}{%
    \newif\ifGPblacktext
    \GPblacktextfalse
  }{}%
  % define a \g@addto@macro without @ in the name:
  \let\gplgaddtomacro\g@addto@macro
  % define empty templates for all commands taking text:
  \gdef\gplbacktext{}%
  \gdef\gplfronttext{}%
  \makeatother
  \ifGPblacktext
    % no textcolor at all
    \def\colorrgb#1{}%
    \def\colorgray#1{}%
  \else
    % gray or color?
    \ifGPcolor
      \def\colorrgb#1{\color[rgb]{#1}}%
      \def\colorgray#1{\color[gray]{#1}}%
      \expandafter\def\csname LTw\endcsname{\color{white}}%
      \expandafter\def\csname LTb\endcsname{\color{black}}%
      \expandafter\def\csname LTa\endcsname{\color{black}}%
      \expandafter\def\csname LT0\endcsname{\color[rgb]{1,0,0}}%
      \expandafter\def\csname LT1\endcsname{\color[rgb]{0,1,0}}%
      \expandafter\def\csname LT2\endcsname{\color[rgb]{0,0,1}}%
      \expandafter\def\csname LT3\endcsname{\color[rgb]{1,0,1}}%
      \expandafter\def\csname LT4\endcsname{\color[rgb]{0,1,1}}%
      \expandafter\def\csname LT5\endcsname{\color[rgb]{1,1,0}}%
      \expandafter\def\csname LT6\endcsname{\color[rgb]{0,0,0}}%
      \expandafter\def\csname LT7\endcsname{\color[rgb]{1,0.3,0}}%
      \expandafter\def\csname LT8\endcsname{\color[rgb]{0.5,0.5,0.5}}%
    \else
      % gray
      \def\colorrgb#1{\color{black}}%
      \def\colorgray#1{\color[gray]{#1}}%
      \expandafter\def\csname LTw\endcsname{\color{white}}%
      \expandafter\def\csname LTb\endcsname{\color{black}}%
      \expandafter\def\csname LTa\endcsname{\color{black}}%
      \expandafter\def\csname LT0\endcsname{\color{black}}%
      \expandafter\def\csname LT1\endcsname{\color{black}}%
      \expandafter\def\csname LT2\endcsname{\color{black}}%
      \expandafter\def\csname LT3\endcsname{\color{black}}%
      \expandafter\def\csname LT4\endcsname{\color{black}}%
      \expandafter\def\csname LT5\endcsname{\color{black}}%
      \expandafter\def\csname LT6\endcsname{\color{black}}%
      \expandafter\def\csname LT7\endcsname{\color{black}}%
      \expandafter\def\csname LT8\endcsname{\color{black}}%
    \fi
  \fi
  \setlength{\unitlength}{0.0500bp}%
  \begin{picture}(3960.00,2880.00)%
    \gplgaddtomacro\gplbacktext{%
      \csname LTb\endcsname%
      \put(1020,640){\makebox(0,0)[r]{\strut{}\fnt $0$}}%
      \put(1020,1050){\makebox(0,0)[r]{\strut{}\fnt $5000$}}%
      \put(1020,1460){\makebox(0,0)[r]{\strut{}\fnt $10000$}}%
      \put(1020,1869){\makebox(0,0)[r]{\strut{}\fnt $15000$}}%
      \put(1020,2279){\makebox(0,0)[r]{\strut{}\fnt $20000$}}%
      \put(1201,440){\makebox(0,0){\strut{}\fnt $1$}}%
      \put(2339,440){\makebox(0,0){\strut{}\fnt $16$}}%
      \put(3553,440){\makebox(0,0){\strut{}\fnt $32$}}%
      \put(320,1459){\rotatebox{-270}{\makebox(0,0){\strut{}\fnt $\lambda_k$}}}%
      \put(2369,140){\makebox(0,0){\strut{}\fnt $k$}}%
      \put(2369,2579){\makebox(0,0){\strut{}\fnt (a)}}%
    }%
    \gplgaddtomacro\gplfronttext{%
    }%
    \gplbacktext
    \put(0,0){\includegraphics{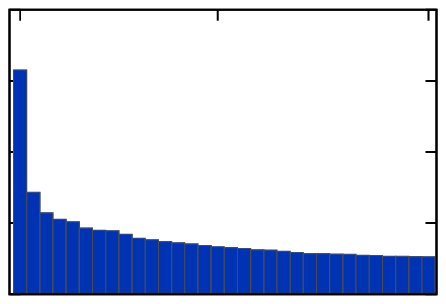}}%
    \gplfronttext
  \end{picture}%
\endgroup

%% file: frank70s_coords_gplt.tex
% GNUPLOT: LaTeX picture with Postscript
\begingroup
\newcommand{\fnt}[0]{\footnotesize}
  \makeatletter
  \providecommand\color[2][]{%
    \GenericError{(gnuplot) \space\space\space\@spaces}{%
      Package color not loaded in conjunction with
      terminal option `colourtext'%
    }{See the gnuplot documentation for explanation.%
    }{Either use 'blacktext' in gnuplot or load the package
      color.sty in LaTeX.}%
    \renewcommand\color[2][]{}%
  }%
  \providecommand\includegraphics[2][]{%
    \GenericError{(gnuplot) \space\space\space\@spaces}{%
      Package graphicx or graphics not loaded%
    }{See the gnuplot documentation for explanation.%
    }{The gnuplot epslatex terminal needs graphicx.sty or graphics.sty.}%
    \renewcommand\includegraphics[2][]{}%
  }%
  \providecommand\rotatebox[2]{#2}%
  \@ifundefined{ifGPcolor}{%
    \newif\ifGPcolor
    \GPcolortrue
  }{}%
  \@ifundefined{ifGPblacktext}{%
    \newif\ifGPblacktext
    \GPblacktextfalse
  }{}%
  % define a \g@addto@macro without @ in the name:
  \let\gplgaddtomacro\g@addto@macro
  % define empty templates for all commands taking text:
  \gdef\gplbacktext{}%
  \gdef\gplfronttext{}%
  \makeatother
  \ifGPblacktext
    % no textcolor at all
    \def\colorrgb#1{}%
    \def\colorgray#1{}%
  \else
    % gray or color?
    \ifGPcolor
      \def\colorrgb#1{\color[rgb]{#1}}%
      \def\colorgray#1{\color[gray]{#1}}%
      \expandafter\def\csname LTw\endcsname{\color{white}}%
      \expandafter\def\csname LTb\endcsname{\color{black}}%
      \expandafter\def\csname LTa\endcsname{\color{black}}%
      \expandafter\def\csname LT0\endcsname{\color[rgb]{1,0,0}}%
      \expandafter\def\csname LT1\endcsname{\color[rgb]{0,1,0}}%
      \expandafter\def\csname LT2\endcsname{\color[rgb]{0,0,1}}%
      \expandafter\def\csname LT3\endcsname{\color[rgb]{1,0,1}}%
      \expandafter\def\csname LT4\endcsname{\color[rgb]{0,1,1}}%
      \expandafter\def\csname LT5\endcsname{\color[rgb]{1,1,0}}%
      \expandafter\def\csname LT6\endcsname{\color[rgb]{0,0,0}}%
      \expandafter\def\csname LT7\endcsname{\color[rgb]{1,0.3,0}}%
      \expandafter\def\csname LT8\endcsname{\color[rgb]{0.5,0.5,0.5}}%
    \else
      % gray
      \def\colorrgb#1{\color{black}}%
      \def\colorgray#1{\color[gray]{#1}}%
      \expandafter\def\csname LTw\endcsname{\color{white}}%
      \expandafter\def\csname LTb\endcsname{\color{black}}%
      \expandafter\def\csname LTa\endcsname{\color{black}}%
      \expandafter\def\csname LT0\endcsname{\color{black}}%
      \expandafter\def\csname LT1\endcsname{\color{black}}%
      \expandafter\def\csname LT2\endcsname{\color{black}}%
      \expandafter\def\csname LT3\endcsname{\color{black}}%
      \expandafter\def\csname LT4\endcsname{\color{black}}%
      \expandafter\def\csname LT5\endcsname{\color{black}}%
      \expandafter\def\csname LT6\endcsname{\color{black}}%
      \expandafter\def\csname LT7\endcsname{\color{black}}%
      \expandafter\def\csname LT8\endcsname{\color{black}}%
    \fi
  \fi
  \setlength{\unitlength}{0.0500bp}%
  \begin{picture}(3960.00,2880.00)%
    \gplgaddtomacro\gplbacktext{%
      \csname LTb\endcsname%
      \put(780,640){\makebox(0,0)[r]{\strut{}\fnt $0$}}%
      \put(780,1077){\makebox(0,0)[r]{\strut{}\fnt $200$}}%
      \put(780,1514){\makebox(0,0)[r]{\strut{}\fnt $400$}}%
      \put(780,1951){\makebox(0,0)[r]{\strut{}\fnt $600$}}%
      \put(900,440){\makebox(0,0){\strut{}\fnt $-500$}}%
      \put(1575,440){\makebox(0,0){\strut{}\fnt $-250$}}%
      \put(2250,440){\makebox(0,0){\strut{}\fnt $0$}}%
      \put(2924,440){\makebox(0,0){\strut{}\fnt $250$}}%
      \put(3599,440){\makebox(0,0){\strut{}\fnt $500$}}%
      \put(320,1459){\rotatebox{-270}{\makebox(0,0){\strut{}\fnt Count}}}%
      \put(2249,140){\makebox(0,0){\strut{}\fnt $\coordest_s^{(1)}$}}%
      \put(2249,2579){\makebox(0,0){\strut{}\fnt (c)}}%
    }%
    \gplgaddtomacro\gplfronttext{%
    }%
    \gplbacktext
    \put(0,0){\includegraphics{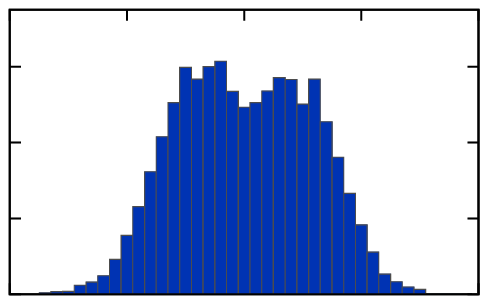}}%
    \gplfronttext
  \end{picture}%
\endgroup

%% file: frank70s_coords_2d_gplt.tex
% GNUPLOT: LaTeX picture with Postscript
\begingroup
\newcommand{\fnt}[0]{\footnotesize}
  \makeatletter
  \providecommand\color[2][]{%
    \GenericError{(gnuplot) \space\space\space\@spaces}{%
      Package color not loaded in conjunction with
      terminal option `colourtext'%
    }{See the gnuplot documentation for explanation.%
    }{Either use 'blacktext' in gnuplot or load the package
      color.sty in LaTeX.}%
    \renewcommand\color[2][]{}%
  }%
  \providecommand\includegraphics[2][]{%
    \GenericError{(gnuplot) \space\space\space\@spaces}{%
      Package graphicx or graphics not loaded%
    }{See the gnuplot documentation for explanation.%
    }{The gnuplot epslatex terminal needs graphicx.sty or graphics.sty.}%
    \renewcommand\includegraphics[2][]{}%
  }%
  \providecommand\rotatebox[2]{#2}%
  \@ifundefined{ifGPcolor}{%
    \newif\ifGPcolor
    \GPcolortrue
  }{}%
  \@ifundefined{ifGPblacktext}{%
    \newif\ifGPblacktext
    \GPblacktextfalse
  }{}%
  % define a \g@addto@macro without @ in the name:
  \let\gplgaddtomacro\g@addto@macro
  % define empty templates for all commands taking text:
  \gdef\gplbacktext{}%
  \gdef\gplfronttext{}%
  \makeatother
  \ifGPblacktext
    % no textcolor at all
    \def\colorrgb#1{}%
    \def\colorgray#1{}%
  \else
    % gray or color?
    \ifGPcolor
      \def\colorrgb#1{\color[rgb]{#1}}%
      \def\colorgray#1{\color[gray]{#1}}%
      \expandafter\def\csname LTw\endcsname{\color{white}}%
      \expandafter\def\csname LTb\endcsname{\color{black}}%
      \expandafter\def\csname LTa\endcsname{\color{black}}%
      \expandafter\def\csname LT0\endcsname{\color[rgb]{1,0,0}}%
      \expandafter\def\csname LT1\endcsname{\color[rgb]{0,1,0}}%
      \expandafter\def\csname LT2\endcsname{\color[rgb]{0,0,1}}%
      \expandafter\def\csname LT3\endcsname{\color[rgb]{1,0,1}}%
      \expandafter\def\csname LT4\endcsname{\color[rgb]{0,1,1}}%
      \expandafter\def\csname LT5\endcsname{\color[rgb]{1,1,0}}%
      \expandafter\def\csname LT6\endcsname{\color[rgb]{0,0,0}}%
      \expandafter\def\csname LT7\endcsname{\color[rgb]{1,0.3,0}}%
      \expandafter\def\csname LT8\endcsname{\color[rgb]{0.5,0.5,0.5}}%
    \else
      % gray
      \def\colorrgb#1{\color{black}}%
      \def\colorgray#1{\color[gray]{#1}}%
      \expandafter\def\csname LTw\endcsname{\color{white}}%
      \expandafter\def\csname LTb\endcsname{\color{black}}%
      \expandafter\def\csname LTa\endcsname{\color{black}}%
      \expandafter\def\csname LT0\endcsname{\color{black}}%
      \expandafter\def\csname LT1\endcsname{\color{black}}%
      \expandafter\def\csname LT2\endcsname{\color{black}}%
      \expandafter\def\csname LT3\endcsname{\color{black}}%
      \expandafter\def\csname LT4\endcsname{\color{black}}%
      \expandafter\def\csname LT5\endcsname{\color{black}}%
      \expandafter\def\csname LT6\endcsname{\color{black}}%
      \expandafter\def\csname LT7\endcsname{\color{black}}%
      \expandafter\def\csname LT8\endcsname{\color{black}}%
    \fi
  \fi
  \setlength{\unitlength}{0.0500bp}%
  \begin{picture}(3960.00,2880.00)%
    \gplgaddtomacro\gplbacktext{%
      \csname LTb\endcsname%
      \put(900,640){\makebox(0,0)[r]{\strut{}\fnt $-500$}}%
      \put(900,1050){\makebox(0,0)[r]{\strut{}\fnt $-250$}}%
      \put(900,1460){\makebox(0,0)[r]{\strut{}\fnt $0$}}%
      \put(900,1869){\makebox(0,0)[r]{\strut{}\fnt $250$}}%
      \put(900,2279){\makebox(0,0)[r]{\strut{}\fnt $500$}}%
      \put(1020,440){\makebox(0,0){\strut{}\fnt $-500$}}%
      \put(1507,440){\makebox(0,0){\strut{}\fnt $-250$}}%
      \put(1995,440){\makebox(0,0){\strut{}\fnt $0$}}%
      \put(2482,440){\makebox(0,0){\strut{}\fnt $250$}}%
      \put(2969,440){\makebox(0,0){\strut{}\fnt $500$}}%
      \put(320,1459){\rotatebox{-270}{\makebox(0,0){\strut{}\fnt $\coordest_s^{(1)}$}}}%
      \put(1994,140){\makebox(0,0){\strut{}\fnt $\coordest_s^{(2)}$}}%
      \put(1994,2579){\makebox(0,0){\strut{}\fnt (d)}}%
    }%
    \gplgaddtomacro\gplfronttext{%
      \csname LTb\endcsname%
      \put(3235,640){\makebox(0,0)[l]{\strut{}\fnt $0$}}%
      \put(3235,874){\makebox(0,0)[l]{\strut{}\fnt $20$}}%
      \put(3235,1108){\makebox(0,0)[l]{\strut{}\fnt $40$}}%
      \put(3235,1342){\makebox(0,0)[l]{\strut{}\fnt $60$}}%
      \put(3235,1576){\makebox(0,0)[l]{\strut{}\fnt $80$}}%
      \put(3235,1810){\makebox(0,0)[l]{\strut{}\fnt $100$}}%
      \put(3235,2044){\makebox(0,0)[l]{\strut{}\fnt $120$}}%
      \put(3235,2279){\makebox(0,0)[l]{\strut{}\fnt $140$}}%
    }%
    \gplbacktext
    \put(0,0){\includegraphics{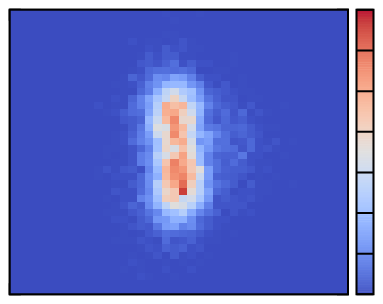}}%
    \gplfronttext
  \end{picture}%
\endgroup

%% file: arxiv.bbl
\begin{thebibliography}{10}

\bibitem{ammar1987generalized}
{\sc G.~S. Ammar and W.~B. Gragg}, {\em The generalized {S}chur algorithm for
  the superfast solution of {T}oeplitz systems}, in Rational approximation and
  its applications in mathematics and physics, J.~Gilewicz, M.~Pindor, and
  W.~Siemaszko, eds., vol.~1237 of Lecture Notes in Mathematics, Springer,
  1987, pp.~315--330, \url{https://doi.org/10.1007/BFb0072474}.

\bibitem{amunts2014structure}
{\sc A.~Amunts, A.~Brown, X.-c. Bai, J.~L. Ll{\'a}cer, T.~Hussain, P.~Emsley,
  F.~Long, G.~Murshudov, S.~H.~W. Scheres, and V.~Ramakrishnan}, {\em Structure
  of the yeast mitochondrial large ribosomal subunit}, Science, 343 (2014),
  pp.~1485--1489, \url{https://doi.org/10.1126/science.1249410}.

\bibitem{isbi15}
{\sc J.~And\'{e}n, E.~Katsevich, and A.~Singer}, {\em Covariance estimation
  using conjugate gradient for {3D} classification in {CRYO-EM}}, in Proc.
  ISBI, April 2015, pp.~200--204,
  \url{https://doi.org/10.1109/ISBI.2015.7163849}.

\bibitem{sampta}
{\sc J.~And\'en and A.~Singer}, {\em Factor analysis for spectral estimation},
  in Proc. SampTA, July 2017, pp.~169--173,
  \url{https://doi.org/10.1109/SAMPTA.2017.8024447}.

\bibitem{axelsson1996iterative}
{\sc O.~Axelsson}, {\em Iterative solution methods}, Cambridge university
  press, 1996.

\bibitem{marina}
{\sc A.~Barnett, L.~Greengard, A.~Pataki, and M.~Spivak}, {\em Rapid solution
  of the cryo-{EM} reconstruction problem by frequency marching}, SIAM J.
  Imaging Sci., 10 (2017), pp.~1170--1195,
  \url{https://doi.org/10.1137/16M1097171}.

\bibitem{baxter2009determination}
{\sc W.~T. Baxter, R.~A. Grassucci, H.~Gao, and J.~Frank}, {\em Determination
  of signal-to-noise ratios and spectral {SNRs} in cryo-{EM} low-dose imaging
  of molecules}, J. Struct. Biol., 166 (2009), pp.~126--132,
  \url{https://doi.org/10.1016/j.jsb.2009.02.012}.

\bibitem{twicing}
{\sc T.~Bhamre, T.~Zhang, and A.~Singer}, {\em Anisotropic twicing for single
  particle reconstruction using autocorrelation analysis}.
\newblock Submitted, 2017, \url{https://arxiv.org/abs/1704.07969}.

\bibitem{chan1996conjugate}
{\sc R.~H. Chan and M.~K. Ng}, {\em Conjugate gradient methods for {T}oeplitz
  systems}, SIAM Rev., 38 (1996), pp.~427--482,
  \url{https://doi.org/10.1137/S0036144594276474}.

\bibitem{chan1988optimal}
{\sc T.~F. Chan}, {\em An optimal circulant preconditioner for {T}oeplitz
  systems}, SIAM J. Sci. and Stat. Comput., 9 (1988), pp.~766--771,
  \url{https://doi.org/10.1137/0909051}.

\bibitem{xiuyuan}
{\sc X.~Cheng}, {\em Random Matrices in High-dimensional Data Analysis}, PhD
  thesis, Princeton University, 2013,
  \url{http://arks.princeton.edu/ark:/88435/dsp01wh246s26t}.

\bibitem{cheng2015primer}
{\sc Y.~Cheng, N.~Grigorieff, P.~Penczek, and T.~Walz}, {\em A primer to
  single-particle cryo-electron microscopy}, Cell, 161 (2015), pp.~438--449,
  \url{https://doi.org/https://doi.org/10.1016/j.cell.2015.03.050}.

\bibitem{lafon}
{\sc R.~R. Coifman and S.~Lafon}, {\em Diffusion maps}, Appl. Comput. Harmon.
  Anal., 21 (2006), pp.~5--30,
  \url{https://doi.org/https://doi.org/10.1016/j.acha.2006.04.006}.

\bibitem{cooley-tukey}
{\sc J.~W. Cooley and J.~W. Tukey}, {\em An algorithm for the machine
  calculation of complex {F}ourier series}, Math. Comp., 19 (1965),
  pp.~297--301, \url{https://doi.org/10.2307/2003354}.

\bibitem{dashti}
{\sc A.~Dashti, P.~Schwander, R.~Langlois, R.~Fung, W.~Li, A.~Hosseinizadeh,
  H.~Y. Liao, J.~Pallesen, G.~Sharma, V.~A. Stupina, A.~E. Simon, J.~D. Dinman,
  J.~Frank, and A.~Ourmazd}, {\em Trajectories of the ribosome as a {B}rownian
  nanomachine}, Proc. Natl. Acad. Sci. U.S.A., 111 (2014), pp.~17492--17497,
  \url{https://doi.org/10.1073/pnas.1419276111}.

\bibitem{davis-kahan}
{\sc C.~Davis and W.~M. Kahan}, {\em The rotation of eigenvectors by a
  perturbation. {III}}, SIAM J. Numer. Anal., 7 (1970), pp.~1--46,
  \url{https://doi.org/10.1137/0707001}.

\bibitem{dempster-laird}
{\sc A.~P. Dempster, N.~M. Laird, and D.~B. Rubin}, {\em Maximum likelihood
  from incomplete data via the {EM} algorithm}, J. Royal Stat. Soc. B Stat.
  Methol., 39 (1977), pp.~1--38, \url{http://www.jstor.org/stable/2984875}.

\bibitem{dobriban}
{\sc E.~Dobriban, W.~Leeb, and A.~Singer}, {\em Optimal prediction in the
  linearly transformed spiked model}.
\newblock Submitted, 2017, \url{https://arxiv.org/abs/1709.03393}.

\bibitem{donoho-gavish}
{\sc D.~L. Donoho, M.~Gavish, and I.~M. Johnstone}, {\em Optimal shrinkage of
  eigenvalues in the spiked covariance model}.
\newblock 2013, \url{https://arxiv.org/abs/1311.0851}.

\bibitem{druskin1989two}
{\sc V.~Druskin and L.~Knizhnerman}, {\em Two polynomial methods of calculating
  functions of symmetric matrices}, USSR Comput. Math. Math. Phys., 29 (1989),
  pp.~112--121, \url{https://doi.org/10.1016/S0041-5553(89)80020-5}.

\bibitem{nufft}
{\sc A.~Dutt and V.~Rokhlin}, {\em Fast {F}ourier transforms for nonequispaced
  data}, SIAM J. Sci. Comput., 14 (1993), pp.~1368--1393,
  \url{https://doi.org/10.1137/0914081}.

\bibitem{erickson1971measurement}
{\sc H.~Erickson and A.~Klug}, {\em Measurement and compensation of defocusing
  and aberrations by {F}ourier processing of electron micrographs}, Philos.
  Trans. Royal Soc. B, 261 (1971), pp.~105--118,
  \url{https://doi.org/10.1098/rstb.1971.0040}.

\bibitem{fessler-toeplitz}
{\sc J.~A. Fessler, S.~Lee, V.~T. Olafsson, H.~R. Shi, and D.~C. Noll}, {\em
  {T}oeplitz-based iterative image reconstruction for {MRI} with correction for
  magnetic field inhomogeneity}, IEEE Trans. Sig. Process., 53 (2005),
  pp.~3393--3402, \url{https://doi.org/10.1109/TSP.2005.853152}.

\bibitem{frank}
{\sc J.~Frank}, {\em Three-dimensional electron microscopy of macromolecular
  assemblies}, Academic Press, 2006.

\bibitem{donoho-gavish-svd}
{\sc M.~Gavish and D.~L. Donoho}, {\em Optimal shrinkage of singular values},
  IEEE Trans. Inf. Theory, 63 (2017), pp.~2137--2152,
  \url{https://doi.org/10.1109/TIT.2017.2653801}.

\bibitem{golub-vanloan}
{\sc G.~H. Golub and C.~F. Van~Loan}, {\em Matrix computations}, Johns Hopkins
  University Press, 4th~ed., 2013.

\bibitem{greengard2004accelerating}
{\sc L.~Greengard and J.-Y. Lee}, {\em Accelerating the nonuniform fast
  {Fourier} transform}, SIAM Rev., 46 (2004), pp.~443--454,
  \url{https://doi.org/10.1137/S003614450343200X}.

\bibitem{frealign}
{\sc N.~Grigorieff}, {\em {FREALIGN}: {H}igh-resolution refinement of single
  particle structures}, J. Struct. Biol., 157 (2007), pp.~117 -- 125,
  \url{https://doi.org/10.1016/j.jsb.2006.05.004}.

\bibitem{guerquin-kern}
{\sc M.~Guerquin-Kern, D.~V.~D. Ville, C.~Vonesch, J.-C. Baritaux, K.~P.
  Pruessmann, and M.~Unser}, {\em Wavelet-regularized reconstruction for rapid
  {MRI}}, in Proc. ISBI, IEEE, June 2009, pp.~193--196,
  \url{https://doi.org/10.1109/ISBI.2009.5193016}.

\bibitem{harauz1986exact}
{\sc G.~Harauz and M.~van Heel}, {\em Exact filters for general geometry three
  dimensional reconstruction}, Optik, 73 (1986), pp.~146--156.

\bibitem{herman2009fundamentals}
{\sc G.~T. Herman}, {\em Fundamentals of computerized tomography: {I}mage
  reconstruction from projections}, Springer-Verlag London, 2009,
  \url{https://doi.org/10.1007/978-1-84628-723-7}.

\bibitem{hestenes}
{\sc M.~R. Hestenes and E.~Stiefel}, {\em Methods of conjugate gradients for
  solving linear systems}, J. Res. Natl. Bur. Stand., 49 (1952),
  \url{https://doi.org/10.6028/jres.049.044}.

\bibitem{higham2008functions}
{\sc N.~J. Higham}, {\em Functions of matrices: {T}heory and computation},
  Society for Industrial and Applied Mathematics, 2008,
  \url{https://doi.org/10.1137/1.9780898717778}.

\bibitem{jain2013low}
{\sc P.~Jain, P.~Netrapalli, and S.~Sanghavi}, {\em Low-rank matrix completion
  using alternating minimization}, in Proc. STOC, ACM, 2013, pp.~665--674,
  \url{https://doi.org/10.1145/2488608.2488693}.

\bibitem{jin2014iterative}
{\sc Q.~Jin, C.~Sorzano, J.~{de~la}~{Rosa-Trev\'{i}n}, J.~{Bilbao-Castro},
  R.~{N\'{u}\~{n}ez-Ram\'{i}rez}, O.~Llorca, F.~Tama, and S.~Joni\'{c}}, {\em
  Iterative elastic {3D}-to-{2D} alignment method using normal modes for
  studying structural dynamics of large macromolecular complexes}, Structure,
  22 (2014), pp.~496--506, \url{https://doi.org/10.1016/j.str.2014.01.004}.

\bibitem{johnstone}
{\sc I.~M. Johnstone}, {\em On the distribution of the largest eigenvalue in
  principal components analysis}, Ann. Statist., 29 (2001), pp.~295--327,
  \url{http://www.jstor.org/stable/2674106}.

\bibitem{jonic2017computational}
{\sc S.~Joni\'{c}}, {\em Computational methods for analyzing conformational
  variability of macromolecular complexes from cryo-electron microscopy
  images}, Curr. Opin. Struct. Biol., 43 (2017), pp.~114--121,
  \url{https://doi.org/10.1016/j.sbi.2016.12.011}.

\bibitem{gene}
{\sc E.~Katsevich, A.~Katsevich, and A.~Singer}, {\em Covariance matrix
  estimation for the cryo-{EM} heterogeneity problem}, SIAM J. Imaging Sci., 8
  (2015), pp.~126--185, \url{https://doi.org/10.1137/130935434}.

\bibitem{khoshouei2017cryo}
{\sc M.~Khoshouei, M.~Radjainia, W.~Baumeister, and R.~Danev}, {\em Cryo-{EM}
  structure of haemoglobin at 3.2 {$\AA$} determined with the volta phase
  plate}, Nat. Commun., 8 (2017), \url{https://doi.org/10.1038/ncomms16099}.

\bibitem{kimanius}
{\sc D.~Kimanius, B.~O. Forsberg, S.~H. Scheres, and E.~Lindahl}, {\em
  Accelerated cryo-{EM} structure determination with parallelisation using
  {GPUs} in {RELION}-2}, eLife, 5 (2016),
  \url{https://doi.org/10.7554/eLife.18722}.

\bibitem{klug}
{\sc A.~Klug and R.~A. Crowther}, {\em Three-dimensional image reconstruction
  from the viewpoint of information theory}, Nature, 238 (1972), pp.~435--440,
  \url{https://doi.org/10.1038/238435a0}.

\bibitem{kuhlbrandt}
{\sc W.~K\"{u}hlbrandt}, {\em The resolution revolution}, Science, 343 (2014),
  pp.~1443--1444, \url{https://doi.org/10.1126/science.1251652}.

\bibitem{kunis-potts-sph}
{\sc S.~Kunis and D.~Potts}, {\em Fast spherical {Fourier} algorithms}, J.
  Comput. Appl. Math., 161 (2003), pp.~75--98,
  \url{https://doi.org/10.1016/S0377-0427(03)00546-6}.

\bibitem{roy}
{\sc R.~R. Lederman and A.~Singer}, {\em A representation theory perspective on
  simultaneous alignment and classification}.
\newblock Submitted, 2016, \url{https://arxiv.org/abs/1607.03464}.

\bibitem{lederman2017continuously}
{\sc R.~R. Lederman and A.~Singer}, {\em Continuously heterogeneous
  hyper-objects in cryo-{EM} and {3-D} movies of many temporal dimensions}.
\newblock Submitted, 2017, \url{https://arxiv.org/abs/1704.02899}.

\bibitem{frank70s_10k}
{\sc H.~Liao and J.~Frank}, {\em Classification by bootstrapping in single
  particle methods}, in Proc. ISBI, IEEE, April 2010, pp.~169--172,
  \url{https://doi.org/10.1109/ISBI.2010.5490386}.

\bibitem{liao-kaczmarz}
{\sc H.~Y. Liao, Y.~Hashem, and J.~Frank}, {\em Efficient estimation of
  three-dimensional covariance and its application in the analysis of
  heterogeneous samples in cryo-electron microscopy}, Structure, 23 (2015),
  pp.~1129--1137, \url{https://doi.org/10.1016/j.str.2015.04.004}.

\bibitem{liao2013structure}
{\sc M.~Liao, E.~Cao, D.~Julius, and Y.~Cheng}, {\em Structure of the {TRPV1}
  ion channel determined by electron cryo-microscopy}, Nature, 504 (2013),
  pp.~107--112, \url{https://doi.org/10.1038/nature12822}.

\bibitem{liu1995estimation}
{\sc W.~Liu and J.~Frank}, {\em Estimation of variance distribution in
  three-dimensional reconstruction. {I}. {T}heory}, J. Opt. Soc. Am. A, 12
  (1995), pp.~2615--2627, \url{https://doi.org/10.1364/JOSAA.12.002615}.

\bibitem{lloyd}
{\sc S.~Lloyd}, {\em Least squares quantization in {PCM}}, IEEE Trans. Inf.
  Theory, 28 (1982), pp.~129--137,
  \url{https://doi.org/10.1109/TIT.1982.1056489}.

\bibitem{stephane-book}
{\sc S.~Mallat}, {\em A wavelet tour of signal processing}, Academic Press,
  3rd~ed., 2008.

\bibitem{marcenko-pastur}
{\sc V.~A. Mar{\v{c}}enko and L.~A. Pastur}, {\em Distribution of eigenvalues
  for some sets of random matrices}, Math. USSR Sb., 1 (1967), p.~457,
  \url{https://doi.org/10.1070/SM1967v001n04ABEH001994}.

\bibitem{michielssen1996multilevel}
{\sc E.~Michielssen and A.~Boag}, {\em A multilevel matrix decomposition
  algorithm for analyzing scattering from large structures}, IEEE Trans.
  Antennas Propag., 44 (1996), pp.~1086--1093,
  \url{https://doi.org/10.1109/8.511816}.

\bibitem{milne2013cryo}
{\sc J.~L. Milne, M.~J. Borgnia, A.~Bartesaghi, E.~E. Tran, L.~A. Earl, D.~M.
  Schauder, J.~Lengyel, J.~Pierson, A.~Patwardhan, and S.~Subramaniam}, {\em
  Cryo-electron microscopy--{A} primer for the non-microscopist}, FEBS Journal,
  280 (2013), pp.~28--45, \url{https://doi.org/10.1111/febs.12078}.

\bibitem{mindell2003accurate}
{\sc J.~A. Mindell and N.~Grigorieff}, {\em Accurate determination of local
  defocus and specimen tilt in electron microscopy}, J. Struct. Biol., 142
  (2003), pp.~334--347, \url{https://doi.org/10.1016/S1047-8477(03)00069-8}.

\bibitem{musicus1984levinson}
{\sc B.~Musicus}, {\em {L}evinson and fast {C}holesky algorithms for {T}oeplitz
  and almost {T}oeplitz matrices}, Tech. Report 538, MIT Research Laboratory of
  Electronics, 1988.

\bibitem{natterer}
{\sc F.~Natterer}, {\em The mathematics of computerized tomography}, Society
  for Industrial and Applied Mathematics, 2001,
  \url{https://doi.org/10.1137/1.9780898719284}.

\bibitem{paul2007asymptotics}
{\sc D.~Paul}, {\em Asymptotics of sample eigenstructure for a large
  dimensional spiked covariance model}, Statist. Sinica, 17 (2007),
  pp.~1617--1642, \url{http://www.jstor.org/stable/24307692}.

\bibitem{penczek-pca}
{\sc P.~Penczek, M.~Kimmel, and C.~Spahn}, {\em Identifying conformational
  states of macromolecules by eigen-analysis of resampled cryo-{EM} images},
  Structure, 19 (2011), pp.~1582--1590,
  \url{https://doi.org/10.1016/j.str.2011.10.003}.

\bibitem{penczek-variance}
{\sc P.~A. Penczek}, {\em Variance in three-dimensional reconstructions from
  projections}, in Proc. ISBI, 2002, pp.~749--752,
  \url{https://doi.org/10.1109/ISBI.2002.1029366}.

\bibitem{penczek2010chapter}
{\sc P.~A. Penczek}, {\em Image restoration in cryo-electron microscopy}, in
  Cryo-EM, Part B: 3-D Reconstruction, G.~J. Jensen, ed., vol.~482 of Methods
  Enzymol., Academic Press, 2010, pp.~35--72,
  \url{https://doi.org/10.1016/S0076-6879(10)82002-6}.

\bibitem{penczek-classification}
{\sc P.~A. Penczek, J.~Frank, and C.~M. Spahn}, {\em A method of focused
  classification, based on the bootstrap {3D} variance analysis, and its
  application to {EF-G}-dependent translocation}, J. Struct. Biol., 154 (2006),
  pp.~184--194, \url{https://doi.org/10.1016/j.jsb.2005.12.013}.

\bibitem{penczek-bootstrap}
{\sc P.~A. Penczek, C.~Yang, J.~Frank, and C.~M. Spahn}, {\em Estimation of
  variance in single-particle reconstruction using the bootstrap technique}, J.
  Struct. Biol., 154 (2006), pp.~168--183,
  \url{https://doi.org/10.1016/j.jsb.2006.01.003}.

\bibitem{brubaker}
{\sc A.~Punjani, J.~L. Rubinstein, D.~J. Fleet, and M.~A. Brubaker}, {\em
  {cryoSPARC}: algorithms for rapid unsupervised cryo-{EM} structure
  determination}, Nat. Methods, 14 (2017), pp.~290--296,
  \url{https://doi.org/10.1038/nmeth.4169}.

\bibitem{radon1917uber}
{\sc J.~Radon}, {\em \"{U}ber die {B}estimmung von {F}unktionen durch ihre
  {I}ntegralwerte l\"{a}ngs gewisser {M}annigfaltigkeiten}, Berichte
  S\"{a}chsishen Akad. Wissenschaft., Math. Phys. Klass, 69 (1917),
  pp.~262--277.

\bibitem{saad1992analysis}
{\sc Y.~Saad}, {\em Analysis of some {K}rylov subspace approximations to the
  matrix exponential operator}, SIAM J. Numer. Anal., 29 (1992), pp.~209--228,
  \url{https://doi.org/10.1137/0729014}.

\bibitem{saad2003iterative}
{\sc Y.~Saad}, {\em Iterative methods for sparse linear systems}, SIAM,
  2nd~ed., 2003, \url{https://doi.org/10.1137/1.9780898718003}.

\bibitem{scheres-relion}
{\sc S.~Scheres}, {\em {RELION}: {I}mplementation of a {B}ayesian approach to
  cryo-{EM} structure determination}, J. Struct. Biol., 180 (2012),
  pp.~519--530, \url{https://doi.org/10.1016/j.jsb.2012.09.006}.

\bibitem{scheres}
{\sc S.~H. Scheres}, {\em A {B}ayesian view on cryo-{EM} structure
  determination}, J. Mol. Biol., 415 (2012), pp.~406--418,
  \url{https://doi.org/10.1016/j.jmb.2011.11.010}.

\bibitem{schur1917potenzreihen}
{\sc J.~Schur}, {\em {\"U}ber potenzreihen, die im innern des einheitskreises
  beschr{\"a}nkt sind.}, J. Reine Angew. Math., 147 (1917), pp.~205--232.

\bibitem{shatsky}
{\sc M.~Shatsky, R.~Hall, E.~Nogales, J.~Malik, and S.~Brenner}, {\em Automated
  multi-model reconstruction from single-particle electron microscopy data}, J.
  Struct. Biol., 170 (2010), pp.~98--108,
  \url{https://doi.org/10.1016/j.jsb.2010.01.007}.

\bibitem{shi-malik}
{\sc J.~Shi and J.~Malik}, {\em Normalized cuts and image segmentation}, IEEE
  Trans. Pattern Anal. Mach. Intell., 22 (2000), pp.~888--905,
  \url{https://doi.org/10.1109/34.868688}.

\bibitem{shkolnisky2012viewing}
{\sc Y.~Shkolnisky and A.~Singer}, {\em Viewing direction estimation in
  cryo-{EM} using synchronization}, SIAM J. Imaging Sci., 5 (2012),
  pp.~1088--1110, \url{https://doi.org/10.1137/120863642}.

\bibitem{fred-ml}
{\sc F.~J. Sigworth}, {\em A maximum-likelihood approach to single-particle
  image refinement}, J. Struct. Biol., 122 (1998), pp.~328--339,
  \url{https://doi.org/10.1006/jsbi.1998.4014}.

\bibitem{singer2010detecting}
{\sc A.~Singer, R.~R. Coifman, F.~J. Sigworth, D.~W. Chester, and
  Y.~Shkolnisky}, {\em Detecting consistent common lines in cryo-{EM} by
  voting}, J. Struct. Biol., 169 (2010), pp.~312--322,
  \url{https://doi.org/10.1016/j.jsb.2009.11.003}.

\bibitem{slepian}
{\sc D.~Slepian}, {\em Prolate spheroidal wave functions, {F}ourier analysis
  and uncertainty--{IV}: {E}xtensions to many dimensions; generalized prolate
  spheroidal functions}, Bell Syst. Tech. J., 43 (1964), pp.~3009--3057,
  \url{https://doi.org/10.1002/j.1538-7305.1964.tb01037.x}.

\bibitem{strang1986proposal}
{\sc G.~Strang}, {\em A proposal for {T}oeplitz matrix calculations}, Stud.
  Appl. Math., 74 (1986), pp.~171--176,
  \url{https://doi.org/10.1002/sapm1986742171}.

\bibitem{tagare}
{\sc H.~D. Tagare, A.~Kucukelbir, F.~J. Sigworth, H.~Wang, and M.~Rao}, {\em
  Directly reconstructing principal components of heterogeneous particles from
  cryo-{EM} images}, J. Struct. Biol., 191 (2015), pp.~245--262,
  \url{https://doi.org/10.1016/j.jsb.2015.05.007}.

\bibitem{trefethen-bau}
{\sc L.~N. Trefethen and D.~Bau}, {\em Numerical linear algebra}, SIAM, 1997.

\bibitem{tygert-sph-3}
{\sc M.~Tygert}, {\em Fast algorithms for spherical harmonic expansions,
  {III}}, J. Comput. Phys., 229 (2010), pp.~6181--6192,
  \url{https://doi.org/10.1016/j.jcp.2010.05.004}.

\bibitem{circulant}
{\sc E.~E. Tyrtyshnikov}, {\em Optimal and superoptimal circulant
  preconditioners}, SIAM J. Matrix Anal. Appl., 13 (1992), pp.~459--473,
  \url{https://doi.org/10.1137/0613030}.

\bibitem{van2000single}
{\sc M.~van Heel, B.~Gowen, R.~Matadeen, E.~V. Orlova, R.~Finn, T.~Pape,
  D.~Cohen, H.~Stark, R.~Schmidt, M.~Schatz, and A.~Patwardhan}, {\em
  Single-particle electron cryo-microscopy: {T}owards atomic resolution}, Q.
  Rev. Biophys., 33 (2000), pp.~307--369,
  \url{https://doi.org/10.1017/S0033583500003644}.

\bibitem{vinothkumar2016single}
{\sc K.~R. Vinothkumar and R.~Henderson}, {\em Single particle electron
  cryomicroscopy: {T}rends, issues and future perspective}, Q. Rev. Biophys.,
  49 (2016), \url{https://doi.org/10.1017/S0033583516000068}.

\bibitem{vonesch}
{\sc C.~Vonesch, L.~Wang, Y.~Shkolnisky, and A.~Singer}, {\em Fast
  wavelet-based single-particle reconstruction in cryo-{EM}}, in Proc. ISBI,
  IEEE, March 2011, pp.~1950--1953,
  \url{https://doi.org/10.1109/ISBI.2011.5872791}.

\bibitem{vulovic2013image}
{\sc M.~Vulovi{\'c}, R.~B. Ravelli, L.~J. van Vliet, A.~J. Koster,
  I.~Lazi\'{c}, U.~L\"{u}cken, H.~Rullg\aa{}rd, O.~\"{O}ktem, and B.~Rieger},
  {\em Image formation modeling in cryo-electron microscopy}, J. Struct. Biol.,
  183 (2013), pp.~19--32, \url{https://doi.org/10.1016/j.jsb.2013.05.008}.

\bibitem{wade1992brief}
{\sc R.~Wade}, {\em A brief look at imaging and contrast transfer},
  Ultramicroscopy, 46 (1992), pp.~145--156,
  \url{https://doi.org/10.1016/0304-3991(92)90011-8}.

\bibitem{wajer-pruessmann}
{\sc F.~T. A.~W. Wajer and K.~P. Pruessmann}, {\em Major speedup of
  reconstruction for sensitivity encoding with arbitrary trajectories}, in
  Proc. ISMRM, 2001, p.~767.

\bibitem{lanhui-firm}
{\sc L.~Wang, Y.~Shkolnisky, and A.~Singer}, {\em A {F}ourier-based approach
  for iterative {3D} reconstruction from cryo-{EM} images}.
\newblock Unpublished, 2013, \url{https://arxiv.org/abs/1307.5824}.

\bibitem{lanhui-LUD}
{\sc L.~Wang, A.~Singer, and Z.~Wen}, {\em Orientation determination of
  cryo-{EM} images using least unsquared deviations}, SIAM J. Imaging Sci., 6
  (2013), pp.~2450--2483, \url{https://doi.org/10.1137/130916436}.

\bibitem{xu2018allosteric}
{\sc N.~Xu, D.~Veesler, P.~C. Doerschuk, and J.~E. Johnson}, {\em Allosteric
  effects in bacteriophage {HK97} procapsids revealed directly from covariance
  analysis of cryo {EM} data}, J. Struct. Biol,  (2018),
  \url{https://doi.org/10.1016/j.jsb.2017.12.013}.
\newblock In press.

\bibitem{zhang2016gctf}
{\sc K.~Zhang}, {\em {Gctf}: Real-time {CTF} determination and correction}, J.
  Struct. Biol., 193 (2016), pp.~1--12,
  \url{https://doi.org/10.1016/j.jsb.2015.11.003}.

\bibitem{zhao}
{\sc Z.~Zhao and A.~Singer}, {\em {F}ourier--{B}essel rotational invariant
  eigenimages}, J. Opt. Soc. Am. A, 30 (2013), pp.~871--877,
  \url{https://doi.org/10.1364/JOSAA.30.000871}.

\bibitem{zhao2014rotationally}
{\sc Z.~Zhao and A.~Singer}, {\em Rotationally invariant image representation
  for viewing direction classification in cryo-{EM}}, J. Struct. Biol., 186
  (2014), pp.~153--166, \url{https://doi.org/10.1016/j.jsb.2014.03.003}.

\bibitem{zheng2012three}
{\sc Y.~Zheng, Q.~Wang, and P.~C. Doerschuk}, {\em Three-dimensional
  reconstruction of the statistics of heterogeneous objects from a collection
  of one projection image of each object}, J. Opt. Soc. Am. A, 29 (2012),
  pp.~959--970, \url{https://doi.org/10.1364/JOSAA.29.000959}.

\end{thebibliography}
